%% file: thesis1.tex
\documentclass[12pt,oneside,aps,english,a4paper]{report}

\usepackage{amssymb}
\usepackage{graphics}
\usepackage{epsfig}
\usepackage{a4wide}
\usepackage{latexsym}
\usepackage[Sonny]{fncychap} 
\usepackage{fancyhdr}
\usepackage{graphicx}
\usepackage{amsmath}
\usepackage{bm}
\usepackage{xspace}
\usepackage{bbold}
\numberwithin{equation}{section}
\usepackage{enumerate}
\usepackage{doublespace}

\def\baselinestretch{1.2}

\renewcommand{\theequation}
             {\arabic{chapter}.\arabic{section}.\arabic{equation}}
\newcommand{\clearemptydoublepage}
             {\newpage{\pagestyle{empty}\cleardoublepage}}
\renewcommand{\baselinestretch}{1.2}

\begin{document}


\pagestyle{empty}

\include{frontpage}

\clearemptydoublepage

\pagestyle{plain}
\pagenumbering{roman}
\include{preamble}

\clearemptydoublepage


\pagestyle{fancy}

\addtolength{\headheight}{\baselineskip}

\renewcommand{\chaptermark}[1]
        {\markboth{#1}{}}
\renewcommand{\sectionmark}[1] 
        {\markright{\thesection\ #1}}

\fancyhf{}
\fancyhead[LE,RO]{\it\bfseries\thepage}
\fancyhead[LO]{\it\bfseries\rightmark}
\fancyhead[RE]{\it\bfseries\leftmark}

\tableofcontents
\clearemptydoublepage
\listoffigures
\clearemptydoublepage
\listoftables
\clearemptydoublepage

\fancyhf{}
\fancyhead[LE,RO]{\it\bfseries\thepage}
\fancyhead[LO]{\it\bfseries\rightmark}


\pagenumbering{arabic}
\setcounter{page}{1}

\include{chapter01}
\clearemptydoublepage
\include{chapter21}

\clearemptydoublepage
\include{chapter31}

\clearemptydoublepage
\include{chapter41}

\clearemptydoublepage
\include{chapter53}
\clearemptydoublepage
\include{conclusion}
\clearemptydoublepage


\appendix
\renewcommand{\theequation}
             {\Alph{chapter}.\arabic{section}.\arabic{equation}}

\clearemptydoublepage

\fancyhf{}
\fancypagestyle{plain}{\fancyhead{}\renewcommand{\headrulewidth}{0pt}}
\fancyhead[LE,RO]{\bfseries\thepage}

{\small
\addcontentsline{toc}{chapter}{Bibliography}
\bibliographystyle{unsrt}
\bibliography{references}
}


\end{document}

%% file: frontpage.tex
\renewcommand{\baselinestretch}{1}

\begin{titlepage}
\topskip1.3in
\begin{center}
\vspace{2cm}

\begin{center}
\begin{tabular}{l}
\hline
  \\ \\
\huge{{\bfseries\textsc{\qquad Variations of the}}}\\ \\
\huge{{\bfseries\textsc{Fine Structure Constant}}}\\ \\
\huge{{\bfseries\textsc{\qquad in Space and Time}}}\\
\\\\  
\hline
\end{tabular}
\end{center}

\vspace{2.5cm}
{\Large
David Fonseca Mota\\
\vspace{0.5cm}
Corpus Christi College\\}
\vspace{2.3cm}
\parbox{20mm}{
    \epsfxsize=20mm
    \epsfbox{CU.arms.colour}
}\\
\vspace{2.3cm}
{\large
Dissertation submitted for the degree of Doctor of Philosophy\\
\vskip5pt
Department of Applied Mathematics and Theoretical Physics\\
University of Cambridge\\ 
September 2003}
\end{center}

\clearemptydoublepage

\begin{flushright}
{\it {\small A descoberta dessa impossibilidade,}}\\ 
{\it {\small a de  sermos  j\'a  no caminhar   e  s\'o  no caminhar o sermos;}}\\
{\it {\small  n\~ao podendo ser de outra maneira,}}\\
{\it {\small \'e a  possibilidade de criar o nosso modo de caminhar}}\\
{\it {\small -- Ernesto Mota --}}
\end{flushright}

\bigskip

\vspace{4cm}
\begin{center}
\large{{This thesis is dedicated to my parents:\\ Irene Fonseca (M\~{a}ezitas) and
    Joaquim Mota (Paizolas)}}
\end{center}

\clearemptydoublepage

\end{titlepage}

%% file: preamble.tex
\begin{center}
\bfseries{\textsc{DECLARATION}}
\end{center}

The research presented in this thesis was performed in the Department
of Applied Mathematics and Theoretical Physics at the University of
Cambridge between October 1999 and August 2003. 
This dissertation is the
result of my own work, except as stated below or where
explicit reference is made to the results of others. 
Chapters \ref{qualitative}, \ref{gaugeinvariant}, \ref{realistic} and \ref{darkenergy} of this thesis are respectively based on
the papers (Refs.~\cite{mota1, mota2, mota3,mota4}): 
\begin{itemize}
\item
J.~D.~Barrow and D.~F.~Mota,
``Qualitative analysis of universes with varying alpha'',
Class.\ Quant.\ Grav.\  {\bf 19}, 6197 (2002) 
[arXiv:gr-qc/0207012].
\item
J.~D.~Barrow and D.~F.~Mota,
``Gauge-invariant perturbations of varying-alpha cosmologies'',
Class.\ Quant.\ Grav.\  {\bf 20}, 2045 (2003) 
[arXiv:gr-qc/0212032].
\item
D. F. Mota and J. D. Barrow,
``Varying Alpha in a More Realistic Universe'', Submitted to
Phys. Lett. B, 
[arXiv:astro-ph/0306047]
\item
D. F. Mota and J. D. Barrow,
``Local and Global Variations of The Fine Structure
Constant'',  Submitted to
Mon. Not. Roy. Astron. Soc. , [arXiv:astro-ph/0309273]

\end{itemize}

Much of the work of chapters~\ref{qualitative}, ~\ref{gaugeinvariant},
~\ref{realistic} and \ref{darkenergy} was a result of a close 
collaborative effort with my 
supervisor John D. Barrow. In chapter~\ref{qualitative}, my supervisor
has set up the approximation method, the validity of the approximation
and the linearisation of the instability. He has also found the exact
solution in the $n>2/3$ case and has performed the stability analysis
of the $n=2/3$ asymptote. In chapters~\ref{gaugeinvariant},
~\ref{realistic} and \ref{darkenergy}, my supervisor has contributed to
the interpretation, discussion and clarification of the results.
 
This dissertation is not substantially the same as any that I
have submitted, or am submitting, for a degree, diploma or other 
qualification at any other university.

\vspace{3cm}
Signed: .......................................
\hspace{1.2cm}  
Dated: ........................................
\clearemptydoublepage

\vspace{5cm}
\begin{center}
\bfseries{\textsc{ACKNOWLEDGEMENTS}}
\end{center}

I would like to thank my supervisor John D. Barrow for his constant
advice, encouragement, enthusiasm and support over my years here in
Cambridge.   

During the course of this work I have had helpful discussions with
many people for which I am very grateful, including  Carsten van de
Bruck, Martin Bucher, Jo\~{a}o Lopes Dias, Yves Gaspar,
Kaviland Moodley, Michael Murphy, Fernando Quevedo and Constantinos Skordis. 

I would like to thank Pedro Ferreira and Joseph Silk for
hospitality at the Oxford University Physics Department  during the
time when part of this thesis was written. 

I also would like to thank Orfeu Bertolami for his  encouragement and
support to apply to Cambridge while still an undergraduate student in Lisbon.

For the financial support, I would like to thank  Funda\c{c}\~{a}o para
a Ci\^{e}ncia e a Tecnologia, through the research grant BD/15981/98
and  Funda\c{c}\~{a}o Calouste Gulb\^{e}nkian, through the research grant Proc.50096.

On a personal note I would like to thank the unforgettable and unique
friendship of Milind, John, Pearson and Al  during all these years in
Cambridge. Many others have contribute to my enjoyable stay in Cambridge
they are Andre, Caroline, Jose, Khaled,  Maria Jo\~ao, Miguel, Nidhi,
Nuno, Pedro, Saghir, Shaun and Toto.

Although not living in Cambridge, I would like to thank Marisa and my
'new family' Adriano Couto, Maria de Lurdes, Marie-Th\'er\`ese and
Marl\`ene for their friendship. I also would like to thank my
grandmother, Maria da Concei\c{c}\~ao,
and my aunt, Zeza, to whom I will be always grateful for taking
care of me, and for all their love, since I was a baby. 

In a very special way, I would like to thank my parents, 
for their love and support  over all these years, of which,
as far as I can remember, have always   encourage me to ``learn more''
and to ``widen my horizons''. 

To my brother, Ernesto, any word of gratitude would be too small, 
I may only say: Thank you for existing.

Finally it is an honour to thank my wife and soul-mate Elisabeth
for her understanding and patience throughout the process of
writing the thesis. Most specially, thank you for all your  love and
encouragement when I most need it.

\clearemptydoublepage

\begin{center}
{\Large \bfseries\textsc{Variations of the Fine Structure Constant in
    Space and Time}\\}
\vspace{.5cm}
{\large David Fonseca Mota\\}
\vspace{1cm}
\bfseries{\textsc{Summary}}
\end{center}

This thesis describes a detailed investigation of the effects of matter
inhomogeneities on the cosmological evolution of the fine structure
constant.

In chapter \ref{introduction}, we briefly describe the observational
and theoretical motivations to this work. and we  review the standard
cosmological model. We also review the Bekenstein-Sandvik-Barrow-Magueijo (BSBM) theory
for a varying fine structure constant, $\alpha$. 

Assuming a Friedmann universe which evolves with a power-law scale
factor, $a=t^{n}$, in chapter \ref{qualitative}, we analyse the phase space of the system of
equations that 
describes a time-varying $\alpha$, in a homogeneous and isotropic background universe.
We classify all the possible behaviours of $\alpha (t)$ in
ever-expanding universes with different $n$ and find exact solutions for 
$\alpha (t)$. In general, $\alpha $ will be a non-decreasing function of time that
increases logarithmically in time during a period when the expansion
is dust-dominated ($n=2/3$), but becomes constant when $n>2/3$. 
$\alpha $ tends rapidly
to a constant when the expansion scale factor increases exponentially. A
general set of conditions is established for $\alpha $ to become
asymptotically constant at late times in an expanding universe.

In chapter \ref{gaugeinvariant}, using a gauge-invariant formalism, we
derive and solve the linearly perturbed Einstein
cosmological equations for the BSBM theory. 
We calculate the time evolution of inhomogeneous perturbations
of the fine structure constant, $\frac{\delta \alpha }{\alpha }$ on small
and large scales with respect to the Hubble radius. In a radiation-dominated
universe small inhomogeneities in $\alpha $ decay on large scales
but on scales smaller than the Hubble radius they undergo stable
oscillations. In a dust-dominated universe, small inhomogeneous perturbations
in $\alpha $ approach a constant on large scales and on small scales they
grow as $t^{2/3}$, and $\frac{\delta \alpha }{\alpha }$ tracks $%
\frac{\delta \rho _{m}}{\rho _{m}}$ . If the expansion accelerates, as in
the case of a $\Lambda $ or quintessence-dominated phase, inhomogeneities in 
$\alpha $ decay on both large and small scales. The amplitude of
perturbations in $\alpha $ are much smaller than that of matter or
radiation perturbations. We also present a numerical study of the non-linear
evolution of spherical inhomogeneities in radiation and dust universes by
means of a comparison between the evolution of flat and closed Friedmann
models with time-varying $\alpha .$Various limitations of these simple
models are also discussed.

Chapter \ref{realistic} is dedicated to the effects of non-linear
structure formation in the evolution of $\alpha$.  
We study the space-time evolution of the fine structure constant inside
evolving spherical overdensities in a $\Lambda$ Cold Dark Matter ($CDM$) Friedmann
universe, using the spherical infall model. We show that its value inside
virialised regions will be significantly larger than in the low-density
background universe. The consideration of the inhomogeneous evolution of the
universe is therefore essential for a correct comparison of extragalactic
and solar system limits on, and observations of, possible time
variation of   $\alpha$ and other constants. Time variation of   $\alpha$ 
in the cosmological background can give rise to no
locally observable variations inside virialised overdensities like the one
in which we live, explaining the discrepancy between astrophysical and
geochemical observations. 

In chapter \ref{darkenergy}, using the BSBM varying-alpha theory, and
the spherical  collapse model for cosmological structure formation, we
study the effects of the dark-energy equation of state and the
coupling  of $\alpha$ to the matter fields on the space and time
evolution of $\alpha$.  
We compare its evolution inside virialised
overdensities  with that in the cosmological background, using the
standard ($\Lambda =0$)  $CDM$ model of structure formation and the
dark-energy modification,  $wCDM$. We find that, independently of the
model of structure  formation one considers, there is always a
difference between the value  of alpha in an overdensity and in the
background.  In a $SCDM$ model, this difference is the same,
independent of  the virialisation redshift of the overdense region. In
the  case of a $wCDM$ model, especially at low redshifts, the
difference  depends on the time when virialisation occurs and the
equation  of state of the dark energy. At high redshifts, when the
$wCDM$  model becomes asymptotically equivalent to the $SCDM$ one, the
difference is constant. At low redshifts, when dark energy starts to
dominate  the cosmological expansion, the difference between $\alpha$ in
a cluster  and in the background grows.

The last chapter contains a summary of the results obtained in the
thesis and a discussion of some open problems that require further investigation.

\cleardoublepage

%% file: chapter01.tex
\chapter{Introduction}
\label{introduction}

\begin{flushright}
{\it {\small  ... contudo fazemos da nossa  angustia,}}\\ {\it {\small
desta quietude  irriquieta que nos atormenta a  alma,}}\\ {\it {\small
as mil  e uma  perguntas sem resposta...}}\\  {\it {\small  -- Ernesto
Mota --}}
\end{flushright}

\bigskip

\section{Constants of Nature}

Once upon a  time, humankind started to describe  the universe we live
in using  mathematics.  When a  mathematical formalism is  proposed in
order to describe the different  phenomena that occur in our universe,
a theory in  physics is set. Any theory which claims  to explain a set
of  events  in  nature  has   to  be  confronted  with  the  empirical
observations.  Hence, having the  formal structure of a given physical
theory, there  is the need to quantify it  in order to test  it. The
quantification of  a given physical model consist  in the introduction
of  numbers. But this  raises a  question: How  shall we  choose these
numbers and to which quantities  shall we attribute them?  Theories in
physics  have  several  free  parameters  of  which  value  cannot  be
calculated or deduced from its formal structure. These free parameters
are the  fundamental constants of the  theory. We need  to attribute a
value to  these quantities, and the  value is chosen in  order to match
the quantitative  predictions of the theory with  the observations and
experiments performed.  Why are these quantities fundamental?  Because
their  values  cannot  be  calculated, only  measured.  Why  constant?
Because every experiment yields  the same value independently of time,
position, temperature, pressure, etc.

But not all the constants  in physics are fundamental. If one conducts
an  experiment with greater  precision and/or  in an  environment more
extreme than  previously considered,  one might observe  variations of
the constants.  This may precipitate  physicists to search for  a more
basic  theory which explains  the origin  and the  value of  the (then
non-fundamental and variable) constant.  The status of a constant
depends then on the  considered theory and on the  observer measuring them,
that is, on  whether this observer belongs to  the world of low-energy
quasi-particles or to the high  energies one. For example, the density
of  a given  material  was once  regarded  as constant,  but was later
found by experimenters to vary with temperature and pressure.

\subsection{Why study variations of the fundamental constants?}
 
By definition,  the constants of  nature are quantities which  have no
dependence  on other measured  parameters, it  is not  surprising that
despite the  enormous advances  in particle physics  we still  have no
idea why the  constants of nature take the values  they do. This might
be something we  will never be able to answer, but  we have learnt from the
past,  that what  we call  today a  constant might  not be  so  in the
future.

The hypothesis  of the constancy  of the fundamental constants plays  an important
role  in  astronomy and  cosmology  where the  redshift
measures  the  look-back time.  Ignoring  the  possibility of  varying
constants could lead to a distorted view of our universe and if such 
variations are  established corrections should  be applied. It  is thus
important   to  investigate  that   possibility,  especially   as  the
measurements become more precise.

The study of  the variations of the constants of  nature also offers a
new  link  between  astrophysics,  cosmology and  high-energy  physics
complementary  to   early  universe  cosmology.   In  particular,  the
observation   of  the   variability  of   the   fundamental  constants
constitutes  one of the  few ways  to test  directly the  existence of
extra-dimensions and to test high energy physics models.

\bigskip

From all the possible fundamental constants of nature we will focus on
the so  called fine structure  constant, $\alpha$. The  fine structure
constant   is  also  regarded   as  the   coupling  constant   of  the
electromagnetic  interactions.   The fine  structure  constant can  be
derived from other constants as follows \cite{particlebook}:
\begin{equation}
\alpha= \frac{e^2}{4\pi \epsilon_0 \hbar c}
\label{alpha}
\end{equation} 
where $c$ is the speed of light in vacuum, $\hbar\equiv h/2\pi$ is the
reduced  Planck constant,  $e$ is  the electron  charge  magnitude and
$\epsilon_0$ is the  permitivity of free space. The  value of $\alpha$
measured     today    on    earth     is    $\alpha_0\approx1/137.035$
\cite{particlebook}.

In this thesis we will study the theoretical possibilities of time and
spatial variations of the fine structure constant along the history of
the universe.

Time  variations of  $\alpha$ can  be measured  using the  'time shift
density parameter':
\begin{equation}
\frac{\Delta \alpha}{\alpha}\equiv \frac{\alpha(z)-\alpha_0}{\alpha_0}
\label{timeshift}
\end{equation}
where $\alpha(z)$  is the  value of the  fine structure constant  at a
redshift $z$. Similarly, spatial variations of $\alpha$ can be studied
using the 'spatial shift density parameter':
\begin{equation}
\frac{\delta
\alpha}{\alpha}\equiv
\frac{\alpha_{x_1}(z)-\alpha_{x_2}(z)}{\alpha_{x_2}(z)}
\label{spatialshift}
\end{equation}
where $\alpha_{x_{1,2}}$  is the value of the  fine structure constant
in a  two different  regions in the  universe, for instance  $x_1$ and
$x_2$ can represent two different galaxies.

But  why shall  we look  at  possible variations  of the  fine
structure constant?  The next two sections  describe both observational
and theoretical motivations for this work.

\section{Observations and Constraints on Variations of 
the Fine Structure Constant}
\label{constraints}

Many  methods were  used to  observe possible  variations of  the fine
structure  constant \cite{uzan}.   A  general conclusion  of those  experiments and
observations, is that there is no variation of $\alpha$, or at least
the  results  are consistent  with  the hypothesis of a  null  variation. 
If there are  any time or  spatial variations of
$\alpha$, these  will be  very small.
Hence,  up  to
nowadays, we basically have upper limits that constraint variations of
the  fine   structure  constant.   Nevertheless, there is at least one
observation
which   explicitly  gives   a  non-null   result  for   variations  of
$\alpha$. As we will see,  this observation has a very unique feature:
It  has  observed variations  in  the  fine  structure constant  in  a
redshift range of $z=0.5-3.5$.

When   taking  conclusions  from   experiments  and   observations  on
variations (or none)  of the constants of nature  we should always bear
in mind that  those measurements are set on  a given time-scale. Hence
we need  to be very careful  when extrapolating those  results into an
absolute conclusion. For  instance, in the case of  the fine structure
constant, one expect  to be able to constrain  a relative variation of
$\alpha$  in  a  time  scale  of $10^9$  years  ($z\approx0.3$)  using
geochemical constraints \cite{fuj}; $10^9-10^{10}$ years ($z\approx3$)
using astrophysical  (quasars) methods \cite{murphy};  $10^{10}$ years
($z\approx1000$)   using   cosmological   methods  (Cosmic   Microwave
Background   Radiation)  \cite{avelino};   and  $1-12$   months  using
laboratory methods  \cite{clocks}.  It is then not  a straight forward
task  to interpret  and take  absolute  conclusions from  most of  the
observations performed until today. For example, if we observe no time
variations of $\alpha$ using geochemical methods, we can only conclude
that the fine structure constant  did not vary from a redshift $z=0.3$
up to today.

Another common  feature to the observational  and experimental results
is the  presence of degeneracies. Usually, there  are other parameters
in  the  model,  used  to  study the  physical  phenomenon,  that  may
influence the  results. Due  to that, the  conclusions taken  are then
based on  some assumptions, which usually  lead to an  extra error bar
besides  the  systematic  experimental  errors.  The  problem  of  the
degeneracies is not an easy one, since such analyses are also dependent
on    our    understanding    of    the    fundamental    interactions
\cite{landau,sisterna}.  For instance, grand unifying theories predict
that  all the  constants  should vary  simultaneously.   One needs  to
further study the systematic errors of the observations and to propose
and   realise  new   experiments  which   are  less   model  dependent
\cite{karshenboim,braxmaier}.

In this section, we shall not attempt to give a full review, deferring
the reader to \cite{uzan} and the references therein.

\subsection{Geochemical constraints}

All the geological  studies are on time scales of order  of the age of
the  Earth,  meaning  that  constraint  variations  of  the  fine
structure  constant in  a  range of  redshifts of  $z\approx0.1-0.15$,
depending on the values of the cosmological parameters.

Atomic  clocks are one  of the  principal methods  we have  to measure
possible  variations of  the fine  structure constant  on  Earth.  The
latest  constraint on  possible variations  of $\alpha$,  using atomic
clocks, is \cite{marion}:
\begin{equation}
\label{atmbound}
\left| \frac{\dot{\alpha}}{\alpha}\right| <  4.2 \times 10^{-15} ~{\rm
yr}^{-1} \,.
\end{equation}
Notice that, although the constraint is very tight, it is only valid
for a  very short period of time.  Usually a few months,  the time the
experience lasted. 

\bigskip

One  of the oldest  observations, with  non-null results,  looking for
variations of  the fine structure  constant came from the  analysis of
the Oklo phenomenon - a natural nuclear fission reactor which operated
in Gabon, Africa, at approximately $1.8$ billion years ago \cite{book,
petrov,naudet,naudet1}.   The first analysis  were done  by Shlyakhter
\cite{shly} and an upper limit on time variations of $\alpha$ resulted
to be:  $\Delta\alpha/\alpha< 10^{-7}$ over a period  of $1.8$ billion
years. However that study provided two possible results:
\begin{eqnarray}
\frac{\dot\alpha}{\alpha}=\left(-0.2\pm0.8\right)\times10^{-17}
\qquad\text{per                        year}                        \\
\frac{\dot\alpha}{\alpha}=\left(4.9\pm0.4\right)\times10^{-17}
\qquad\text{per year}
\end{eqnarray}
which come from two possible physical branches \cite{book,shly,olive1}.
Note that only the first  solution is consistent with a null variation
of the  fine structure  constant, whereas the  other implies  a higher
value of $\alpha$ in the past.

Recently many others analysis were performed using the Oklo phenomenon
\cite{fuj,damour1,olive1}. Up  to today, the  tighter constraint using
new samples from the Oklo reactor is:
\begin{equation}
\frac{\Delta\alpha}{\alpha}=\left(-0.04\pm0.15\right)\times10^{-7}
\end{equation} 
These limits  should be  considered with care,  since in general  there are
considerable uncertainties regarding  the complicated Oklo environment
and any limits on $\alpha$ derived from it can be weakened by allowing
other interaction  strengths and mass ratios  to vary in  time as well
\cite{flambaum,olive1}.

\bigskip

\subsubsection{Constraints at slightly higher redshifts}

Another geochemical process, which  is sensitive to changes in $\alpha$,
is  the $\beta$-decay  rate of  $^{187}Re$ \cite{dicke}.  Updating the
work of  Dyson \cite{dyson,dyson1},  Olive et al.  \cite{olive1} claim
that,  improved  measurements  of  the $^{187}Re$  $\beta-$decay  rate
(derived from meteoritic $^{187}Re/^{187}Os$ abundances ratios), imply
$\vert\Delta\alpha/\alpha\vert<3\times  10^{-7}$ over  the  $4.65$ Gyr
history   of  the  solar   system  ($z\approx0.45$).    However,  Uzan
\cite{uzan} notes that the above result assumes that variations of the
$\beta$-decay rate  depend entirely on variations of  $\alpha$ and not
on possible  variations of  the weak coupling  constant, see also
\cite{fujiinew}. 
Very recently, Olive et
al. \cite{olivenew}, have rederived this constraint, in models where
all gauge and Yukawa couplings vary independently, as would be
expected in grand unification theories. The new constraint is
$\vert\Delta\alpha/\alpha\vert<(-8\pm 8)\times  10^{-7}$. 
In spite of the potential problems of these experiments, 
the  study  of  this  phenomenon  is  quite  important.   It  provides
constraints to variations of $\alpha$ at slightly higher redshift than
the usual geochemical constraints, and at  the same time, it give us a
hint on the  possible evolution of the fine  structure constant inside
our local system of gravity. 

\subsection{Astrophysical observations}

Astrophysical observation are probably one of the most important types
of  experiments  which  can  be   performed  in  order  to  study  the
possibility  of   variations  of   the  fine  structure   constant  in
cosmological scales.

Observing the quasar absorption  spectra give us information about the
atomic levels at their position  and time of emissions.  It is usually
assumed,  in those observations,  that all  transitions have  the same
$\alpha$ dependence.  So any variation of the  fine structure constant
will  affect  all  the  wave  lengths by  the  same  factor.   Another
assumption is  the independence of  this uniform shift of  the spectra
from the  Doppler effect  due to the  motion of  the source or  to the
gravitational field where  it sits.  The idea is  to compare different
absorption lines  from different species to  extract information about
different combinations  of the constants  at the time of  emission and
then   to    compare   them   with   the    laboratory   values   (see
\cite{murphythesis} for  a detailed description).   A quite surprising
result  was  obtained  from   these  observations:  an  explicit  time
variation  of the  fine structure  constant  was found.  The value  of
$\alpha$, in  a redshift  range of $z=0.5-3.5$,  was smaller  than the
standard          value          on          earth          $\alpha_0$
\cite{murphy,murphy1,webb,webb1,drink} (see however \cite{alternative}
for another interpretation of the  results).  The latest value for the
time   shift   parameter   of   the   fine   structure   constant   is
\cite{murphylast}:
\begin{equation}
\frac{\Delta\alpha}{\alpha}=-0.543\pm 0.116\times 10^{-5}
\end{equation}
for a redshift range of $z=0.5-3.5$.

Many     other    astrophysical     observations     were    performed
\cite{bahcall,bahcall1,bahcall2,potekhin,savedoff,chiba},  but none  of them
has  explicitly  indicate the  existence  of  variations  of the  fine
structure constant, that is, they are just upper bounds.  This show us
how relevant and important are the series of results obtained from the
quasar absorption  spectra.  Such a non-zero  detection, if confirmed,
would bring  tremendous implications on our  understanding of physics,
for instance,  it would be interesting  to investigate if such  a variation is
compatible   with  the  test   of  the   universality  of   free  fall
\cite{nordtvedt}.

Note  that,  a feature  of  the  results  from the  quasar  absorption
spectra,  is  that  they   are  'incompatible'  with  the  geochemical
observations if  the variation  of $\alpha$ is  linear with  time.  If
both the geochemical and  the quasar's absorption spectra observations
are  confirmed,  then  the  theoretical  models  proposed  to  explain
variations  of $\alpha$ will  need to  explain the  reason for  such a
discrepancy   between  astrophysical   observations   and  geochemical
ones.  In  particular,  the  constraint  of  $\Delta\alpha/\alpha\leq3
\times 10^{-7}$  at redshift  $z=0.45$, coming from  the $\beta$-decay
experiments,  is very  difficult  to be  compatible  with an  explicit
variation of the fine structure constant  of $\Delta\alpha/\alpha\approx 5\times 10^{-6}$
at $z=0.5$, coming from the quasar's absorption spectra.

\subsection{Early universe constraints}

\subsubsection{Cosmic Microwave Background Radiation}

Time or spatial  variations of the fine structure  constant affect the
power-spectrum  of the  Cosmic Microwave  Background  Radiation (CMBR)
anisotropies \cite{hannestad,kaplinghat}. This is due to the fact that
a  different value  of  $\alpha$  would affect  the  biding energy  of
hydrogen, the Thompson scattering  cross section and the recombination
rates. A smaller value of $\alpha$ would postpone the recombination of
electrons  and protons,  i.e., the  last scattering  would occur  at a
lower  redshift.  It would  also alter  the baryon-to-photon  ratio at
last-scattering,  leading  to  changes  in  both  the  amplitudes  and
positions of  features in the  power spectrum.  The  strongest current
constraints at $z\approx1000$ from the  CMBR power spectrum are at the
$\Delta\alpha/\alpha\approx10^{-2}$   level  if   one   considers  the
uncertainties  in,  and  degeneracies  with,  the  usual  cosmological
parameters   (for    instance   $\Omega_m,   \Omega_{\Lambda}$,   etc)
\cite{avelino,avelino1,martins,martins1,martins2,landau1}.
However, note there is a critical degeneracy between variations of
$\alpha$ and the    electron   mass    $m_e$    \cite{kujat,battye},
since    the relativistic-quantum corrections  of the  electron mass
depend  on the strength of  the electromagnetic interaction and consequently  on
$\alpha$. This degeneracy dramatically weakens the current constraints.

\subsubsection{Big Bang Nucleosynthesis}

The  theory  of   Big  Bang  Nucleosynthesis  (BBN)  is   one  of  the
corner-stones of  the hot Big Bang  cosmology, successfully explaining
the  abundances of  the light  elements, $D$,  $^{3}He$,  $^{4}He$ and
$^{7}Li$.  Assuming a simple scaling between the value of $\alpha$ and
the  proton-neutron   mass  difference  a  limit  can   be  placed  on
$\Delta\alpha/\alpha$  \cite{kolb}.  A  similar  method  can  be  used
considering   simultaneous  variations   of  the   weak,   strong  and
electromagnetic  couplings   on  BBN  \cite{campbell}.    However,  in
general, estimates based on BBN abundances of $^{4}He$ suffer from the
crucial   uncertainty   of   electromagnetic   contribution   to   the
proton-neutron mass  difference.  Much  weaker limits are  possible if
attention  is restricted  to the  nuclear interaction  effects  on the
nucleosynthesis  of   $D$,  $^{3}He$  and   $^{7}Li$  \cite{bergstrom,
nollett}.  Given the present  observational uncertainties in the light
element    abundances,    the    most    conservative    limits    are
$\Delta\alpha/\alpha=(3\pm7)\times10^{-2}$.

\section{Theoretical Motivations}

In the previous section we have described the empirical motivations to
study variations of the fine structure constant. In reality, they were
not very  convincing, since  most of the  results are  basically upper
bounds to variations  of $\alpha$ and the only  explicit results, that
claimed a lower value of $\alpha$ in the past, was presented by a sole
group    of    collaborators    in    a    series    of    experiments
\cite{murphylast,murphy,murphy1,webb,webb1,drink}.
 
The situation  is completely different  from the theoretical  point of
view.  We will see  that there  are several  important reasons  why we
should seriously take into account  the possibility of a varying-$\alpha$. Most of
them come  from theoretical models which  try to give  a more complete
description of the universe.

These theories have strong observational predictions, for instance, in
the case  of a  time-varying $\alpha$, they  allow us to  evaluate the
effects of a varying $\alpha$  on free fall which leads to potentially
observable violations  of the weak  equivalence principle \cite{dvali,
olive, damour,nordtvedt}. They also allow us to investigate whether or
not  other   cosmological  observations  like   the  Cosmic  Microwave
Background Radiation are consistent with the variations of $\alpha$ to
fit the quasar observations \cite{martins, avelino,battye}.

Once again we shall not attempt  to give a full review here, deferring
the  reader to  \cite{uzan}  and  \cite{ porto}  for  a more  thorough
discussion.

\subsection{Multi-dimensional theoretical models}

There  are  very strong  theoretical  motivations  to  study time  and
spatial variations  of the fine  structure constant. The  reasons come
essentially from  novel theories which claim to  be more 'fundamental'
and  so are candidates  to be  the 'Theory  of Everything'.   The best
candidates  for unification  of  the  forces of  nature,  in a  single
unified theory of  quantum gravity, only seem to  exist in finite form
if there are more dimensions of space than the common four that we are
familiar  with. This  means  that  the true  constants  of nature  are
defined in a higher dimensional world and the constants we observe are
merely  effective  values,  that   is,  they  are  a  four-dimensional
projection  of  the  real   fundamental  constants.  So  the  coupling
constants we  observe and measure may  not be fundamental  and so do
not need to be constant.

Since we  do not  observe any other  dimensions than the  common four,
there is  a strong indication  that, if those  extra-dimensions exist,
they   are   indeed   very   small.    A  general   feature   of   the
multi-dimensional theories,  is that the size  of the extra-dimensions
is associated to a scalar  field.  This field is a dynamical quantity,
which has  the desired property  of making the  extra-dimensions small
and stable (in the sense that their size will not change).  The reason
why we desire the stability of the extra-dimensions, is related to the
fact that  in most of the cases  that scalar field will  be coupled to
the matter fields.  Any time  or spacial variations of this field will
then  be   'seen'  as  a  variation  of   the  interactions'  coupling
constants.  Up to nowadays,  no mechanisms  of stabilising  the scalar
field,  was found.   Hence, until  this mechanism  is to  be  found, a
common feature  of the multi-dimensional theories is  the existence of
slow  changes in  the scale  of  the extra-dimensions  which would  be
revealed by measurable changes in our four-dimensional ``constants''.

Multi-dimensional  theories also  predict relations  between different
constants,  due to  relations  between the  coupling  of dilaton  type
fields and matter, see for instance \cite{chodos,banks,carroll,ranada}
for  more complex  effects then just  the variation  of the  coupling
constants.

\bigskip

One of  the earliest  attempts to create  a single unified  theory was
presented  by  Kaluza   \cite{kaluza}  and  Klein  \cite{klein}.  They
considered   a  five-dimensional   model   with  the   aim  to   unify
electromagnetism and  gravity (for a review  see \cite{overduin}). The
fifth  dimension is  compactified  in  order to  turn  it small.   The
compactification  results into  a coupling  between the  scalar field,
which is related to the radius  of the fifth dimension, and the matter
fields.   Variations  of  the   scalar  field  will  be  reflected  as
variations of  all the  coupling constants.  Various  functional forms
for  monotonic time  variations have  been derived  where  usually the
four-dimensional  gauge couplings vary  as the  inverse square  of the
mean   scale    of   the    extra   dimension   (for    instance   see
\cite{chodos,freund,wu}).   Non-monotonic variations of  $\alpha$ with
time  where analysed  in \cite{marciano}  and the  requirements  for a
self-consistent relations, if there  are simultaneous variations of the
different  coupling constants,  were discussed.  Other five-dimensional
theories, based on the so called Brane-World models, were also used to
study  variations of  the fine  structure constant.  See  for instance
\cite{uzan,porto,palma,brax} and references therein.

\bigskip

A much more recent and popular candidate to a theory of unification is
string theory.  Superstrings theories offer a theoretical framework to
discuss  the value  of  the fundamental  constants  since they  become
expectation values of some  fields \cite{uzan}.  One of the definitive
predictions  of superstring  theories  is the  existence  of a  scalar
field, the dilaton, that  couples directly to matter \cite{taylor} and
whose vacuum expectation value determines the string coupling constant
\cite{witten}.  The four-dimensional couplings are determined in terms
of  a string  scale and  various dynamical  fields, for  instance, the
dilaton and the  volume of compact space. Due  to the coupling between
the  dilaton  and all  the  matter  fields,  we expect  the  following
effects: a  scalar mixture of  a scalar component  inducing deviations
from general relativity, a variation  of the coupling constants, and a
violation  of  the  weak  equivalence principle  \cite{uzan}.  Various
string  theories will  exhibit different  variations of  the effective
couplings;  in  particular, the  cosmological  evolution  of the  fine
structure constant will depend on the form of the coupling between the
dilaton and the matter fields which interact electromagnetically.  The
undefined structure  of the  coupling gave origin  to many  low energy
models  and approaches  to study  time and  spatial variations  of the
fundamental constants and its relation to the extra-dimension radius
(see for instance \cite{kolb,ichikawa,damour2,damour3,vayonakis}).  In
\cite{damour} it  was shown  that cosmological variations  of $\alpha$
may proceed at different rates  at different points in space-time (see
also \cite{forgacs,barrowextra,li}). This  mean that variations of the
fine structure constant may not only be in time but in space as well.


\subsection{The Bekenstein model for varying-$\alpha$}
\label{bek}

Independently    of    any    extra-dimensional   model,    Bekenstein
\cite{bek1,bek2,bek3} formulated a  framework to incorporate a varying
fine structure constant. Working in units in which $\hbar$ and $c$ are
constants, he  adopted a classical description  of the electromagnetic
field  and  made   a  set  of  assumptions  to   obtain  a  reasonable
modification of Maxwell  equations to take into account  the effect of
the variation of the elementary  charge, $e$. His eight postulates are: (i)
For  a  constant $\alpha$  electromagnetism  is  described by  Maxwell
theory and the coupling of  the potential vector $A_{\mu}$ to mater is
minimal.  (ii) The variation of $\alpha$ results from dynamics.  (iii)
The dynamics of electromagnetism and  $\alpha$ can be obtained from an
invariant action. (iv) The action  is locally invariant (v) The action
is time  reversal invariant.  (vi) Electromagnetism  is causal.  (vii)
The  shortest  length is  the  Plank  length.   (viii) Gravitation  is
described by a metric which satisfies Einstein equations.

Assuming that the  charges of all particles vary in  the same way, one
can  set $e=e_0 \epsilon  (x^{\mu})$ where  $\epsilon (x^{\mu})$  is a
dimensionless  universal field and  $e_0$ is  a constant  denoting the
present  value   of  $e$.  This  means  that   some  well  established
assumptions,    like    charge    conservation,   must    give    away
\cite{landau2}. Still,  the principles  of local gauge  invariance and
causality  are maintained,  as is  the scale  invariance of  the field
$\epsilon$.

Since  $e$  is  the  electromagnetic coupling,  the  $\epsilon$  field
couples to the gauge field as $\epsilon A_{\mu}$ in the Lagrangian and
the  gauge  transformation  which   leaves  the  action  invariant  is
$\epsilon   A_{\mu}\rightarrow  \epsilon  A_{\mu}+\partial_{\mu}\chi$,
rather       than      the       usual       $      A_{\mu}\rightarrow
A_{\mu}+\partial_{\mu}\chi$.   The electromagnetic  tensor generalises
to
\begin{equation}
F_{\mu\nu}=  \epsilon^{-1}  \left(  \partial_{\mu}(\epsilon  A_{\nu})-
\partial_{\nu}(\epsilon A_{\mu})\right)
\end{equation}
which reduces  to the  usual form when  $\epsilon$ is a  constant. The
electromagnetic action is given by
\begin{equation}
S_{EM}=-\frac{1}{16\pi}\int dx^4\sqrt{-g}F_{\mu\nu}F^{\mu\nu}
\end{equation}
and  the dynamics  of $\epsilon$  are controlled  by the  kinetic term
action
\begin{equation}
S_{\epsilon}=-\frac{1}{2}\frac{\hbar
c}{l^2}\int
dx^4\sqrt{-g}\frac{\partial_{\mu}\epsilon\partial^{\mu}\epsilon}{\epsilon^2}
\end{equation}
where $l$ is a length  scale of the theory, introduced for dimensional
reasons , and which needs to be small enough to be compatible with the
observed  scale invariance of  electromagnetism. This  constant length
scale gives the scale down to  which the electric field around a point
charge  is accurately  Coulombic ($l_{pl}<  l<10^{-15}-10^{-16}cm$) to
avoid conflict with experiments \cite{bek1,barrowomega}.

Olive and  Pospelov \cite{olive}  generalised the Bekenstein  model to
allow additional coupling of  a scalar field $\epsilon^2=B_F(\phi)$ to
non-baryonic dark matter (as  first proposed by Damour \cite{gibbons})
and  cosmological constant, arguing  that in  certain classes  of dark
matter models, and particularly  in supersymmetric ones, it is natural
to expect that  $\phi$ would couple more strongly  to dark matter than
to baryons.

The  formalism developed  by Bekenstein  was also  
in the braneworld context
\cite{youm,palma}  and   Magueijo  \cite{magueijokibble}  studied  the
effect of a varying fine  structure constant on a complex scalar field
undergoing  an  electromagnetic   $U(1)$  symmetry  breaking  in  this
framework (see also \cite{bertolami1,bertolami2,bertolami3}). Also, inspired by the Bekenstein model,
Armend\'ariz-Pic\'on  \cite{armendariz}  derived the  most
general low energy action including a real scalar field that is local,
invariant  under  space inversion  and  time reversal,  diffeomorphism
invariant and with a $U(1)$ gauge invariance.

\subsection{Other investigations}

When  discussing  variations  of  the fundamental  constants,  careful
distinction  should  be  made  between dimensional  and  dimensionless
constants. The reason is  the fact that every experimental measurement
consist  in  comparing  the   value  of  two  quantities.  Hence,  any
measurement  of a  dimensional constant  needs to  be followed  by the
units  chosen \cite{dicke}.  This fact  has important  consequences in
theoretical  models  which  describe  variations  of  a  dimensionless
quantity,  if  the  later  is  obtained via  a  combination  of  other
(dimensional) quantities. For instance, this degeneracy, among
combination of dimensional constants,  lead Dirac \cite{dirac,dirac1} to propose
the 'Large Numbers Hypothesis'. It sates that the existence of certain
large dimensionless numbers which arise in combinations of some
cosmological numbers and physical constants was not a coincidence but
a consequence of an underlying relationship between them.

Speaking of  variations  of  dimensional
constants  is then somewhat ambiguous,  since observations  and experiments
can only  measure dimensionless quantities.  This means that  we would
not  be able  to  experimentally distinguish  variations  of the  fine
structure constant resulting form a variation of the speed of light or
in the electron charge (see however \cite{bsm4}).
Bekenstein's  model  assumed that  variations  of  the fine  structure
constant  were due  to  variations of  the  electron charge.   Another
approach is to attribute the  change in the fine structure constant to
a          varying           light          propagation          speed
\cite{am,moffat,moffatal,ba,mof,magueijoreview}.   The  motivation for
such  theories are their  ability to  solve the  standard cosmological
problems,                usually                solve               by
inflation\cite{barrowmagueijo,barrowmagueijo1,barrowmagueijo2,jbvsl}.
For instance, the horizon  problem is trivially solved by  claiming a
much higher light speed in the early universe.

\section{Summary}

Observations of  quasar absorption-line spectra  has been found  to be
consistent with  a time variation of  the value of  the fine structure
constant  between redshifts  $z=0.5-3.5$ and  the present.  The entire
data set  of 128 objects gives  spectra consistent with a  shift of 
$\alpha$ with respect to the present
value \cite{murphylast,murphy,webb,webb1}.  Nevertheless, extensive  analysis  has yet  to
find a  selection effect that can  explain the sense  and magnitude of
the relativistic line-shifts underpinning these deductions. Adding to
that, astronomical probes of the constancy of the electron-proton mass ratio
have reported  possible evidence for time  variation, \cite{ivan}, but
as yet the statistical significance is low.

Motivated   by  these  observations,   there  has   been  considerable
theoretical investigation of  the cosmological consequences of varying
$\alpha $, \cite {book}$.$ In particular, in the
theoretical  predictions  of  gravity  theories which  extend  general
relativity to incorporate space-time  variations of the fine structure
constant.  These  have been primary formulated  as Lagrangian theories
with  explicit  variation  of   the  velocity  of  light,  $c$,  \cite
{moffatal,am,ba,mof, jbvsl},  or  of  the  charge  on the  electron, 
 $e,$  \cite{bek2, bsbm,bsm1,bsm2,bsm3}.  
A range of  variant theories have been investigated  with attention to
the possible particle physics motivations and consequences for systems
of    grand   and    partial   unification  
\cite{banks,lang,olive, marciano,damour,jdb,calmet,armendariz,chacko,correia}.

Varying-$\alpha$  theories  offer  the  possibility  of  matching  the
magnitude and trend  of the quasar observations and to  investigate
whether or not other cosmological observations are consistent with the
small  variations of $\alpha  $.  
They
are  also of particular interest because they  predict that violations
of the weak  equivalence principle \cite{olive,damour} 
should be observed  at a level that
is within about an order  of magnitude of existing experimental bounds
\cite{bsm4,dvali,mof}.

\section{Thesis Aim}

 As we  saw in the previous  sections, studies on the  time and spatial
variations of the fine  structure constant, $\alpha$, are motivated by
two main aspects:

\begin{itemize}

\item The  first, are the  recent observations of small  variations of
relativistic  atomic  structure in  quasar  absorption spectra,  which
suggest  that the fine  structure constant,  was smaller  at redshifts
$z=0.5-3.5$    than   the    current    terrestrial   value    $\alpha
_{0}=7.29735308\times  10^{-3}$, with $  \Delta \alpha  /\alpha \equiv
\{\alpha (z)-\alpha _{0}\}/\alpha _{0}=-0.543\pm 0.116\times 10^{-5}$. 

\item  The second  is the  fact that  the current  theoretical models,
candidates to a  grand unified theory of quantum  gravity, require the
existence  of extra-dimensions  beyond  the common  four  we are  used
to. In these models, the  radius of the extra-dimensions is associated
to a scalar field which is coupled to the matter fields. Any change in
this scalar  field reflects variations  of the coupling  constants, in
particular the fine structure constant.

\end{itemize}

We also saw in the  previous sections, that several theories have been
proposed to  investigate the implications of a  varying fine structure
constant. A  common feature to  all of them  is the existence  of a
field responsible for variations of $\alpha$. This field is coupled to
the   matter  fields,   at   least,  to   the   ones  which   interact
electromagnetically.   Due  to  the  coupling between  the  field  and
matter, it is then natural to wonder if the existence and evolution of
inhomogeneities of  the later will  affect the evolution  of $\alpha$.
This will be the main objective of this thesis: to investigate how the
evolution of the inhomogeneities in  the matter fields affect the time
and spatial evolution of the fine structure constant along the history
of the universe.





\bigskip


In order  to investigate  the effects of  the evolution of  the matter
inhomogeneities,  on  the time  and  spatial  variations  of the  fine
structure  constant, we  need to  first to  study the  evolution  of a
homogeneous and isotropic universe and of its components.  In the next
section we briefly review the  the Standard Cosmological Model and the
Friedmann-Robertson-Walker    spacetime,   assuming    a   non-varying
$\alpha$. This  will give  us the usual  background behaviour  for the
usual  matter   components  in  the   universe:  pressureless  matter,
radiation  and cosmological  constant. In  section \ref{bsbm}  we will
then introduce the varying-$\alpha$  model we will use throughout this
thesis.


\section{The Friedmann-Robertson-Walker Universe}

The  standard Friedmann-Robertson-Walker  (FRW) cosmological  model is
based on three main theoretical assumptions: The first is that General
Relativity is  the correct description  of gravitational interactions,
which  implies  that  the   model  is  four  dimensional.  The  second
assumption  concerns the particle  content of  the universe,  which is
assumed  to   be  described  by   the  Standard  Model   of  Particles
(SM). Finally, it is based  on the Cosmological Principle, which tells
us that our universe is  spatially maximally symmetric at any constant
time and  so, isotropic and  homogeneous in space.   These assumptions
are based upon three robust observational facts:



\begin{itemize}

\item The observation by Hubble  and Slipher that all the galaxies are
separating from each other, at  a rate that is roughly proportional to
their separation, is  realised for a Hubble  parameter, $H$ (see
later) being  approximately constant at present, $H=  H_0>0$. This has
been highly verified during the past few decades.

\item The  relative abundance of  the elements with  approximately $75
 \%$  of Hydrogen, $24  \%$ Helium  and other  light elements  such as
 Deuterium and  Helium-4 with small fractions  of a percent,  is a big
 success of nucleosynthesis,  and at present, it is  the farthest away
 in the past that we have been able to compare theory and observation.


\item  The  discovery of  the  Cosmic  Microwave Background  Radiation
(CMBR) by Penzias and Wilson  in $1964$, signalling the time of photon
last  scattering. This  is one  of  the strongest  evidences that  our
universe started in a stage of a hot big bang.
The temperature of the CMBR  across the sky is remarkably uniform: the
deviations from isotropy, $i.e.$ differences in the temperature of the
blackbody spectrum measured in different directions of the sky, are of
order  of   $20\mu$K  on   large  scales,  or   one  part   in  $10^5$
\cite{cobe,cobe1,kamkos,kamkos1,kamkos2,kamkos3,cmbexp,cmbexp1,cmbexp2,wmap,wmap1}.
The observed high degree of isotropy not only provides strong evidence
for the present  level of large-scale isotropy and  homogeneity of our
Hubble volume, but  also provides an important probe  of conditions in
the universe at  red  shifts\footnote{The red shift of an object,
$z$, is  defined in terms of  the ratio of the  detected wavelength to
the emitted wavelength as
\begin{equation}
1+z \equiv \frac{\lambda_0}{\lambda_e} = \frac{a(t_0)}{a(t_e)}\,,
\end{equation}
where $a(t)$ is the scale  factor (see later). Any increase (decrease)
in $a(t)$ leads to a red  shift (blue shift) of the light from distant
sources. Since today observed distant galaxies have red shift spectra,
we can conclude  that the universe is expanding.}  of order 1100.  The
primeval  density  inhomogeneities  necessary  to  initiate  structure
formation result in predictable  temperature fluctuations in the CMBR,
so it can be used to probe theories of structure formation.

\end{itemize}


\bigskip

Now, a mathematical consequence of the cosmological principle, is that
the metric of the cosmological spacetime, takes the Friedman Robertson
Walker (FRW) form
\begin{equation}\label{FRW4}
ds^2\equiv g_{\mu\nu}dx^\mu dx^\nu=dt^2-a^2(t)dx^2\,,
\end{equation}
where $\mu=0,1,2,3$ and $dx^2$ is  the metric of the three dimensional
maximally symmetric  space with constant  curvature, $^{(3)}R= \frac{6
k}{a^2(t)}$, with  parameter $k=\pm 1,0$, corresponding  to a(n) open,
closed or flat three slices. This can be written as:
\begin{equation}\label{metric1}
dx^2={\gamma}_{ij}dx^idx^j =  \frac{dr^2}{1-kr^2} + r^2  [d \theta^2 +
\sin^2 \theta  d \phi^2 ] =  d\chi^2 + f^2(\chi) [d  \theta^2 + \sin^2
\theta d \phi^2 ]\,,
\end{equation}
where
\begin{equation}
d\chi = \frac{dr}{\sqrt{1-kr^2}}\,,
\end{equation}
and
\begin{eqnarray}\label{rchi}
f(\chi) = \left\{
\begin{array}{llll}
\sin \chi  \; \; \;  \; &{\rm  if}\; \; \;  & k =1  \; \; \;  \; &{\rm
closed}  \\ \nonumber  \chi &{\rm  if}& k=0  & {\rm  flat}\\ \nonumber
\sinh \chi &{\rm if}&k = -1 & {\rm open}
\end{array}
\right.
\end{eqnarray}
Spatially open,  flat and closed universes  have different geometries:
Light geodesics on these  universes behave differently, and thus could
in principle be distinguished observationally.

The   scale  factor  $a(t)$,  can be  determined  by solving  the
equations of motion coming from  an action which contains gravity, and
some  matter  content  that  we   describe  as  a  perfect  fluid,  in
consistency  with  the assumptions  of  homogeneity  and isotropy  (an
observer comoving with the fluid would see the universe around her/him
as isotropic).  The action for this system takes the form:
\begin{equation}\label{action1}
S  =  \frac{1}{2\kappa^2}\int{d^4  x\sqrt{-g}\left[  R  +  2\Lambda  -
  2\kappa^2 \,{\mathcal L}_{m} \right]}\,,
\end{equation}
where $\kappa^2  = 8\pi G =  M_P^{-2}$ with $G$  Newton's constant and
$M_P=  2.4  \times 10^{18}{\rm  GeV}$  is  the  Planck mass.  We  have
introduced a possible  cosmological constant $\Lambda$, and ${\mathcal
L}_{m}$  is  the  Lagrangian  describing  the matter  content  in  the
universe.

The  equations of  motion derived  from (\ref{action1})  give  rise to
Einstein's  equations  (where we  will  assume  throughout the  thesis
$\hbar =c\equiv 1$),
\begin{equation}\label{einstein1}
G_{\mu\nu}\equiv  R_{\mu\nu} -  \frac{1}{2} R  g_{\mu\nu}  = -\kappa^2
T_{\mu\nu} - \Lambda g_{\mu\nu}\,.
\end{equation}
Here the stress-energy momentum tensor  for a perfect fluid is defined
by
\begin{equation}\label{tmunudef}
T^{\mu\nu} = \frac{2}{\sqrt{-g}}\,  \frac{ \delta (\sqrt{-g} {\mathcal
        L}_m) }{\delta g_{\mu\nu}}\,,
\end{equation}
and can be written as
\begin{eqnarray}\label{tmunu4}
T^{\mu\nu}\! &=& (\rho +p)  u^\mu u^\nu - \!p\, g^{\mu\nu}\nonumber \\
     \!&=&\! {\rm diag} (\rho, -p,-p,-p)\,,
\end{eqnarray}
where $u^\mu$  is the comoving four-velocity,  which satisfies, $u^\mu
u_\mu  =-1$  and $\rho(t)$  and  $p(t)$  are  the energy  density  and
pressure  of  the  fluid,  respectively,   at  a  given  time  in  the
expansion. They are related by the  equation of state\footnote{In
general, the parameter $w$, which gives the speed of sound, can depend
on time, but throughout this work, we take it to be constant.}:
\begin{equation}\label{eos}
p= w \rho\,.
\end{equation}
%
From Einstein's equations (\ref{einstein1}), one obtains the following
two (not independent)  equations of motion. The first  one is the 
Friedmann equation
\begin{equation}\label{Fried4}
H^{2}=\frac{\kappa^2}{3}\rho             -            \frac{k}{a^{2}}+
\frac{\Lambda}{3}\,,\,\,\,
\end{equation}
where $H=  \frac{\dot a}{a}$ is the  Hubble parameter and  a dot means
derivative with respect to the cosmic time $t$. the second is the 
Raychaudhuri equation
\begin{equation}\label{Ray}
 \frac{\ddot           a}{a}=-\frac{\kappa^2}{6}\,(\rho+3p)          +
 \frac{\Lambda}{3}\,.
\end{equation}
Now,  the  $\mu=0$ component  of  the  conservation  equation for  the
stress-energy momentum tensor, $T^{\mu\nu}_{\,\,\,\,\,\,;\nu}=0$ gives
the conservation of energy equation:
\begin{equation}\label{conserva}
\dot \rho + 3 H (\rho + p) = 0\,,
\end{equation}
which is implied by equations. (\ref{Fried4}, \ref{Ray}).
Therefore, after using  the equation of state (\ref{eos})  we are left
with only  two equations  that we  can take as  the Friedmann  and the
energy  conservation, for $\rho(t)$  and $a(t)$,  which can  be easily
solved. Equation (\ref{conserva}) gives immediately
\begin{equation}
\rho\sim a^{-3(1+w)}\,;
\end{equation}
introducing  this into Friedmann's  equation gives  us a  solution for
$a(t)$ as
\begin{equation}
a(t) \propto t^{\frac{2}{3(1+w)}} \qquad ({\rm for}\,\,\, \Lambda=k=0)
\end{equation}
 The  behaviour of $\rho$  and $a(t)$  for the  typical values  of the
equation of state are shown in Table \ref{table1} for a flat universe,
that is $k=0$. The non flat cases can be found straightforwardly.

\bigskip

\begin{table}[htbp]
\begin{center}
\begin{tabular}{|c|c|c|c|}
\hline Stress Energy  & $w$ & Energy Density &  Scale Factor $a(t)$ \\
\hline \hline Matter & $w=0$ & $\rho\sim a^{-3} $ & $ a(t)\sim t^{2/3}
$  \\ \hline  Radiation &  $ w=\frac{1}{3}$  & $  \rho\sim a^{-4}  $ &
$a(t)\sim t^{1/2}$\\ \hline Vacuum ($\Lambda$)  & $ w=-1$ & $ \rho\sim
\frac{\Lambda}{8\pi  G}$  & $a(t)\sim  \exp(\sqrt{\frac{\Lambda}{3}}\,
t)$\\ \hline
\end{tabular}
\end{center}
\caption{The behaviour of the scale  factor and the energy density for
matter, radiation and vacuum dominated universes for $k=0$.}
\label{table1}
\end{table}

\bigskip

\smallskip
For the more general cases,  we can say several things without solving
the  equations  explicitly.   First  of all,  note  that  Raychaudhuri
equation   (\ref{Ray})   for   a   vanishing   cosmological   constant
$\Lambda=0$,  implies  $\ddot  a/a<0$,  if  $(\rho+3p)$  is  positive.
Moreover, $a>0$  by definition and  $\dot a/a>0$ since we  observe red
shifts, not  blue shifts. So  one can conclude  that $a$ is  a growing
function of time  at present and that the curve  $a(t)$ $vs.$ $t$ hits
the axis at some finite time  in the past, if Einstein's equations are
valid at all times. Then $t=0$ is defined by $a(0)=0$, where the model
has a singularity.  The age of the universe $t_0$,  is the time passed
since then and has a finite value.

\begin{figure}[htbp!]
\centering \epsfig{file=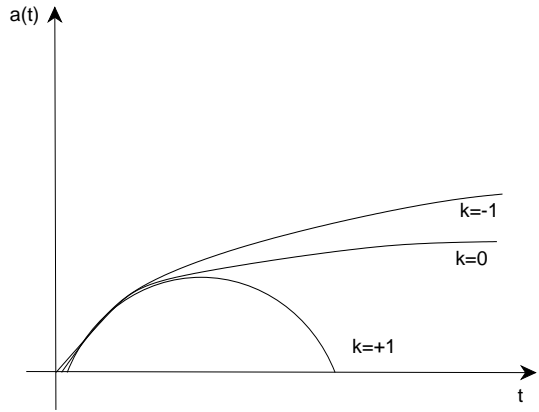,height=6cm} %
\caption{{\protect\small   \textit{Three   Universes:   Closed($K=1$),
Flat($K=0$) and Open ($K=-1$)}}}
\label{ofc}
\end{figure}

On the  other hand,  conservation of energy  (\ref{conserva}) requires
that $\rho$ decrease at least as fast as $a^{-3}$ as $a$ increases, if
$p$ remains positive. Then  equation (\ref{Fried4}) implies that $\dot
a^2\rightarrow a^{-1}-k$ (at  least) as $a\rightarrow\infty$. Thus the
behaviour of  the expansion depends on  the value of  $k$: for $k=-1$,
$\dot  a^2$  remains  positive  definite, so  $a(t)\rightarrow  t$  as
$t\rightarrow\infty$  and the  universe  expands for  ever. If  $k=0$,
$\dot a^2$ also remains  positive definite, and $a(t)\rightarrow t^c$,
with  $c<1$,  as   $t\rightarrow\infty$.  Then  the  universe  expands
indefinitely, but  slower than in  the previous case.  For  $k=1$, the
expansion stops at some point, $\dot a=0$, when
\begin{equation}
\frac{\kappa^2\rho}{3}-\frac{1}{a^2}=0
\end{equation}
and  $\dot  a$ takes  negative  values  since  $\ddot a<0$.  Then  the
universe begin to decrease again, reaching the singularity in a finite
time in the future, see figure \ref{ofc} .

Moreover, using  (\ref{Fried4}) and (\ref{conserva}) we  can write the
derivative of the Hubble parameter as
\begin{equation}\label{hpunto}
\dot H = -\frac{\kappa^2}{2} (\rho + p) + \frac{k}{a^2}\,.
\end{equation}
This equation is already telling us something very important. For flat
models, $k=0$,  this relation  implies that reversal  from contraction
($H<0$) to expansion ($H>0$) of  the scale factor $a(t)$ is impossible
since  $(\rho +  p)  >0$ by  the  weak  energy condition  (see
\cite{hawking}  for  a review  on  the  energy  conditions in  general
relativity).

\bigskip

Let  us now  introduce a  useful concept  that will  allow us  to make
        general comments  about the cosmological evolution  of the FRW
        universe.  This is,  the  critical  density, which  is
        defined by:
\begin{equation}\label{rocritica}
\rho_c  \equiv\frac{3\,H_0^2}{8\pi G}  \sim 1.7  \times \,10^{-29}{\rm
               g}/{\rm cm}^3\,,
\end{equation}
where $H_0$  is the  present value of  the Hubble  parameter, $H_0\sim
65$Km/s/Mpc$^{-1}$~\footnote{1  Mpc   =  $3\times  10^{22}$  meters.}.
Using  this  relation, we  can  also  define  the dimensionless  
density parameter $\Omega_i$ as

\begin{equation}\label{omega}
\Omega_i \equiv \frac{\rho_i}{\rho_c}\,,
\end{equation}
where  the subindex  $i$  stands for  matter, radiation,  cosmological
constant and curvature, today, that is

\begin{eqnarray}
&& \Omega_m = \frac{8\pi  G \rho_m}{3 H_0^2}\,, \qquad \qquad \Omega_m
         =    \frac{8\pi   G    \rho_r}{3   H_0^2}\nonumber    \\   &&
         \Omega_{\Lambda} = \frac{\Lambda}{3  H_0^2} \,, \qquad \qquad
         \quad \,\,\, \Omega_k = -\frac{k}{3 H_0^2} \,.
\end{eqnarray}
Using these relations, we can rewrite Friedmann's equation in the form
\begin{equation}
\Omega_k = \Omega_m +\Omega_{\Lambda} - 1 = \Omega_{tot} - 1
\end{equation}
where we have neglected $\Omega_r$ today.  Then, we can write down the
following equivalences:
\begin{equation}
\begin{array}{lllllllll}
 k=1\; \; \; \; &\Leftrightarrow\; \; \; & \Omega_{tot}>1\; \; \; \; &
\Leftrightarrow\; \; \; & \rho>\rho_c  \; \; \; \; & \Leftrightarrow\;
\; \; &  {\rm closed} \\ \nonumber k=0 \;  \; \; \; &\Leftrightarrow\;
\;  \;  &  \Omega_{tot}=1\; \;  \;  \;  &  \Leftrightarrow\; \;  \;  &
\rho=\rho_c  \; \;  \;  \;  & \Leftrightarrow\;  \;  \; &{\rm  flat}\\
\nonumber k=-1\; \; \;  \; &\Leftrightarrow\; \; \; & \Omega_{tot}<1\;
\;  \; \;  &  \Leftrightarrow\; \;  \; &  \rho<\rho_c  \; \;  \; \;  &
\Leftrightarrow\; \; \; & {\rm open}
\end{array}
\end{equation}

\smallskip
At  this  point, we  have  all the  information  needed  to trace  the
evolution  of  the universe  with  the  assumption  that the  universe
corresponds to an expanding gas of particles described by the Standard
Model of particle  physics. This gas is assumed  to be in equilibrium.
There are  two ways in which  the particles leave  equilibrium. One is
when the  mass threshold of the  particle is reached  by the effective
temperature of the universe, and so,  it is easier for the particle to
annihilate  with its  antiparticle, rather  than being  produced again
since,  as the  universe cools  down, there  is not  enough  energy to
produce such  a heavy  object. The other  way is when  the interaction
rate of the relevant reactions $\Gamma$, is smaller than the expansion
rate of the universe, measured by  $H$. At this time the particles get
out  of equilibrium.  For instance  at  temperatures above  1 MeV  the
reactions  that keep  neutrinos  in equilibrium  are  faster than  the
expansion rate but at this temperature $H\leq\Gamma$ and they decouple
from the hot  plasma, leaving then an observable,  in principle, trace
of the very  early universe. However, at the present  time, we are far
from being able to detect such radiation (for a detailed analysis, see
\cite{kolb,bookpeebles}).







At  a temperature  of about  $10^5$K,  corresponding to  $10$ eV,  the
non-relativistic  matter  content in  the  universe  reached the  same
density as the  relativistic one, and the universe  changed from being
radiation dominated to being matter dominated. When this happened, the
expansion  rate increased from  $a(t)\propto t^{1/2}$  to $a(t)\propto
t^{2/3}$. From  that point on, the temperature  decreased more quickly
than the  (fourth root of  the) energy density. The  first significant
event of this  epoch occurred at around 3000K,  when decoupling of the
radiation from matter occurred. The universe was cold enough for atoms
to be formed,  and so radiation decoupled completely  from the matter,
and the photons propagated freely, giving rise to the cosmic microwave
background radiation.   After this, structure must have  formed in the
universe,  including the  first large  gravitationally  bound systems,
such as clusters and galaxies, probably due to quantum fluctuations of
the early universe, leading to our present time.

\section{The Bekenstein-Sandvik-Barrow-Magueijo Theory}
\label{bsbm}

In  the  previous  section  we  have  described  the  evolution  of  a
Friedmann-Robertson-Walker universe and its components.  Now, in order
to  study  the time  and  spatial  variations  of the  fine  structure
constant during the history of  the evolution of the universe, we need
to  consider  a theoretical  model  which  incorporates variations  of
$\alpha$ into the previous cosmological model.

Due  to  the amazing  theoretical  and  observational  success of  the
standard cosmological model, we  will choose a varying-$\alpha$ theory
which  will preserve  as  much as  possible  the behaviour  of an  FRW
universe.

As  we  saw  in  the  previous  sections,  a  common  feature,  to  all
theoretical  varying-$\alpha$ models,  is  the existence  of a  scalar
field, responsible for the variations of $\alpha$, which is, at least,
coupled  to  the  matter  fields which  interact  electromagnetically.
Since  we  are  interested   to  investigate  the  effects  of  matter
inhomogeneities on  the evolution of the fine  structure constant, and
we  want  the preserve  Einstein  gravity  formulation,  we will  then
consider  a four-dimensional  model  where the  scalar  field is  only
coupled to  the matter fields which  interact electromagnetically.


With  all  these  desired  features   in  mind,  we  will  choose  the
Bekenstein-Sandvik-Barrow-Magueijo  (BSBM) theory  for a  varying fine
structure constant.

\bigskip

Bekenstein  did  not  take  into  account  the  effect  of  the  field
$\epsilon$  in  the  Einstein  equations  and studied  only  the  time
variation of $\epsilon$ in a matter dominated universe.
Sandvik,  Barrow and Magueijo  have generalised  the scalar  theory by
Bekenstein in order to include the gravitational effects of the field,
$\epsilon$,  responsible   for  the  variations  of   the  fine  structure
constant \cite{bsbm}. 
In that sense, they  have extended the analysis by Bekenstein
by  solving the  coupled  system  of the  Einstein  equations and  the
Klein-Gordon equation of $\epsilon$ \cite{bsbm}.

In order to simplify
Bekenstein's  theory,   they  introduced  an   auxiliary  gauge  field
$a_{\mu}=\epsilon   A_{\mu}$   \cite{bsbm,magueijokibble},   and   the
electromagnetic field tensor can be written now as
\begin{equation}
f_{\mu\nu}=\epsilon
F_{\mu\nu}=\partial_{\mu}a_{\nu}-\partial_{\nu}a_{\mu}
\end{equation}
The  covariant derivative  then becomes  $D_{\mu}=\partial_{\mu}+i e_0
a_{\mu}$.  To  simplify  further  another transformation  is  possible
\cite{magueijokibble}:  $\epsilon\rightarrow  \psi=\ln \epsilon$.  The
field $\psi$  will then be the  responsible for the  variations of the
fine structure constant. The scalar  field $\psi$ plays a similar role
to the  dilaton in the low  energy limit of string  and M-theory, with
the  important  difference that  it  couples  only to  electromagnetic
energy. Since  the dilaton field  couples to all the  matter (although
generally to different sectors  with different powers) then the strong
and electroweak  charges, as  well as particle  masses, can  also vary
with  the  time-position   coordinate  $x_{\mu}$.  These  similarities
highlight the deep  connections between effective fundamental theories
in     higher     dimensions     and     varying-constant     theories
\cite{damour,barrowextra,forgacs,marciano}.

The total action describing  the dynamics of
the Universe with a varying-$\alpha$ and  
including the  Einstein-Hilbert action for  gravity and
normal matter takes the form:
\begin{equation}
S=\int
d^4x\sqrt{-g}\left(
\mathcal{L}_{grav}+\mathcal{L}_{matter}+\mathcal{L%
}_\psi +\mathcal{L}_{em}e^{-2\psi }\right) , \label{S}
\end{equation}
where $\mathcal{L}_\psi  ={\frac \omega 2}\partial  _\mu \psi \partial
^\mu \psi  $, $\omega =  \frac{\hbar c}{l^2}$ is a  coupling constant,
and  $\mathcal{L}_{em}=-\frac   14f_{\mu  \nu  }f^{\mu   \nu  }$.  The
gravitational  Lagrangian is  the  usual $  \mathcal{L}_{grav}=-\frac
1{16\pi    G}R$,    with     $R$    the    curvature    scalar,    and
$\mathcal{L}_{matter}$ is the matter fields Lagrangian.

\bigskip

To obtain the cosmological equations we vary the action (\ref{S}) with
respect to the metric to give the generalised Einstein equations
\begin{equation}
G_{\mu\nu}=8                           \pi                           G
\left(T^{mat}_{\mu\nu}+T^{\psi}_{\mu\nu}+T^{em}_{\mu\nu}\right)
\label{einstein}
\end{equation}
which are similar to the one derived in equation (\ref{einstein1}) but
now  we have  included the  energy momentum  of $\psi$,  which  can be
obtained using  equation (\ref{tmunudef}). The equation  of motion for
$\psi$ comes, varying the action (\ref{S}) with respect to it
\begin{equation}
\square\psi=\frac{2}{\omega}e^{-2\psi}{\mathcal{L}}_{em}
\label{psiKG}
\end{equation}

The  right-hand-side  (RHS)  of  equation (\ref{psiKG})  represents  a
source term for $\psi$, which includes all the matter fields which are
coupled  to  it. These  include  not  only  relativistic matter  (like
photons),  but   as  well   as  non-relativistic  one   that  interact
electromagnetically.

It  is  clear  that  ${\cal{L}}_{em}$,  vanishes for  a  sea  of  pure
radiation  since  ${\cal{L}}_{em}=\left(E^2-B^2\right)/2=0$. The  only
significant contribution  to a variation  of $\psi$ comes  from nearly
pure   electrostatic    or   magnetostatic   energy    associated   to
non-relativistic particles.  In order to make quantitative predictions
we  then  need  to  know  how  much  of  the  non-relativistic  matter
contributes    to    the    right-hand-side    (RHS)    of    equation
(\ref{psiKG}).    This   can    be   parametrised    by    the   ratio
$\zeta={\cal{L}}_{em}/\rho_m$, where $\rho_m$ is the energy density of
the non-relativistic matter \cite{havard}.

For  protons   and  neutrons,  $\zeta$  can  be   estimated  from  the
electromagnetic corrections to the nucleon mass, $0.63 MeV$ and $-0.13
MeV$, respectively \cite{dvali}.  This correction contains the $E^2/2$
contribution,  which is  always positive,  and also  terms of  the form
$j_{\mu}a^{\mu}$,   where    $j_{\mu}$   is   the    quarks'   current
\cite{bsm1}. Hence  we take a guidance  value of $\zeta\approx10^{-4}$
for protons and neutrons.

\bigskip

Using  the parameter $\zeta$,  the fraction  of electric  and magnetic
energies may then be written as:
\begin{equation}
\zeta^{E}=\frac{E^2}{\rho_m} \qquad \zeta^{B}=\frac{B^2}{\rho_m}
\label{zetas}
\end{equation}
where  $E^2$  and  $B^2$   are  the  electric  and  magnetic  energies
respectively.    Using   equation    (\ref{zetas}),    then   equation
(\ref{psiKG}) becomes
\begin{equation}
\square\psi=\frac{2}{\omega}e^{-2\psi}\rho_m
\left(\zeta^{E}-\zeta^{B}\right)
\end{equation}

\bigskip

 

Since  we are interested  in the  cosmological evolution  of $\alpha$,
instead of using both  parameters $\zeta^{E}$ and $\zeta^{B}$, we will
use  throughout  this  thesis,  the cosmological  parameter,  $\zeta$,
defined as $\zeta\equiv\zeta^{E}-\zeta^{B}$,  which in the limit where
$\zeta^{E}\gg\zeta^{B}$ is  positive, and when $\zeta^{E}\ll\zeta^{B}$
is negative.

Note that, the  cosmological value of $\zeta$ has  to be weighted, not
only by  the electromagnetic-interacting baryon fraction,  but also by
the fraction of matter that  is non-baryonic, for instance dark
matter. 
The coupling to  dark matter is motivated by  two aspects. The first
is the fact that dark matter might be electromagnetically charged, for
instance superconducting cosmic  strings, and if that is  the case the
scaler field is necessarily coupled  to it. The second is the Olive and
Pospelov generalisation of the Bekenstein model, where they claim that
the  field responsible  for the  variations  in $\alpha$  may be  more
strongly  coupled to  dark matter  than to  baryons (depending  on the
nature of dark  matter) \cite{olive}. Hence the value and sign
of $\zeta$ depends also on the  nature of dark matter to which the
fields $\psi$ might be coupled. For instance, $\zeta=-1$  for superconducting
cosmic       strings,       where      ${\cal{L}}_{em}\approx-B^{2}/2$
\cite{bsm1,bsm2}, and in the case of neutrinos $\vert\zeta\vert\ll1$.

\bigskip

The  universe, we  will  be studying,  will  be described  by a  flat,
homogeneous  and  isotropic Friedmann  metric.  The universe  contains
pressure-free matter, of density  $\rho _{m}$, a cosmological constant
$\Lambda $, of  density $\rho _{\Lambda }\equiv $  $\Lambda /(8\pi G)$
and  radiation, of density  $\rho_r$.  The  Friedmann equation  can be
obtained from the Einstein equations (\ref{einstein}),
\begin{equation}
H^{2}=\frac{8\pi   G}{3}\left(  \rho  _{m}\left(   1+\left\vert  \zeta
\right\vert e^{-2\psi }\right) + \rho_r e^{-2\psi} +\rho _{\psi }+\rho
_{\Lambda }\right)
\label{fried}
\end{equation}
where  $\rho  _{\psi}   =  \frac{\omega}{2}  \dot\psi^2$.   Using  the
cosmological  parameter  $\zeta$, the  evolution  equation for  $\psi$
comes from \ref{psiKG}
\begin{equation}
\ddot{\psi}+3H\dot{\psi}=-\frac{2}{\omega  }\ e^{{-2\psi  }}\zeta \rho
_{m}.
\label{psidot}
\end{equation}
The  conservation  equations  for  the non-interacting  radiation  and
matter densities, $\rho _r$ and $\rho _m$ respectively, are:
\begin{eqnarray}
\dot{\rho    _m}+3H\rho    _m    &=&0   \label{mat}    \nonumber    \\
\dot{\rho}_r+4H\rho _r &=&2\dot{\psi}\rho _r \label{dotrho}
\label{matter}
\end{eqnarray}
so   $\rho  _m\propto   a^{-3}$.   The  last   relation  in   equation
(\ref{matter}) can be written as
\begin{equation}
\dot{\tilde{\rho}_r}+4H\tilde{\rho}_r=0,
\end{equation}
with $\tilde{\rho}_r\equiv \rho _re^{-2\psi }\propto a^{-4}$.

Equation  (\ref{psidot}) may  be  expressed in  terms  of the  kinetic
energy  density  of   the  $  \psi  $  field,   $\rho  _\psi  =\omega
\dot{\psi}^2/2,$ to give
\begin{equation}
\dot{\rho  _\psi  }+6H\rho  _\psi =2\sqrt{{\frac  2\omega  }}e^{-2\psi
}\zeta_m \rho _m\sqrt{\rho _\psi }.  \label{dotrhopsi}
\end{equation}
The  $\psi $ field  behaves like  a stiff  Zeldovich fluid  with $\rho
_\psi \propto a^{-6}$ when the RHS vanishes.

\bigskip

\begin{figure}[p]
\centering
\includegraphics[height=7.0cm,width=10cm]{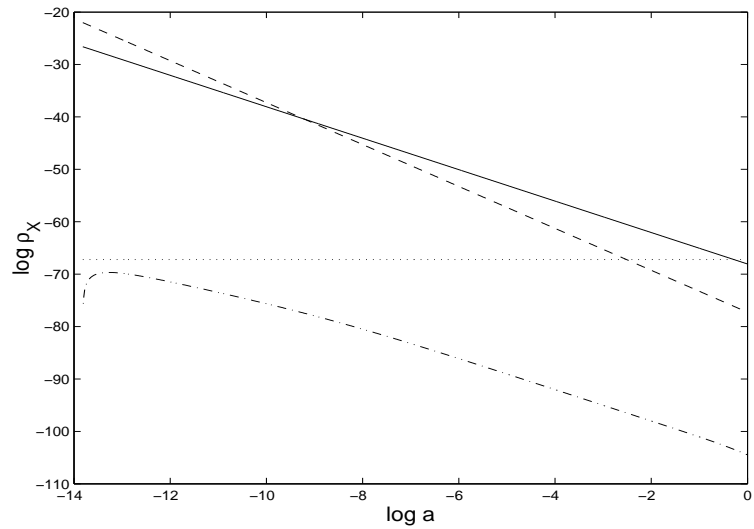}
\caption{{\protect {\it  Evolution of $\log \rho_{m}$  (solid lines) ,
$\log \rho_{\psi}$  (dash-dotted line), $\log  \rho_{\lambda}$ (dotted
line)  and  $\log \rho_{r}$  (dashed  line)  as  a function  of  $\log
(a)$.}}}
\label{backgroundrho}
\end{figure}

\begin{figure}[p]
\centering
\includegraphics[height=7.0cm,width=10cm]{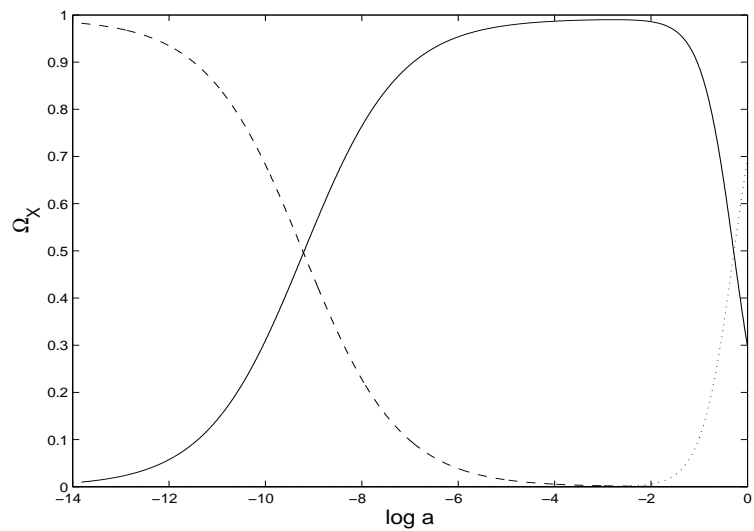}
\caption{{\protect {\it  Evolution of $\log \rho_{m}$  (solid lines) ,
$\log \rho_{\psi}$  (dash-dotted line), $\log  \rho_{\lambda}$ (dotted
line)  and  $\log \rho_{r}$  (dashed  line)  as  a function  of  $\log
(a)$.}}}
\label{backgroundomega}
\end{figure}

The    Bekenstein-Sandvik-Barrow-Magueijo    theory,    enables    the
cosmological consequences of a varying fine structure constant to be
analysed self-consistently. Equations (\ref{fried}), (\ref{psidot} and
(\ref{matter}), govern the Friedman universe with time-varying $\alpha$ . These
equation  can  be  evolved  numerically from  early  radiation  epoch,
through the  matter era  and into vacuum  domination.
Notice from Figure  \ref{backgroundrho} and \ref{backgroundomega} that
the energy  density associated  to the scalar  field $\psi$  is always
negligible at all the stages of the universe history.

 The inclusion of
a  varying-$\alpha$  will then  affect  very  little the  cosmological
expansion of the universe. Hence, the Hubble diagram will be precisely
the  same  as  that  of  a  universe,  with  a  constant  $\alpha$,  with
$\Omega_m\approx0.3$,    $\Omega_{\Lambda}\approx0.7$   and   $h=0.65$
\cite{bsbm}.  This is the effect that we desired in a varying-$\alpha$
model: the less possible deviations from the standard FRW universe.

%% file: chapter21.tex
\chapter{Background Solutions for the Fine Structure Constant}
\label{qualitative}

\begin{flushright}
{\it {\small Finjo viver num tempo sem fim}}\\
{\it {\small e de tanto fingir...}}\\
{\it {\small esqueci-me que s\'o sei viver assim}}\\
{\it {\small -- Ernesto Mota --}}
\end{flushright}

\bigskip

\section{Introduction}

In  the  last chapter  we  have  described  the BSBM  varying-$\alpha$
cosmological  model we  will use  to study  the effect  of  the matter
inhomogeneities in  the cosmological  evolution of the  fine structure
constant.  Before starting to investigate those effects, we first need
to study  the evolution of  $\alpha$ in the homogeneous  and isotropic
background  universe  and to  calculate  analytical  solutions of  the
equation  of  motion   (\ref{psidot})  for  $\psi$.  These  analytical
solutions  will be  useful when  we  study the
existence and behaviour of the inhomogeneities.

We  will then  provide a  complete analysis  of the  behaviour  of the
solutions  of the non-linear  propagation equation  (\ref{psidot}) for
appropriate  behaviours of $a(t)$,  for each  epoch the  universe went
through:   radiation dominated,  dust dominated  and   dark  energy
dominated.

The aim  of this  chapter brings  a question: How  shall we  solve the
non-linear  differential equation (\ref{psidot})?  One cannot  hope to
obtain exact solutions to  most non-linear differential equations. The
reason  is  that  there  are  only  a  limited  number  of  systematic
procedures for solving them, and these apply only to a very restricted
class  of equations.  Moreover, even  when a  closed-form  solution is
known, it  may be so  complicated that its qualitative  properties are
obscure. Thus, for  most non-linear equations it is  necessary to have
reliable  techniques to  determine  the approximate  behaviour of  the
solutions. In order to calculate and to understand the exact solutions
of equation (\ref{psidot}), we  will use a phase-space analysis.  From
this  analysis, we  will be  able to  find stable  asymptotic solution
which can  be considered  to describe the  behaviour of  $\psi$ during
the different eras the universe went through.



\section{An Approximation Method}

We have seen in the last chapter, that the energy density of the field
associated to variations of $\alpha$ is always negligible with respect
to    all    the    other    energy   content    of    the    universe
($\rho_{\psi}\ll\rho_m,   \rho_r,\rho_{\Lambda}$).   This   means  the
Friedmann  models with  varying $\alpha$  have the  property  that the
homogeneous motion of  the $\psi $ does not  create significant metric
perturbations  at late  times \cite{bsm1}.  So, far  from  the initial
singularity,  we can  safely assume  that the  expansion  scale factor
follows the form for the  Friedmann universe containing the same fluid
when  $\alpha $ does  not vary  $(\zeta =0=\psi  )$. The  behaviour of
$\psi $  then follows  from a solution  of equation  (\ref{psidot}) in
which $a(t)$ has the form for a Friedmann universe for matter with the
same equation  of state in  general relativity when $\zeta  =0=\psi .$
This  behaviour  is natural.  We  would  not  expect that  very  small
variations of  the coupling to  electromagnetically interacting matter
would  have large  gravitational  effects upon  the  expansion of  the
universe. Thus, in this chapter we will provide a complete analysis of
the behaviour of the  solutions of  equation
(\ref{psidot}) for appropriate behaviours of $a(t).$

Before  starting the analysis  and to  calculate the  asymptotic exact
solutions  let  us show  that  the  assumptions  we claim  are  indeed
right. Hence, we  will show how good is  the approximation of assuming
that the scale factor, $a(t)$, will not deviate from the the usual FRW
universe that  we have calculated  in the previous chapter,  see table
\ref{table1}.

\bigskip

We  consider spatially  flat  universes ($k=0$)  and  assume that  the
expansion scale  factor is  that of the  Friedmann model  containing a
perfect fluid:
\begin{equation}
a=t^n \label{a}
\end{equation}
where  $n$  is  a  constant.  The  late stages  of  an  open  universe
containing fluid with  density $\rho $ and pressure  $p$ obeying $\rho
+3p>0$ can  be studied by considering  the case $n=1.$  We rewrite the
wave equation (\ref {psidot}) as
\[
\ (\dot{\psi}a^3\dot{)}=-\frac{2\zeta }\omega \rho _ma^3\exp [-2\psi ]
\]
Therefore, since $\rho _ma^3$ is  constant this reduces to a Liouville
equation of the form
\begin{equation}
(\dot{\psi}a^{3})^{.}=\ N\exp [-2\psi ] \label{psi}
\end{equation}
where $N$ is a constant, defined by
\begin{equation}
N\equiv -\frac{2\zeta }{\omega }\rho _{m}a^{3}\ .  \label{N}
\end{equation}

We shall  consider first the  cosmological models that arise  when the
defining  constant $N$  is  negative. This  arises  when the  constant
$\zeta  <0,$ indicating  that the  matter content  of the  universe is
dominated by  magnetic rather than electrostatic energy.  The value of
$\zeta   $  for   baryonic   and  dark   matter   has  been   disputed
\cite{dvali,olive,bsbm}. It  is the difference  between the percentage
of mass in electrostatic and  magnetostatic forms. As explained in the
previous  chapter and  \cite{bsbm}, we  can at  most \textit{estimate}
this quantity for neutrons  and protons, with $\zeta _{n}\approx \zeta
_{p}\sim 10^{-4}$. We may expect  that for baryonic matter $\zeta \sim
10^{-4}$, with composition-dependent variations of the same order. The
value  of $\zeta $  for the  dark matter,  for all  we know,  could be
anything  between  -1  and  1.   Superconducting  cosmic  strings,  or
magnetic  monopoles, display  a \textit{  negative} $\zeta  $, unlike
more conventional dark matter. It is clear that the only way to obtain
a cosmologically increasing $\alpha $  in BSBM is with $\zeta <0$, i.e
with  unusual dark  matter, in  which magnetic  energy  dominates over
electrostatic energy.  In \cite{bsbm} it  was showed that  fitting the
Webb et  al results requires  $\zeta /\omega =-2\pm  1\times 10^{-4}$,
where  $\zeta $ is  weighted by  the necessary  fractions of  dark and
baryonic matter required by  observations of the gravitational effects
of dark  matter and the  calculations of Big Bang  nucleosynthesis. We
note  also that in  practise $  \zeta $  might display  a significant
spatial variation because of the  change in the nature of the dominant
form  of matter over  different  length scales.  For example,  a
magnetically dominated form of dark matter might contribute a negative
value  of $\zeta  $ on  large scales  while domination  of  the matter
content  by  baryons  on  small   scales  would  lead  to  $\zeta  >0$
locally. We  will not investigate the  effects of such  variations in this
thesis.

\subsection{The validity of the approximation}

\bigskip We  have assumed that  the scale factor  is given by  the FRW
model and then  solved the $\psi $ evolution equation.  This is a good
approximation up  to logarithmic corrections. Here is  what happens to
higher order.

We take the leading order behaviour in (\ref{fried})
\begin{equation}
\left(   \frac{\dot{a}}{a}\right)   ^{2}\approx  \frac{8\pi   }{3}\rho
_{m}\left( 1+\left| \zeta \right| e^{-2\psi }\right) \label{next}
\end{equation}
Now if we take $a=t^{2/3}$ so
\[
\frac{8\pi }{3}\rho _{m}=\frac{4}{9t^{2}}
\]
and solve equation (\ref{psi}) we get asymptotically,
\begin{equation}
\psi =\frac{1}{2}\ln [2N\ln t]\rightarrow \ln [\ln t] \label{psisol}
\end{equation}
to leading order  at late times. Suppose we  now re-solve (\ref{next})
with the $\left| \zeta \right| e^{-2\psi }$ correction included
\begin{equation}
\left(  \frac{\dot{a}}{a}\right) ^{2}\approx  \frac{4}{9t^{2}}\ \left(
1+\left|     \zeta      \right|     e^{-2\psi     }\right)     \approx
\frac{4}{9t^{2}}\left(   1+  \frac{\left|   \zeta  \right|   }{\  \ln
t}\right) \label{newfried}
\end{equation}
Note that the kinetic term which we neglected is of order
\begin{equation}
\omega \dot{\psi}^{2}\sim \frac{1}{4t^{2}\ln ^{2}t}
\end{equation}
and so  is smaller  than the term  we have retained.  Solving equation
(\ref {next}) we have
\begin{equation}
\ln   a=\frac{2\left|    \zeta   \right|   }{3}\left[   sh^{-1}\left\{
\sqrt{\frac{\ln t }{\left|  \zeta \right| }}\right\} +\sqrt{\frac{\ln
t}{\left| \zeta \right|  }} \sqrt{1+\frac{\ln t}{\left| \zeta \right|
}}\right]
\end{equation}
Note that  when $\left|  \zeta \right| \rightarrow  0$ this  gives the
usual $ \ln  a=\frac{2}{3}\ln t$. When $\left| \zeta  \right| \neq 0$
we have
\[
\ln   a\rightarrow  \frac{2\left|   \zeta   \right|  }{3}\left\{   \ln
[2\sqrt{\frac{  \ln t}{\left|  \zeta \right|  }}]+\frac{\ln t}{\left|
\zeta \right| }\right\}
\]
and
\begin{equation}
a=t^{2/3}(\ln t)^{\left| \zeta \right| /3} \label{scale}
\end{equation}
where $\left| \zeta  \right| $ is small and so  the corrections to the
$ a=t^{2/3}$ ansatz are small. In terms of the Hubble rate:
\[
H=\frac{2}{3t}+\frac{\left| \zeta \right| }{3t\ln t}
\]

If we include the kinetic corrections to equation (\ref{next}) then as
$\dot{ \psi}=\frac{1}{2t\ln t}$
\begin{equation}
\left(  \frac{\dot{a}}{a}\right) ^{2}\approx  \frac{4}{9t^{2}}\ \left(
1+\left|  \zeta  \right|   e^{-2\psi  }\  \right)  +\frac{4\pi  \omega
}{3}\dot{\psi}   ^{2}\approx   \frac{4}{9t^{2}}\left(  1+\frac{\left|
\zeta \right| }{\ \ln t}+ \frac{S}{\ \ln ^{2}t}\right)
\end{equation}
where
\[
S\equiv \frac{3\pi \omega }{4}
\]

So, if $x=\ln t,$ we have
\begin{eqnarray}
\frac{3}{2}\ln   a  &=&\sqrt{x^{2}+\left|  \zeta   \right|  x+S}+S\int
\frac{dx}{x  \sqrt{x^{2}+\left|  \zeta  \right|  x+S}}  \nonumber  \\
&&+\frac{\left|  \zeta  \right| }{2}\int  \frac{dx}{\sqrt{x^{2}+\left|
\zeta  \right| x+S}} \\  \frac{3}{2}\ln a  &=&\sqrt{x^{2}+\left| \zeta
\right|    x+S}-\ln   \left[    \frac{   2S+\left|    \zeta   \right|
x+2\sqrt{S}\sqrt{x^{2}+\left|   \zeta    \right|   x+S}}{x}   \right]
\nonumber   \\   &&+\frac{\left|    \zeta   \right|   }{2}\ln   \left[
2\sqrt{x^{2}+\left| \zeta \right| x+S}+2x+\left| \zeta \right| \right]
\end{eqnarray}
Again, as $t\rightarrow  \infty $ the leading order  behaviour is that
found in equation (\ref{scale}).

In the  radiation era, where  $a(t)\propto t^{1/2}$, we have  an exact
solution of equation (\ref{psi}) with
\begin{equation}
2\psi =\ln (8N)+\frac{1}{2}\ln t
\end{equation}
so the corrections to the Friedmann equation look like
\begin{equation}
\left(  \frac{\dot{a}}{a}\right) ^{2}\approx  \frac{1}{4t^{2}}\ \left(
1+\left|     \zeta      \right|     e^{-2\psi     }\right)     \approx
\frac{1}{4t^{2}}\left(   1+   \frac{\left|   \zeta   \right|   }{8N\
t^{1/2}}\right)
\end{equation}
and  these  corrections  fall  off   much  faster  than  in  the  dust
case.  Again,  our  basic  approximation  method holds  good  to  high
accuracy.

\subsection{A linearisation instability}

Despite the robustness of  the basic test-motion approximation that we
are employing to  analyse the evolution of $\psi  (t)$ as the universe
expands, there  is a subtle feature the  non-linear evolution equation
(\ref{psi}) which must be noted in order that spurious conclusions are
not drawn  from an  approximate analysis. We  see that  the right-hand
side of equation  (\ref{psi} ) is always positive.  Therefore $\psi $
can never  experience an  expansion maximum (where  $\dot{\psi}=0$ and
$\ddot{\psi}<0$)    and    therefore     $\psi    (t)$    can    never
oscillate.  However, if  we  were to  linearise equation  (\ref{psi}),
obtaining
\[
(\dot{\psi}a^{3})^{.}=\  N\exp   [-2\psi  ]\approx  N(1-2\psi  +O(\psi
^{2}))
\]
then for  $\psi >1/2$  the right-hand side  takes negative  values and
pseudo-oscillatory solutions for $\psi $ would appear that are not the
linearised  approximation  to  any  true solution  of  the  non-linear
equation (  \ref{psi}). Care must  therefore be taken to  ensure that
analytic approximations  are not  extended to large  $\psi $  and that
numerical  analyses are  not creating  spurious spirals  in  the phase
plane by virtue of a  linearisation procedure; for a fuller discussion
see ref. \cite{bsm3}.

These considerations  can be taken  further. It is possible  for $\psi
(t)$ to  decrease, reach a  minimum and then  increase. But it  is not
possible for  $ \psi  (t)$ to  decrease if it  has ever  increased. A
second interesting consequence of this feature of equation (\ref{psi})
is that it holds true even  if $a(t)$ reaches an expansion maximum and
begins to  contract. Thus in a  closed universe we expect  $\psi $ and
$\alpha  $ to  continue to  increase  slowly even  after the  universe
begins  to contract.  This will  have important  consequences  for the
expected  variation  of  $\psi   $  and  $\alpha  $  in  realistically
inhomogeneous universes.

\section{Two-Dimensional Non-Linear Autonomous Systems}

 
Autonomous  systems  of  equations,   when  they  are  interpreted  as
describing the motion of a  point in the phase space, are particularly
susceptible to  some very beautiful  techniques of local  analysis. By
performing  a local  analysis  of the  system  near what  are know  as
critical points,  one can  make remarkably accurate  predictions about
the global behaviour of the solution.

In this section,  we shall not attempted to  give a proper description
of the the  dynamical systems field, but we will  only state the basic
results needed  to understand  this chapter. For  a better  and proper
study of the topic see for instance \cite{andronov, wiggins,bautin}.
 
\bigskip

Differential equations  which do not contain  the independent variable
explicitly  are said to  be autonomous.  Any differential  equation is
equivalent to an autonomous equation of one higher order.

It is convenient  to study the approximate behaviour  of an autonomous
equation  of order  $2$ when  it is  in the  form of  a system  of $2$
coupled first-order differential equations. The general form of such a
system is
\begin{eqnarray}
\dot y_1 &=& f_1(y_1,y_2) \nonumber\\ \dot y_2 &=& f_2(y_1,y_2)
\label{systemgeneral}
\end{eqnarray}
where  the  dots  indicate  a  differentiation  with  respect  to  the
independent variable, for instance $t$.

The  solution  of the  system  (\ref{systemgeneral})  is  a curved  or
trajectory  in  a  two-dimensional   space  called  phase  space.  The
trajectory is parametrised in terms of $t:y_1=y_1(t)$, $y_2=y_2(t)$.

We assume that $f_1$, $f_2$ are continuous differentiable with respect
to  each of  their arguments.  Thus, by  the existence  and uniqueness
theorem   of  differential   equations  \cite{andronov}   any  initial
condition   $y_1(0)=a_1$,  $y_2(0)=a_2$,  gives   rise  to   a  unique
trajectory  through  the  point   $(a_1,  a_2)$.  To  understand  this
uniqueness  property  geometrically,  note  that every  point  on  the
trajectory $[y_1(t),y_2(t)]$, the system (\ref{systemgeneral}) assigns
a unique  velocity vector $[\dot y_1(t),\dot y_2(t)]$  which is tangent
to  the trajectory  at that  point.  It immediately  follows that  two
trajectories  cannot  cross;  otherwise,  the tangent  vector  at  the
crossing point would not be unique.

\subsection{Critical points in phase space}

If  there are  any  solutions  to the  set  of simultaneous  algebraic
equations
\begin{eqnarray}
f_1(y_1,y_2)  &=&0 \qquad  (\Leftrightarrow \dot  y_1  =0) \nonumber\\
f_2(y_1,y_2) &=&0 \qquad (\Leftrightarrow \dot y_2 =0) \
\label{systemcritical}
\end{eqnarray}
then there  are special degenerate  trajectories in phase  space which
are  just  points. The  velocity  at these  points  is  zero so  the
position  vector  does not  move.  These  points  are called  critical
points.

Studying  the  phase space  it  is  possible  to make  elegant  global
analyses of the  system. The possible behaviours of  a trajectory in a
two dimensional system \cite{andronov} are:
\begin{itemize}
\item The  trajectory may approach  a critical point  as $t\rightarrow
+\infty$.
\item The trajectory may approach $\infty$ as $t\rightarrow +\infty$.
\item The trajectory may remain motionless at a critical point for all
$t$.
\item The trajectory may describe a closed orbit or a cycle.
\item The trajectory may  approach a closed orbit (by spiralling inward
  or outward toward the orbit) as $t\rightarrow +\infty$.
\end{itemize}

The possible  local behaviours for trajectories near  a critical point
are:
\begin{itemize}
\item{1.}  All  trajectories may  approach  the  critical point  along
  curves  which  are asymptotically  straight  lines as  $t\rightarrow
  +\infty$. We call such a critical point a stable node.
\item{2.}  All  trajectories may  approach  the  critical point  along
  spiral curves  as $t\rightarrow +\infty$.  Such a critical  point is
  called a stable spiral point.
\item{3.}  All time reversed  trajectories, that  is, $y(t)$  with $t$
  decreasing, may move toward the critical point along paths which are
  asymptotically  straight  lines as  $t\rightarrow  -\infty$. Such  a
  critical  point   is  an  unstable  node.  As   $t$  increases,  all
  trajectories that  start near  an unstable node  move away  from the
  node along paths that are approximate straight lines, at least until
  the trajectory gets far from the node.
\item{4.} All time-reversed trajectories may move forward the critical
  point along spiral curves as $t\rightarrow -\infty$. Such a critical
  point  is called  an  unstable spiral  point. As  $t$ increases,  all
  trajectories  move   away  from  an  unstable   spiral  point  along
  trajectories that are, at least initially, spiral shaped.
\item{5.}  Some trajectories  may  approach the  critical point  while
  others move away from it  as $t\rightarrow +\infty$. Such a critical
  point is called a saddle point.
\item{6.} All trajectories may form  a closed orbit about the critical
  point. Such a critical point is called a centre.
\end{itemize}

\subsection{Matrix methods} 

\subsubsection{Linear autonomous systems}

Since two-dimensional linear autonomous systems can exhibit any of the
critical  point  behaviours  that  we  have  described  above,  it  is
appropriate  to  study  linear  systems  before  going  to  non-linear
ones. With this in mind we introduce an easy method for solving linear
autonomous systems.

A two dimensional linear autonomous system $\dot y_1 = a y_1 + b y_2$,
$\dot y_2 = c y_1 + d y_2$ may be re-written in matrix form as
\begin{equation}
{\bf{\dot Y}}= M {\bf{Y}}
\label{matrix}
\end{equation}
where
$M=  \bordermatrix{  & &\cr  &  a &  b  \cr  & c  &  d  \cr} $  \qquad
and \qquad 
${\bf{Y}}= \bordermatrix{& \cr & y_1 \cr & y_2\cr} $

It  is  easy  to  verify  that  if  the  eigenvalues  $\lambda_1$  and
$\lambda_2$  of  the matrix  $M$  are  distinct  and ${\bf{V_1}}$  and
${\bf{V_2}}$  are eigenvectors  of $M$  associated to  the eigenvalues
$\lambda_1$  and $\lambda_2$,  then the  general solution  to equation
(\ref{matrix}) has the form
\begin{equation}
{\bf{Y}}(t)=      c_1     {\bf{V_1}}e^{\lambda_1     t}      +     c_2
{\bf{V_2}}e^{\lambda_2 t}
\label{solution}
\end{equation}
where  $c_1$  and  $c_2$   are  constants  of  integration  which  are
determined by the initial position ${\bf{Y}}(0)$.

The linear  system (\ref{matrix}) has  a critical point at  the origin
$(0,0)$. It is  easy to classify this critical  point once $\lambda_1$
and  $\lambda_2$ are  known.  Note that,  $\lambda_1$ and  $\lambda_2$
satisfy the eigenvalue condition

$det[M-I\lambda]=det \bordermatrix{ &  &\cr & a-\lambda & b  \cr & c &
d-\lambda \cr} $ = $\lambda^2-\lambda (a+d)+ad-bc=0
\label{eigenvalues}
$

If  $\lambda_1$  and  $\lambda_2$  are  real and  negative,  then  all
trajectories approach  the origin as  $t\rightarrow\infty$ and $(0,0)$
is a stable node. Conversely,  if $\lambda_1$ and $\lambda_2$ are real
and  positive,  then  all  trajectories  move  away  from  $(0,0)$  as
$t\rightarrow\infty$  and  $(0,0)$  is  an  unstable  node.  Also,  if
$\lambda_1$ and  $\lambda_2$ are real but $\lambda_1$  is positive and
$\lambda_2$ is negative, then  $(0,0)$ is a saddle point; trajectories
approach the origin  in the ${\bf{V_2}}$ direction and  move away from
the origin in the direction ${\bf{V_1}}$.

Solutions  $\lambda_1$ and $\lambda_2$  of (\ref{eigenvalues})  may be
complex. However,  when the matrix  $M$ is real, then  $\lambda_1$ and
$\lambda_2$  must be  a  complex conjugate  pair.  If $\lambda_1$  and
$\lambda_2$  are   pure  imaginary,  then   the  vector  ${\bf{Y}}(t)$
represents a  closed orbit  for any $c_1$  and $c_2$ and  the critical
point  at $(0,0)$  is a  centre.  If  $\lambda_1$ and  $\lambda_2$ are
complex with non-zero real part, then the critical point at $(0,0)$ is
a spiral. When $Re$ $\lambda_{1,2}<0$, then ${\bf{Y}}(t)\rightarrow0$,
$(0,0)$   is   a   stable   spiral  point;   conversely,   when   $Re$
$\lambda_{1,2}>0$, $(0,0)$ is an unstable spiral point.

\subsubsection{Two-dimensional non-linear systems}

The  analysis  of  a  non-linear  system is  a  much  more  complicate
case. Nevertheless non-linear system can be analysed near its critical
points, and  the result of that  study can be indeed  very accurate if
the system is almost linear. In our particular case, the condition for
almost-linearity  can  be  checked  comparing  our  results  with  the
numerical integrations performed in  \cite{bsm1}. We will see that our
results  are indeed  very good  approximation to  the  solutions found
numerically, with  exception to one  case where one of  the eigenvalue
solution  of  equation  (\ref{eigenvalues})  is  zero.  The  stability
analysis of this case will be described bellow in \ref{zeroeigen}.

The  local  analysis of  a  non-linear  system,  to which  the  linear
approximation  works  well,  consist  into  linearise  the
equations and then procedure as usually done for a linear system.  We first
identify the critical points. Then we perform, a local analysis of the
system very near  them.  Using matrix methods, we  identify the nature
of the critical points of  the linear system. Finally, we assemble the
results  of our  local analysis  and synthesise  a  qualitative global
picture of the solution to the non-linear system.

\subsubsection{Critical point with one zero eigenvalue}
\label{zeroeigen}

In  this  section  we  will  only  present  the  results  obtained  in
\cite{sonoda} and we refer the reader to that reference for a detailed
explanation.
 
If  one  of the  eigenvalue  solutions,  $\lambda_{1,2}$, of  equation
(\ref{eigenvalues})   has,  for   instance,   $Re(\lambda_{1})=0$  and
$Re(\lambda_{2})<0$ then the stability cannot be decided by the linear
approximation.

In general, a non-linear autonomous system can be written in the form
\begin{equation}
{\bf \dot Y}= M {\bf Y} + N({\bf Y})
\end{equation}
where  $M$  is a  constant  matrix $2\times  2$  and  $N({\bf Y})$  is
non-linear  in  ${\bf Y}$.  The  linearisation  of  this system  gives
(\ref{matrix}).  Consider  now that, the two eigenvalues  of $M$, have
$Re(\lambda_{1})=0$  and  $Re(\lambda_{2})<0$,  corresponding  to  the
eigenvectors ${\bf{V_1}}$  and ${\bf{V_2}}$.  In order to  perform the
stability analysis of the  critical point $(0,0)$, we should procedure
as follows (For other alternatives to this method see \cite{wiggins}).
\begin{itemize}
\item First, we  apply a linear transformation to  ${\bf Y}$ using the
matrix formed by the  eigenvectors ${\bf{V_1}}$ and ${\bf{V_2}}$. This
will split  the system into  critical and non-critical  variables; the
critical    variable    is    the   eigenvector    corresponding    to
$Re(\lambda_{1})=0$,  in  our   case  ${\bf{V_1}}$.  Hence,  ${\bf{Y}}
\rightarrow {\bf{X}}=(x_1, x_2)$. In  the new coordinates, we now have
that the following system
\begin{eqnarray}
\dot  x_1 &=& n_1  (x_1,x_2)\nonumber \\  \dot x_2  &=& p_{21}x_1+\bar
p_2x_1+ n_2 (x_1,x_2)
\end{eqnarray}
where $n_1$ and $n_2$ are at least quadratic in both $x_1$ and $x_2$.
\item  Secondly, if  there are  linear terms  in $x_1$,  in  the above
system, then they must be eliminated. That can be achieved solving the
equation  $\dot x_2=0$, which  will give  us $x_2$  has a  function of
$x_1$  (the critical variable).  Say, for  instance, that  solution give
$x^{*}_{2}(x_1)$.
\item    At   last,   we    perform   the    following   transformation
  $(x_1,x_2)\rightarrow(z_1,z_2)$ defined as
\begin{equation}
(z_1,z_2)=\left(x_1,x_2-x_{2}^{*}\left(x_1\right)\right)
\end{equation}
\end{itemize}

\bigskip

Having performed  this steps we end  up with an  autonomous system for
the coordinates $(z_1,z_2)$
\begin{eqnarray}
\dot  z_1 &=&  h_1(z_1,z_2)\nonumber \\  \dot z_2  &=&  p_{21}x_1+ h_2
(z_1,z_2)
\label{last}
\end{eqnarray}
where the  functions $h_1$ and $h_2$  are at least  quadratic in their
arguments.  the stability  of the  critical point  ${\bf{Y}}=(0,0)$ is
determined by the stability of ${\bf{Z}}\equiv(z_1,z_2)=(0,0)$.

Since  the  non-null  eigenvalue  of  $M$ has  a  negative  real  part
($Re(\lambda_{2})<0$,   by  assumption),   the  coordinate   $z_2$  is
asymptotically     stable    about     the    origin     $(0,0)$    as
$t\rightarrow\infty$.   The  asymptotic   stability  of  the  critical
variable   is  therefore   determined   as  $t\rightarrow\infty$   and
$z_2\rightarrow0$  by the  leading term  of $h_1(z_1,0)$.   It  can be
shown \cite{sonoda} that when the  first power of $h_1(z_1,0)$ is even
(recall  that   $h_1$  is  at   least  a  quadratic   function),  then
$(z_1,z_2)=(0,0)$ is unstable.

\section{Phase-Plane Analysis}

We   are  ready   now  to   analyse  the   phase  space   of  equation
(\ref{psi}). We  first transform it  into an autonomous system  of two
first  order differential  equations.  Then we  identify the  critical
points and we perform, a local  analysis of the system very near them.
The  exact system  will be  well approximated  by a  linear autonomous
system near the critical points. This will be seen when we compare our
results  with the  numerical results  performed in  \cite{bsm1}. Using
matrix methods  we identify the nature  of the critical  points of the
linear system. Finally, we assemble  the results of our local analysis
and synthesise  a qualitative  global picture of  the solution  to the
non-linear system.


\subsection{A transformation of variables}

In  this section  we  will look  at  the $\psi  $  equation of  motion
(\ref{psi} ) when the expansion scale factor takes the power-law form
(\ref{a}). The evolution equation for the field then becomes:
\begin{equation}
\frac d{dt}\left( \dot{\psi}t^{3n}\right) =N\exp [-2\psi ] \label{n}
\end{equation}
with $N>0.$ We introduce the following variables :
\begin{eqnarray}
x=ln(t)\qquad  y=\psi -Ax-B\qquad  A=1-\frac{3n}2\qquad  B=\frac 12\ln
(N)
\label{coord}
\end{eqnarray}
and rewrite (\ref{psi}) as:
\begin{equation}
y^{\prime \prime }+(3n-1)(y^{\prime }+1-\frac{3n}2)=e^{-2y}, \label{y}
\end{equation}
where  $^{\prime  }$  $\equiv  d/dx$. This  second-order  differential
equation  can be  transformed into  an autonomous  system  by defining
$v=\frac{dy}{dx}$ and $u=y$:
\begin{eqnarray}
\frac{dv}{dx}    &=&e^{-2u}+(1-3n)(v+1-\frac{3n}2)    \label{nl}    \\
\frac{du}{dx} &=&v \nonumber
\end{eqnarray}

\bigskip

With the  aim to obtain asymptotic  exact solution, we  have to search
for the critical points of equation (\ref{nl}). The critical points of
a system are defined as the values of $u$ and $v$ that give $v'=0$ and
$u'=0$,  where  the prime  represents  a  derivative  with respect  to
$x$. The  importance of the  critical points is  the fact that  we can
linearise the system around then, and if they are stable, the solution
of the system near them, can be used as an asymptotically stable
solution.

Looking at equation (\ref{nl}), it is straight forward to see that the
system has a finite critical point at
\begin{equation}
(u_c,v_c)=(-\frac 12\ln {[(1-3n)(\frac{3n}2-1)]},0) \label{crit}
\end{equation}
when $n\in (\frac  13,\frac 23)$ and it has  a infinite critical point
at $ (u_c,v_c)=(+\infty  ,0)$ when $n=\frac 13$ or  $n=\frac 23$. The
finite critical points correspond to  the family of exact solutions of
the form  $ \psi =B+A\ln (t)$  found in \cite{bsm1}.  In the original
variables these solutions are
\[
\psi =\frac 12\ln [\frac{2N}{(2-3n)(3n-1)}]+\left( \frac{2-3n}2\right)
\ln (t).
\]

\bigskip

In order  to analyse the system  fully, we will  study finite critical
points and  the critical points  at infinity separately. We  will also
distinguish  several  domains  of  behaviour for  $n$:  $n=\frac  13$,
$n=\frac 23$,  $n\in (\frac 13,\frac  23)$, $n<\frac 13$  and $n>\frac
23$. In some cases we  will supplement our investigations by numerical
study of the system (\ref{nl}).

\subsection{The finite critical points}

\subsubsection{The cases $n\in (\frac 13,\frac 23)$}

In the domain  where $n\in (\frac 13,\frac 23)$  the system (\ref{nl})
does not appear  to be exactly integrable. So in  this section we will
study the behaviour of the system near the critical points and at late
times.

There   is   a   finite   critical  point   $(u_c,v_c)=(-\frac   12\ln
{[(1-3n)(\frac{3n}   2-1)]},0)$    for    $n\in   (\frac    13,\frac
23)$. Linearising the system \ref{nl} about it we obtain:
\begin{eqnarray}
\frac{dV}{dx} &=&(3n-1)(3n-2)U+(1-3n)V \label{almost} \\ \frac{dU}{dx}
&=&V \nonumber
\end{eqnarray}
where $V=v-v_c$ and $U=u-u_c$.  Since the characteristic matrix is non
singular  the  critical  point  is simple.  Hence  the  non-linearised
version  of   this  system  has   the  same  phase  portrait   at  the
neighbourhood of the critical point.

The  eigenvalues  $\xi _{1,2}$  and  corresponding eigenvectors  $\chi
^{\xi _{1,2}}$ of the system (\ref{almost}) are:
\begin{eqnarray}
\xi  _{1,2} &=&{\frac{1-3n\pm  \sqrt{9-42n+45{n^2}}}2}  \\ \chi  ^{\xi
_{1,2}} &=&({\frac{1-3n\pm \sqrt{9-42n+45{n^2}}}2},1) \nonumber
\end{eqnarray}
These eigenvalues  will be complex  numbers for $n\in  (\frac 13,\frac
35)$ and  they will  be pure real  numbers when $n\in  (\frac 35,\frac
23)$. Notice that  for both cases the real part  of the eigenvalues is
always  negative, so  the critical  point is  a stable  attractor. The
general  solution  of  the  linearised system  (\ref{almost})  can  be
expressed as:
\[
(U(x),V(x))=A{e^{\xi _1x}}\chi _1+B{e^{\xi _2x}}\chi _2
\]
where  $A$, $B$  are arbitrary  constants  and $x=\ln  (t)$. From  the
transformations  (\ref{psi}) we can  obtain explicitly  the expression
for $  \psi (t)$  near the critical  point. There are  three possible
behaviours of the solutions near the critical point:

\subsubsection{Pseudo-oscillatory $\protect\psi $ behaviour}

\begin{figure}[htbp!]
\centering       
\epsfig{file=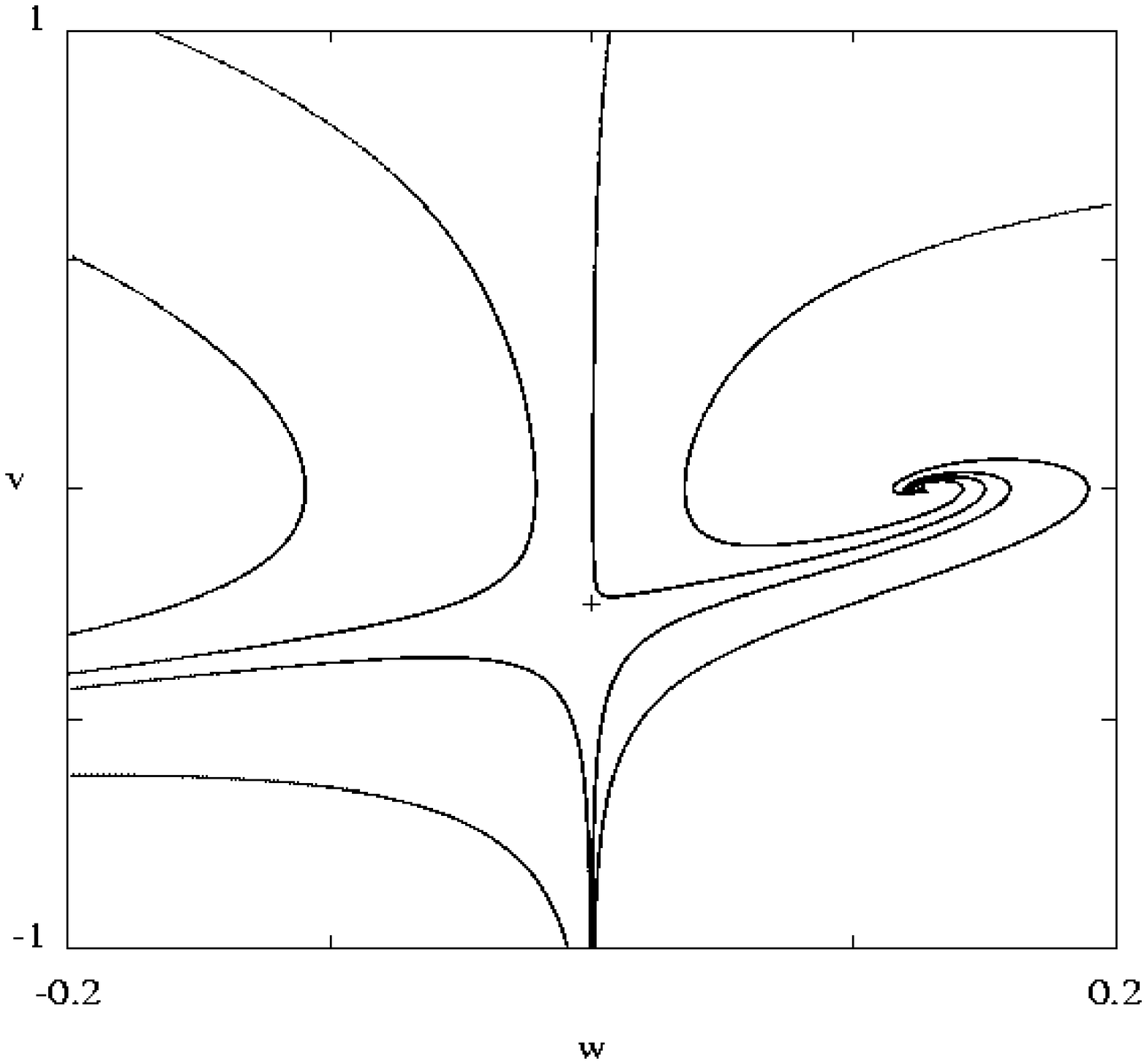,height=6.5cm,width=7.4cm}
\epsfig{file=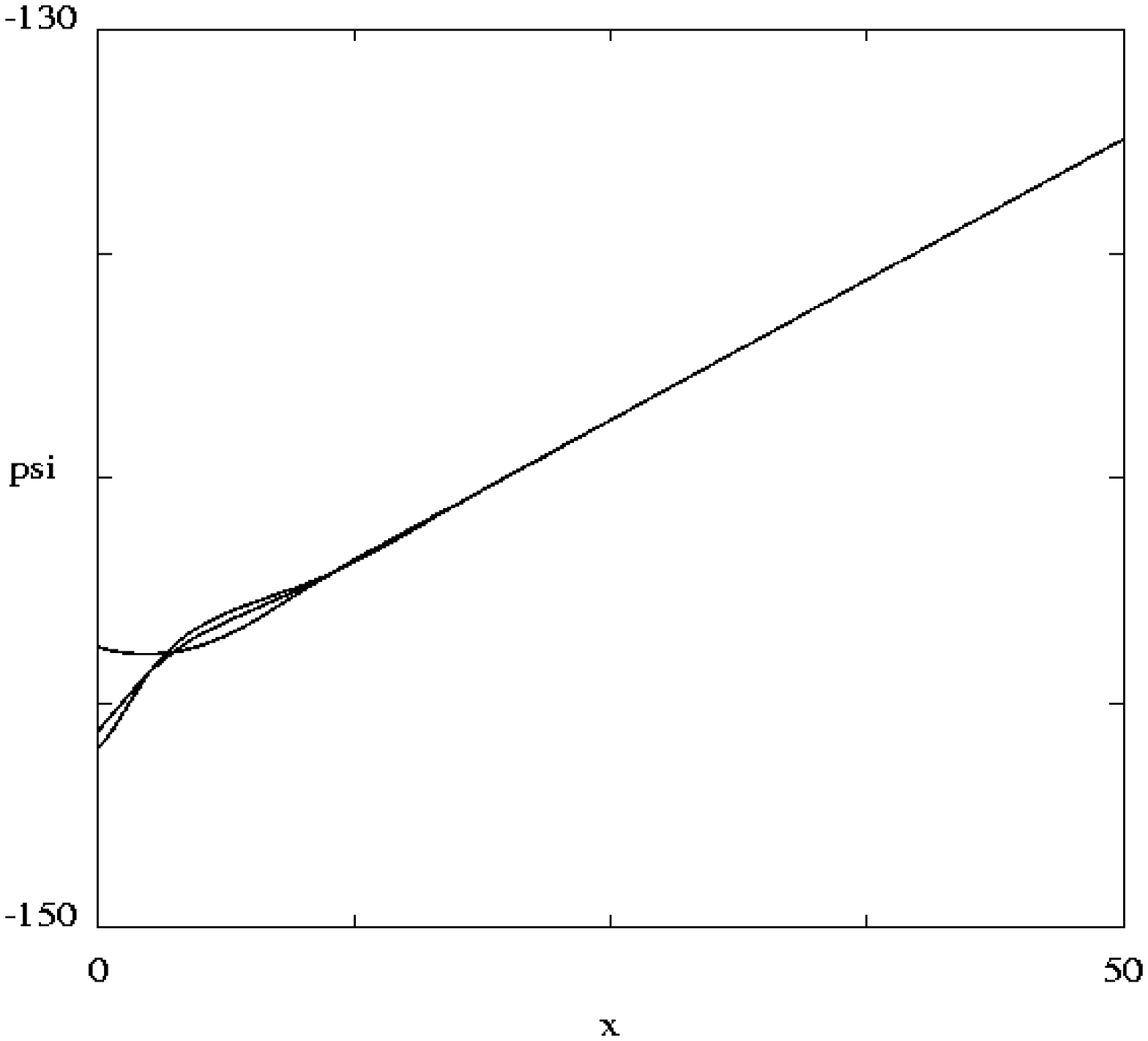,height=6.5cm,width=7.4cm}
\caption{{\protect \textit{Numerical  plots of the phase  space in the
$   (w,v)$  coordinates,  and   the  $\protect\psi$   evolution  with
$x=log(t)$ for $n= \frac{1}{2}$ .The '$+$' sign is a saddle point and
the triangle is a stable node. }}}
\label{n12}
\end{figure}

When $n\in (\frac{1}{3},\frac{3}{5})$ the linearised evolution for the
$\psi  $  field  exhibits  damped oscillations  about  its  asymptotic
solution as $ t\rightarrow \infty $:
\begin{eqnarray}
\psi   (t)  &=&\frac   12\ln   [\frac{2N}{\left(  1-3n\right)   \left(
3n-2\right) }  ]+\frac 12\left(  2-3n\right) \ln (t)  \label{comp} \\
&&+{t^\alpha   }\left(  B\cos   [\beta  \ln   (t)]+A\sin   [\beta  \ln
(t)]\right) \nonumber
\end{eqnarray}
where
\begin{eqnarray}
\alpha       &=&{\frac{1-3n}{2}}      \label{alph}       \\      \beta
&=&{\frac{\sqrt{9+42n-45{n^{2}}}}{2}} \label{beta}
\end{eqnarray}
We note that this family of solutions and asymptotes includes the case
of the radiation-dominated universe  ($n=1/2$). The phase portrait and
the $\psi $ vs. $\ln t$ for this case are shown in figure \ref{n12}.
But in  the case of  a universe containing  the balance of  matter and
radiation, displayed  by our  own, it  need not be  the case  that the
asymptotic behaviour, displayed by the exact solution for the critical
point, is reached before the radiation-dominated expansion is replaced
by matter-domination,  see references \cite{bsbm}  and \cite{bsm1} for
further discussion of this point.  However as we discussed above these
oscillations are an artifact of the linearisation process and the part
of the solution ( \ref{comp}) controlled by the constants $A$ and $B$
is  only  valid  for  small  times, hence  we  called  this  behaviour
pseudo-oscillatory.

\subsubsection{Non-oscillatory behaviour}

When $n\in (\frac  35,\frac 23)$, the $\psi $  field will approach its
asymptotic  behaviour in  a non-oscillatory  fashion  as $t\rightarrow
\infty $:
\begin{eqnarray}
\psi   (t)  &=&\frac   12\ln   [\frac{2N}{\left(  1-3n\right)   \left(
3n-2\right) }]
\label{real} \\
&&+\frac 12\left( 2-3n\right)  \ln (t)+t^{\delta -\gamma }\left( A+B{%
t^{2\gamma }}\right) \nonumber
\end{eqnarray}
where
\begin{eqnarray}
\delta   &=&{\frac{1-3n}2}    \label{del}   \\   \gamma    &=&   \frac
12\sqrt{9-42n+45{n^2}} \label{gam}
\end{eqnarray}

Note    that   $\delta    +\gamma    <0$   in    the   domain    $n\in
(\frac{1}{3},\frac{2}{3})$ ,  so at late  times ($t\rightarrow \infty
$),  both the  pseudo-oscillatory and  non-oscillatory  solutions case
will approach the asymptotic solution defined by the appropriate value
of $n$.

\subsubsection{Intermediate behaviour}

\begin{figure}[htbp!]
\centering   \epsfig{file=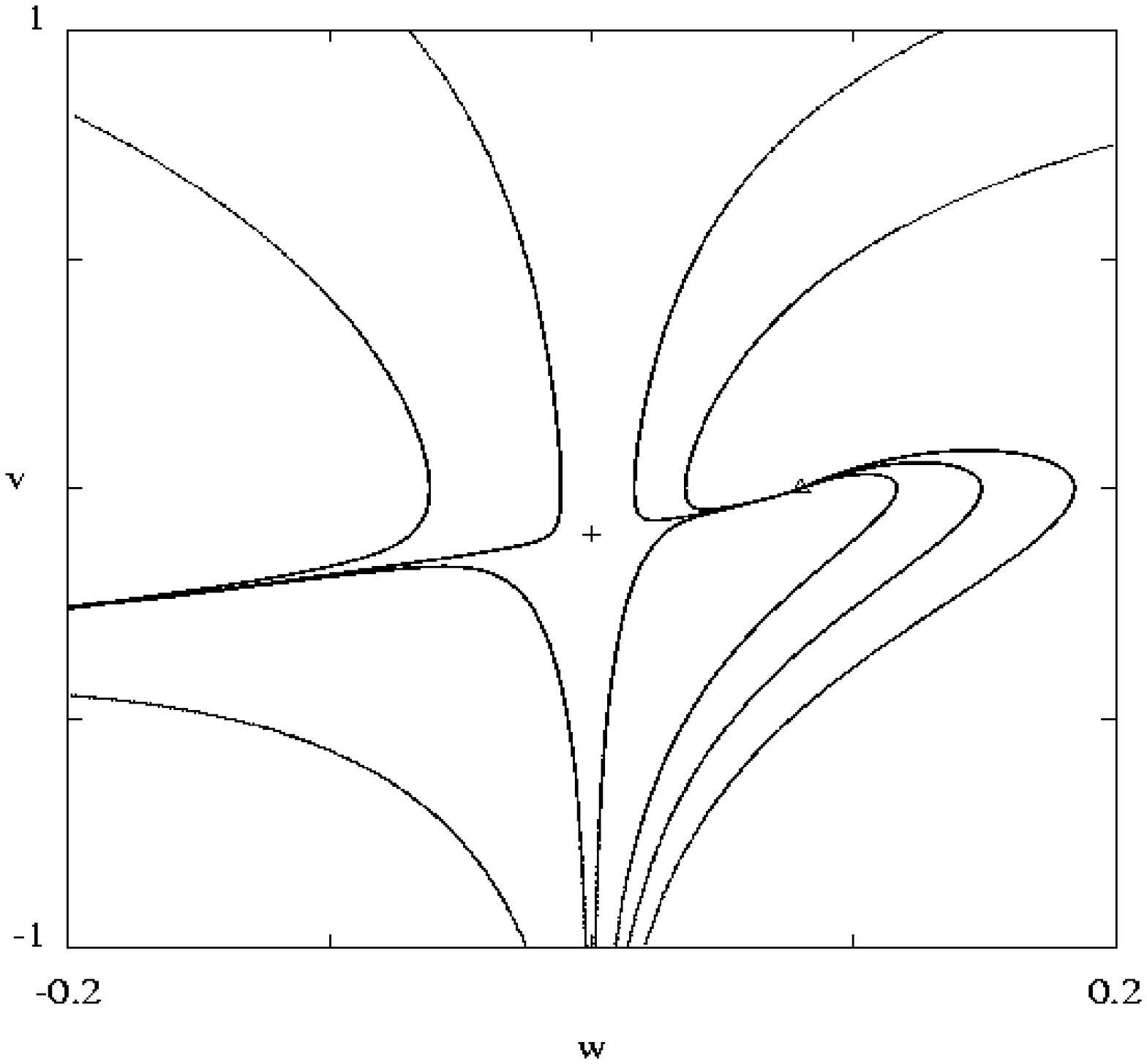,height=6.5cm,width=7.4cm}  %
\epsfig{file=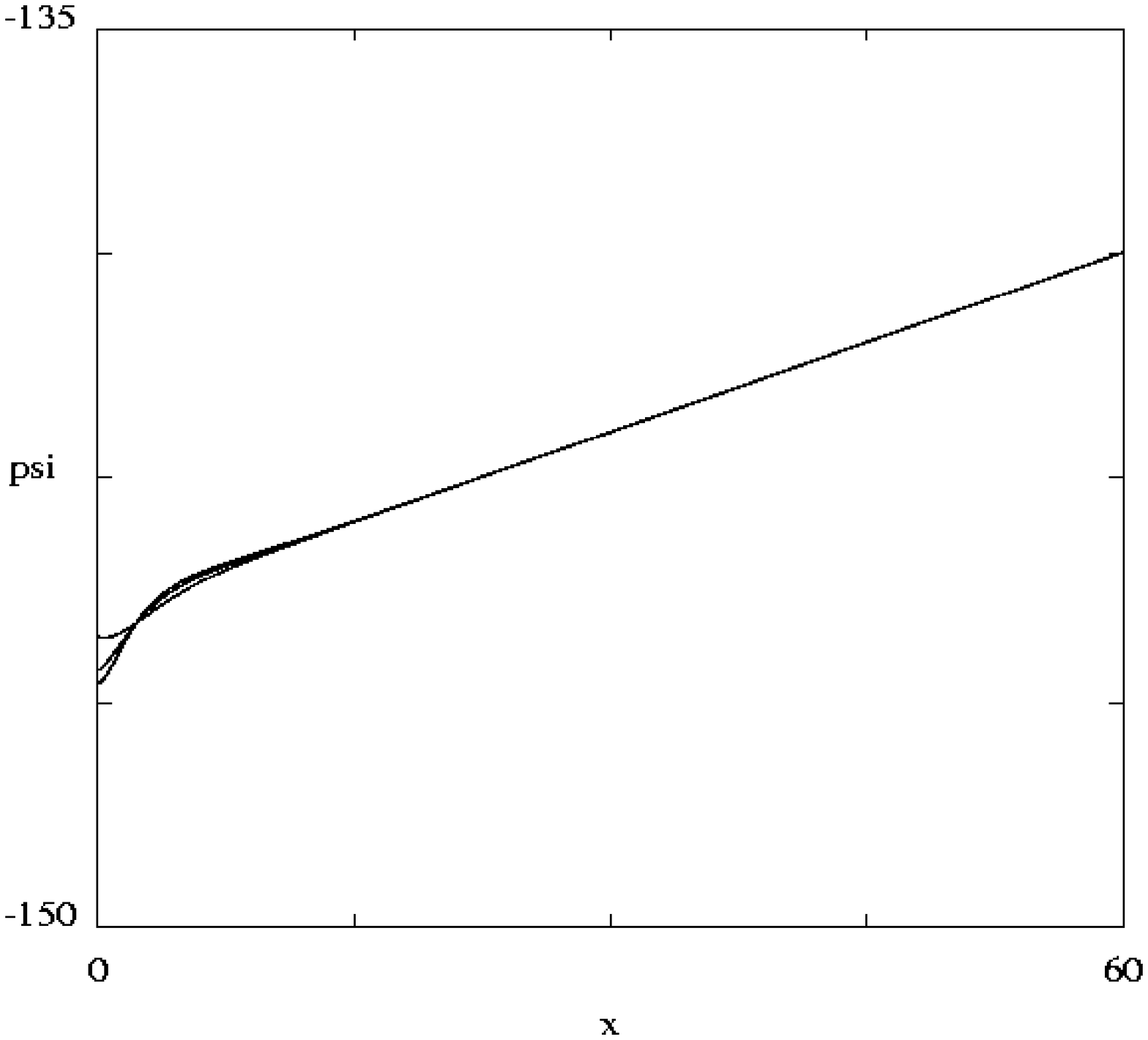,height=6.5cm,width=7.4cm}
\caption{{\protect \textit{Numerical  plots of the phase  space in the
$   (w,v)$  coordinates,  and   the  $\protect\psi$   evolution  with
$x=log(t)$ for $n= \frac{3}{5}$. The '$+$' sign is a saddle point and
the triangle is a stable node. }}}
\label{n35}
\end{figure}

A  transition  between   the  pseudo-oscillatory  and  non-oscillatory
regimes happens when $\beta =0$ and $n=\frac{3}{5}$; the $\psi $ field
then has the following solution in the vicinity of the critical point:
\begin{equation}
\psi (t)=\frac{1}{5}\ln (N)+\frac{1}{10}\ln (t)+At^{-{\frac{2}{5}}}.
\label{n3/5}
\end{equation}
The phase  plane structure and  evolution of $\psi  $ vs. $\ln  t$ for
this case are shown in figure \ref{n35}.


\subsubsection{Overview}

The  late-time solution  of the  $\psi $  field solution  is  given by
asymptotic behaviour of  equations (\ref{comp}) or (\ref{real}), which
is:
\[
\psi  \rightarrow  \frac  12\ln [\frac{2N}{\left(  1-3n\right)  \left(
3n-2\right) }]+\frac 12\left( 2-3n\right) \ln (t)
\]
This shows that the solutions (\ref{comp}) and (\ref{real}) generalise
the  ones found  in \cite{bsm1},  since they  can be  obtained setting
$A=B=0$   in  (   \ref{comp}).   In  particular,   the   case  of   a
radiation-dominated   Universe  ($   n=\frac  12$),   we   have  from
(\ref{comp}):
\begin{eqnarray}
\psi  (t)  &=&\frac{1}{2}\left(   \log  (8)+\log  (N)+\log  (t)\right)
\label{rad}        \\        &&+{t^{{-\frac{1}{4}}}}\left(       B\cos
[{\frac{5\sqrt{3}\log    (t)}{4}}]+A\sin    [{   \frac{5\sqrt{3}\log
(t)}{4}}]\right) \nonumber
\end{eqnarray}

A full  mathematical summary of the  change in structure  of the phase
space with changing $n$ for the system (\ref{psi}) is given in section
\ref{overview}. There  we include cosmologically  unphysical values of
$n$  and show  how the  critical point  structure bifurcates  with the
change in value of $n$.

\subsection{The critical points at infinity}

In order to describe the  qualitative evolution of the system, we must
determine  the behaviour of  the system  (\ref{nl}) near  the critical
point $ (u_c,v_c)=(+\infty ,0)$. In order to bring the critical point
to a finite value, we define:
\[
u=-\frac 12\ln (w)
\]
Using the new coordinate $w$ we can re-write the system (\ref{nl}) as:
\begin{eqnarray}
\frac{dw}{dx}   &=&-2wv  \label{nllog}   \\   \frac{dv}{dx}  &=&\left(
1-3n\right) \left( 1-{\frac{3n}2}+v\right) +w \nonumber
\end{eqnarray}

This system  has critical  points on the  $(w,v)$ plane when  $(0,-1 +
{\frac{3n     }{2}})$      and     $(\left(1-     3n\right)     \left(
{\frac{3n}2}-1\right),0).$ Note that the second critical point is just
the same as the one we  have analysed in the previous section. In this
subsection   we   will   then   only  analyse   the   critical   point
$(w_c,v_c)=(0,-1 +  {\frac{3n}{2}})$ since it corresponds  to the case
where $u \rightarrow +\infty$.

Proceeding   as  before,  and   linearising  (\ref{nllog})   about  $
(w_{c},v_{c})=(0,-1+{\frac{3n}{2}})$ we obtain:
\begin{eqnarray}
\frac{dW}{dx}  &=&(2-3n)W \label{llog}  \\  \frac{dV}{dx} &=&W+(1-3n)V
\nonumber
\end{eqnarray}
where $V=v-v_{c}$ and $W=w-w_{c}$. Again, the characteristic matrix of
the system is  non singular, and the critical  point is simple. Hence,
system   (\ref  {llog})  will   have  the   same  phase   portrait  as
(\ref{nllog}) in the neighbourhood of the critical point.

The  eigenvalues  $\xi _{1,2}$  and  corresponding eigenvectors  $\chi
^{\xi _{1,2}}$ of the system (\ref{llog}) are:
\begin{eqnarray}
\xi  _1  &=&1-3n,\hspace{1.0in}\chi  ^{\xi _1}=(0,1)  \label{inf1}  \\
\qquad \xi _2 &=&2-3n,\hspace{1.0in}\chi ^{\xi _2}=(1,1) \label{inf2}
\end{eqnarray}
These eigenvalues are always real. The critical point is an attractive
node for $n>\frac 23$, a saddle point when $n\in (\frac 13,\frac 23)$,
and  it will  be  an unstable  point  when $n<\frac  13$. The  general
solution  of the  system  (\ref{nllog}) in  the  neighbourhood of  the
critical point $ (w_c,v_c)=(0,-1+{\frac{3n}2})$, can be expressed as:
\[
(w(x)-w_c,v(x)-v_c)=A{e^{\xi _1x}}\chi _1+B{e^{\xi _2x}}\chi _2
\]
where  $A$, $B$ are  arbitrary constants  and $x=\ln  (t)$. Therefore,
near the critical point $(u_c,v_c)=(+\infty ,0)$:
\begin{equation}
\psi (t)=\frac 12\ln (\frac NB) \label{psisolution}
\end{equation}
so the scalar field is constant.

\bigskip

Note that in the  domain $n\in (\frac{1}{3},\frac{2}{3})$ this is just
a transitory solution  since the critical point is  a saddle point. In
the domain $n<\frac{1}{3}$ it  is an unstable critical point, possibly
relevant as  an early-time solution  to braneworld cosmologies  in the
high-density regime where the Hubble expansion rate of the universe is
linearly  proportional to  the density,  so $n=1/4$  for  a braneworld
containing  radiation,  $n=1/12$  for  a massless  scalar  field,  and
$n=1/6$  for  dust.   However,  in  the  $t\rightarrow  0$  limit  the
assumption  that the  $\dot{\psi} ^{2}$  and $\zeta  \exp  [-2\psi ]$
terms can  be neglected in  the Friedmann equation  (\ref{fried}) will
break down.

\subsubsection{The cases of $n>2/3$ and de Sitter expansion}

An interesting case exists when $n>\frac 23$, since the critical point
is  a stable attractor  and so  this means  that the  constant-$\psi $
behaviour  (\ref   {psisolution})  is  the   late-time  attractor,  in
agreement   with  the  conclusions   of  references   \cite{bsbm}  and
\cite{bsm1}. This is an important feature of a universe which exhibits
accelerated expansion  in its late  stages ($n>1$). It means  that the
present value of  $\alpha $ is the asymptotic one.  It also means that
variations  of $\alpha  $  are turned  off  by the  domination of  the
expansion dynamics by negative  curvature or by any quintessence field
\cite {bsbm},  \cite{bsm1}. This property may  provide important clues
to explaining why our universe possesses small but finite curvature or
quintessence  energy today:  if it  did  not then  the fine  structure
constant would continue to increase  until it was impossible for atoms
and molecules to exist \cite{bsm2}

\begin{figure}[htbp!]
\centering   \epsfig{file=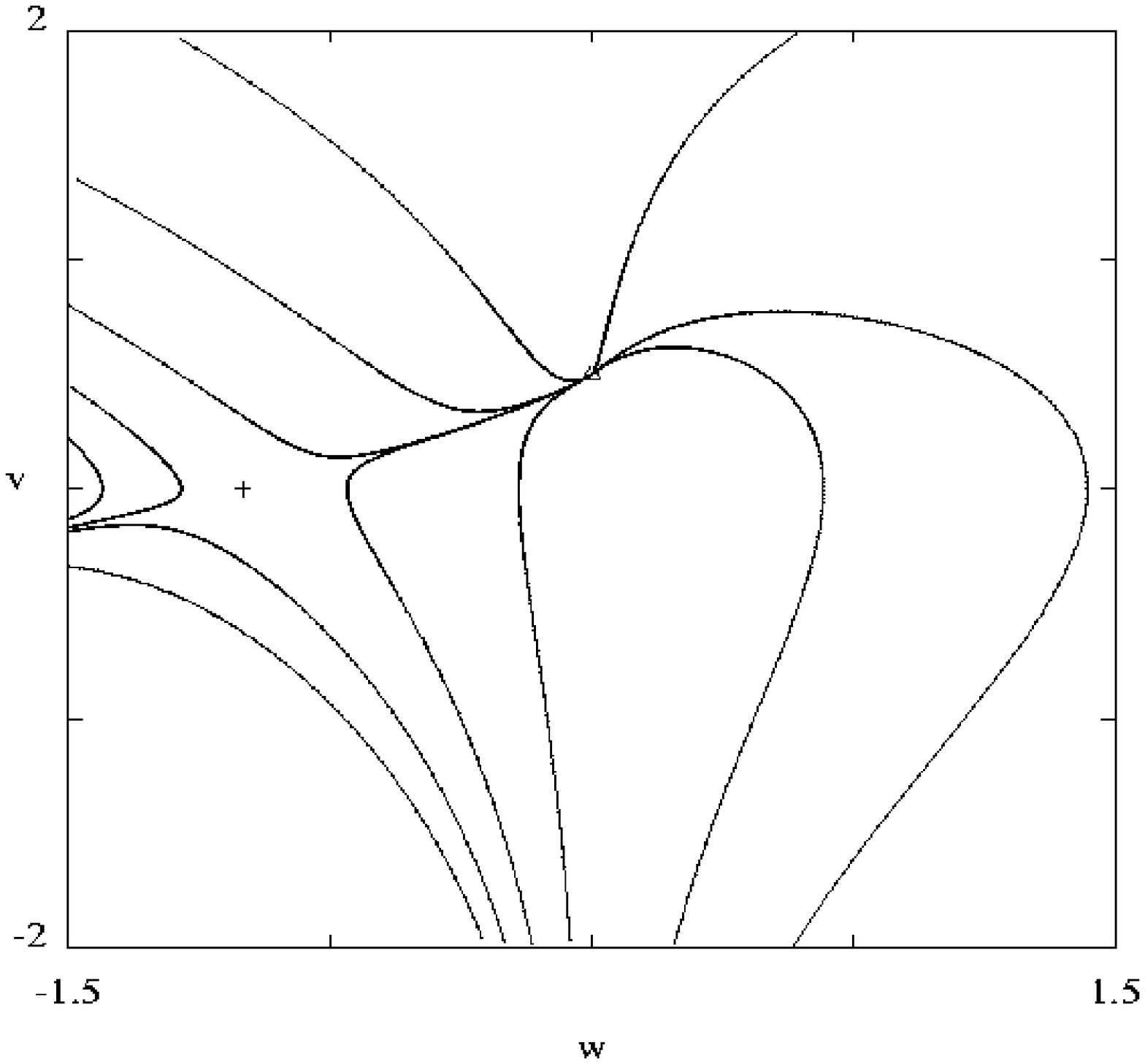,height=6.5cm,width=7.4cm}   %
\epsfig{file=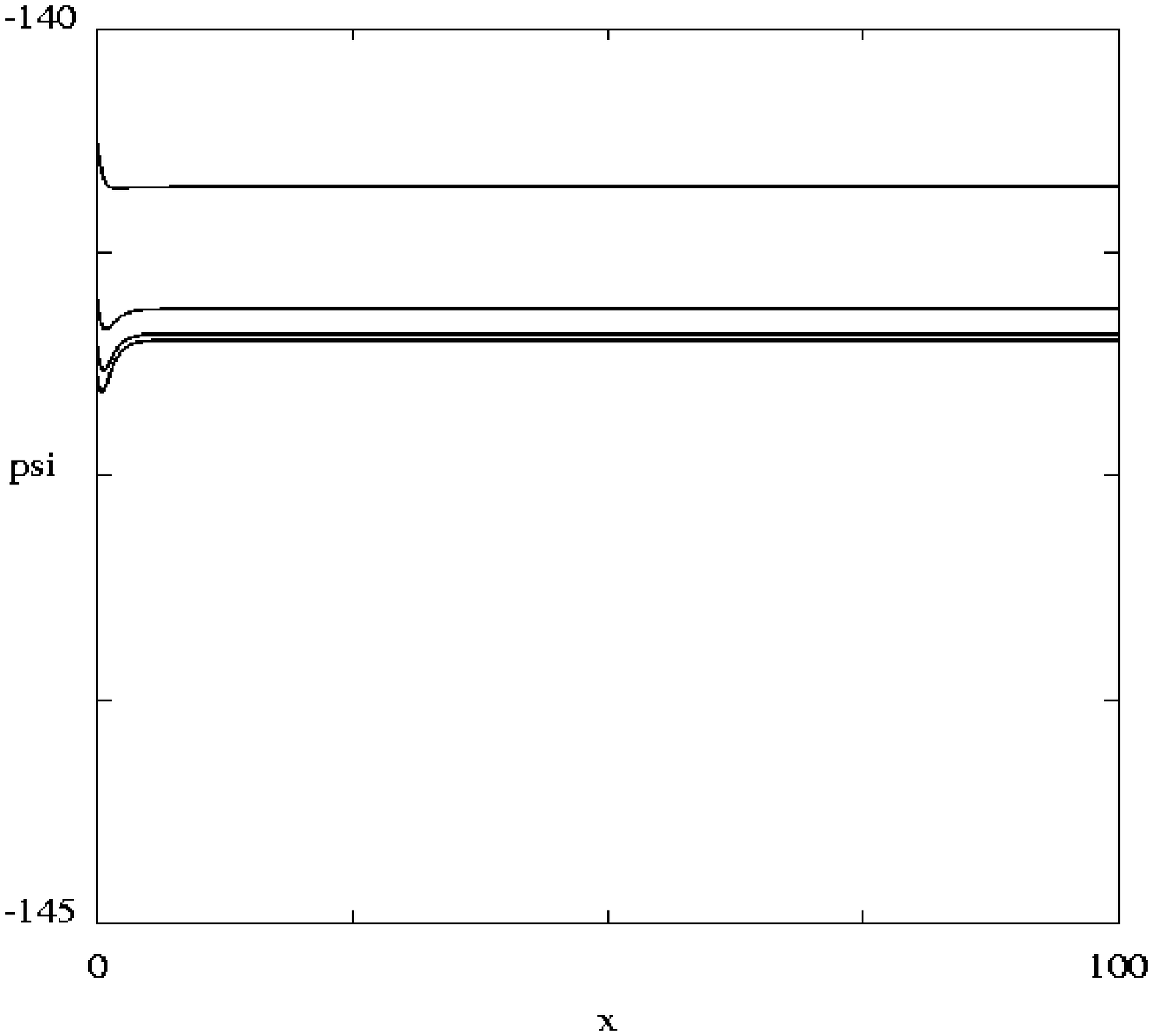,height=6.5cm,width=7.4cm}
\caption{{\protect \textit{Numerical  plots of the phase  space in the
$   (w,v)$  coordinates,  and   the  $\protect\psi$   evolution  with
$x=log(t)$  for $n=1  $. The  '$+$'  sign is  a saddle  point and  the
triangle is a stable node. }}}
\label{n1}
\end{figure}

In the  $n>2/3$ case  we can find  a detailed asymptotic  solution for
equation ( \ref{n}) which has the form
\[
\psi                              =C+B(t+t_{0}^{{}})^{1-3n}+\frac{N\exp
[-2C]}{2-3n}(t+t_{0}^{{}})^{2-3n}
\]
with $C,t_{0}$  constants. We see  immediately that for  this solution
$\psi $ approaches a constant as $t\rightarrow \infty $. Of particular
interest is the case of a curvature-dominated open universe, which has
$n=1$. The phase plane structure and the evolution of $\psi $ vs. $\ln
t$ for this case are shown in figure \ref{n1}.


We  see  that this  approach  to  constant  behaviour occurs  for  all
universes that  accelerate ($n>1$) and so  we would expect  to find it
also in the case of a de Sitter background universe with
\[
a(t)=\exp [\lambda t]
\]
where $\lambda >0$ is constant.

Substituting  this  in  equation   (\ref{psi})  we  find  a  late-time
asymptotic solution
\[
\psi   =C+B\exp   [-3\lambda   t]-\frac{N\left(  3\lambda   t+1\right)
}{9\lambda ^2} \exp [-2C-3\lambda t]\rightarrow C
\]
as $t\rightarrow \infty .$

\bigskip

Notice that  we were  not able  to fully describe  the nature  of this
critical point  in the cases where  $n=\frac{1}{3}$ or $n=\frac{2}{3}$
since one  of the eigenvalues of the  systems is zero. In  order to do
so, we will study these two cases individually, in particular the case
$n=\frac{2}{3}$  is  important since  it  describes  the scale  factor
evolution on a dust dominated universe.

\subsubsection{The $n=\frac 13$, $a(t)=t^{\frac 13}$ case: an exact solution}

The $n=\frac 13$ case can be  exactly integrated. It is of interest as
an exact solution in its own right but it corresponds to the case of a
universe whose  expansion dynamics are  dominated by the effects  of a
fluid  with a  $ p=\rho  $ equation  of state,  or a  massless scalar
field. It  also describes  the behaviour  of the $\psi  $ field  in an
anisotropic universe of the simple Kasner type.

We see when $n=1/3$ the system (\ref{nl}) has the form:
\begin{eqnarray}  \label{0333system}
\frac{dv}{dx}& =& e^{-2u} \\ \frac{du}{dx}& =& v \nonumber
\end{eqnarray}

\begin{figure}[htbp!]
\centering  \epsfig{file=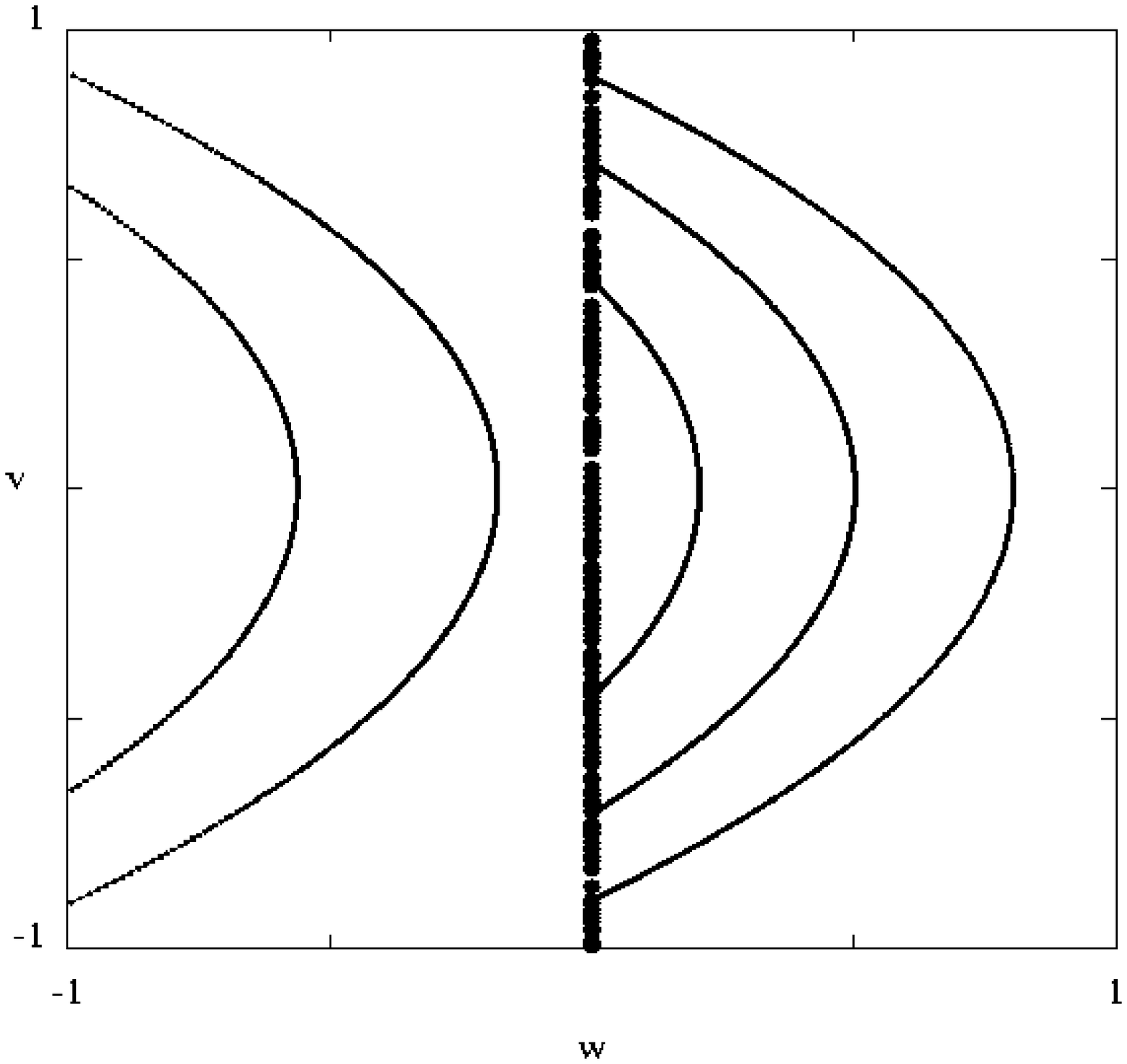,height=6.5cm,width=7.4cm} %
\epsfig{file=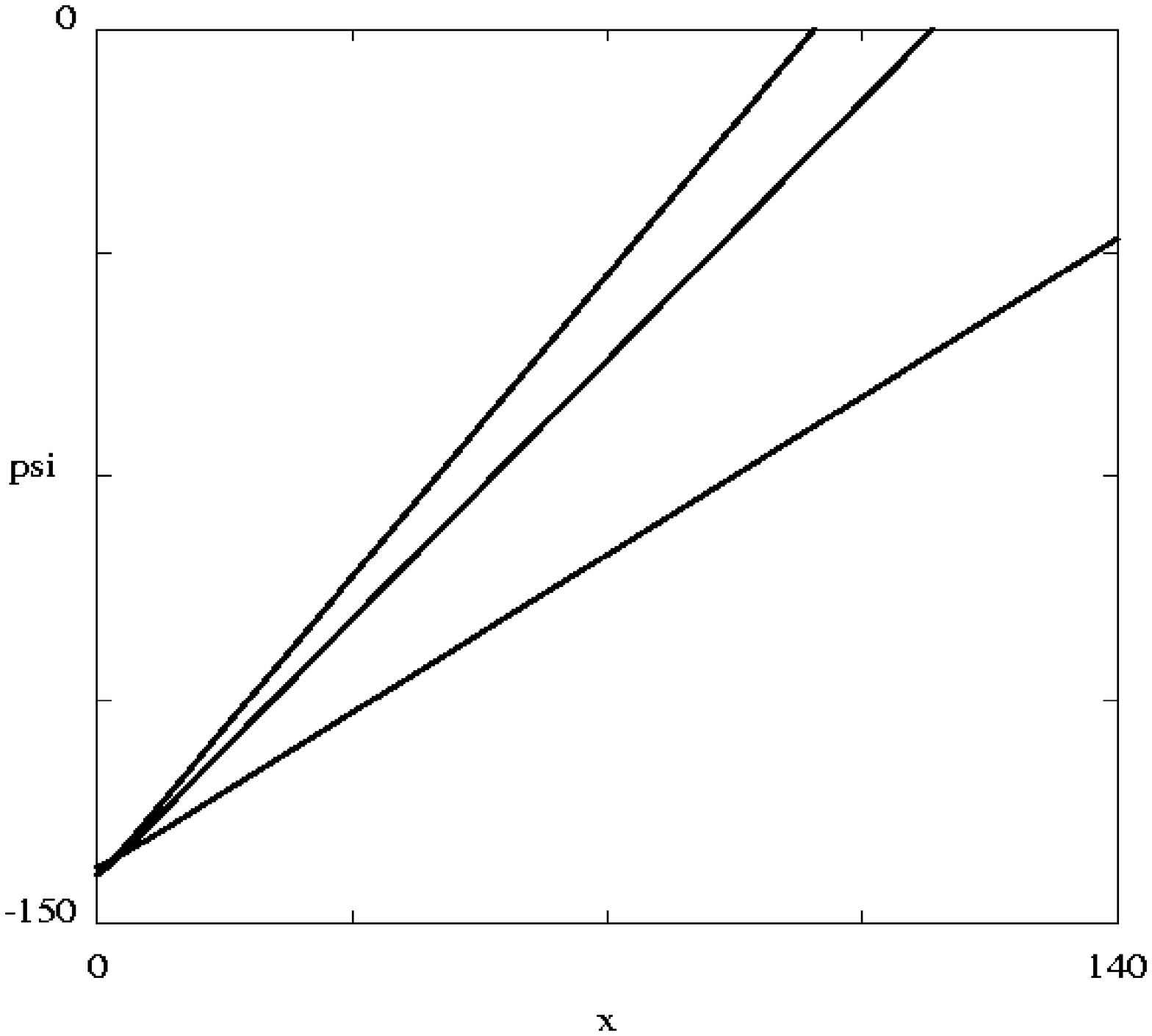,height=6.5cm,width=7.4cm}
\caption{{\protect \textit{Numerical  plots of the phase  space in the
$  (w,v)$   coordinates  and  time   evolution  of  $\protect\psi(x)$
$n=\frac{1}{3}$.  The central line  which passes through the origin is
an attractor for $w>0$, and an unstable line for $w<0$. }}}
\label{n13}
\end{figure}

This integrates to give
\[
v^2+e^{-2u}=E^2
\]
where  $E$  is  a  constant.  Hence, we  have  two  possible  solution
branches:  $  v(u)=\pm \sqrt{E-e^{-2u}}$.  Both,  positive ($+$)  and
negative ($-$) solutions for $v$ lead to the same result. Choosing the
positive   branch  of   the  $v(u)$   solution  and   using  equations
(\ref{0333system}) we obtain:
\[
u(x)=-\ln (E)+\ln [\cosh \left( E\left( C_1-x\right) \right) ]
\]
where $C_1$ is  an integration constant. From (\ref{coord})  we have a
solution for $\psi (t)$:
\[
\psi (t)={\frac{\ln (\frac N4)\  }2}-\ln (E)+\frac 12\ln (t)+\ln [\exp
[EC_1]t^{-E}+t^E\exp [-EC_1]]
\]
and the asymptotic limit when $t\rightarrow \infty $ gives
\[
\psi (t)\rightarrow (\frac{1}{2}+E)\ln (t+t_{0}).
\]
with  $t_{0}$ constant. The  phase space  and $\psi  (t)$ vs.  $\ln t$
evolution is shown in figure \ref{n13}.


\subsubsection{The $n=\frac 23$ , $a=t^{\frac 23}$ case}

The  case  of  a  dust-dominated universe  is  mathematically  special
because  of  the  presence  of  a zero  eigenvalue  in  the  stability
analysis. It is illuminating to  consider this case separately with an
asymptotic analysis that extends the earlier study in \cite{bsm1}.

Consider  equation (\ref{n})  with $n=2/3$.  If we  introduce  the new
variable $x=\ln (t),$ then
\begin{equation}
\psi ^{\prime \prime }+\psi ^{\prime }=N\exp [-2\psi ] \label{dash}
\end{equation}

At large $x$, the asymptotic form of this equation has the form:
\begin{equation}
\psi    =\frac{1}{2}\ln    [2N(x+x_{0})]-\frac{1}{2}\sum_{n=1}^{\infty
}\frac{ (n-1)!}{(x+x_{0})^{n}}+C\exp [-x] \label{asym}
\end{equation}
where  $C$ and $x_{0}>0$  are arbitrary  constants. The  leading order
behaviour as $t\rightarrow \infty $ is therefore (cf.(\ref{psisol}))
\[
\psi \sim \frac{1}{2}\ln [2N\ln (t)]
\]

\begin{figure}[htbp!]
\centering   \epsfig{file=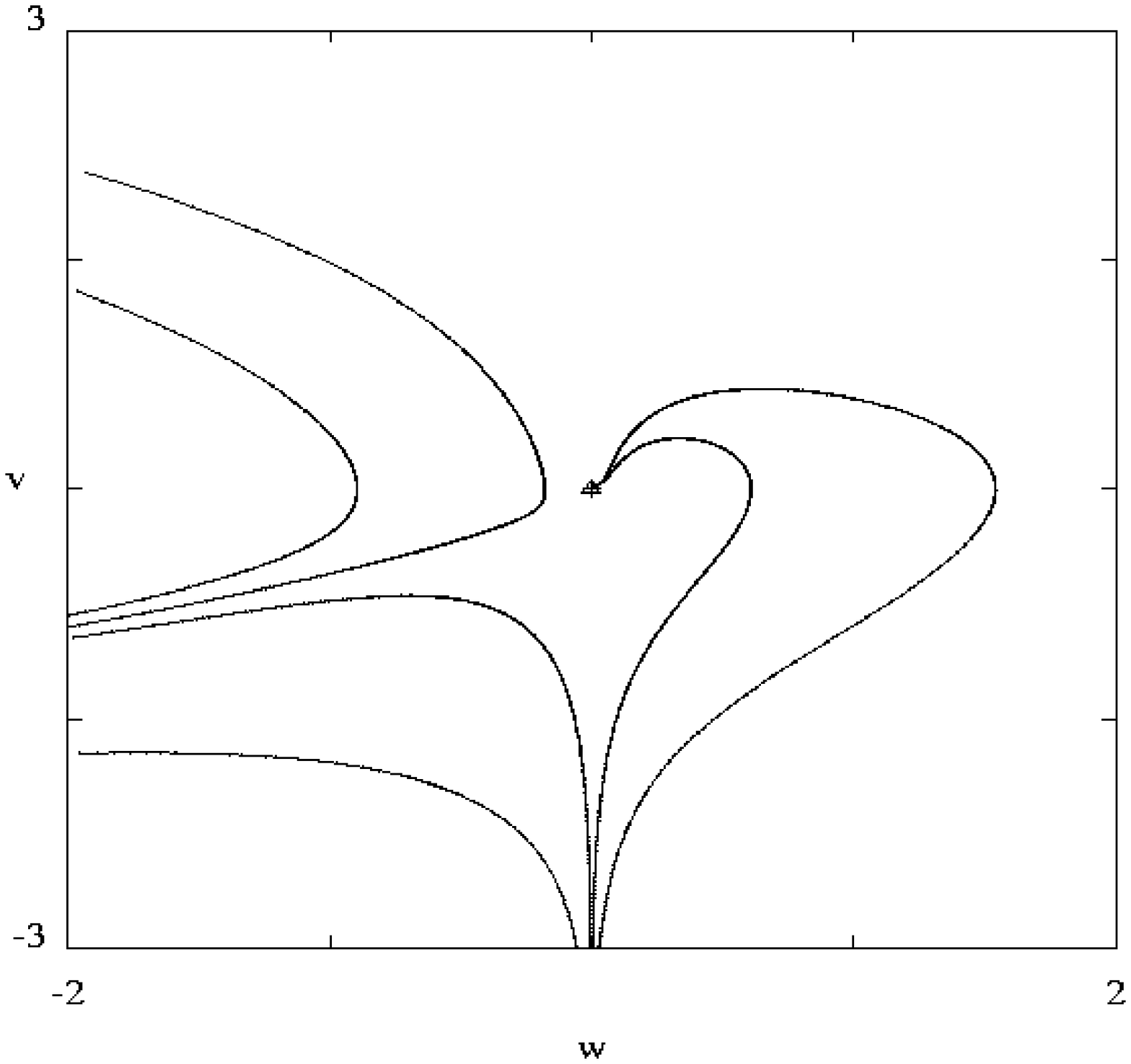,height=6.5cm,width=7.4cm}  %
\epsfig{file=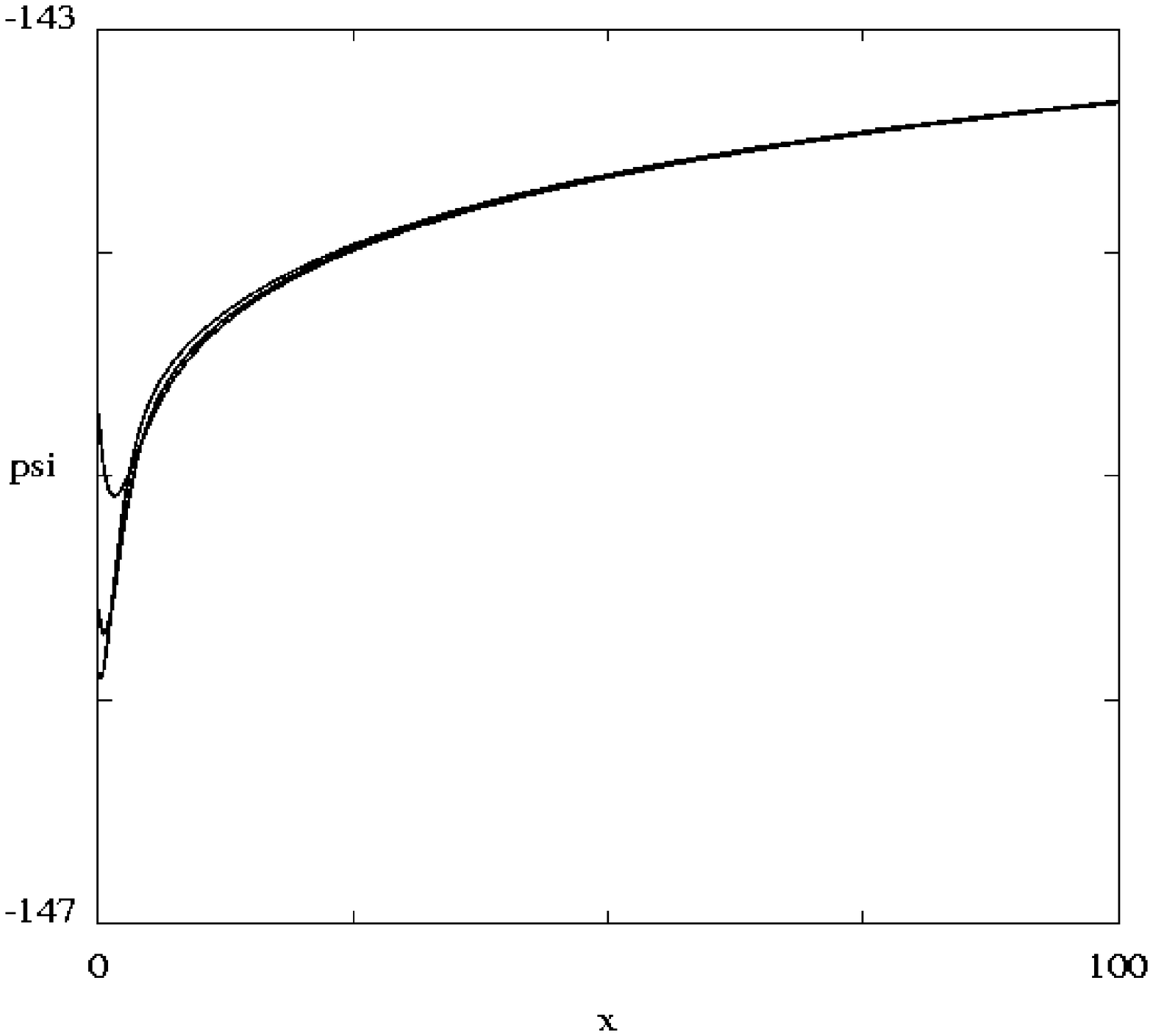,height=6.5cm,width=7.4cm}
\caption{{\protect \textit{Numerical  plots of the phase  space in the
$   (w,v)$  coordinates,  and   the  $\protect\psi$   evolution  with
$x=log(t)$ for $n=  \frac{2}{3}$. The triangle is a  saddle point for
$w<0$ and a stable node for $w>0$. }}}
\label{n23}
\end{figure}

\paragraph{Stability analysis of the $n=2/3$ asymptote}:

Performing the coordinate transformation $u=-\frac 12\ln (w)${\ on the
system (\ref{nl}) when }$n=2/3${, we have: }
\begin{eqnarray}
\frac{dv}{dx}   &=&w-v   \label{0667non}   \\  \frac{dw}{dx}   &=&-2vw
\nonumber
\end{eqnarray}
This  system has a  critical point  at the  origin of  the $(w,v)$
plane. It corresponds to the  case where $u\rightarrow \infty $. The
linearisation  about  this   critical  point  gives  two  eigenvalues:
$\lambda  _1=-1$ with   the  eigenvector  $\chi  _1=(0,1)$ and
$\lambda _2=0 $  with the eigenvector $ \chi  _2=(1,1)$. Since
we  have  a  zero  eigenvalue,  the stability  is  determined  by  the
non-linear  behaviour  and  is  one  of  Lyapunov's  'critical'  cases
\cite{sonoda},\cite{bendixon}   \cite{bautin}.  We   apply   a  linear
transformation  to split  the  system into  critical and  non-critical
variables; where  the critical  variables are those  eigenvectors with
zero eigenvalue and the non-critical  variable are the others. Then we
apply a  non-linear transformation which will  eliminate the influence
of the  critical variables upon  the non-critical ones at  the leading
order.\ This  gives $  (w,v)\rightarrow (W,W+V)$, and  the system
(\ref{0667non}) becomes: \ 
\begin{eqnarray}
W^{\prime  }  &=&-2W(W+V)  \label{WV}  \\  V^{\prime  }  &=&-V+2W(W+V)
\nonumber
\end{eqnarray}
{with the critical point at }$(W,V)=(0,0).${The Lyapunov procedure for
the system (\ref{WV}) is to  set the linearly stable variable, }$V,${\
equal to zero in the }$W^{\prime }${\ equation so,}
\[
W^{\prime }=-2W^{2}
\]
{and  so   the  second-order   analysis  shows  that   critical  point
}$(W,V)=(0,0)${\ is unstable. In fact  this critical point is a saddle
point  as can  be seen  from the  numerical-phase-plane plot  figure \
\ref{n23}. The unstable part correspond to a non-physical range of the
cosmological  variables,  since  it   gives  $w<0$.  The  stable  part
corresponds to  the range of values  where $w\geq 0$.  With respect to
our (approximate) exact solution,  the unphysical case will correspond
to the range of $x_{0}<0$, since  they lead to $\alpha <0$.  Hence the
asymptotic solution (\ref{asym}) is  the stable late-time behaviour of
the dust universes with small }$\zeta .$ The phase plane structure and
the evolution of $\psi $ vs. $\ln t$ is shown in figure \ref{n23}.


\section{Some General Asymptotic Features}

\subsection{Models with asymptotically constant $\protect\psi $ and $\protect%
\alpha $}

We  have seen that  $\psi $,  and hence  the fine  structure constant,
$\alpha $, tends to a constant at late times in accelerating universes
with power-law  and exponential increase  of the scale factor.  We can
establish a useful general  criterion for this asymptotic behaviour to
occur for general $a(t)$.  Suppose that as $t\rightarrow \infty $ both
sides of equation  (\ref{psi}) tend to a constant  (which may be equal
to zero). Thus
\begin{eqnarray*}
(\dot{\psi}a^3)^{\cdot }  &=&A \\ \psi  &=&A\int \frac{tdt}{a^3}+B\int
\frac{dt}{a^3}+C\rightarrow C
\end{eqnarray*}
as $t\rightarrow \infty $ if
\begin{eqnarray}
A\int   \frac{tdt}{a^3}   &\rightarrow   &0  \label{lim1}   \\   B\int
\frac{dt}{a^3} &\rightarrow &0 \label{lim2}
\end{eqnarray}
Then for consistency we also require, as $t\rightarrow \infty $, that
\[
N\exp [-2C]\rightarrow A
\]
so $A$  cannot be chosen to  be zero. Thus in  all cosmological models
for  which  (\ref{lim1})  and   (\ref{lim2})  hold  there  will  be  a
self-consistent asymptotic  solution as $t\rightarrow \infty  $ of the
form
\[
\psi =N\exp [-2C]\int \frac{tdt}{a^{3}}+B\int \frac{dt}{a^{3}}+C
\]
where $B,C$  are positive  constants. We see  that the $n>2/3$  and de
Sitter  cases  satisfy the  conditions  (\ref{lim1}) and  (\ref{lim2})
hence $\psi $ is asymptotically constant. The dust ($n=2/3$) case, for
which $\psi  \rightarrow \infty $ as $t\rightarrow  \infty $ satisfies
(\ref{lim2}) but violates (\ref{lim1}).

\section{ General Analysis of the Phase Plane Bifurcations}
\label{overview}

In previous sections  we have analysed the $\psi  $ evolution
equation (\ref {psi}) for  a range of variables which  are
physically realistic and correspond to  expanding universes. We
will now  analyse the whole range for variables of the system
(\ref{llog}). As before we see there are    two    critical
points    in    the    $(w,v)$   plane,    at
$(0,-1+{\frac{3n}{2}})$                     and
$ ((1-3n)(\frac{3n}{2}-1),0)$.   Linearising   (\ref{llog})
about   $ (w_{c_{1}},v_{c_{1}})=(0,-1+\frac{3n}{2})$
and           $ (w_{c_{2}},v_{c_{2}})=((1-3n)(\frac{3n}{2}-1),0)$
we   obtain   the following characteristics matrices:

$M_1= \bordermatrix{ & &\cr  & 2-3n & 0 \cr & 1  & 1-3n \cr} $
\qquad $M_2= \bordermatrix{& & \cr & 0 & (3n-1)(2-3n) \cr & 1 &
1-3n \cr} $

\bigskip

The   characteristic   matrices   are   non   singular   except   when
$n=\frac{1}{3}$ or  $ n=\frac{2}{3}$.  In the non-singular  cases the
critical  points  will be  simple  and  the  system defined  by  these
differential  equations is  structurally  stable \cite{andronov},  and
there will be no 'strange' chaotic behaviour outside the neighbourhood
of the  critical points.  Hence, the linearised  system will  have the
same phase portrait as non-linearised  one in the neighbourhood of the
critical points.

The  evolution, with  respect to  changing $n$,  of the  signs  of the
determinant and the trace of these  two matrices is given in the table
\ref{criticalpoints}. This show us  that there are always two critical
points  in our  system, an  unstable  saddle and  an attractor  (which
changes from a spiral to a node).
\begin{table}[htb!]
\centering
\begin{tabular}[htb!]{ccc}
\hline  \textbf{$n$}  &  \multicolumn{2}{c}{\textbf{Critical  Points
$(w_{c},v_{c})$}} \\  \cline{2-3} & $\left(  \left( 3n-1\right) \left(
1-\frac{3n}{2}\right) ,0\right) $  & $ \left( 0,1-\frac{3n}{2}\right)
$ \\  \hline $(-\infty ;\frac{1}{3})$ & Saddle  Point (non-physical) &
Unstable Node  \\ &  det $M_{1}<0$, Tr  $M_{1}>0$ & det  $M_{2}>0$, Tr
$M_{2}>0$  \\ \hline  $\frac{1}{3}$ &  Origin  & Axis  \\ &  det
$M_{1}=0$, Tr $M_{1}=0$ &  det $M_{2}=0$, Tr $M_{2}>0$ \\ \hline
$(\frac{1}{3};\frac{1}{2})$ &  Stable Spiral &  Saddle Point \\  & det
$M_{1}>0$, Tr $M_{1}<0$ &  det $M_{2}<0$, Tr $M_{2}>0$ \\ \hline
$\frac{1}{2}$ &  Stable Spiral & Saddle  Point \\ &  det $M_{1}>0$, Tr
$M_{1}<0$   &   det    $M_{2}<0$,   Tr   $M_{2}=0$   \\   \hline
$(\frac{1}{2};\frac{3}{5})$ &  Stable Spiral &  Saddle Point \\  & det
$M_{1}>0$, Tr $M_{1}<0$ &  det $M_{2}<0$, Tr $M_{2}<0$ \\ \hline
$\frac{3}{5}$  &  Stable  Spiral  (node)  &  Saddle  Point  \\  &  det
$M_{1}>0$, Tr $M_{1}<0$ &  det $M_{2}<0$, Tr $M_{2}<0$ \\ \hline
$(\frac{3}{5};\frac{2}{3})$  & Stable  Node &  Saddle Point  \\  & det
$M_{1}>0$, Tr $M_{1}<0$ &  det $M_{2}<0$, Tr $M_{2}<0$ \\ \hline
$\frac{2}{3}$  & Stable  Axis &  Stable Axis  \\ &  det  $M_{1}=0$, Tr
$M_{1}<0$   &   det    $M_{2}=0$,   Tr   $M_{2}<0$   \\   \hline
$(\frac{2}{3};\infty )$ & Saddle Point (non-physical) & Stable Node \\
& det $M_{1}<0$, Tr $M_{1}<0$ & det $M_{2}>0$, Tr $M_{2}<0$ \\ \hline
\end{tabular}
\caption{Stability  of all  the  critical points  of  the equation  of
motion for $\psi$.}
\label{criticalpoints}
\end{table}

\begin{figure}[htbp!]
\centering  \epsfig{file=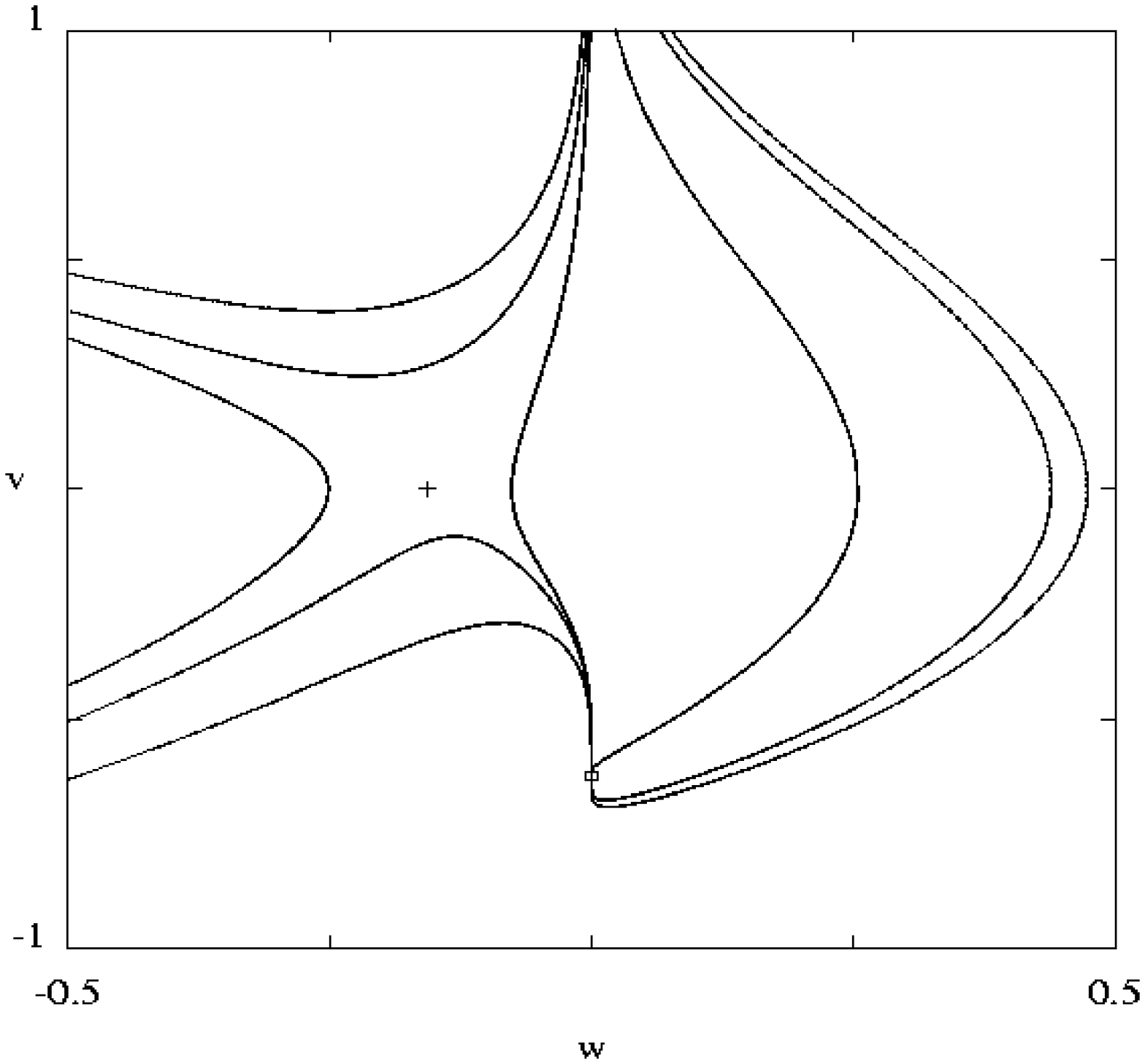,height=6.5cm,width=7.4cm}  %
\epsfig{file=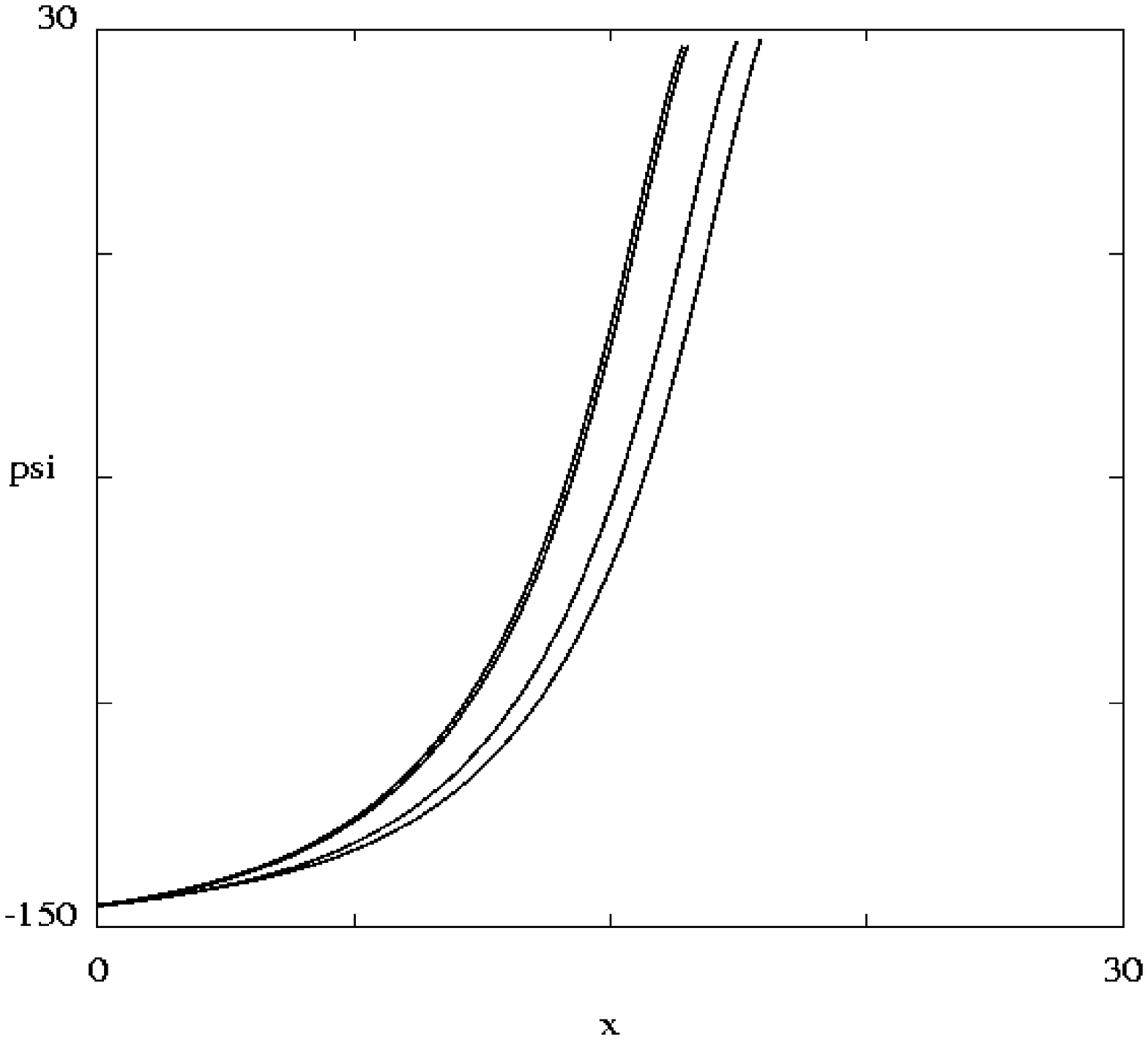,height=6.5cm,width=7.4cm}
\caption{{\protect \textit{Numerical  plots of the phase  space in the
$   (w,v)$  coordinates,  and   the  $\protect\psi$   evolution  with
$x=log(t)$ for $n=\frac{1}{4}$.  The '$+$' sign is a  saddle point and
the square is an unstable source. }}}
\label{n025}
\end{figure}

The cases $n=\frac{1}{3}$ or $n=\frac{2}{3}$, where the determinant of
the  characteristic  matrices  vanishes,  lead  to  a  bifurcation  of
codimension  $1$, in particular,  of Saddle-Node  type \cite{wiggins},
since  they  correspond  to  points  where  the  determinants  of  the
characteristic matrices change sign, det $M_{1}$ $=$ det$M_{2}=0$. At
these   values    of   $n$   the    nature   of   the    system   will
change.  Cosmologically,  these  points  represent  a  change  in  the
behaviour of  the time evolution  of the fine structure  constant as
can  be  seen  from  the figures:  \ref{n025},  \ref{n13},  \ref{n12},
\ref{n35}, \ref{n23} , \ref{n1},  which display the time evolution of
$\psi$. When $n$\  starts to grows from $-\infty  $\ to $\frac{1}{3}$\
the    two   critical    points   slowly\emph{    \    }converge   at
$n=\frac{1}{3}$\emph{. }For  example, in the  $n=0$ case where  we may
without  loss of generality  set $a=1$,  equation (\ref{psi})  has the
exact solution
\[
\alpha =\exp [2\psi ]=A^{-2}\cosh ^{2}[AN^{1/2}(t+t_{0})]
\]
where $A,t_{0}$ are constants. This is an unrealistically rapid growth
asymptotically, $\psi  \varpropto t,$ caused  by the  absence  of the
inhibiting effect  of the cosmological expansion. The  case of $n=1/4$
is  shown   in  figure  \ref{n025},   which  shows  the   phase  space
trajectories and the evolution of $\psi $ vs.  $\ln t.$

At  $n=\frac{1}{3}$ we  are in  the situation  where the  two critical
points  collapse  into  a  unique   one  at  the  origin,  creating  a
saddle-node bifurcation and a  concomitant change in the behaviour and
evolution of $\psi  $. As $n$ keeps growing  the single critical point
splits  into two  critical points  again.  They move  apart until  the
radiation value  is reached,  $n=\frac{1}{2}$ (Tr $M_{2}=0$).  In this
case, $\psi $ is a  asymptotically monotonic growing function of time,
with  some small oscillations  near the  Planck epoch.   However, note
that  in our  universe the  asymptote giving  an increase  of  $\psi $
behaviour  with  time  is  never  reached  before  the  dust-dominated
evolution takes over \cite{bsbm}, \cite{bsm1}, \cite{bsm2}.

As  the  universe  evolves   to  the  dust-dominated  epoch,  and  $n$
approaches  the   intermediate  behaviour  $n=\frac{3}{5}$,   the  two
critical  points start  to coalesce  again into  a single  point. When
$n=\frac{3}{5}$ is  reached, $\psi $ becomes  a strictly monotonically
growing function of time. When  $n$ reaches the value corresponding to
a   dust-dominated  universe,  $n=\frac{2}{3}$,   another  saddle-node
bifurcation  occurs. The two  critical points  collapse into  a single
one. Again there will be change in the behaviour of $\psi $ for larger
values  of $n$.  Accordingly, when  $n>2/3,$ the  two  critical points
reappear  once again and  $\psi $  becomes asymptotically  constant in
value.

Notice,  that although a  bifurcation is  something that  'spoils' the
smooth behaviour of  a system, in our case, that  won't happen, due to
the physical  constraints of  our variables. In  reality due  to those
constraints, the  physical system will never 'feel'  the abrupt change
at $n=\frac{1}{3}$ and $ n=\frac{2}{3}$. This is also due to the fact
that the attracting  critical point always lies in  the physical range
of the variables, while the  unstable one disappears form the physical
system  when the bifurcations  occur, as  can be  seen from  the phase
plane plots.

%% file: chapter31.tex
\chapter{Gauge-Invariant Linear Perturbations of Varying-Alpha Cosmologies}
\label{gaugeinvariant}

\begin{flushright}
{\it  {\small Ser\'a este espelho a porta de entrada para a casa da
    verdade  }}\\ 
{\it  {\small em  que  ainda
n\~ao fomos convidados?}}\\ 
{\it {\small -- Ernesto Mota --}}
\end{flushright}

\bigskip

\section{Introduction}

In the previous chapter, using  a phase space analysis of the equation
of motion of the field  $\psi$ (\ref{psidot}), we have shown that BSBM
theories have a number of  appealing properties. They predict that, in
a homogeneous and isotropic universe,  there should be no variation of
$\alpha  $ during the  radiation era  and none  during any  present or
late-time curvature or cosmological constant dominated era. During the
dust era $\alpha $ should  grow (leading to $\Delta \alpha /\alpha <0$
as observed) but only as  $\log (t).$ Those results were also obtained
by  Barrow, Sandvik, and  Magueijo \cite{bsbm,bsm1,bsm2,bsm3}  using a
different approach.

This  behaviour allows  the  quasar data  to  be accommodated  without
producing conflict with recent geonuclear limits on allowed variations
of $\alpha $,  like the Oklo natural reactor  limits of $\Delta \alpha
/\alpha  \lesssim 10^{-7}$  \cite{shly,  sisterna,landau,damour1, fuj}
because they are imposed at a very low effective redshift of $z\approx
0.15,$  at which  time the  universe  has begun  accelerating and  all
variations  of  $\alpha $  are  damped  out.  Recent deductions  of  a
possible  upper  limit  of  $\Delta \alpha  /\alpha  \lesssim  3\times
10^{-7}$ at $z=0.45$ from nuclear $\beta $ decays are potentially more
restrictive  \cite{peebles,dyson, olive1}.  These observations  may be
indeed problematic,  since there is,  at least, an order  of magnitude
difference between  $\Delta \alpha /\alpha$  from the quasar  data and
the $\beta $-decay rate.

A possible approach to solve the  discrepancy of the two results is to
claim that one of the  measurements has a non-identified problem which
was not  taken into account in  the observations. However,  it must be
remembered  that both  the nuclear  limits  are derived  from a  local
solar-system  environment. In  the absence  of a  theory  relating the
value of, and rate of change  of, $\alpha $ on the cosmological scales
where  quasar lines  form  to  their values  in  the virialised  local
inhomogeneities where galaxies, stars  and planets form, one should be
wary of ignoring the possible corrections that must be introduced when
comparing planetary and quasar bounds:  the density of the Earth would
not, for example, be a good  indicator of the density of the universe.
Adding to that,  we know that any varying-$\alpha$  theory implies the
existence of a field, responsible for variations of $\alpha$, which is
coupled  to  the  matter  fields.  It is  then  natural,  that  strong
variations of the  density of matter in the  universe, lead to spatial
variations of  $\psi$.  Thus, inhomogeneity is an  important factor in
the study of varying-constant cosmological theories.

\bigskip

The first  approach to  address this problem,  and the  possibility of
variations of   $\alpha$ be  influenced by matter inhomogeneities, is to
consider   the   so  called   linear   regime   of  the   cosmological
perturbations.  In this regime,  all the  perturbations in  the matter
fields are very  small. Much smaller than the  background value of the
fields.

In this  chapter, we are then  going to study the  evolution of small,
gauge-invariant perturbations  to the exact Friedmann-Robertson-Walker
solutions of the  BSBM theory. The results of  this investigation will
reveal  whether one  must worry  about  fast growth  of small  initial
inhomogeneities in the value of $\alpha $, which would lead to spatial
variations of the  value of the fine structure  constant that might be
more significant than the time variations at late times.

\bigskip

Our discussion  is organised as follows.  In section 2 of  the we give
the gauge-invariant  perturbation equations following  the development
used in general relativity. In section 3 we specify the BSBM cosmology
with  varying  $ \alpha  $  and  derive  the gauge  invariant  linear
perturbation   equations  which  couple   the  perturbations   in  the
gravitational field to those in  $\alpha $ and the density. In section
4 we solve for the time-evolution of small inhomogeneities in the fine
structure constant on large and  small scales for radiation, dust, and
cosmological   constant   dominated   expansion  of   the   background
universe. In  section 5  we extend these  studies into  the non-linear
regime  by  means of  numerical  solutions  for  flat and  closed  FRW
universes.  The evolution  of spherical  curvature  inhomogeneities in
density and  in $\alpha $ is  followed by computing  the difference in
time evolution  between the FRW  models of different  curvature. These
studies also reveal  for the first time the behaviour  of $\alpha $ in
closed FRW models in the BSBM theory.


\bigskip

Units will be  used in which $c=\hbar =1$; Greek  indices run form $0$
to $3$ and Latin indices only  over the spatial degrees of freedom $1$
to $3$.  The Einstein summation  convention is assumed; $a(t)$  is the
scale factor of the background Friedmann-Robertson-Walker (FRW) metric
and $G$  is Newton's  gravitation constant.  Recall  that in  the BSBM
varying-$\alpha $ theory, the quantities $c$ and $\hbar $ are taken to
be constant,  while $e$ varies  as a function  of a real  scalar field
$\psi ,$ with $\alpha =e^{2},$ hence
\begin{eqnarray}
e &=&e_{0}\exp [\psi ]  \label{e3} \\ \alpha &=&\alpha _{0}\exp [2\psi
]
\end{eqnarray}

\section{Linear Theory of Cosmological Perturbations}

Observations  indicate that  the  universe is  nearly homogeneous  and
isotropic on large  scales, in particular at early  times where metric
perturbations  to  the   cosmic  microwave  background  radiation  are
observed to  be small $\approx  10^{-5}$ $<<1$. It is  usually assumed
that there exist small  primordial perturbations which slowly increase
in amplitude  due to gravitational instability to  form the structures
we observe at the present time  on the scales of galaxies and clusters
of  galaxies.    The  problem  of  describing  the   growth  of  small
perturbations  can  be  tackled  using the  cosmological  perturbation
theory, which is  based on expanding the Einstein  equations to linear
order about the background  metric. The theory was initially developed
by   Lifshitz   \cite{lifshitz},  but   many   other  approaches   and
improvements    were   done    until   today,    see    for   instance
\cite{kodama,bruni,hwang,durrer} for reviews.
  
Here  we  will  follow  the  gauge invariant  formalism  developed  in
\cite{bardeen, mukhanov}  and we refer  the reader to  this references
for more details  in this particular approach. Also,  we will a priori
assume the existence of small  inhomogeneities at some initial time in
the early universe. Hence we  will not discuss any particular model to
create  the primordial  fluctuations  in the  matter  fields, see  
\cite{khoury,lyth,gasperini,hindmarsh,guth}   for  different  proposed
models and hypothesis.

\subsection{The background and metric perturbations}

Assuming that the metric of  spacetime deviates only by a small amount
from a  homogeneous, isotropic FRW  spacetime, which is defined  to be
the background universe. It is convenient to split the metric into two
parts: the first is the  background metric, which represents the ideal
background  spacetime,  and  the  second is  its  perturbation,  which
describes  how  the  'real'  spacetime  deviates  from  the  idealised
background. This approach is  mathematically well defined since it was
shown in \cite{d'eath}  that in the case of  a FRW universe, solutions
of the linearised  field equations can be viewed  as linearisations of
solutions of the full non-linear ones.

The background line element is:
\begin{equation}
ds^{2}=^{(0)}g_{\mu  \nu   }dx^{\mu  }dx^{\nu  }=dt^{2}-a^{2}(t)\gamma
_{ij}dx^{i}dx^{j}=a^{2}(\eta )(d\eta ^{2}-\gamma _{ij}dx^{i}dx^{j}),
\label{backline}
\end{equation}%
where $\eta $ is the conformal time,
\[
d\eta =a^{-1}dt.
\]

We choose the background metric to be the FRW metric, so
\[
\gamma _{ij}=\delta _{ij}[1+\kappa (x^{2}+y^{2}+z^{2})]^{-2},
\]%
where  $\kappa  =0,1,-1$ depending  on  whether the  three-dimensional
hypersurfaces  of constant  $\eta  $  are flat,  closed  or open.  The
Einstein equations are:
\begin{equation}
G_{\nu }^{\mu }=R_{\nu }^{\mu }-\frac{1}{2}\delta _{\nu }^{\mu }R=8\pi
GT_{\nu }^{\mu },
\end{equation}%
where $R_{\nu }^{\mu }$ is the Ricci tensor, $R\equiv R_{\mu }^{\mu }$
is the Ricci scalar and $T_{\nu }^{\mu }$ is the total energy-momentum
tensor.  We  shall exploit the fact  that the BSBM  theory for varying
$\alpha $  can be  expressed as general  relativity with  a particular
linear combination of energy-momentum tensors. For the moment consider
the presence of  a single energy-momentum tensor. In  the next section
this will be decomposed appropriately.

For the  background metric in equation  (\ref{backline}), in conformal
time, the Einstein equations reduce to the $0-0$ equation
\begin{equation}
\mathcal{H}^{2}=\frac{8\pi G}{3}a^{2}T_{0}^{0}-\kappa , \label{00}
\end{equation}%
where $\mathcal{H}\equiv \frac{a^{\prime  }}{a}$ using conformal time,
and the $i-i$ equation:
\begin{equation}  \label{ii}
\mathcal{H^{\prime}}+ \mathcal{H}^2 =  \frac{4\pi G}{3}a^2 T - \kappa,
\qquad T\equiv T^{\mu}_{\mu}
\end{equation}

For the  background metric  (\ref{backline}), the space-space  part of
the   Ricci   tensor    $R_{j}^{i}$   is   proportional   to   $\delta
_{j}^{i}$.   Thus,   for  an   isotropic   background  universe,   the
energy-momentum    tensor   must    also   be    spatially   diagonal,
$T_{j}^{i}\propto  \delta  _{j}^{i}$,   in  order  that  the  Einstein
equations  are satisfied. Differentiating  (\ref{00}) with  respect to
$\eta $ and subtracting $2a^{\prime  }$ we get the continuity equation
for matter $\triangledown _{\mu }T_{0}^{\mu }=0$ :
\begin{equation}  \label{conserv}
dT^{0}_{0}=-(4T^{0}_{0}-T)d\ln{a}.
\end{equation}

\bigskip

We  now introduce small  perturbations around  the FRW  background and
follow   the   gauge-invariant   approach   of  \   Mukhanov   et   al
\cite{mukhanov}. The full line element may be expressed as:
\begin{eqnarray}  \label{fullline}
ds^2  &=&   ^{(0)}g_{\mu\nu}  dx^{\mu}dx^{\nu}  +   \delta  g_{\mu\nu}
dx^{\mu}dx^{\nu},
\end{eqnarray}
where $\delta g_{\mu\nu}$ describes  the perturbation. The full metric
has been decoupled into its background parts and perturbation parts:
\begin{eqnarray}  \label{metric}
g_{\mu\nu} &=& ^{(0)}g_{\mu\nu}+ \delta g_{\mu\nu}.
\end{eqnarray}

The   metric   tensor  has   $10$   independent   components  in   $4$
dimensions. For  linear perturbations, the metric  perturbation can be
divided  into  three  distinct   types:  scalar,  vector  and  tensor,
according to their  transformation properties on spatial hypersurfaces
\cite{bardeen,stewart}.    The  reason   for   splitting  the   metric
perturbation into these three types  is that they are decoupled in the
linear perturbations  equations and so  evolve independently.  Neither
of the  vector and tensor perturbations  exhibit growing instabilities
in  dust   and  radiation  universes  .   Vector  perturbations  decay
kinematically  in an expanding  universe whereas  tensor perturbations
lead  to gravitational  waves  that  do not  couple  to the  isotropic
energy-density and pressure inhomogeneities \cite{mukhanov}.  However,
scalar  perturbations may  lead to  growing inhomogeneities  which, in
turn, have an  important effect on the dynamics  of matter and thereby
on the time and space variations of the fine structure constant in the
BSBM\ theory.   In this thesis we  will then consider  only the scalar
perturbation modes.

It is shown  in the literature, see for  instance \cite{bardeen}, that
scalar perturbations can always be constructed from a scalar quantity,
or  its  derivatives,  and   any  background  quantity,  such  as  the
$3$-metric $\gamma_{ij}$. We can then construct any first-order metric
perturbation in  terms of  four scalars $\phi$,  $\chi$, $B$  and $E$,
which are functions of space and time coordinates \cite{mukhanov}:
\begin{eqnarray}
\delta  g_{00}&=& - 2  a^2 \phi,  \nonumber\\ \delta  g_{0i}&=& \delta
g_{i0}=   a^2  B_{|i}   ,  \\   \delta  g_{ij}&=&   -  2   a^2  \left(
\chi\gamma_{ij}-E_{|ij} \right) \nonumber
\end{eqnarray}
where $|i$ represents the three-dimensional covariant derivative.  The
most  general  form  of   the  scalar  metric  perturbations  is  then
constructed using  the four  scalar quantities$\phi$, $\chi$,  $B$ and
$E$:

$\delta g_{\mu \nu }= \bordermatrix{ &  & \cr & 2\phi & -B_{|i} \cr &
-B_{|i} & 2(\chi \gamma_{ij}-E_{|ij}) \cr},$

\bigskip

From the above equation and equation (\ref{metric}), the line element
for the background and for the scalar metric perturbations is
\begin{equation}
ds^{2}=a^{2}(\eta     )\{(1+2\phi    )d\eta    ^{2}-2B_{|i}dx^{i}d\eta
-[(1-2\chi           )\gamma           _{ij}+2E_{|ij}]dx^{i}dx^{j})\}.
\label{perturbedline}
\end{equation}

\subsection{Gauge-invariant variables}

The  homogeneity  of  a  FRW  spacetime  gives  a  natural  choice  of
coordinates. But in  the presence of first order  perturbations we are
free  to make  a first  order change  in the  coordinates, that  is, a
gauge-transformation.   This   gauge    freedom   leads   to   several
complications.  For example,  not all  perturbed metrics  correspond to
perturbed spacetimes.  It is possible to obtain  an inhomogeneous form
for the metric  in a homogeneous and isotropic  spacetime, by a choice
of  coordinates. Hence,  it is  important  to be  able to  distinguish
between physical inhomogeneities and mere coordinate artifacts.

The  gauge problem  of  the cosmological  perturbations is  intimately
related to  the fact  that we actually  deal with two  spacetimes, the
physical  space  time,  corresponding  to  the  real  universe  and  a
fictitious one, which defines the unperturbed background. We need then
to   establish   an   one-to-one   correspondence  between   the   two
spacetimes. For  instance, when a  coordinate system is  introduced in
the background  we need to  have a correspondent coordinate  system in
the real universe and vice-versa. This is necessary in order for us to
be able to define a background spacetime into the real universe and to
compare quantities between  them. The need for that  is clear from the
definition of a perturbation in a given spacetime: a perturbation in a
given quantity  is the difference between  its value at  some event in
the real  spacetime and  its value at  the corresponding event  in the
background. This  means that  even if a  quantity behaves as  a scalar
under  coordinate  transformations,   its  perturbation  will  not  be
invariant under  gauge transformations.  As a result  we might  end up
with  spurious gauge  modes in  the solutions,  which have  no meaning
whatsoever.

There are two essential ways to deal with the gauge problem. The first
is to choose a particular gauge and compute all the quantities in that
gauge. If the gauge choice  is well motivated, the perturbed variables
will be  easy to  interpret. However, the  task of selecting  the best
gauge for any  given situation is not always  trivial. See for example
\cite{lifshitz,kolb} for more details  on this approach. The second is
to  describe perturbations using  gauge-invariant variables.   We will
follow here the second approach.

\bigskip 

Gauge-invariant   variables  are  unchanged  under  all
infinitesimal   scalar  coordinate   transformations,   so  they   are
independent  of the  background  coordinates. Such  quantities can  be
constructed out of the four scalar functions $\Phi $, $\Psi $, $E$ and
$B$,   see  \cite{bardeen}.   The   simplest  gauge-invariant   linear
combinations which  span the  space of gauge-invariant  variables that
can   be   constructed   from   the   metric   variables   alone   are
\cite{mukhanov}:
\begin{equation}
\Phi =\phi  +\frac{[(B-E^{\prime })a]^{\prime }}{a},\qquad  \Psi =\chi
-\frac{a }{a^{\prime }}\left( B-E^{\prime }\right) .  \label{gauge}
\end{equation}

Note that  any combination of  gauge-invariant variables will  also be
gauge   invariant.    In    general,   a   scalar   quantity   $f(\eta
,\mathbf{{x})}$  defined  in  the  spacetime  can be  split  into  its
background value and  a perturbation $f(\eta ,
)+\delta f(\eta ,\mathbf{x})$. Since, in  general, $ \delta f$ is not
gauge  invariant,   we  cannot   use  this  scalar   quantity  without
modification if  we want to have gauge-invariant  equations. Hence, we
consider the following gauge-invariant combination:
\begin{equation}
\delta     f^{(gi)}=\delta    f+f_{0}^{\prime     }(B-E^{\prime    }).
\label{gaugeinv}
\end{equation}

\bigskip

The freedom  of gauge choice can  be used to impose  two conditions on
the four scalar functions.  The \textit{longitudinal gauge} is defined
by  the conditions  $E=B=0$. This  gauge choice  has the  advantage of
ruling out  the complications of  residual gauge modes. Also,  in this
gauge $\phi $ and $\chi  $ coincide with the gauge-invariant variables
$\Phi  $ and  $\Psi $  respectively. In  this longitudinal  gauge, the
metric takes the form:
\[
ds^{2}=a^{2}(\eta    )[(1+2\Phi     )d\eta    ^{2}-(1-2\Psi    )\gamma
_{ij}dx^{i}dx^{j})],
\]%
and the gauge invariants $\Phi $  and $\Psi $ become the amplitudes of
the metric perturbations in the longitudinal coordinate system. In the
case where there are  no space-space components in the energy-momentum
tensor, so  $ T_{j}^{i}\propto \delta  _{j}^{i}$, we have  that $\Phi
=\Psi $ and  there remains only one free  metric perturbation variable
which is a generalisation of the Newtonian gravitational potential. As
can be seen  from the above equation, the  gauge invariants $\Phi$ and
$\Psi$ have a simple  physical interpretation: they are the amplitudes
of the metric perturbations in the longitudinal coordinate system.

\subsection{The linearly perturbed Einstein equations}

For  small perturbations  of the  metric, the  Einstein tensor  can be
written  as $G_{\nu  }^{\mu  }=\ ^{(0)}G_{\nu  }^{\mu }+\delta  G_{\nu
}^{\mu }$,  and the energy-momentum tensor  can be split  in a similar
way. The fully perturbed Einstein equations can be obtained for scalar
perturbation    modes    with    the    line    element    given    by
(\ref{perturbedline}).   However,  these   equations  are   not  gauge
invariant, since they contain non gauge-invariant quantities. In order
to have gauge-invariant equations we need to replace $\phi $ and $\chi
$ by the  gauge-invariant variables $\Phi $, $\Psi  $ and $B-E^{\prime
}$, and to construct  the gauge-invariant equivalents of $\delta G_{\nu
}^{\mu }$  and $\delta  T_{\nu }^{\mu  }$, we need  to rewrite  then as
\cite{mukhanov}:
\begin{eqnarray}
\delta   G_{0}^{(gi)\   0}&=&\delta  G_{0}^{0}+(^{(0)}G_{0}^{0})^{\prime
}(B-E^{\prime   }),    \nonumber\\   \delta   G_{j}^{(gi)\   i}&=&\delta
G_{j}^{i}+(^{(0)}G_{j}^{i})^{\prime   }(B-E^{\prime   }),  \\   \delta
G_{i}^{(gi)\     0}&=&\delta     G_{i}^{0}+(^{(0)}G_{0}^{0}-\frac{1}{3}\
^{(0)}G_{k}^{k})(B-E^{\prime })_{|i}, \nonumber
\end{eqnarray}
and analogously for $\delta T_{\nu }^{\mu }$,
\begin{eqnarray}
\delta   T_{0}^{(gi)\   0}&=&\delta  T_{0}^{0}+(^{(0)}T_{0}^{0})^{\prime
}(B-E^{\prime }),
\\
\delta   T_{j}^{(gi)\   i}&=&\delta  T_{j}^{i}+(^{(0)}T_{j}^{i})^{\prime
}(B-E^{\prime }),
\\
\delta  T_{i}^{(gi)\ 0}&=&\delta T_{i}^{0}+(^{(0)}T_{0}^{0}-\frac{1}{3}\
^{(0)}T_{k}^{k})(B-E^{\prime })_{|i}.
\end{eqnarray}

The components  of the perturbed Einstein  equations linearised around
small perturbations of the  background are $\delta G_{\nu }^{(gi)\ \mu
}=8\pi G\delta T_{\nu }^{(gi)\ \mu }$:
\begin{eqnarray}  \label{genpertein}
\delta  G_{0}^{0}&=&\triangledown  ^{2}\Phi  -3\mathcal{H}\Phi  ^{\prime
}-3\Phi  ( \mathcal{H}^{2}-\kappa  )=4\pi  Ga^{2}\delta T_{0}^{(gi)0}
\nonumber \\ \delta  G_{i}^{0}&=&\partial _{i}(a\Phi )^{^{\prime }}=4\pi
Ga\delta  T_{i}^{(gi)0}  \\   \delta  G_{i}^{j}&=&\Phi  ^{\prime  \prime
}+3\mathcal{H}\Phi    ^{\prime    }+\Phi    (2    \mathcal{H}^{\prime
}+\mathcal{H}^{2}-\kappa )=-4\pi Ga^{2}\delta T_{j}^{(gi)i} \nonumber
\end{eqnarray}
where   we   have  already   simplified   the   equations  since   the
energy-momentum tensor that we will be considering in section 3 has no
space-space components and so $\Phi =\Psi $.

In order to close our system of equations, we need equations of motion
for  the matter  formulated in  a gauge-invariant  way.  This requires
explicit expressions for the energy-momentum tensor and so we must now
specify the BSBM theory.

\section{Cosmological Perturbations and the BSBM Model}


%


It was shown in chapter \ref{introduction} that, in the context of the
BSBM  model, the  homogeneous evolution  of  $\psi $  does not  create
significant metric  perturbations at  late times and  the cosmological
time-evolution of the expansion scale-factor is very well approximated
by the usual power-laws found in  $\kappa =0$ FRW models filled with a
perfect fluid (see also\cite{bsm3} ).

Therefore,  we will  assume there  are no  major modifications  in the
perturbed spacetime  which would lead  to changes of the  behaviour of
the  perturbed variables  of  a perturbed  FRW  spacetime filled  with
perfect  fluid.  That  is,  we  will assume  that  the  energy-density
perturbations  and the  metric  potential are  the  same as  in a  FRW
universe with  no variation of $\alpha  $.  Physically, this  is to be
expected  for most of  the evolution,  although this  assumption might
break down  (along with  much else) on  approach to initial  and final
cosmological singularities.  It is a  reflection of the fact  that the
changes in $\alpha $ have  negligible feedback into the changes in the
expansion,  which  are  governed  to  leading order  by  gravity.  The
principal effects are those of perturbations in the matter density and
expansion rate on the evolution of $\alpha .$ This simplification will
allow us to  write the time and space variations  of the scalar field,
$\delta \psi  $, as a functions  of $\rho _{m}$,  $\rho _{r}$, $\delta
\rho _{m}$, $\delta  \rho _{r}$ , $\Phi $ and  $\psi $, where $\delta
\rho _{m}$,  $\delta \rho  _{r}$ and $  \Phi $  will be given  by the
solutions  found  in  ref.   \cite{mukhanov}  for  universes  with  no
variation  of $\alpha  ;$  the field  $\psi  $ will  be  given by  the
solutions  found  previously in  chapter \ref{qualitative}.  In  order  to find  an
expression for $\delta \psi $ in  terms of these quantities we need to
write  the gauge-invariant linearly  perturbed Einstein  equations for
the BSBM model.

\subsection{The background equations}

We vary the action (\ref{S}) with respect to the metric to obtain the
generalised Einstein equations:
\[
G_{\nu }^{\ \mu }=8\pi G\left( T_{\nu }^{\mathit{matter}\ \mu }+T_{\nu
}^{\psi \ \mu }+T_{\nu }^{\mathit{em}\ \mu }\right)
\]%
where   $T_{\mu  \nu  }^{\mathrm{mat}}=\frac{2}{\sqrt{-g}}\frac{\delta
(\sqrt{-g} \mathcal{L}_{\mathrm{mat}})}{\delta g^{\mu  \nu }}$ is the
energy-momentum tensor for perfect-fluid matter fields, and
\[
T_{\mu   \nu   }^{\mathrm{matter}}=(\rho   _{m}+p_{m})u_{\mu   }u_{\nu
}-p_{m}g_{\mu \nu },
\]%
where  $u^{\mu  }$  $=\delta   _{0}^{\mu  }$  is  the  comoving  fluid
4-velocity; $ T_{\nu }^{\psi \ \mu }$ and $T_{\nu }^{\mathit{em}\ \mu
}$ are the energy-momentum tensors for the kinetic energy of the field
$\psi $ and the electromagnetic field respectively:
\[
T_{\mu \nu }^{\psi }=\omega  \partial _{\mu }\psi \partial _{\nu }\psi
-{  \frac{\omega  }{2}}g_{\mu \nu  }\partial  _{\beta }\psi  \partial
^{\beta }\psi  ,\quad T_{\mu \nu  }^{\mathrm{em}}=F_{\mu \beta }F_{\nu
}^{\beta  }e^{-2\psi  }-  {\frac{1}{4}}g_{\mu  \nu  }F_{\sigma  \beta
}F^{\sigma \beta }e^{-2\psi }.
\]%
Note, the total energy density of the electromagnetic field is the sum
of  the Coulomb  energy density  $\zeta \rho  _{m}$ and  the radiation
energy  density $\rho  _{r}$  , where  $-1\leq  \zeta \leq  1$ is  the
fraction of  mass density  $\rho _{m}$  of matter in  the form  of the
Coulomb energy.  We will then consider $T_{\nu  }^{\mathit{em}\ \mu }$
as a perfect fluid:
\[
T_{\mu      \nu      }^{\mathrm{em}}=(|\zeta      |\rho      _{m}+\rho
_{r}+p_{m}+p_{r})e^{-2\psi }u_{\mu }u_{\nu }-\left( p_{m}+p_{r}\right)
e^{-2\psi }g_{\mu \nu }
\]

The  propagation  equation for  $\psi  $  comes  from the  variational
principle as:
\begin{equation}
{\partial  _{\mu  }\left[ \sqrt{-g}g^{\mu  \nu  }\partial _{\nu  }\psi
\right]         =-        \frac{2}{\omega        }}\sqrt{-g}e^{-2\psi
}\mathcal{L}_{\mathrm{em}}.
\label{psieq}
\end{equation}%
This equation  determines how  $e,$ and hence  $\alpha $,  varies with
time.

The background equations can now be explicitly obtained:
\begin{eqnarray}
3{{\mathcal{H}}^{2}}   &=&8\pi  Ga^{2}\left(  \rho   _{m}+\left(  \rho
_{r}+|\zeta     |\rho     _{m}\right)     e^{-2\psi     }+\frac{\omega
}{2}a^{-2}{{\psi  ^{\prime   }}^{2}}  \right)  +a^{2}\Lambda  -3\kappa
\label{fried3} \\ 3{\mathcal{H^{\prime  }}} &=&-4\pi Ga^{2}\left( \rho
_{m}\left( 1+|\zeta |e^{-2\psi }\right) +2\rho _{r}e^{-2\psi }+2\omega
a^{-2}{{\psi ^{\prime }} ^{2}}\right) +a^{2}\Lambda \nonumber
\end{eqnarray}%
where $\Lambda $ is the  cosmological constant. The equation of motion
for the $\psi $ field is:
\begin{equation}
2\mathcal{H}\psi  ^{\prime   }+\psi  ^{\prime  \prime  }=\frac{2|\zeta
|{{a}^{2}}}{ \omega }\rho _{m}e^{-2\psi } \label{psiddot3}
\end{equation}

The conservation equations for the  matter fields, $\rho _r$ and $\rho
_m$ respectively, are:
\begin{eqnarray}
\rho      _m^{\prime}+3{\mathcal{H}}     \rho      _m      &=&0     \\
\rho_r^{\prime}+4{\mathcal{H}} \rho_r^{\prime}&=&2\psi^{\prime}\rho_r.
\label{dotrho3}
\end{eqnarray}

\subsection{Linear perturbations in the BSBM theory}

The  perturbed components  of the  total energy-momentum  ($T_{\mu \nu
}^{total}=T_{\mu \nu }^{mat}+T_{\mu  \nu }^{\psi }+T_{\mu \nu }^{em}$)
arise from perturbations of the different matter fields which are time
and space dependent. In particular, we have $\rho _{m}\rightarrow \rho
_{m}+\delta  \rho _{m}$, $\rho  _{r}\rightarrow \rho  _{r}+\delta \rho
_{r}$ and  $\psi \rightarrow  \psi +\delta \psi  $. Note also  that we
have   to   perturb  the   fluid   4-velocity   field,   so  we   have
$u_{i}\rightarrow u_{i}+\delta u_{i}$, where $i$ $=1,2,3$.

\bigskip

In  order to  have gauge-invariant  equations we  need to  express the
perturbed  energy-momentum  tensor  in  terms of  the  gauge-invariant
energy   density,   pressure,   scalar   field  and   velocity   field
perturbations  .  The  gauge  invariants  $\delta  \rho  _{m}^{(gi)}$,
$\delta \rho _{r}^{(gi)}$ and $\delta \psi ^{(gi)}$ are defined in the
same way as the gauge-invariant perturbation of a general four-scalar,
see equation (\ref{gauge}), so:
\[
\delta    \rho   _{m}^{(gi)}=\delta   \rho    _{m}+\rho   _{m}^{\prime
}(B-E^{\prime    }),\quad     \delta    p^{(gi)}=\delta    p+p^{\prime
}(B-E^{\prime }),\quad \delta  \psi ^{(gi)}=\delta \psi +\psi ^{\prime
}(B-E^{\prime }),
\]%
where $\delta  p$ is the  perturbed pressure for a  specific component
and the gauge-invariant  three-velocity $\delta u_{i}^{(gi)}$ is given
by \cite {mukhanov}:
\[
\delta u_{i}^{(gi)}=\delta u_{i}+a(B-E^{\prime })_{|i}.
\]%
The physical meaning of the quantities which enter the gauge-invariant
equations  is  very  simple:  they  coincide  with  the  corresponding
perturbations in the longitudinal gauge.  From now on we will drop the
superscript  $(gi)$  since  we   will  always  be  dealing  only  with
gauge-invariant quantities.

We   can   now    write   the   gauge-invariant   linearly   perturbed
energy-momentum tensor:
\begin{eqnarray}
\delta    T_{0}^{\mathit{matter}\     0}    &=&\delta    \rho    _{m},
\label{emmatter} \\  \delta T_{i}^{\mathit{matter}\ 0}  &=&\left( \rho
_{m}+p_{m}\right)       a^{-1}\delta       u_{i},      \\       \delta
T_{j}^{\mathit{matter}\ i} &=&-\delta p_{m}\delta _{j}^{i},
\end{eqnarray}
\begin{eqnarray}
\delta T_{0}^{\psi  \ 0} &=&\omega a^{-2}\left(  \psi ^{\prime }\delta
\psi  ^{\prime  }-\psi  ^{{\prime   }^{2}}\Phi  \right)  ,  \\  \delta
T_{i}^{\psi  \ 0} &=&\omega  a^{-2}\psi ^{\prime  }\partial _{i}\delta
\psi  ,  \\  \delta  T_{j}^{\psi  \  i}  &=&\omega  a^{-2}\left(  \psi
^{{\prime  }^{2}}\Phi -\psi  ^{\prime }\delta  \psi  ^{\prime }\right)
\delta _{j}^{i},
\end{eqnarray}
\begin{eqnarray}
\delta   T_{0}^{\mathit{em}\   0}   &=&\left(  |\zeta   |\delta   \rho
_{m}+\delta  \rho _{r}\right)  e^{-2\psi  }-2e^{-2\psi }\left(  |\zeta
|\rho  _{m}+\rho  _{r}\right) \delta  \psi  ,  \\  \delta
T_{i}^{\mathit{em}\  0} &=&e^{-2\psi  }\left(  |\zeta |\rho  _{m}+\rho
_{r}+p_{m}+p_{r}\right)     a^{-1}\delta      u_{i},     \\     \delta
T_{j}^{\mathit{em}\     i}    &=&\left(     2\delta     \psi    \left(
p_{m}+p_{r}\right) e^{-2\psi }-\left( \delta p_{m}+\delta p_{r}\right)
e^{-2\psi }\right) \delta _{j}^{i}.
\label{emem}
\end{eqnarray}%
We   have  assumed   there  are   no  anisotropic   stresses   in  the
energy-momentum  tensors   and  we  have   considered  only  adiabatic
perturbations;  that is,  we  consider the  pressure perturbations  to
depend  only on  the  energy-density perturbations.  In  the dust  and
radiation   cases   this  means   $p_{m}=0$,   $\delta  p_{m}=0$,   or
$p_{r}=\frac{1}{3}\rho  _{r}$  and  $\delta p_{r}=\frac{1}{3}
\rho _{r}$, respectively.

\bigskip

The  fully  perturbed   gauge--invariant  Einstein  equations  can  be
obtained  from  (\ref{genpertein})  using  the expressions  above  for
$\delta T_{\nu }^{\mu }$ ( \ref{emmatter}-\ref{emem}). We have
\begin{eqnarray}  \label{einspert00}
\Phi  \left( -3\mathcal{H}^2 +  3k +  4G\pi \omega  {\psi ^{\prime}}^2
\right)    +    \triangledown^2    \Phi    -   4G\pi    \omega    \psi
^{\prime}\delta\psi^{\prime}-  3  \mathcal{H}\Phi^{\prime}= \nonumber
\\ 4G\pi {a}^2  e^{-2\psi }\left[ \left( e^{2\psi }  + |\zeta| \right)
\delta\rho_m   +   \delta\rho_r   -   2\delta\psi  \left(   \rho_r   +
|\zeta|\rho_m \right) \right]
\end{eqnarray}
\begin{equation}  \label{einspert01}
\mathcal{H} \triangledown^2  \Phi + \triangledown^2  \Phi^{\prime}= \\
4G\pi \omega \psi ^{\prime}\triangledown^2 \delta\psi -\frac{4G\pi}{3}
a e^{-2\psi  } \left[ 4\rho_r +  3\left( e^{2\psi }  + |\zeta| \right)
\rho_m  \right] \triangledown \delta u \nonumber
\end{equation}
\begin{eqnarray}  \label{einspert11}
\Phi \left( {\mathcal{H}}^2 +  2\mathcal{H}^{\prime}- k + 4G\pi \omega
{\psi     ^{\prime}}^2     \right)    +     3\mathcal{H}\Phi^{\prime}+
\Phi^{\prime\prime}=   \nonumber  \\   \frac{4G\pi}{3}   \left[  {a}^2
e^{-2\psi}\left(  \delta\rho_r -  2\rho_r\delta\psi \right)  + 3\omega
\psi ^{\prime}\delta\psi^{\prime}\right]
\end{eqnarray}

It  is  useful to  write  the  perturbed energy-momentum  conservation
equations for each component:
\begin{equation}  \label{psiperteq}
\delta  \psi ^{^{\prime  \prime }}=\frac{2|\zeta  |}{\omega }e^{-2\psi
}{a}^{2} \left[  \delta \rho _{m}+2\rho _{m}\left(  \Phi -\delta \psi
\right) \right] -2 \mathcal{H}\delta \psi ^{^{\prime }}+\triangledown
^{2}\delta \psi +4\psi ^{\prime }\Phi ^{^{\prime }},
\end{equation}
\begin{equation}  \label{radpert}
\delta   \rho   _{r}^{^{\prime   }}=-\frac{2}{3}\left[   \delta   \rho
_{r}\left(  6 \mathcal{H}-3\psi  ^{\prime  }\right) +\rho  _{r}\left(
2\triangledown  \delta u-3\delta  \psi ^{^{\prime  }}-6\Phi ^{^{\prime
}}\right) \right] ,
\end{equation}
\begin{equation}  \label{dustpert}
\delta  \rho   _{m}^{^{\prime  }}=-3\mathcal{H}\delta  \rho  _{m}-\rho
_{m}\left( \triangledown \delta u-3\Phi ^{^{\prime }}\right) .
\end{equation}

From these expressions it is clear that perturbations in $\alpha $ are
sourced by  perturbations in  the dust, but  not vice versa.  Hence we
expect that, in a dust-dominated  universe with varying $\alpha $, the
cold  dark matter  perturbations will  behave as  in  a dust-dominated
universe  with no  varying $\alpha  $.  Notice however  that the  same
cannot  be concluded so  easily for  a radiation-dominated  era, since
there  is a source  term in  equation (\ref{radpert})  proportional to
$\delta  \psi ^{\prime }$.  We expect  this term  to be  negligible at
large scales, but that might not be the case on small scales.

\bigskip

The   gauge-invariant  perturbation   for  $\alpha   $  is   given  by
equation (\ref{e3}) as
\[
\frac{\delta \alpha }{\alpha }=2\delta \psi .
\]%
From       equations      (\ref{einspert00}),      (\ref{einspert01}),
(\ref{einspert11}),  using (\ref{fried3}) to  simplify, we  obtain the
general form for $\delta \psi  ,$ the perturbation to the scalar field
which drives variations of the fine structure constant, as
\begin{eqnarray}
\delta \psi  &=&\frac{1}{8G\pi {a}^{2}\left( 2\rho  _{r}+3|\zeta |\rho
_{m}\right) }\left( 4G\pi {a}^{2}\left[ 2\left( \delta \rho _{r}-4\rho
_{r}\Phi \right) +3\left( e^{2\psi }+|\zeta |\right) \right. \right.
\label{psipert} \\
&&\left. \left. \left( \delta \rho _{m}-2\rho _{m}\Phi \right) \right]
+3e^{2\psi }\left[ 2\Phi  \left( 3\mathcal{H}^{2}-k-4G\pi \omega {\psi
^{\prime   }}^{2}\right)  -\triangledown   ^{2}\Phi  +6\mathcal{H}\Phi
^{^{\prime }}+\Phi ^{^{\prime \prime }}\right] \right) \nonumber
\end{eqnarray}%
This is a gauge-invariant expression  for $\delta \psi $, written as a
function  of  the gauge-invariant  quantities  $\Phi  $, $\delta  \rho
_{m}$, and $ \delta \rho _{r}$.

\section{Evolution of the Perturbations}

In a spatially flat universe, $\kappa =0$, the general gauge-invariant
expression for $\delta \psi $ becomes:
\begin{eqnarray}
\delta \psi  &=&\frac{1}{8G\pi {a}^{2}\left( 2\rho  _{r}+3|\zeta |\rho
_{m}\right) }\left( 4G\pi {a}^{2}\left[ 2\left( \delta \rho _{r}-4\rho
_{r}\Phi \right) +3\left( e^{2\psi }+|\zeta |\right) \right. \right.
\label{deltapsiflat} \\
&&\left. \left. \left( \delta \rho _{m}-2\rho _{m}\Phi \right) \right]
+3e^{2\psi  }\left[ \Phi  \left(  6\mathcal{H}^{2}-8G\pi \omega  {\psi
^{\prime  }  }^{2}\right)  -\triangledown ^{2}\Phi  +6\mathcal{H}\Phi
^{^{\prime }}+\Phi ^{^{\prime \prime }}\right] \right) .  \nonumber
\end{eqnarray}

It  was shown in  chapter \ref{qualitative} that  the cosmological
evolution of  the metric scale factor  is unchanged (up  to very small
logarithmic  corrections) to leading  order by  the time  evolution of
$\psi $.   The dominant effect is  that of the evolution  of the scale
factor on the evolution of $\psi $ through its propagation equation.

It is  known that perturbations  of massless scalar fields,  or scalar
fields  with  a  very  small  mass  are  negligible  with  respect  to
perturbations   in   the   matter   fields   and   the   gravitational
potential.  Guided  by  this,  in  order to  obtain  the  evolutionary
behaviour for  $\frac{\delta \alpha }{\alpha  }$, we will  assume that
the matter field perturbations, $\delta  \rho _{m}$ and $ \delta \rho
_{r}$, and  the metric perturbations,  $\Phi $, are unaffected  by the
$\psi  $ perturbations  to leading  order. We  will assume  that these
three quantities will therefore be the  same to this order as they are
in a flat  FRW universe filled with barotropic  matter and a minimally
coupled  scalar  field. These  assumptions  are  valid  if the  energy
density of  $\psi $  is much  smaller than the  energy density  of the
matter  fields,  so  $\Phi  $  will  be  driven  only  by  the  matter
perturbations.  If  we examine  the  perturbations  in the  non-linear
regime   it  is   confirmed  that   $\frac{\delta  \psi   }{\psi  }\ll
\frac{\delta         \rho         }{\rho         }$ (see         figure
(\ref{dustdeltaalpha})).\textbf{\ }

\subsection{Radiation-dominated universes}

In any radiation-dominated era in  which the expansion of the universe
is  dominated by  relativistic  particles with  an  equation of  state
$p=\frac{1}{3} \rho  $, we can neglect  the non-relativistic stresses
in  the  universe,  in  particular,  the  cold  dark  matter  and  the
cosmological constant  since $\rho _{r}>>\rho  _{m}>>\rho _{\Lambda }$
to  a good  approximation. If  we assume  that $\rho  _{\Lambda }=\rho
_{\Lambda  }=0=\kappa  $  and  $\delta \rho  _{m}=0$,  the  background
equations of  motion give the  usual conformal time evolution  for the
scale factor and the energy density of the radiation:
\[
\rho  _{r}=\rho   _{r_{0}}a^{-4}e^{2\psi  }\qquad  a=\sqrt{8\pi  G\rho
_{r_{0}}} \eta
\]%
where $\rho _{r_{0}}$ is a constant.

The  perturbations in the  barotropic matter  fluid and  the potential
$\Phi    $    come     from    the    equations    (\ref{einspert00}),
(\ref{einspert01}). The usual flat FRW solutions with constant $\alpha
$ are obtained  by setting the terms proportional to  $\psi $ to zero,
(\ref{einspert11}).  The general  solution of  these equations  can be
obtained  by  expanding  the  physical  quantities  in  terms  of  the
eigenfunctions of the operator $\triangledown ^{2} $ (where $ -k^{2}$
denotes the  eigenvalue of  this operator) and  solving for  each mode
separately. Resuming the terms,  the general solutions of the linearly
perturbed  equations for  the  potential $\Phi  $  and the  barotropic
matter are:
\begin{equation}
\Phi  =\eta ^{-3}\{\left[  w\eta  \cos (w\eta  )-\sin (w\eta  )\right]
C_{1}+   \left[   w\eta    \sin   (w\eta   )+\cos   (w\eta   )\right]
C_{2}\}e^{i\mathbf{kx}}
\label{radbackgroundphi}
\end{equation}%
and
\begin{eqnarray}
\frac{{\delta  \rho  _{r}}}{\rho  _{r}}  &=&\frac{4}{\eta  ^{3}}\left(
C_{1}\{\eta w\left[ 1-\frac{1}{2}\left( \eta w\right) ^{2}\right] \cos
(\eta  w)+\left[   \left(  \eta  w\right)   ^{2}-1\right]  \sin  (\eta
w)\}\right.
\label{radbackgroundrho} \\
&&+\{\left[  1-\left(  \eta w\right)  ^{2}\right]  \cos (\eta  w)+\eta
w\left[  1- \frac{1}{2}\left(  \eta w\right)  ^{2}\right]  \sin (\eta
w)\}C_{2})e^{i\mathbf{ kx}}, \nonumber
\end{eqnarray}%
where we  have expanded the general  solution in plane  waves since we
are  assuming  a spatially  flat  universe;  $C_{1}$  and $C_{2}$  are
arbitrary functions of the spatial coordinates; $k$ is the wave vector
mode  and  $w=k/  \sqrt{3}$.  Note  that  these  quantities  are  all
expressed in  their gauge-invariant format. Finally,  to calculate the
explicit  time  dependence of  $\delta  \psi $,  we  need  to use  the
background solution for $\psi $:
\begin{equation}
\psi   =\frac{1}{2}\log  {(8N)}+\frac{1}{4}\log  {(\frac{a_{0}}{2}\eta
^{2}).}
\label{radol}
\end{equation}%
This   was    found   in chapter \ref{qualitative} and  \cite{bsm2, mota1}   for   a
radiation-dominated  universe, where  $N=-\frac{2\zeta  }{\omega }\rho
_{m}a^{3}>0$  since  $\zeta  <0$  in  the  magnetic  energy  dominated
theories  considered  by BSBM.  It  is  important  to notice  that  in
universes with  an entropy to  baryon ratio ($S\sim 10^{9}$)  like our
own, $\psi $  does not experience any growth  in time \cite{bsm1}. The
constant term  on the right-hand  side of (\ref{radol})  dominates the
solution for  $\psi (\eta )$  throughout the radiation  era. Numerical
solutions confirm this freezing in of $\psi $, and hence of $\alpha $,
during the radiation era,\ \cite{bsbm}.

\subsubsection{Large-scale perturbations in a radiation-dominated era}

In  the long-wavelength  limit ($w\eta  <<1$) where  the scale  of the
perturbation  exceeds  the  Hubble  radius,  we  can  neglect  spatial
gradients.   So,  in  this  limit, $\frac{\delta  \alpha  }{\alpha  }$
becomes:
\[
\delta   \psi   =\frac{1}{2}\frac{\delta   \alpha  }{\alpha   }\propto
\frac{1}{2\eta      }e^{i\mathbf{kx}}\triangledown      ^{2}C_{2}-4\pi
Gwe^{i\mathbf{kx}}C_{2}\eta ^{-3}
\]

From  this  expression  we  can  see  at  that  on  large  scales  the
inhomogeneous perturbations in $\alpha  $ will decrease as a power-law
in time.  This behaviour agrees with  the one found  in \cite{bsm3} by
other methods.

\subsubsection{Small-scale perturbations in a radiation-dominated era}

On scales  smaller than the  Hubble radius ($w\eta >>1$)  the dominant
terms are proportional to $\triangledown ^{2}C_{1}$ and $\triangledown
^{2}C_{2}$,  so  the  asymptotic  behaviour for  $\frac{\delta  \alpha
}{\alpha }$ will be: \emph{\ }
\begin{eqnarray}
\delta   \psi  &=&\frac{1}{2}\frac{\delta  \alpha   }{\alpha  }\propto
-\frac{1}{2}w \left[ \cos  (\eta w)\triangledown ^{2}C_{1}+\sin (\eta
w)\triangledown           ^{2}C_{2}\right]           e^{i\mathbf{kx}}+
\label{deltaradsmall}   \\   &&+\frac{1}{2\eta   }\left[  \sin   (\eta
w)\triangledown ^{2}C_{1}-\cos  (\eta w)\triangledown ^{2}C_{2}\right]
e^{i\mathbf{kx}} \nonumber
\end{eqnarray}

On small  scales we  can see  the perturbations on  $\alpha $  will be
oscillatory. This behaviour is new and does not coincide with the ones
found in \cite{bsm3}.

In  reality, we expect  dissipation of  the adiabatic  fluctuations to
occur by Silk  damping \cite{silk} and small-scale fluctuations in  $\alpha $ will
also undergo decay as their driving terms damp out. However, while the
exact solution  for the evolution  of $\alpha $  is a linear sum  of a
constant and a slow power-law  growth, the power-law evolution does not
become dominant by the end of  the radiation era in universes like our
own with entropy per baryon $ O(10^{9}).$

\subsection{Dust-dominated universes}

In the  case of  a flat dust-dominated  universe, filled with  a $p=0$
fluid,  we  can assume  $\rho  _{r}=\rho  _{\Lambda  }=0=\kappa $  and
$\delta  \rho  _{r}=0$.   Again,   in  order  to  obtain  an  explicit
expression of  $\delta \psi  $ from equation  (\ref{deltapsiflat}), we
will  assume  that the  matter-field  perturbations,  $\rho _{m}$  and
$\delta  \rho _{m}$,  and the  metric perturbation,  $\Phi $,  are not
affected by the $\psi $  perturbations to leading order. Thus, we will
assume that  these functions  behave as in  a perturbed flat  FRW dust
universe.

In general, for a flat  dust universe we have the following background
solutions:
\[
\rho     _{m}=\rho     _{m_{0}}a^{-3},\qquad    a=\frac{2\pi     G\rho
_{m_{0}}}{3}\eta ^{2},
\]%
where $\rho _{m_{0}}$ is constant.

As before, we can calculate the most general gauge-invariant solutions
for the energy-momentum perturbations  $\delta \rho _{m},$ and for the
potential $ \Phi  $, assuming that the $\psi $  field does not affect
them to leading order, so their time dependences are \cite{mukhanov}:
\begin{equation}
\Phi =C_{1}+C_{2}\eta ^{-5} \label{dustbackground}
\end{equation}%
and
\begin{equation}
\frac{\delta  \rho   _{m}}{\rho  _{m}}=\frac{1}{6}\left[  \left(  \eta
^{2}\triangledown      ^{2}C_{1}-12C_{1}\right)      +\left(      \eta
^{2}\triangledown ^{2}C_{2}+18C_{2}\right) \eta ^{-5}\right] ,
\end{equation}%
where  $C_{1}$ and  $C_{2}$  are arbitrary  functions  of the  spatial
coordinates.

Once again, we  need the background solution for $\psi  $, in order to
calculate  $\delta \psi  $ as  an explicit  function of  the conformal
time. We use the asymptotic solution
\[
\psi =\frac{1}{2}\log {[2N\log {(\frac{a_{m}}{6}\eta ^{3})}],}
\]%
which was found  in \cite{bsm3}, as an asymptotic  approximation for a
dust-dominated  universe which  is  in good  agreement with  numerical
solutions,  and  where,  as  above,  $N=-\frac{2\zeta  }{\omega  }\rho
_{m}a^{3}>0$ is a constant.

\subsubsection{Large-scale perturbations in a dust-dominated era}

On  scales larger  than the  Hubble radius  we can  neglect  the terms
proportional  to  $\partial  _{i}C_{1}$,$\triangledown ^{2}C_{1}$,  $
\triangledown ^{2}C_{2}$ and $\partial _{i}C_{2}$; so in this limit we
have the following asymptotic behaviour for the non-decaying mode:
\begin{equation}
\delta   \psi   =\frac{1}{2}\frac{\delta   \alpha  }{\alpha   }\propto
-2C_{1}-\frac{    2\pi     G\rho    _{m_{0}}}{\ln    (\eta    )}C_{1}
\label{deltadustlarge}
\end{equation}%
Therefore, on large scales  the inhomogeneous perturbations of $\alpha
$ will not  grow in time by gravitational  instability. This behaviour
can be understood with reference to the general evolution equation for
$\psi  $  in  Friedmann  universes.  On large  scales,  where  spatial
gradients  in  $\psi $  can  be neglected  with  respect  to its  time
derivatives, we may view inhomogeneities  in density and in $\psi $ as
if  they  are  separate  Friedmann  universes  of  non-zero  curvature
($\kappa  \neq 0$).  The growth  of  inhomogeneity can  be deduced  by
comparing  the  evolution  of $\alpha  $  in  the  $ \kappa  \neq  0$
universes  with  those in  the  $\kappa  =0$  model (a  more  detailed
numerical  study  of  this  model  will  be  presented  in  section  5
below).  In   effect,  this  uses  the   Birkhoff-Newton  property  of
gravitational  fields  with  spherical  symmetry \cite{wald}.  We  note  that  the
(\ref{psiddot3}) evolution equation has the simple property that $\psi
$ cannot have a  maximum because $ \ddot{\psi}>0$ when $\dot{\psi}>0$
because  $N>0$ , \cite{bsm3}.  This result  holds irrespective  of the
value of  $\kappa \lesseqqgtr 0.$  Thus $\psi $  and $ \alpha  $ will
continue their  slow increase in  both over-densities, under-densities
and the flat background until we reach scales small enough for spatial
derivative  to come significantly  into play.  This has  the important
consequence  that we do  not expect  large spatial  inhomogeneities in
$\alpha  $ to have  developed. However,  it should  be noted  that the
sensitivity of the observations of varying-$\alpha $ effects in quasar
spectra is sufficient to  discern variations of redshift space smaller
than $  O(10^{-5})$, which is of  the same order as  the amplitude of
density fluctuation on very large scales in the universe. \emph{\ }

\subsubsection{Small-scale perturbations in a dust-dominated era}

On scales smaller than the  Hubble radius the dominant terms are those
proportional   to   $\triangledown   ^{2}C_{1}$   and   $\triangledown
^{2}C_{2}$, so the asymptotic behaviour for the growing mode will be:
\begin{equation}
\delta \psi \propto \frac{1}{12}\eta ^{2}\triangledown ^{2}C_{1}
\label{deltadustsmall}
\end{equation}%
This shows that  perturbations of $\alpha  $ will grow  on small
scales.   This result is  a product  of the  assumption that  on small
scales  the  universe  can  be   considered  as  being  filled  by  an
homogeneous  and isotropic  fluid, however  we know  this is  not true
below the scale where  gravitational clustering becomes non-linear. On
these small  scales we  also have to  worry about new  consequences of
inhomogeneity  which  have not  been  included  in  our analysis.  For
example,  the  constant parameter  $N=-\frac{2\zeta  }{\omega }
_{m}a^{3}\varpropto  \zeta \Omega  _{m}$  will vary  in  space due  to
inhomogeneity in the background matter density parameter $\Omega _{m}$
and  in the  dark  matter parameter  $-1\leq  \zeta \leq  1$. We  have
assumed that $ \zeta <0$ for  the cold dark matter on large scales in
order for  the cosmological consequences of time-varying  $\alpha $ to
be a small perturbation to  the standard cosmological dynamics. But on
small scales the dark matter will  be baryonic in nature and so $\zeta
>0$  there.  Hence, we  expect  $\zeta  $  to be  significantly  scale
dependent  as  we   go  to  small  scales.

\subsection{Accelerated expansion}

In  an era  of  accelerated expansion,  $\ddot{a}>0$,  as would  arise
during  inflation or  during a  $\Lambda $-  or quintessence-dominated
epoch at late  times, we can consider the scale factor  to evolve as a
power-law of the conformal time as $a=\eta ^{-n}$, where $n\geq 1$ and
$\eta $ \  runs from $ -\infty  $ to $0$. The case  of $a=\eta ^{-1}$
corresponds to a $\Lambda $ -dominated epoch.

As in the previous sections, we  will assume that all the other matter
components  which  fill the  universe  will  behave  exactly as  in  a
perturbed FRW universe with no variations of $\alpha $. We also assume
that neither of  the dust and radiation perturbations  will affect the
behaviour  and  evolution  of  $   \delta  \psi  $,  and  that  these
perturbations  are negligible with  respect to  the $\Lambda  $ stress
driving the  expansion, so we  will consider $\delta \rho  _{m}=0$ and
$\delta \rho _{r}=0$.

\bigskip

We saw in chapter \ref{qualitative}, that a universe which is
undergoing accelerated expansion, the  asymptotic solution for $\psi $
is a constant, so we will  assume that $\psi =\psi _{\infty }$, in the
background,  where $\psi _{\infty  }$ is  a constant.  Thus, equations
(\ref {einspert00}) and (\ref{einspert11}) become:
\[
\frac{8G\pi  |\zeta |\delta \psi  \rho _{m}}{e^{2\psi  _{\infty }}\eta
^{2n}}  +\triangledown  ^{2}\Phi  +\frac{3n\left(  \eta  \Phi  -n\Phi
^{^{\prime }}\right) }{\eta ^{2}}=0
\]%
\[
n\left(  2+n\right)  \Phi +\eta  \left(  \eta  \Phi ^{^{\prime  \prime
}}-3n\Phi ^{^{\prime }}\right) =0
\]%
where we  have also  considered $\rho _{r}=0$,  but $\rho  _{m}\neq 0$
because of the coupling with $\psi  $ in the equation of motion of the
scalar field.   Note that if  we had also  set $\rho _{m}=0$  here, we
would have imposed a no  $ \alpha $-variation condition: $\delta \psi
=0$.

Integrating the last equation, we obtain
\[
\Phi   =\eta   ^{\frac{3n-\sqrt{1+n\left(   5n-2\right)   }}{2}}\left(
\sqrt{\eta  }  C_{1}+\eta  ^{\frac{1}{2}+\sqrt{1+n\left(  5n-2\right)
}}C_{2}\right)
\]%
where  $C_{1}$ and  $C_{2}$  are arbitrary  functions  of the  spatial
coordinates. Note  that since $n>1$  in accelerating universes  we see
that $ \Phi $ will \textbf{decay in time as $\eta\rightarrow0$}. From
this solution for $\Phi $, we obtain:
\begin{eqnarray}
\delta     \psi     &=&\frac{1}{2}\frac{\delta     \alpha     }{\alpha
}=-\frac{e^{2\psi  _{\infty   }}}{16G\rho  _{m_{0}}\pi  |\zeta  |}\eta
^{\frac{-3+n-\sqrt{1+n\left(    -2+5n\right)   }}{2}}\left[   3n\left(
1+n-\sqrt{1+n\left(    5n-2\right)    }\right)    C_{1}+\right. \nonumber    \\
&&\left.   3n\left(  1+n+\sqrt{1+n\left(  5n-2\right)   }\right)  \eta
^{\sqrt{   1+n\left(   5n-2\right)  }}C_{2}+2\eta   ^{2}\triangledown
^{2}C_{1}+2\eta  ^{2+   \sqrt{1+n\left(  5n-2\right)  }}\triangledown
^{2}C_{2}\right] \nonumber
\end{eqnarray}%
which is  a decaying function of  the conformal time when  $n>1$ and a
constant  when $n=1$.  Thus, in  accord with  the expectations  of the
cosmic  no hair  theorem, the  universe approaches  the FRW  model and
$\alpha $ is asymptotically constant at late times \cite{nohair}.

\subsubsection{Large-scale perturbations during accelerated expansion}

On  scales larger  than the  Hubble radius  we can  neglect  the terms
proportional to the spatial derivatives;  so in this limit we have the
following asymptotic behaviour:
\begin{eqnarray}
\delta   \psi   &\propto   &-\frac{3e^{2\psi   _{\infty   }}n}{16G\rho
  _{m_{0}}\pi  |\zeta  |}\eta ^{\frac{n-3-\sqrt{1+n\left(  5n-2\right)
  }}{2}}\left[   \left(   1+n-\sqrt{1+n\left(   5n-2\right)   }\right)
  C_{1}\right.   \nonumber  \\  &&\left.  +\left(  1+n+\sqrt{1+n\left(
    5n-2\right)   }\right)   \eta   ^{\sqrt{  1+n\left(   5n-2\right)
  }}C_{2}\right]
\end{eqnarray}%
Therefore,  as expected,  on large  scales during  an  accelerated era
inhomogeneities in $\alpha $ will decrease on time when $n>1$ and will
be a constant when $n=1$.

\subsubsection{Small-scale perturbations during accelerated expansion}

On scales  smaller than the Hubble  radius the dominant  terms are the
ones  proportional  to  $\triangledown ^{2}C_{1}$  and  $\triangledown
^{2}C_{2}$, so the asymptotic behaviour will be
\begin{equation}
\delta  \psi \propto  -\frac{e^{2\psi _{\infty  }}}{8G\rho _{m_{0}}\pi
|\zeta  |} \eta ^{\frac{1+n-\sqrt{1+n\left(  5n-2\right) }}{2}}\left[
\triangledown     ^{2}C_{1}+\eta     ^{\sqrt{1+n\left(     5n-2\right)
}}\triangledown ^{2}C_{2}\right]
\end{equation}%
This confirms  that perturbations of  $\alpha $, as  $\eta \rightarrow
0$, will decrease on small scales when $n>1$ and will be constant when
$n=1$.

\subsection{\protect\bigskip Summary of behaviour}

In Table \ref{linearperturbations} we summarise the time evolution of small inhomogeneities in
$ \alpha $\ found under different conditions in this section.
%
\begin{table}[hbtp!]
\begin{center}
\begin{tabular}[htbp!]{ccc}
\hline  \textbf{Universal equation  of  state} &  \multicolumn{2}{c}
{\textbf{Time}    \textbf{Evolution     of    the    perturbations}}\\
&\multicolumn{2}{c}{{   $\delta   \psi   =$}$\frac{1}{2}\frac{\delta
\alpha  }{\alpha   }$}  \\  \cline{2-3}  &   \textit{Large  scales}  &
\textit{Small scales}  \\ \hline \textit{Radiation-dominated  epoch} &
Decaying & Oscillatory \\ $p=\frac{1}{3}\rho ,\qquad a\propto \eta $ &
&  \\ \hline  \textit{Dust-dominated epoch}  & Constant  &  Growing \\
$p=0,\qquad a\propto \eta ^{2}$ & & \\ \hline \textit{Accelerated expn
}$\mathit{{a\propto \eta  ^{-n},n\geq 1}}$\textit{\ } &  & \\ $\Lambda
$-dominated &  Constant & Constant  \\ $p=-\rho ,\qquad  a\propto \eta
^{-1}$ & &  \\ \cline{2-3} Power-law acceleration, $n>1$  & Decaying &
Decaying \\  $p=w\rho ,\qquad w<0,\qquad  a\propto \eta ^{-n}$ &  & \\
\hline
\end{tabular}
\caption{\label{linearperturbations} Time evolution of small inhomogeneities in $\alpha$ found for
the different epochs the universe went through.}
\end{center}
\end{table}

We  should  point out  that  those  results  were obtained  under  the
assumption  that  there is  no  back-reaction,  in  the matter  fields
perturbations,  due  to  the   perturbations  in  the  fine  structure
constant.  Back-reactions may have  an important effect, and should be
taken  into  considerations,   specially,  when  there  are  coupled
fluid-perturbations    in    the    perturbed    Einstein    equations
\cite{mukhanov1,abramo,geshnizjani}.  Perturbations in
a given fluid, and so  in its energy-momentum tensor, affect the
dynamics  of  spacetime. These  perturbations  would  then affect  the
evolution of  the different components  in the universe.  Although, if
the perturbations  in a given fluid  are very small and  do not change
much the spacetime  geometry, then we can neglect its  effects up to a
reasonable approximation.

Varying-$\alpha$ models have  a scalar field coupled to  the matter, a
proper  study of  the  possible back-reaction  should  then have  been
done. Although, there are very
good indications that perturbations in $\alpha$ will not affect at all
the matter inhomogeneities. The reason is the fact that the background
universe  is  not  affected  by  the inclusion  of  the  scalar  field
responsible for the variations of $\alpha$, because the energy density
associated to the  scalar field, when compared to  the matter fields, is
very small.

Nevertheless, this assumption should be  checked. In order to do that,
we  will use  a  simple toy  model  to mimic  the  situation when  the
perturbations in $\psi$ are large and so might affect the evolution of
the  other  perturbations. We  will  study  what is  called  the
non-linear regime  of the cosmological perturbations.  In this regime,
the  approximation of  using a  first order  perturbations,  about the
background,  is   no  longer  valid.   Both  the  matter   and  $\psi$
perturbations will be  no longer considered small with  respect to its
background values.

\section{The Non-Linear Regime - A Toy Model}

In order to study the evolution of inhomogeneities in $\alpha $ beyond
the domain  of linear perturbation theory  we need to  use a different
model. The  simplest approach is  to confine attention  to spherically
symmetric inhomogeneities. This will be done by comparing the solution
of the BSBM  theory for $\alpha $ in a  closed ($\kappa =1$) universe,
with the solution  for $\alpha $ in a flat  ($\kappa =0$) universe. We
are assuming  a Birkhoff property for  the BSBM theory so  that we can
treat the  perturbation as an  independent closed universe \cite{wald}. This  is a
standard  technique in  general  relativity which  was  first used  by
Lema\^{\i}tre \cite{lemaitre}.

We  define  the  alpha   'over-density  perturbation'  (which  is  not
necessarily small) by
\[
\frac{\delta \alpha }{\alpha  }\equiv \frac{\alpha _{\kappa =1}-\alpha
_{\kappa =0}}{\alpha _{\kappa =0}}
\]%
where $\alpha  _{\kappa }$ is  the solution of  equation (\ref{psieq})
for a universe with curvature $\kappa $.

\subsection{Radiation-dominated era}

There  is  no  non-linear  structure formation  during  the  radiation
epoch.  Nevertheless, it  is  interesting  to see  what  would be  the
magnitude of the perturbations in  $\alpha$ with respect to the matter
fields.  The  scale factor  for a radiation-dominated  closed ($\kappa
=1$) FRW  universe is given by  $a=\sin (\eta )$; for  a flat ($\kappa
=0$) FRW universe the normalised scale factor is given by $a=\eta $.

\begin{figure}[htbp!]
\centering
\epsfig{file=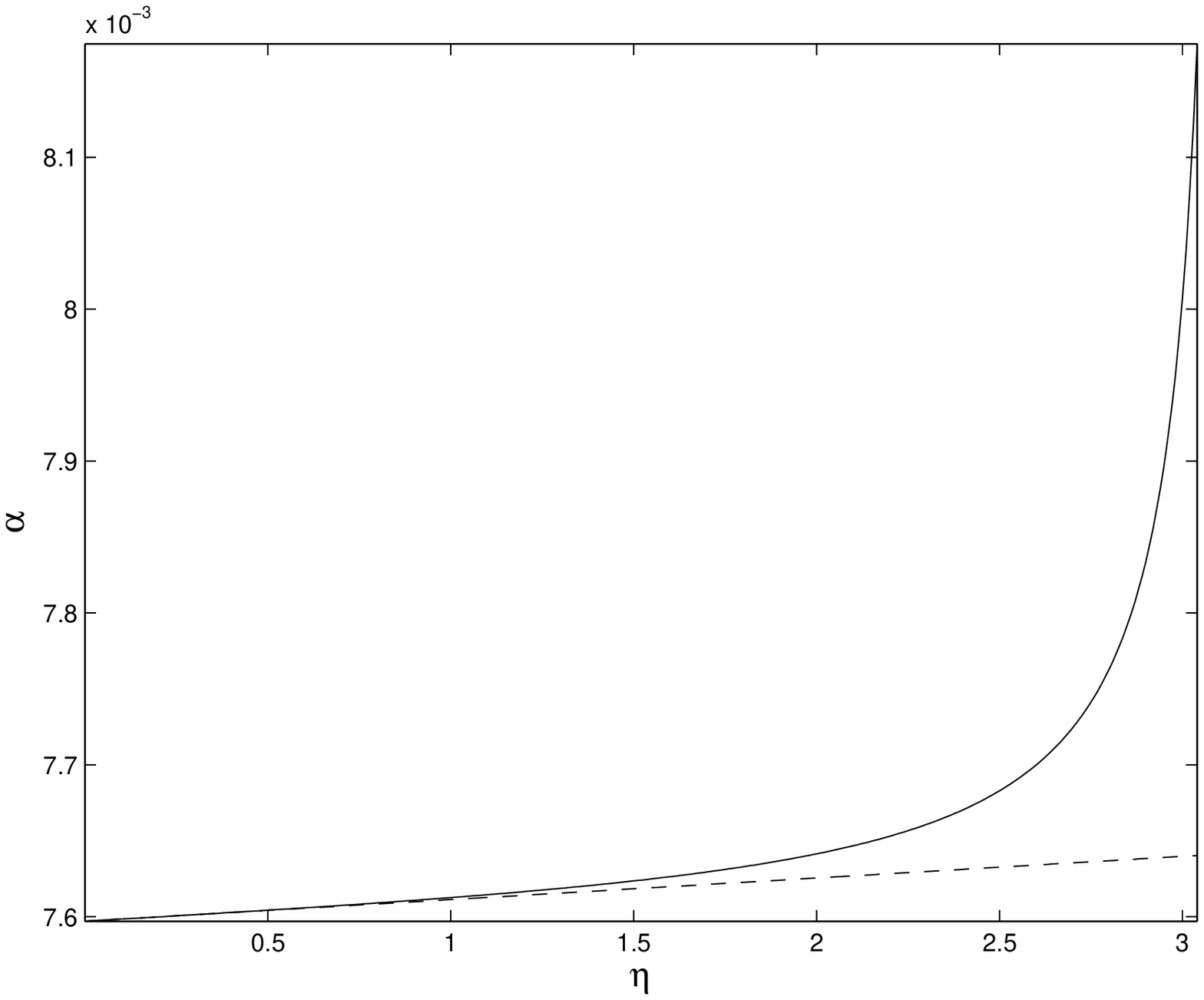,height=6.5cm,width=7.4cm}
\epsfig{file=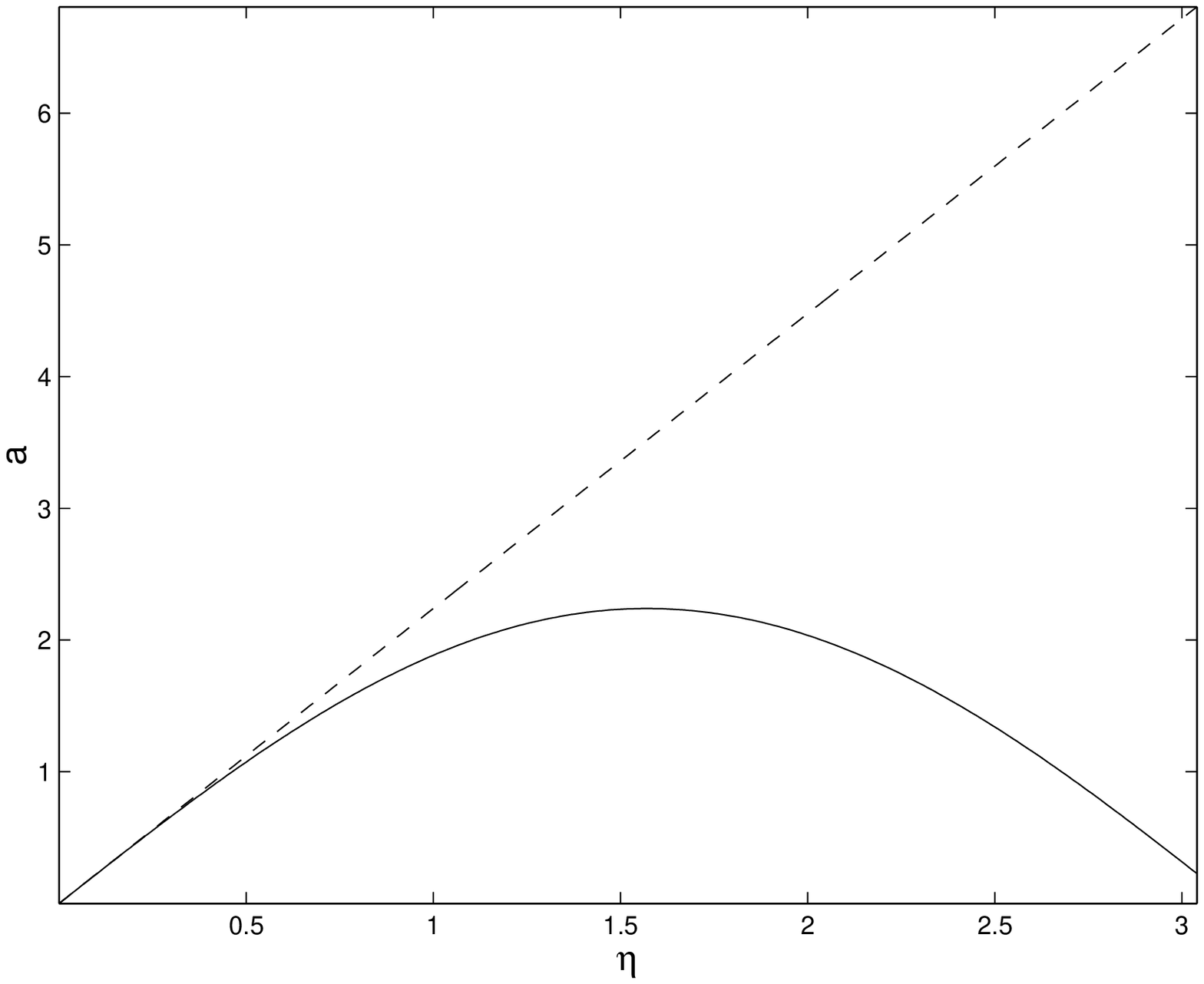,height=6.3cm,width=7.4cm}
\caption{{\protect       \textit{        The       evolution       of
$\protect\alpha(\protect    \eta)$    and    $a(\protect\eta)$    for
radiation-dominated universes  with $\protect\kappa=0$  (dashed) and
$\protect\kappa=1$ (solid).}}}
\label{radalpha}
\end{figure}

The evolution of $\alpha $  can be seen in figure \ref{radalpha} along
with the evolution of $a$ for $\kappa =1,0$.
\clearpage
As expected,  we can see  there is no  difference in the  evolution of
$\alpha $ at early times and $\alpha \propto \eta $ as it was found in
\cite{mota1}. When the difference between the scale factors of the two
universes becomes significant, the  behaviour of $\alpha _{\kappa =1}$
begins to  deviate from  that of $\alpha  _{\kappa =0}$.  We  see that
$\alpha  _{\kappa  =1}$ clearly  grows  faster  than $\alpha  _{\kappa
=0}$. The difference in the  growth rates become very significant near
the  expansion  maximum  of   the  bound  region  ($\eta  \approx  \pi
$). However, after this time our assumption that the background is not
affected by  changes of $ \psi  $ in the  cosmological equations that
describe the background universe breaks down, since the kinetic energy
of the scalar field will diverge and can no longer be neglected in the
Friedmann equation. We expect the behaviour near the final singularity
to  be similar  to the  kinetic-dominated evolution  near  the initial
singularity discussed in ref. \cite{bsm1}.

\begin{figure}[htbp!]
\centering
\epsfig{file=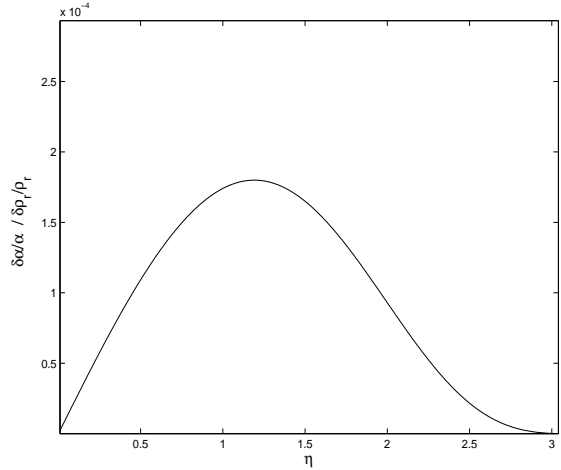,height=6.3cm,width=7.4cm}
\caption{{\protect  \textit{The  evolution  of  $\frac{\protect\delta
\protect\alpha}{\protect\alpha}/
\frac{\protect\delta\protect\rho_r}{\protect        \rho_r}$       vs
$\protect\eta$ in radiation-dominated universes. }}}
\label{raddeltaalpha}
\end{figure}

From figure \ref{raddeltaalpha} we  see that the variations of $\alpha
$ will become increasingly important  as $\eta $ approaches the second
singularity.
%
Notice  that   although  there  is   a  considerable  growth   in  the
perturbations  in  $  \alpha  $,   when  we  compare  them  with  the
perturbations in the radiation, they  are not as significant as can be
observed  from  the evolution  of  the  ratio  $ \frac{\delta  \alpha
}{\alpha  }/\frac{\delta \rho _{r}}{\rho  _{r}}$ versus  $ \eta  $ in
figure \ref{raddeltaalpha}.


\subsection{Dust-dominated era}

During the dust-dominated phase of a closed universe ($\kappa =1$) the
normalised scale  factor is given by $a=2\left(  1-\cos (\eta )\right)
$, while for a flat universe ($\kappa =0$) the normalised scale factor
is given  by $a=\ \eta  ^{2}$. Integrating equation  (\ref{psieq}) for
both cases we obtain the evolution of $\alpha $ for both cases.

\begin{figure}[htbp!]
\centering
\epsfig{file=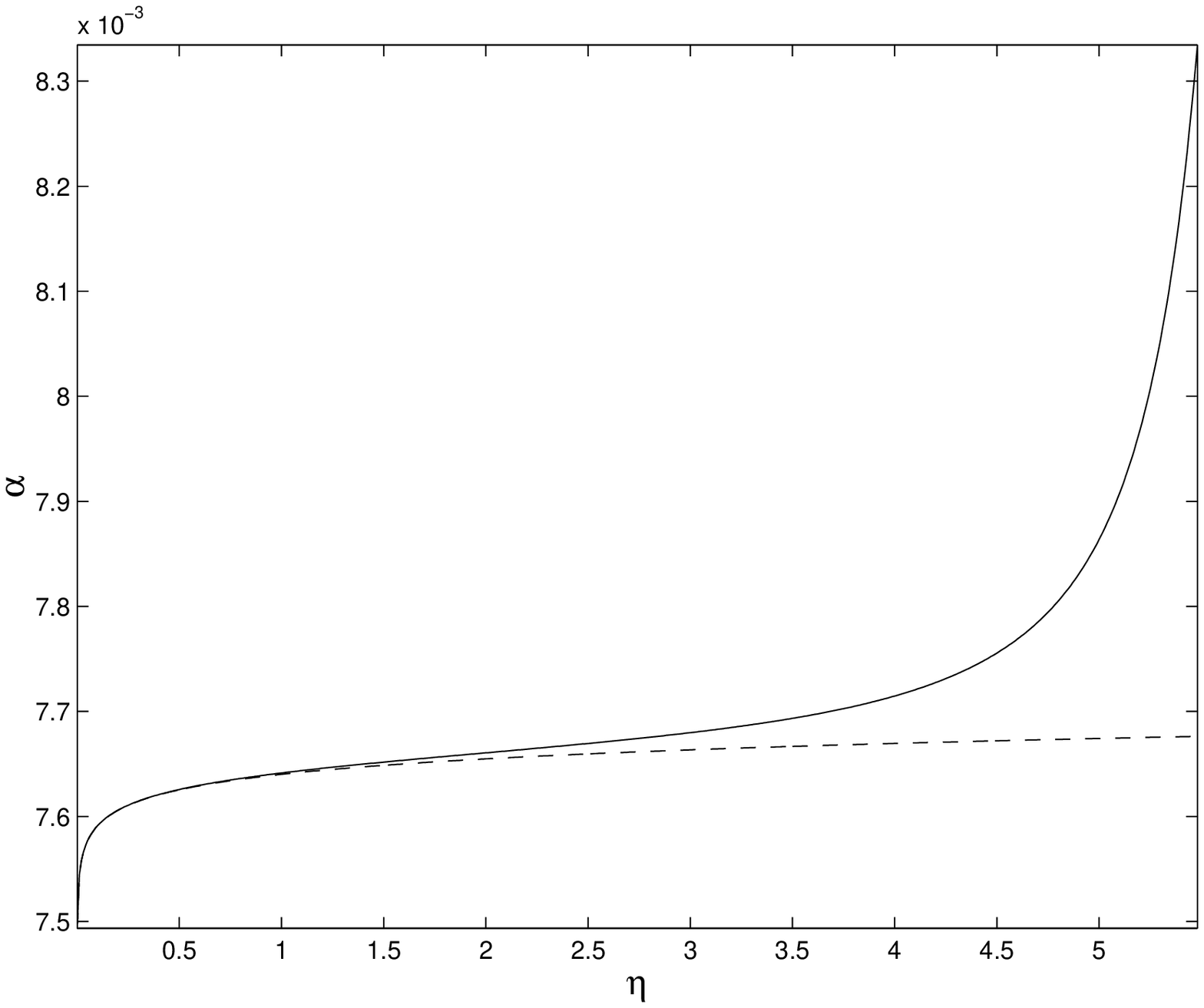,height=6.3cm,width=7.4cm}         
\epsfig{file=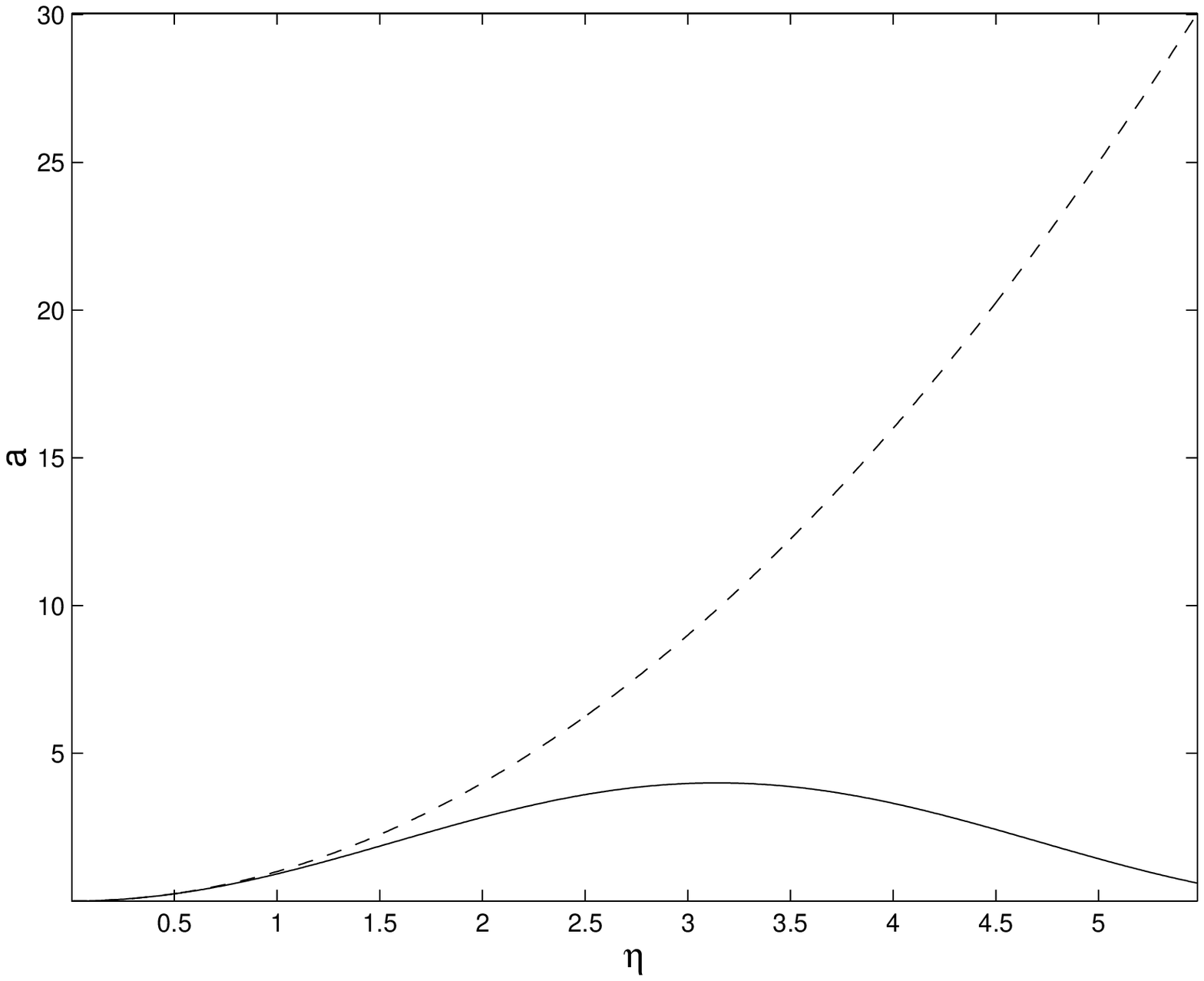,height=6.3cm,width=7.4cm}
\caption{{\protect \textit{The  evolution of $\protect\alpha(\protect
\eta)$  and   $a(\protect\eta)$  for  dust-dominated   universes  with
$\protect \kappa=0$ (dashed) and $\protect\kappa=1$ (solid). }}}
\label{dustalpha}
\end{figure}

The   evolution  of   $\alpha  (\eta   )$  can   be  seen   in  figure
\ref{dustalpha}  along with the  evolution of  $a(\eta )$  for $\kappa
=1,0$.
As in the radiation case, we can see there is little difference in the
evolution of  $ \alpha $ at  early times, since the  scale factor for
the closed model evolves very similarly to the flat one for $\eta <<1$
and $\alpha  \propto \ln  (\eta ) $.  The differences  between $\alpha
_{\kappa =1}$  and $\alpha  _{\kappa =0}$ cases  start to  appear when
$\eta  \approx   1$,  when  the  nonlinear   regime  commences.  These
differences become  more accentuated near the  second singularity, but
once again this is the region where our approximations break down.

\begin{figure}[htbp!]
\centering
\epsfig{file=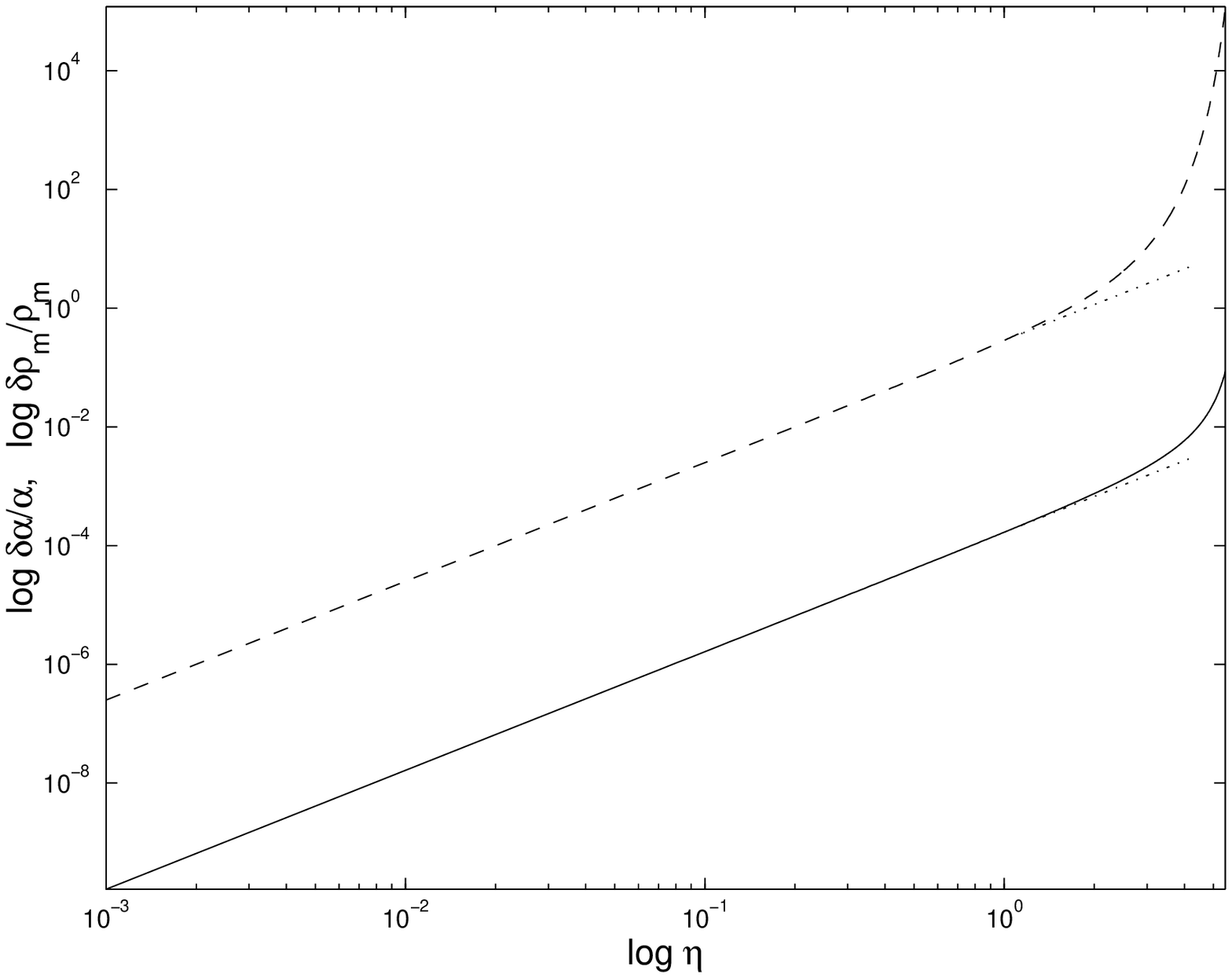,height=6.3cm,width=7.4cm}      %
\epsfig{file=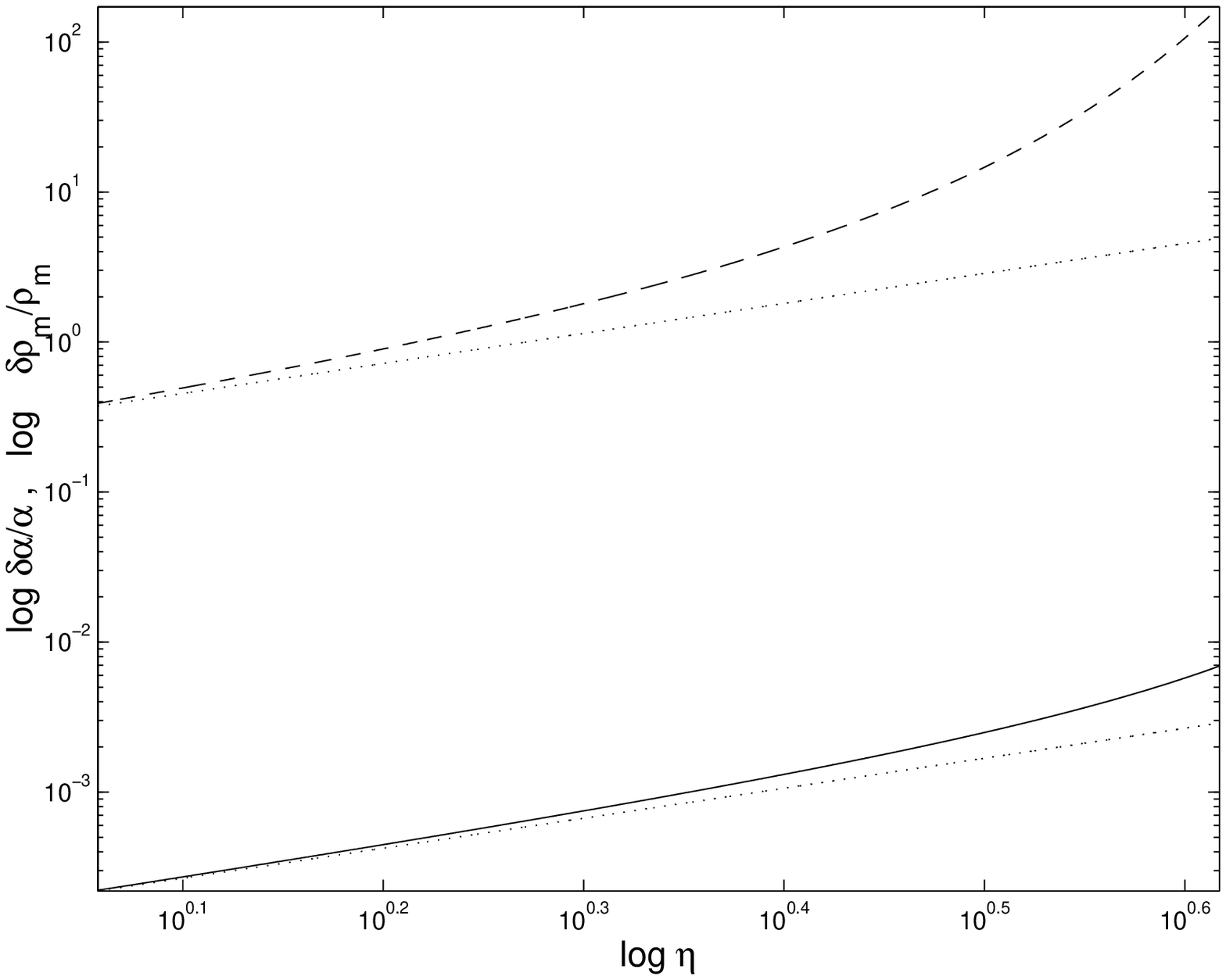,height=6.3cm,width=7.4cm}
\caption{{\protect      \textit{The       evolution      of      $\log
\frac{\protect\delta   \protect\alpha}{\protect\alpha}$  (solid)  and
$\log  \frac{\protect\delta \protect\rho_m}{\protect\rho_m}$ (dashed)
vs  $\log(\protect\eta)$  for a  dust-dominated  universe. The  dotted
lines correspond to evolution in the linear perturbation solution. }}}
\label{dustdeltaalpha}
\end{figure}

Notice  that   although  there  is   a  considerable  growth   in  the
perturbations  in  $  \alpha  $,   when  we  compare  them  with  the
perturbations in the cold dark  matter, they are not as significant as
can  be observed  from  the evolution  of  $\log (\frac{\delta  \alpha
}{\alpha })$  and $\log (\frac{\delta \rho _{m}}{  \rho _{m}})$ $vs.$
$\log (\eta )$ in figure \ref{dustdeltaalpha}.
In this case perturbations in $\alpha $ are even less significant than
in the radiation case. We note that the fact that the linear regime is
a very  good approximation at early  times, since as can  be seen from
figure \ref {dustdeltaalpha}, $\frac{\delta \alpha }{\alpha }$ tracks
$\frac{\delta   \rho  _{m}}{\rho   _{m}}\propto   t^{2/3}$  at   early
times. From  the detail of  figure  \ref{dustdeltaalpha} we  see that
variations of  the cold  dark matter will  start to occur  before than
variations of  $\alpha $, and $\frac{\delta \rho  _{m}}{\rho _{m}}$ is
at least three orders of  magnitude bigger than $\frac{ \delta \alpha
}{\alpha }$.  This is an  indication that perturbations in  the matter
fields will drive perturbations in $\alpha$.

\section{Summary and Comments}

The results presented in this chapter, are a quite  good  approximations when  we
consider large scales, but are expected to break down when extended to
small  scales,  where non-linear  effects  come  into  play and  local
deviations  from isotropy and  homogeneity of  the matter  content are
significant. We note that the  background solutions for $\psi $, about
which we have linearised  the perturbations of the Einstein equations,
are  solutions which  describe  the time  evolution  of $\psi  $ on  a
'standard'  FRW background. Our  neglect of  the back-reaction  of the
$\psi $ perturbations  on the background expansion dynamics  is a good
approximation up to logarithmic corrections.

When we examined the non-linear evolution of spherical inhomogeneities
by means of a comparative numerical study of flat and closed Friedmann
models  we  found  that  perturbations  in  $\alpha  $  remain  almost
negligible with  respect to perturbations in the  fluid that dominates
the energy  density of the universe  at the same  epoch. Comparing the
flat with the closed solution for $\alpha $, we concluded that in both
the  cases of  a radiation  or  dust-dominated epoch,  $\alpha $  will
'feel'  the  change  of  behaviour   in  the  scale  factor,  and  the
perturbations $\frac{\delta \alpha }{\alpha  }$ will grow in time. The
early-time behaviour  of the non-linear solutions  confirms the linear
behaviour  found  previously.  In particular $\frac{\delta  \alpha  }{
\alpha }$  changes in  proportion to $\frac{\delta  \rho }{\rho  }$ at
early times.

We have provided a detailed analysis of the behaviour of inhomogeneous
perturbations  in $\alpha  $ and  its time  variation on  large scales
under the  assumption that the  defining constant of the  BSBM theory,
$\zeta $, is constant and negative in sign. In reality this assumption
will break down on small scales. The negativity of the effective value
of $\zeta  $ requires that  the cold dark  matter is dominated  by the
magnetic  rather  than  the   electric  field  energy  (see  also  the
discussion of reference \cite{bsbm} and by Bekenstein \cite{bek3}). However,
on sufficiently small scales the  dark matter will become dominated by
baryons and  the sign of  the effective $\zeta  $ will have  to change
sign.  Overall,  there  will  also  be  a gradient  in  the  value  of
$\left\vert \zeta \right\vert $ reflecting the scale dependence of the
relative  contribution of  dark matter  to  the total  density of  the
universe.  We   have  not  included  these  effects   in  the  present
analysis. They would need \ to be included in any detailed analysis of
the small-scale behaviour of inhomogeneities  in $\alpha .$ This is an
important challenge for future work because it would enable the quasar
data on varying $\alpha $ to be compared directly with the limits from
the  Oklo natural  reactor  \cite {fuj,  damour1} and  Rhenium-Osmium
abundances in meteorites \cite{dyson, olive1}. At present the relation
between the  cosmological and geonuclear evidences  is unclear because
the latter  are derived from  physical processes occurring  within the
cosmologically non-evolving solar system environment.

Variations of  $\alpha $ also  affect the cosmic  microwave background
radiation spectrum  and anisotropy in different ways,  but the effects
must  be  disentangled  from  allowed changes  in  other  cosmological
parameters that  can contribute similar effects. These  changes in the
microwave  background  with  $  \alpha  $ left  as  a  free  constant
parameter  were   analysed  in  \cite{martins}  using   the  new  WMAP
data. They are far less  sensitive that the many-multiplet analyses of
quasars at  $z=0.5-3.5$ \cite{murphylast,murphy,webb,webb1}, although  they derive
from  higher  redshifts,   $z<1100$.  These  studies  can  accommodate
constant and  varying $\alpha $  but up to  a level that would  be too
large  to   be  consistent   with  the  quasar   data  and   the  slow
time-evolution of the theory with time-varying $\alpha $\ described in
this thesis.

%% file: chapter41.tex
\chapter{Spherical Collapse Model in the Presence of Varying Alpha}
\label{realistic}

\begin{flushright}
{\it  {\small ...\'e  neste horizonte  do  instante que  se espelha  a
liberdade,}}\\ 
{\it {\small um rosto sem fundo sob um sil\^encio mudo:
o \'eco da temporalidade.}}\\  
{\it {\small -- Ernesto Mota --}}
\end{flushright}

\bigskip

\section{Introduction}


In  the last  chapter, using  the  linear theory  of the  cosmological
perturbations, it  was shown that small inhomogeneities  in the matter
fields will induce small perturbations in the fine structure constant.
Those  results just confirmed  our initial  intuition that  one cannot
neglect the  effects of gravity  in the cosmological evolution  of the
fine structure  constant. The  reason is simple:  any varying-$\alpha$
theory implies the existence of  a field responsible for variations of
the fine structure constant. This  field is necessarily coupled to the
matter   fields,    at   least,    to   the   ones    which   interact
electromagnetically.   Variations   of  $\alpha  $,   allowed  by  the
conservation of energy and momentum  will then depend on the expansion
of the  universe and the evolution of  the electromagnetically coupled
matter.   Any  inhomogeneities of  the later  will then
affect the  evolution of $\alpha$. 
In particular,  spatial variations of  the matter
fields  will then  produce spatial  variations of  the  fine structure
constant. In  reality, we will see  in this chapter that  not only the
spatial  homogeneity  of  $\alpha$   is  affected  by  the  growth  of
inhomogeneities  in matter, but  also its  time evolution  will change
according to it.   This effect is model  independent since it is
just a product  of the coupling between the  field responsible for the
variations of $\alpha$ and matter.
The inclusion of electroweak or grand unification theories will create
even  more general  couplings, since  the field,  responsible  for the
$\alpha$-variations, may  also be couple to  all or some  of the other
matter fields, besides the ones that interact electromagnetically. Due
to this, additional consequences will then result \cite{lang,banks}.

\bigskip

It  is useful  to  recall  once again  the  present observational  and
experimental constraints that  a successful varying-$\alpha$ model has
to  satisfy.   We  saw   in  chapter  \ref{introduction},  that
these constraints can be
divided into two main groups: local and astro-cosmological. The local
constraints derive from experiments in our local bound gravitational system:
the Oklo natural reactor $(z=0.14)$, where $\left\vert \frac{\Delta \alpha }{%
\alpha }\right\vert \leq 10^{-7}$ \cite{fuj}; the intra solar-system decay
rate $^{187}\mathrm{Re}~\rightarrow ~^{187}\mathrm{Os}$, $(z=0.45)$, where $%
\left\vert \frac{\Delta \alpha }{\alpha }\right\vert \leq 
10^{-7}$ \cite{olive}; and the stability of terrestrial atomic clocks $(z=0)$%
, where $\left\vert \frac{\dot{\alpha}}{\alpha }\right\vert <4.2\times
10^{-15}~\mathrm{yr}^{-1}$\cite{marion}. Other limits arise from weak
equivalence experiments \cite{nordtvedt} but the limits they provide are more
model dependent. The astro-cosmological constraints are: the quasar
absorption spectra $z=0.5-3.5$, where  $\left\vert \frac{%
\Delta \alpha }{\alpha }\right\vert \approx 10^{-6}$; the
cosmic microwave background radiation $(z=10^{3})$, where $\left\vert \frac{%
\Delta \alpha }{\alpha }\right\vert <10^{-2}$ \cite{martins} and Big Bang
nucleosynthesis $(z=10^{8}-10^{10})$, where $\left\vert \frac{\Delta \alpha 
}{\alpha }\right\vert \leq 2\times 10^{-2}$ \cite{martins}.

There is a potential discrepancy between local and astro-cosmological
constraints; in particular, between the constraint of $\Delta \alpha /\alpha
\leq 10^{-7}$ at redshift $z=0.45$, coming from the $\beta $-decay in
meteoritic samples \cite{olive}, and the explicit variation in $\alpha $ of $%
\Delta \alpha /\alpha \approx 10^{-6}$ at $z=0.5$, coming from the
low-redshift end of the quasar absorption spectra \cite{murphy}. A
successful theory of varying $\alpha $ needs to explain this difference.
This is a challenge. Models which use a very light scalar field, to drive
variations in $\alpha $, need an extreme fine tuning in order to satisfy the
phenomenological constraints coming from geochemical data (Oklo, $\beta $%
-decay), the present equivalent principle tests, and the quasar absorption
spectra, simultaneously \cite{damour1}.

We will see in this chapter,  that just by looking at the evolution of
$\alpha$  in a universe  where non-linear  structures are  formed, the
discrepancy  between  geochemical  constraints  and  the  quasar  data
becomes ``naturally'' justified.


\section{The Spherical Collapse Model}

In order to study the  behaviour of the fine structure constant during
the formation  of non-linear overdensities  we will use  the spherical
infall model \cite{bookpeebles,kolb,carsten}. This will give us an idea on how
the evolution of $\alpha$ in  the overdensity ``breaks away'' from the
corresponding one in the background.

An overdense sphere  is a very useful non-linear  model, as it behaves
in  exactly  the  same  way  as a  closed  sub-universe.  The  density
perturbations  need  not  to  be  a uniform  sphere:  any  spherically
symmetric perturbation will  evolve at a given radius  in the same way
as  a uniform  sphere  containing the  same  amount of  mass \cite{bookpeebles}. In  what
follows,  therefore, density  refers to  {\it{mean}} density  inside a
given sphere.

Let us consider a  spherical perturbation with constant density inside
it which, at an initial time, has an amplitude $\delta_i>0$ and $\vert
\delta_i\vert  \ll 1$.  For the  case of  a flat  matter-dominated FRW
universe it is possible to find an exact solution for the evolution of
the overdensity, although that is not the case for more complex cases,
like ours,  and a numerical  simulation is required.  Nevertheless the
general behaviour of  the overdensity will be similar.  At early times
the sphere  expands together with  the background. For  a sufficiently
large  $\delta_i$ gravity prevents  the sphere  to expand  forever. In
that case, there  are three interesting epochs in  the final stages of
its development.  {\it{Turnaround}}: The  sphere breaks away  from the
general expansion  and reaches  a maximum radius.  {\it{Collapse}}: If
only gravity operates, then the  sphere will collapse to a singularity
where  the densities of  the matter  fields would  go to  infinity. In
practise the  collapse will never happen. As  overdensities break away
from the  background expansion,  {\it Virialisation} occurs,  and they
become gravitationally  bounded systems  through a process  of violent
relaxation  \cite{lynden,shu}.  The  underlying  mechanism of  violent
relaxation  is  the  chaotic  gravitational  field  of  a  collapsing,
non-spherical  self-gravitational system. The  time and  space varying
gravitational field provides the means for individual, non-interacting
particles to change  their energies and to become  well mixed in phase
space. After only a  few dynamical times, ($\tau\sim (G \rho)^{-1/2}$)
the result  is a virialised  distribution of matter whose  phase space
distribution is  roughly Maxwellian and whose density  varies as $\sim
r^{-2}$ \cite{bookpeebles}. Due to its spherical symmetry and isothermal solution, such a
configuration  is  referred  as  an isothermal  sphere.  In  addition,
baryonic matter  can, through  dissipative processes, lose  energy and
condense even further into cores of the isothermal spheres that form.
The process of  structure formation is much more complex and
we  shall not  attempt to  describe it  here, referring  the  reader to
\cite{kolb,bookpeebles}.


With virialisation, a condition of equilibrium is then reached and the
system becomes stationary: There are  no more variations of the radius
of the system, or in  its energy components.  The stationary condition
reached by the process of virialisation is very important, and affects
dramatically the variations of the fine structure constant, as we will
see bellow.

\bigskip

Once again we  will use the BSBM theory in order  to study the effects
of structure formation in the  evolution of $\alpha$. We will assume a
$\Lambda$-Cold-Dark-Matter   ($\Lambda   CDM$)   model  of   structure
formation.

The background evolution  of the universe and its  components is given
by   the   equations   (\ref{fried}),  (\ref{psidot})   and   equation
(\ref{mat}).  This  equations  will  describe  the  evolution  of  the
background scale factor, $a$, the scalar field, $\psi$, responsible for
the variations of $\alpha$  and the energy density of non-relativistic
matter $\rho_m$, respectively.

Just to clarify to notation from  now on, we will label some variables
which ``live  in'' the background  with a $b$. Quantities  which ``live''
inside the overdensities  will be labelled by $c$.  For instance, the
value   of   $\alpha$  inside   a   cluster   will   be  referred   to
$\alpha_c$.  Quantities   which  represent  the  epoch   at  or  after
virialisation,   will   be  labelled   by   a   $v$.  For   instance,
$\Delta\alpha_v/\alpha_v$, represents  the ``time density  contrast'' at
virialisation. This quantity represents a comparison between  the
value of the fine structure constant in
a  virialised  clusters or  in the background, at a given redshift,
and its value on Earth, $\alpha_0$.

\subsection{Dynamics of the overdensities}

Variations  of  $\alpha$ inside  overdense  spherical  regions can  be
studied using the spherical infall model. The evolution of a spherical
overdense  patch of  scale radius  $R(t)$  is given  by the  Friedmann
acceleration equation:
\begin{equation}
\frac{\ddot{R}}{R}=-\frac{4\pi   G}{3}    \left(   \rho   _{cdm}\left(
1+\left\vert  \zeta \right\vert  e^{-2\psi _{c}}\right)  +4\rho _{\psi
_{c}}-2\rho _{\Lambda }\right) \label{rcluster}
\end{equation}%
where  $\rho _{cdm}$  is the  total density  of cold  dark  matter and
baryons in the  cluster, $\psi_c$ is the homogeneous  field inside the
cluster, $\rho_{\Lambda}$ is  the cosmological constant density and 
$\rho _{\psi_{c}}\equiv \dot\psi_c^2/2$. We
have used the equations of state $p_{\psi _{c}}=\rho _{\psi _{c}}$, $%
p_{cdm}=0$ and $p_{\Lambda }=-\rho  _{\Lambda }$, where $p$ represents
the pressure of the fluid.

In the  cluster, the  evolution of $\psi  _{c}$ and $\rho  _{cdm}$ are
given by
\begin{eqnarray}
\ddot{\psi  _{c}}+3\frac{\dot{R}}{R}\dot{\psi _{c}}&=&-\frac{2}{\omega
} e^{-2\psi _{c}}\zeta  \rho _{cdm} \label{psidotcluster} \\ \dot\rho
_{cdm} &=& - 3 \frac{\dot R}{R} \rho_{cdm}
\label{rhocluster}
\end{eqnarray}

\bigskip

According  to the  {\it{Virial theorem}},  a condition  of equilibrium
will be reached when
\begin{equation}
T=\frac{1}{2} R \frac{\partial U}{\partial R}
\label{virialtheorem}
\end{equation}
where, $T$ is the average total  kinetic energy and $U$ is the average
total potential  energy in the sphere.  Note that we  obtain the usual
$T=\frac{n}{2} U$  condition, for any  potential with a  powerlaw form
($U \propto R^n$), which will be our case.

The potential energy for a  given component $x$ can be calculated from
its general expression in a spherical region \cite{landaubook}:
\begin{eqnarray}
U_{x}&=&2  \pi  \int_{0}^{R}  \rho_{tot}  \phi_{x} r^2  dr  \\  \phi_x
\left(r\right)&=& -2 \pi G \rho_x \left(R^2 - \frac{r^2}{3}\right)
\label{generalpotential}
\end{eqnarray}
where  $\rho_{tot}$ is  the total  energy density  inside  the sphere,
$\phi_{x}$ is the gravitational  field potential due to the $\rho_{x}$
density component.

In the case of a $\Lambda CDM$ model the potential energies inside the
cluster would be:
\begin{eqnarray}
U_{G}&=&-\frac{3}{5} G  M^{2} R^{-1}\\ U_{\Lambda  }&=&-\frac{4}{5} \pi G
\rho _{\Lambda  } M R^{2}\\  U_{\psi _{c}}&=&-\frac{3}{5} G  M M_{\psi
_{c}} R^{-4}
\label{potential}
\end{eqnarray}
where  $U_{G}$ is  the potential  energy associated  with  the uniform
spherical overdensity, $U_{\Lambda }$ is the potential associated with
$\Lambda ,$ and $U_{\psi _{c}}$ is the potential associated with $\psi
_{c}.$ $M = M_{cdm} + M_{\psi _{c}}$ is the cluster mass, with
\begin{eqnarray}
M_{cdm}  &=& \frac{4  \pi }{3}  \rho _{cdm}(1  + |\zeta  |  e^{-2 \psi
_{c}}) R^{3}\\  M_{\psi_{c}}& =& \frac{4  \pi }{3} \rho _{  \psi _{c}}
R^{6}
\label{masses}
\end{eqnarray}

\bigskip

The virial theorem would be satisfied when:
\begin{equation}
T_{vir}=-\frac{1}{2}U_{G}+U_{\Lambda }-2U_{\psi _{c}} \label{virial}
\end{equation}
where  $T_{vir}$  $=\frac{1}{2}  M  \bar{v}_{vir}^{2}$  is  the  total
kinetic  energy  at   virialisation  and  $\bar{v}_{vir}^{2}$  is  the
mean-square velocity of masses in the cluster.

Using the virial theorem (\ref{virial}) and energy conservation at the
turn around and when the  cluster virialises, we obtain an equilibrium
condition in term of the potential energies
\begin{equation}
\frac{1}{2}U_{G}(z_{v})+2U_{\Lambda    }(z_{v})-U_{\psi   _{c}}(z_{v})
=U_{G}(z_{ta})+U_{\Lambda }(z_{ta})+U_{\psi _{c}}(z_{ta})
\label{virialcond}
\end{equation}%
where $z_{v}$  is the  redshift of virialisation  and $z_{ta}$  is the
redshift at the turn-around of the over-density at its maximum radius,
when $R=R_{max} $  and $\dot{R}\equiv 0$. In the  case where $\zeta=0$
and $\psi=\dot\psi=0$  we just have the  usual virialisation condition
for $\Lambda CDM$ models \cite{lahav,wang}.

\bigskip

\begin{figure}[Hhtbp!]
\centering
\includegraphics[height=7.0cm,width=10cm]{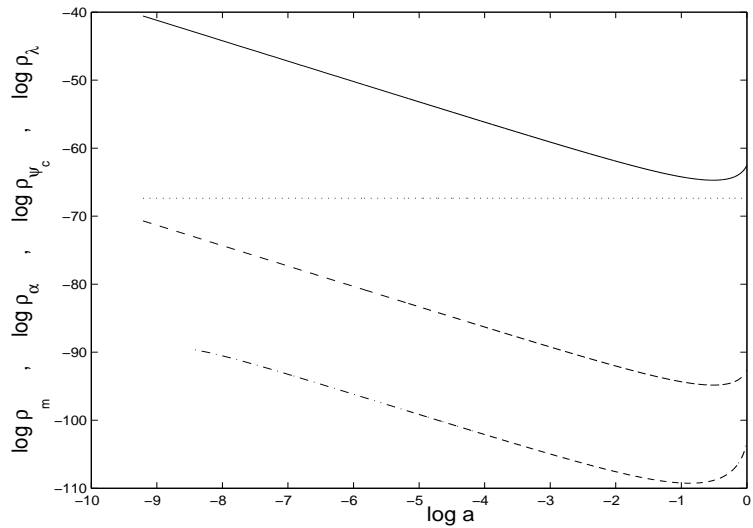}
\caption{{\protect {\it Evolution of $\log \rho_{cdm}$ (solid lines) ,
$\log \rho_{\psi_c}$ (dash-dotted line), $\log \rho_{\lambda}$ (dotted
line) and $\log \rho_{\alpha}$ (dashed line) inside an overdensity, as
a  function   of  $\log  (a)$,   in  a  $\Lambda  CDM$   model.  Where
$\rho_{\alpha}\equiv   \rho_{\psi_c}   +  \vert\zeta\vert   \rho_{cdm}
e^{-2\psi_c}$. }}}
\label{cluster}
\end{figure}

The  inclusion of  a  varying $\alpha  $  at a  level consistent  with
observation          \cite{murphylast,murphy,murphy1,webb,webb1,drink},
($\dot{\alpha}/\alpha _{0}\sim 10^{-6}H$), does not affect the overall
expansion of the universe up to logarithmic corrections, if we are far
from the initial  singularity \cite{bsm1}. The reason for  that is the
negligible  contribution  of  $\psi$  to  the energy  content  of  the
universe,  see  figure \ref{backgroundrho}.   In  a  similar way,  the
dynamical collapse of the overdense  regions is also not affected: The
energy  density  associated with  $\psi_c  $  is  always a  negligible
contribution  to  the energy  content  of  cluster.  Even the  kinetic
energy,  $\rho_{\psi_c}=\frac{\omega}{2}\dot\psi^2$,  which starts  to
grow in the final stages  of the cluster evolution will be negligible,
see Figure \ref{cluster}.

\section{Evolution of $\alpha$ During the Cluster Formation}

The behaviour of the fine structure constant during the evolution of a
cluster  can  now  be  obtained numerically  evolving  the  background
equations (\ref{fried})-(\ref{mat})  and the cluster  equations (\ref
{rcluster})-(\ref{rhocluster})  until  the virialisation  condition
holds, equation (\ref{virialcond}).

\subsection{Setting the initial conditions}

In order to satisfy the constraints imposed by the observations, we
need to set up initial conditions for the evolution. Since the Earth is now
a virialised overdense region, the initial condition for $\psi $ is chosen
so as to obtain our measured laboratory value of $\alpha $ at virialisation, 
$\alpha _{c}(z_{v})\equiv \alpha _{v}=\alpha _{0}$. Since the redshift at
which our cluster has virialised is uncertain, we will choose a
representative example where virialisation occurs over the range $0<z_{v}<10$.

After we have set $z_{v}$ for Earth, another constraint we need to satisfy
is given by the quasar observations \cite{murphylast}. This means that when
comparing the value of the fine structure constant on Earth, at its
virialisation, $\alpha _{v}=\alpha _{0}$, with the value of the fine
structure constant of another region at some given redshift in the range
accessed by the quasar spectra, $3.5\geq z\geq 0.5$, we need to obtain $%
\Delta \alpha /\alpha \equiv (\alpha (z)-\alpha _{v})/\alpha _{v}\approx
-5.4\times 10^{-6}$. This raises the question as to the location of the
clouds where the quasar absorption lines are formed: are they in a region
which should be consider as part of the background or an overdensity with
somewhat lower contrast than exists in our Galaxy? Unfortunately, this
question cannot be answered because we do not know the density of the
clouds, only the column density. Nevertheless, these clouds are much less
dense than the solar system. Because of this, it is a good approximation to
assume that the clouds possess the background density.

Thus, the initial conditions for $\psi $ are chosen so as to obtain our
measured laboratory value of $\alpha $ at virialisation $\alpha
_{c}(z_{v})=\alpha _{0}$ and to match the latest observations \cite%
{murphylast} for background regions at $3.5\geq |z-z_{v}|\geq 0.5$.

\subsection{Differences between the background and the overdensities}

\begin{figure}[htp!]
\centering \epsfig{file=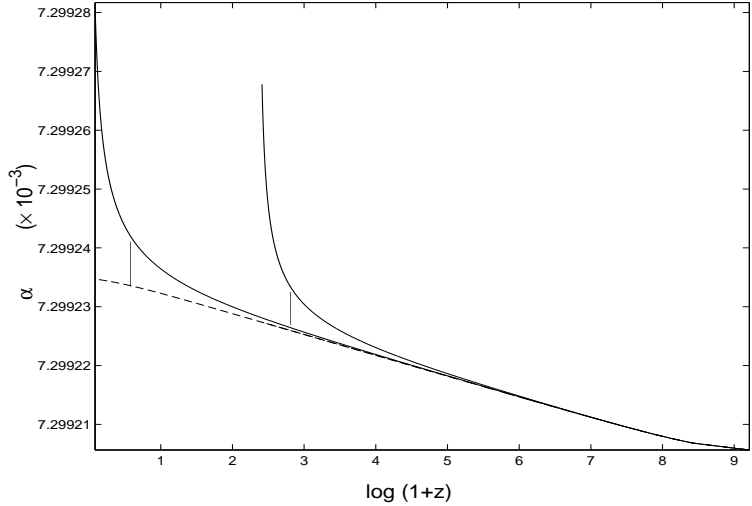,height=7cm,width=10cm} %
\caption{{\protect\small  {\textit{Evolution of  $\protect\alpha  $ in
the background  (dashed line) and  inside clusters (solid lines)  as a
function  of  $\log  (1+z)$.  Initial  conditions were  set  to  match
observations    of    }$\protect\alpha    $    variation    of    ref.
\protect\cite{murphy}. Two clusters  virialise at different redshifts,
one    of    them    in    order    to    have    $
_{c}(z_{v}=1)=\protect\alpha  _{0}$.   Vertical  lines  represent  the
moment of turn-around.}}}
\label{alphabalphac}
\end{figure}

Figure \ref{alphabalphac} show us  the evolution of the fine structure
constant inside an  evolving overdensity and in the  background. It is
clear that  there will  be differences between  the value of  the fine
structure  constant   inside  a   cluster,  $\alpha_c$,  and   in  the
background,  $\alpha_b$. It  is also  shown in  that figure,  that the
value  of $\alpha$ will  also dependent  on the  redshift at  which an
overdensity will virialise.

\begin{figure}[hbp!]
\centering \epsfig{file=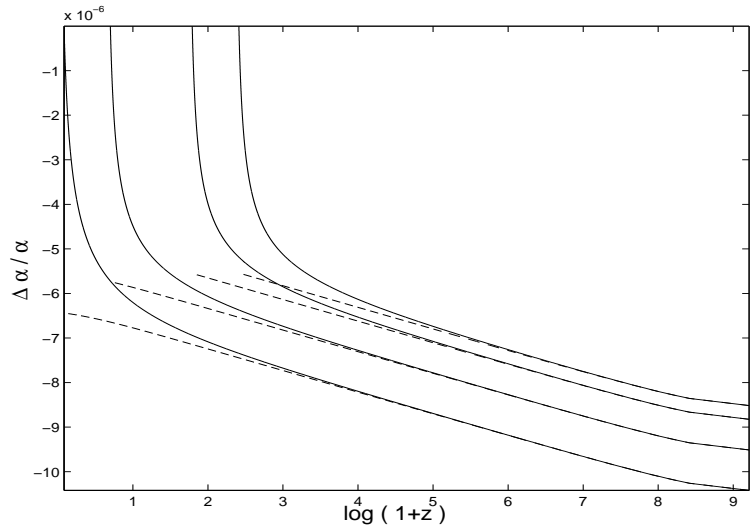,height=7cm,width=10cm}
\caption{{\protect\small {\textit{Evolution  of $\Delta \protect\alpha
/\protect\alpha $ in the background (dashed lines) and inside clusters
(solid lines) as  a function of $\log (1+z)$.  Initial conditions were
set  to match observations  of }$\protect\alpha  $\ variation  of ref.
\protect\cite{murphy}.  Four  clusters  that virialised  at  different
redshifts.  All clusters were  started so  as to  have $\protect\alpha
_{c}(z_{v})=\protect\alpha _{0}$.}}}
\label{deltaalphabalphac}
\end{figure}

In  figure  \ref{deltaalphabalphac}, we  have  evolved four  clusters,
which  have  different initial  conditions  in  order  to satisfy  the
constraints described in the previous section.
 This is just an example, since in reality, the initial condition for $\psi 
$ needs to be fixed only once, for our Galaxy. Hence, $\alpha $
in other clusters  will have a lower or higher  value (with respect to
$\alpha _{0}$)  depending on their $z_{v}$;  as can be  seen in figure
\ref{alphabalphac}.  The   example  given   was  just  to   show  that
independently of the redshift at  which Earth virialises, the order of
magnitude between  virialised regions and the background  is always of
order  $10^{-6}$.  Of  course  this   is  not  a  coincidence,  but  a
consequence of setting the initial  conditions in order to satisfy the
latest quasar observations \cite{murphylast}.
This justifies why our study, although is a qualitative analysis, also
can give  us a good approximation  of the order  of magnitude expected
for   variations  of   the   fine  structure   constant  both   inside
overdensities and in the background. The choice of having assumed that
the clouds, where the quasar observations were made, correspond to the
background, is then a not very stringent condition.


\subsection{Constancy of $\alpha$ after virialisation}

\begin{figure}[hbtp]
\centering \epsfig{file=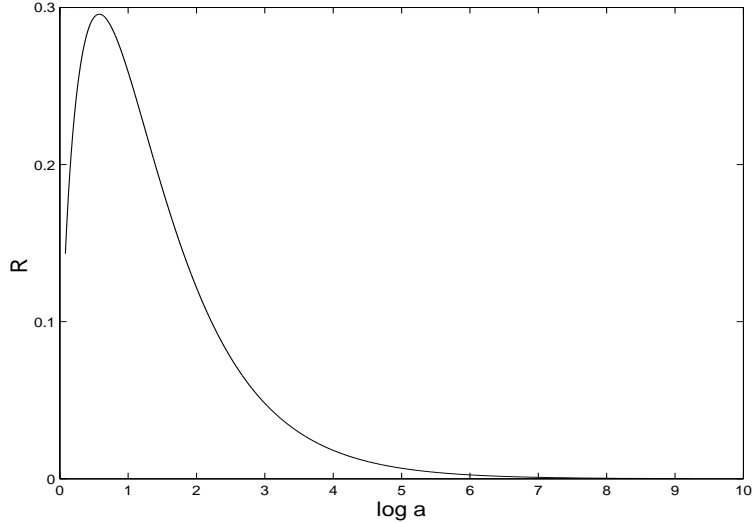,height=7cm,width=10cm}
\caption{{\protect\small  {\textit{Evolution   of  the  radius   of  a
cluster, $R$, as a function of $\log (1+z)$.}}}}
\label{radius}
\end{figure}

After virialisation occurs, the cluster radius, $R$, becomes constant;
time and space  variations of $\alpha $ are  suppressed, and $\alpha $
becomes  constant. If  there were  any variations  of $\alpha  $ after
virialisation, the energy components and radius of the cluster would need to vary
in order  to conserve energy  and momentum. This would  inconsistent with
virialisation.   This   phenomenon   is   not   included   in   figure
\ref{alphabalphac}, since we did follow the evolution to virialisation
with  a many-body  simulation which  would need  to include  the fluid
equations  that describe  the  pressure inside  the  cluster, see  for
instance  \cite{davis,efstathiou}.   In  our  simulation,  the  virial
condition is  a 'stop  condition' and so  we just observe  the typical
behaviour of  the cluster's collapse  as $R\rightarrow 0$,  see figure
\ref{radius},   or   $\rho_{cdm}\rightarrow   +\infty$,   see   figure
\ref{cluster}.

In addition to the stationarity condition that occurs when the cluster
virialises, it can be seen from figure \ref{deltaalphabalphac} that in
all  cases the  variation  of $\alpha  $  since the  beginning of  the
cluster  formation is  of order  $10^{-5}$, and  numerical simulations
give $\dot{\alpha}/\alpha \approx 10^{-22}s^{-1}$, we will see this in
the next chapter.
If  variations of $\alpha  $ are  so small  for such  a wide  range of
virialisation redshifts, we can assume that the difference between the
value   of  $\alpha   _{c}$  at   $z_{v}$   and  at   $z=0$  will   be
negligible. Therefore  it is a  good approximation to assume  that the
time-evolution of  both $\alpha  _{c}$ and of  the cluster  will cease
after  virialisation. Although  this  is not  necessarily  true (  for
instance  the  cluster  could  keep  accreting mass),  it  is  a  good
approximation in respect of the evolution of $\alpha $, especially for
clusters which have virialised at lower redshifts.

\bigskip

The fact  that local  $\alpha $ values  'freeze in'  at virialisation,
means we would observe no time or spatial variations of $ \alpha $ on
Earth,  or elsewhere  in our  Galaxy, even  though  time-variations in
$\alpha  $ might  still be  occurring on  extragalactic scales.  For a
cluster, the value  of $\alpha $ today will be the  value of $\alpha $
at  the   virialisation  time  of  the  cluster.   We  should  observe
significant  differences in  $\alpha  $ only  when comparing  clusters
which virialised at quite  different redshifts. Differences will arise
within  the  same  bound system  only  if  it  has not  reached  viral
equilibrium. Hence, variations of  $\alpha $ using geochemical methods
could easily give a value that is $10-100$ times smaller than is inferred
from quasar spectra. For instance, in spite of the fact that we cannot simulate the
evolution of $\alpha_c$ after 
virialisation, it is  useful to compare the value of $\Delta\alpha/\alpha $ in
clusters that have have virialised at a low redshift and in the
background, to have an order of magnitude of the variations
in the fine structure constant.
Numerical    simulations,   normalised    to   have
$\alpha_c(z_v=0)=\alpha_0$  and to  satisfy  the quasar  observations,
give  $\Delta\alpha/\alpha (z=2)=-6\times10^{-6}$  for  the background
and $\Delta\alpha/\alpha (z_v=0.13)=-1.5\times10^{-7}$ in a
cluster which has virialised at $z_v=0.13$.


\section{Measuring $\alpha$ in Virialised Regions}

\begin{figure}[htbp!]
\centering
\epsfig{file=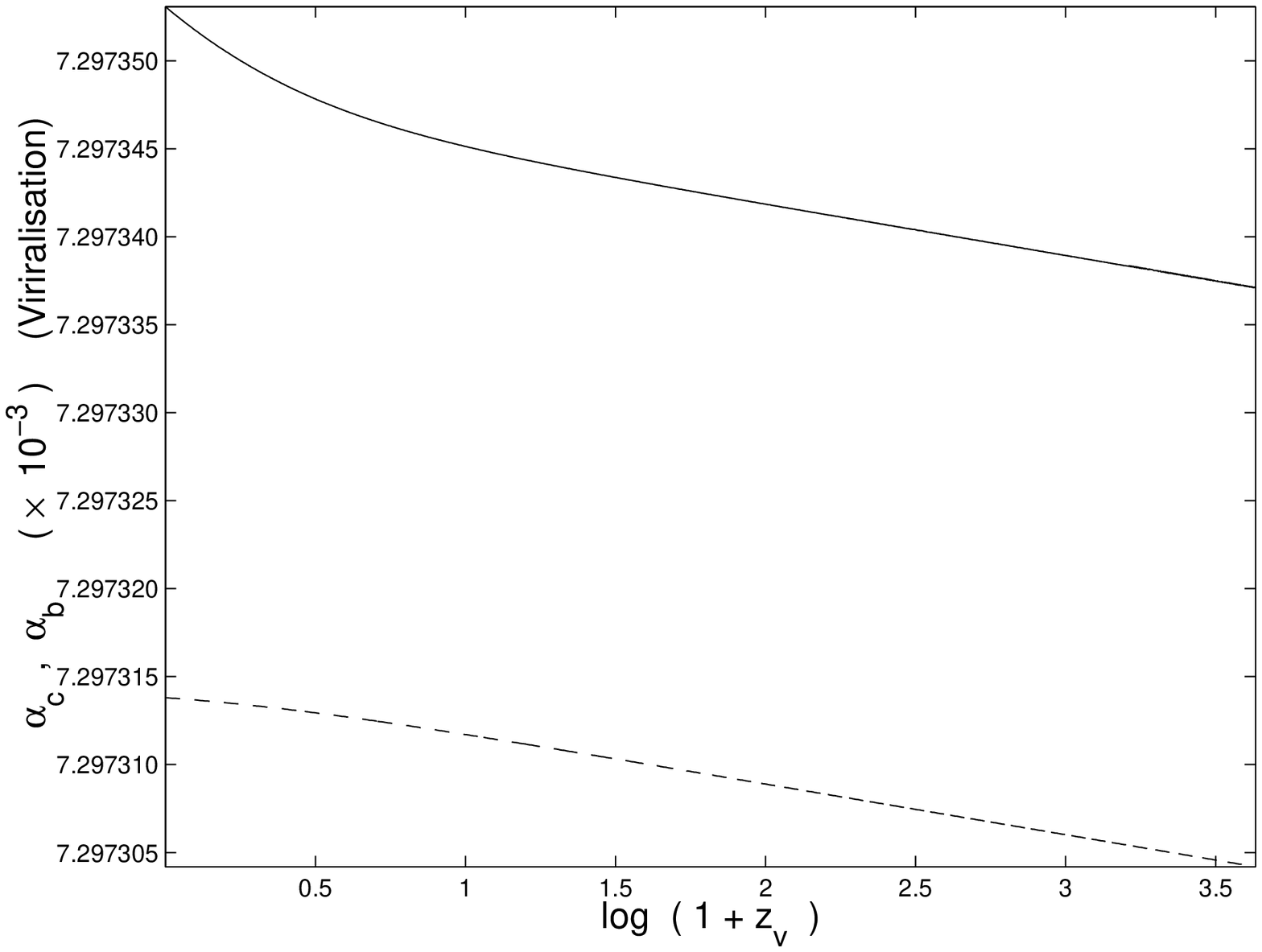,height=7cm,width=10cm}
\epsfig{file=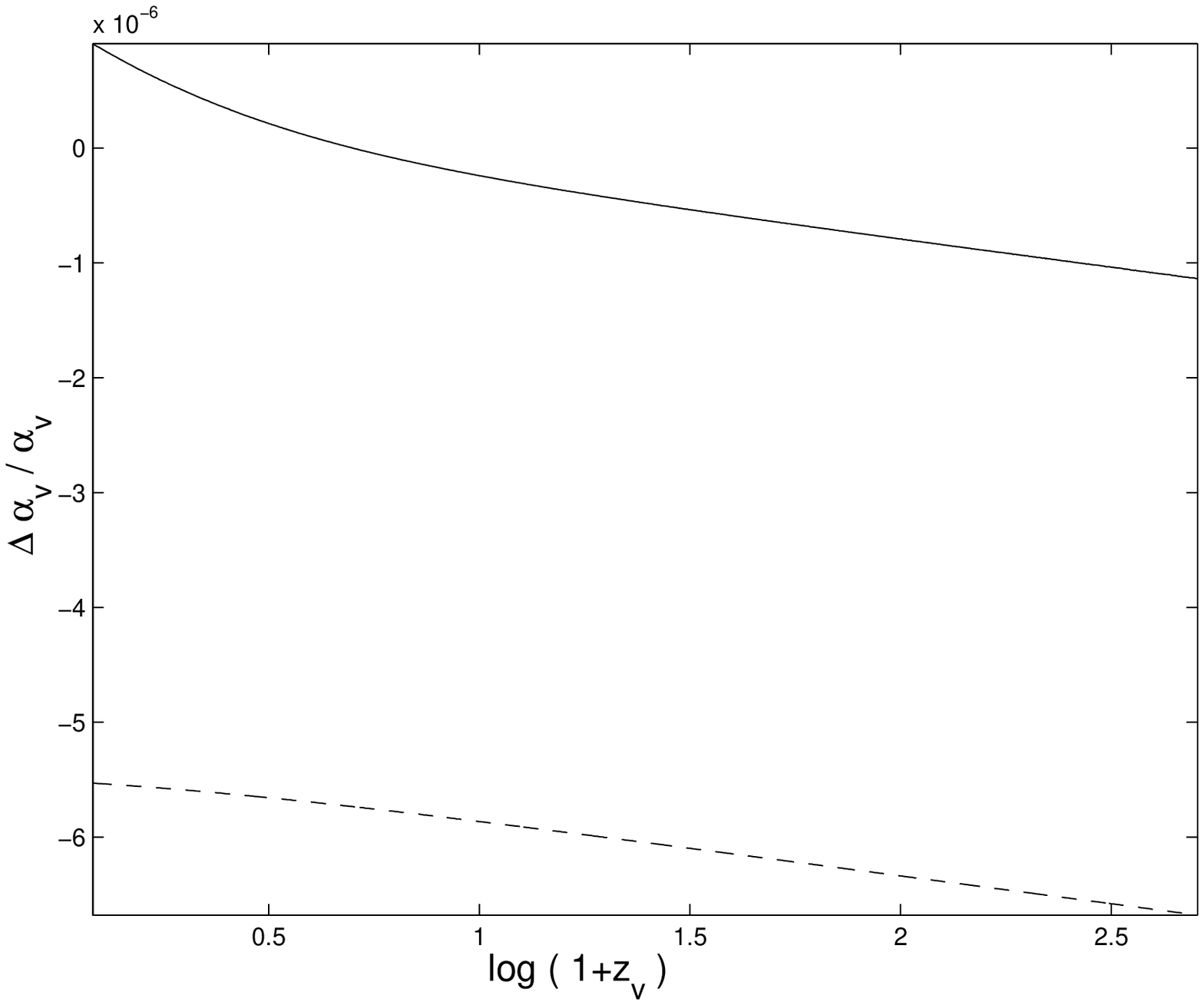,height=7cm,width=10cm}
\caption{{\protect\small  {\textit{Plot  of  $\protect\alpha  $ and
$\Delta  \protect\alpha/ \protect\alpha   $    as  a
function  of $\log  (1+z)$, at  virialisation. Clusters  (solid line),
background (dashed line).}}}}
\label{alphabalphav}
\end{figure}

If  we could  measure the  fine structure  constant  inside virialised
overdensities and  the corresponding value  in the background,  at the
time  of  their  virialisation,  then  figure  \ref{alphabalphav}  
would  give us the evolution  of $\alpha_v$ or
its  time  shift, $\Delta  \alpha_v  /\alpha_v  ,$  as a  function  of
$z_v$.
%

Those figures show us  that differences in $\alpha$ between the
background and the overdensities increase as $z_v\rightarrow0$.
This  is due to  the earlier  freezing of  the value  of $\alpha  $ at
virialisation, and to  our assumption that we live  in a $\Lambda CDM$
universe.  \textit{At  lower   redshifts,  specially  after  }$\Lambda
$\textit{\ starts  to dominate,  variations of }$\alpha  $\textit{\ in
the background  are turned off by the  accelerated expansion. However,
the value of  }$\alpha $\textit{\ in the collapsing  cluster will keep
growing until virialisation occurs}.
At higher redshifts,  $z\gg 1$, both $\alpha _{b}$  and $ \alpha _{c}$
evolve  in expanding  environments: their  increase is  logarithmic in
time before $\Lambda  $ starts to dominate, so  the difference between
them will be much smaller.

These  results mean  that if  we  were able  to compare  the value  of
$\alpha$, in overdensities which have virialised at a similar redshift
as ours, with regions that  have virialised much before or much after,
we would  notice that  there were indeed  differences in  the measured
$\Delta\alpha/\alpha$.


\subsection{Dependence on the matter density of the clusters}

Variations of  $\alpha_c$ with respect to $\alpha_b$  can  be tracked  using the  ``spatial
density contrast'',
\[
\frac{\delta   \alpha   }{\alpha}   \equiv  \frac{   \alpha_{c}-\alpha
_{b}}{\alpha _{b}},
\] 
and computed at virialisation or while the cluster is still evolving.

\bigskip 

\begin{figure}[htbp!]
\centering
\epsfig{file=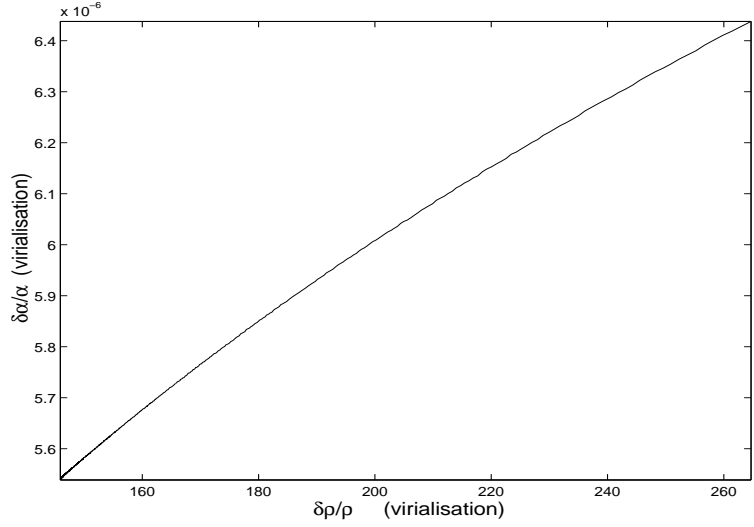,height=7cm,width=10cm}
\caption{{\protect\small    {\textit{Variation    of   $\protect\delta
\protect \alpha  /\protect\alpha $ with  $\protect\delta \protect\rho
/\protect\rho $  at the cluster virialisation  redshift. The evolution
of $\protect\alpha $ inside the clusters was normalised to satisfy the
latest   time-variation    observational   results,   and    to   have
$\protect\alpha     _{c}(z_{v})=\protect     \alpha     _{0}$     for
}${z}_{v}{=1.}$}}}
\label{alpharho}
\end{figure}

A plot of the  ``spatial
density contrast'' in $\alpha $ with respect to the
matter density  inhomogeneity, $\delta \rho /\rho  ,$ at virialisation
is shown in figure. \ref{alpharho}. Note that $\delta \alpha /\alpha $
grows  in   proportional  to  the  density  contrast   of  the  matter
inhomogeneities.  
%

In a  $ \Lambda CDM$  model, the density contrast,  $\Delta _{c}=\rho
_{cdm}(z_{v})/\rho  _{b}(z_{v})$ increases  as the  redshift decreases
\cite{lahav}.   For   high   redshifts,   the  density   contrast   at
virialisation   becomes  the   asymptotically  constant   in  standard
($\Lambda =0$) $CDM$, $\Delta  _{c}\approx 178$, if $\rho_b$ is measured
at collapse or $\Delta _{c}\approx  148$, if $\rho_b$ is measured at
virialisation.   This is  another reason  why at  lower  redshifts the
difference between $\alpha_v$ and $\alpha_b$ increases.

Trends of variation of $\alpha $  can be then predicted from the value
of the matter density contrast of the regions observed. Useful fitting
formulae for  the dependence  of $\alpha $  variation on  $\delta \rho
/\rho $ and the scale factor $a$ $\equiv (1+z)^{-1}$ are:
%
\begin{eqnarray}
\frac{\delta         \alpha        }{\alpha}        &=&(5.56-0.7\sigma
^{\frac{1}{2}}+0.078\sigma     +0.00352\sigma    ^{\frac{3}{2}})\times
10^{-6}  \nonumber  \label{fit2}   \\  \frac{\delta  \alpha  }{\alpha}
&=&(5.37+0.373\theta            ^{\frac{1}{6}            }-0.27\theta
^{\frac{1}{3}}+0.007\theta  ^{\frac{1}{2}})\times 10^{-6} \label{fit3}
\\  \frac{\Delta  \alpha}{\alpha} &=&-(2.47-1.81\sigma  ^{\frac{1}{2}%
}+0.59\sigma -0.094\sigma ^{\frac{3}{2}})\times 10^{-6} \nonumber
\label{fit4}
\end{eqnarray}%
where  $\theta  \equiv \frac{\delta  \rho  }{\rho }/\Delta  _{c_{v}}$,
$\sigma  \equiv a/a_{v}$.  $\Delta _{c_{v}}$  and $a_{v}$  are ``input''
parameters  defined by  our choice  of  the redshift  of when  $\alpha
(z_{v})=\alpha _{0}$ .
These   formulae  are   valid   for  clusters   which  virialised   at
$0<z_{v}<15$. For the case  where $\alpha (z_{v}=1)=\alpha _{0}$ these
formulae are accurate to better than $1\%$.


\section{Comments and Discussion}

Generally,  past  studies of  spatially  homogeneous cosmologies  have
matched  the value  of $\alpha  _{b}$  with the  terrestrial value  of
$\alpha $ measured  today. However, it is clear that  the value of the
fine  structure constant  on Earth,  and  most probably  in our  local
cluster,  will  differ from  that  in  the  background universe.  This
feature  has been  ignored when  comparing observations  of  $\alpha $
variations        from         quasar        absorption        spectra
\cite{murphylast,murphy,webb}  with  local  bounds derived  by  direct
measurement       \cite{prestage,sortais}      or       from      Oklo
\cite{shly,fuj,damour1}    and   long-lived   $    \beta   $-decayers
\cite{olive}.
A similar  unwarranted assumption is  generally made when  using solar
system tests  of general relativity to bound  possible time variations
of $G$  in Brans-Dicke theory  \cite{scharre}: there is no  reason why
$\dot{G}/G$ should be the same in the solar system and on cosmological
scales. Since any varying-constant  model require the existence of a
scalar field coupled to the matter fields, our considerations apply to
all other theories besides BSBM and to variations of other ``constants''
\cite{ivan,uzan}.
The consideration  of the inhomogeneous  evolution of the  universe is
therefore  essential for  a correct  comparison of  extra-galactic and
solar system  limits on, and observations of,  possible time variation
of the fine structure constant and other ``constants''.

\bigskip

There are  many  different
constraints,  a  theoretical   model  of  varying-$\alpha$,  needs  to
satisfy,  in  order to  be  in  agreement  with the  observations  and
experiments.  In  particular, it  needs to 
satisfy    the    geochemical    constraints    at,    $z=0.45$,    of
$\Delta\alpha/\alpha<10^{-7}$  and  the quasar  data,  at $z=0.5$,  of
$\Delta\alpha/\alpha\sim10^{-6}$.

In order to explain, why any experiment made on earth \cite{shly,fuj},
or in  our local system \cite{scharre,olive1},  present constraints on
variations of   $\alpha$  which are, $10$ to $100$  times lower than the
quasar spectra  observations \cite{murphylast},  we need to  take into
account the mechanism of structure formation in the universe.
We have then considered that  today's value of $\alpha$ is measured on
earth  (and  on our  solar  system),  and  that these  are  virialised
overdensities.   We  have  matched   the  quasar   absorption  spectra
observations  \cite{murphylast},   comparing  the  today's   value  of
$\alpha$ on a virialised region (our galaxy) and the background (which
represented the clouds where the quasar data was obtained).
Using these assumption, to set up our initial conditions, we have used
the spherical collapse model to investigate the effects of non-linear
structure formation on the cosmological evolution of the fine
structure constant.   Evolving a tiny overdensity from  the radiation era
up  to today, we  were able  to observe  the differences  between the
evolution of $\alpha$ inside the overdensity and the background.
When the  density of these structures  becomes big enough  for them to
break away from the background expansion, they start to collapse until
its  virialisation.  When  the overdensity  virialises,  variations of
$\alpha$ are  highly suppressed, and  it becomes constant  inside that
region, otherwise the system would not be able to virialise.

The 'freeze in'  effect of the fine structure constant after
virialisation is of major importance.
Any  experiment or  observation made,  in order  to study
variations  of $\alpha$,  on  Earth or  in  our local  system ,  would
measure  no variation  at all.  Or  if they  do, those  would be  much
smaller than the results obtained from the quasar absorption spectra.
So, variations of $\alpha$, would naturally satisfy any experiments on
the  violation of  the  principle of  equivalence \cite{dvali,  olive,
damour}, in  our local  overdensity, and any  geochemical observation.
At the  same time we would be able to  explain the quasars absorption
spectra observations \cite{murphylast,murphy,murphy1,webb}.

The dependence of $\alpha $ on the matter-field perturbations is much less
important when one is studying effects on the early universe, for example on
the CMBR or Big Bang nucleosynthesis \cite{martins}. In the linear regime of
the cosmological perturbations, small perturbations in $\alpha $ will decay
or become constant in the radiation era \cite{mota2}. Also, in the
non-linear regime of structure formation, the evolution of the fine
structure constant in the background and inside the overdensities will not
differ until the overdensities turnaround. Turnaround occurs at
a far lower redshift than that of last scattering, $z=1100$. At these high
redshifts, it was found  that $\Delta \alpha /\alpha \leq
10^{-5}$.  In the background, the fine structure constant, will be a
constant during the radiation epoch \cite{bsbm}. A growth in value of $%
\alpha $ will happen only in the matter dominated era \cite{mota1}. Hence,
the early-universe constraint, coming from the CMBR and Big Bang
Nucleosynthesis, $\Delta \alpha /\alpha \leq 10^{-2}$, are easily
satisfied.

%% file: chapter53.tex
   
\chapter{The Fine Structure Constant Dependence on the Dark-Energy Equation
of State and on the Coupling to Matter}
\label{darkenergy}

\begin{flushright}

{\it {\small O  todo \'e tudo, e  o mim que sou \'e  o resto,}}\\ {\it
{\small n\~ao h\'a resto, h\'a  uma angustia isso sim}}\\ {\it {\small
de saber que o todo que  sou n\~ao o sei dizer...}}\\ {\it {\small mas
o sei que sou todo assim...}}\\
{\it {\small -- Ernesto Mota --}}
\end{flushright}

\bigskip

\section{Introduction}

In the  last chapter,  we saw that  the cosmological evolution  of the
fine  structure  constant  depends  on  the  evolution  of  non-linear
structures in the universe.
There are  two main  features which affect  the evolution of  $\alpha$
in  a universe  where large  scale  structures are
formed. The  most obvious, is  the coupling between the  scalar field,
which  drives variations  in  $\alpha$, and  the  matter fields.   The
second is  the dependence of non-linear models  of structure formation
on  the equation  of state  of the  universe, and in particular that of the dark
energy one   \cite{lahav,wang}.

Looking at the  right-hand-side of equation (\ref{psidot}), we see
that, $\zeta/\omega$ represents the  coupling between the scalar field
$\psi$ an  the matter density.  It  is then natural to  think that the
rate of change of the  fine structure constant, both in the background
and inside overdensities, will be  a function of the absolute value of
$\zeta/\omega $.  Smaller values of  $|\zeta/\omega |$ would lead to a
smaller  dependence  of the  variation  of  $\alpha  $ on  the  matter
inhomogeneities.


Besides the dependence on the coupling $\zeta/\omega$, the fine
structure constant also depends on the equation of state of the
universe. For instance,  several solutions were found for $\alpha_b$, 
for the different epochs the universe went through.
In the case of the  overdensities, the dependence of $\alpha_c$ on the
equation of  state of the universe is related to the 
dependence of  the density contrast of the overdensities on the
structure formation model.

\bigskip

In  resume,  the evolution  of  the  fine  structure constant  in  the
background, and  inside clusters then  depends mainly on  the dominant
equation  of state  of the  universe  and the  value and  sign of  the
coupling constant $\frac{\zeta }{\omega }$. 
We will then investigate the inhomogeneous evolution of the fine structure
constant after  the development of non-linear  cosmic structures, that
is, after they have virialised.
This  allow us  to  predict  variations  of the  fine
structure  constant among  virialised  systems and to investigate the
dependence of $\alpha$ on the equation of state of the
dark energy. In the last section, we show how spatial variations in $\alpha $
may occur due to possible spatial variations in the coupling of $\alpha $ to
the matter fields.

\section{BSBM model and Dark Energy Structure Formation Models}

In order to  study the dependence of the evolution  of $\alpha$ on the
equation of state of dark  energy, we need to generalise the equations
used  in the  previous  chapter both  for  the background  and in  the
overdensities.

The background universe  will be described by a  flat, homogeneous and
isotropic  Friedmann metric  with expansion  scale factor  $a(t)$. The
universe  contains  pressure-free   matter,  of  density  $\rho  _{m}$
$\propto a^{-3}$ and  a dark energy fluid with  a constant equation of
state,   $w_{\phi}$,   and   an   energy-density   $\rho_{\phi}\propto
a^{-3(1+w_{\phi})}$. In the case where dark energy is the cosmological
constant $\Lambda  $, then $\rho_{\phi}\equiv\rho  _{\Lambda }\equiv $
$\Lambda /(8\pi  G)$ and  $w_{\phi}=-1$. Hence the  Friedmann equation
(\ref{fried}) becomes:
\begin{equation}
H^{2}=\frac{8\pi   G}{3}\left(  \rho  _{m}\left(   1+\left\vert  \zeta
\right\vert e^{-2\psi }\right) \ +\rho _{\psi }+\rho _{\phi }\right)
\label{fried1}
\end{equation}

The evolution of a spherical overdense patch of scale radius $R(t)$ is
the  given by  the Friedmann  acceleration  equation (\ref{rcluster}),
which becomes:
\begin{equation}
\frac{\ddot{R}}{R}=-\frac{4\pi     G}{3}\left(    \rho    _{cdm}\left(
1+\left\vert  \zeta \right\vert  e^{-2\psi _{c}}\right)  +4\rho _{\psi
_{c}}+(1+3w_{\phi_c})\rho _{\phi_c }\right) \label{rcluster1}
\end{equation}
where $\rho _{cdm}$ is the density of cold dark matter in the cluster,
$\rho _{\phi_c}$ is  the energy density of the  dark energy inside the
cluster  and  $\rho  _{\psi  _{c}}\equiv\frac{\omega}{2}\dot\psi_c^2$,
where $\psi_c$ represents the scalar field inside the overdensity.  We
have  also used  the  equations of  state  $p_{\psi _{c}}=\rho  _{\psi
_{c}}$, $ p_{cdm}=0$ and $p_{\phi_c}=w_{\phi_c}\rho_{\phi_c}$.

In  the cluster,  the  evolution  of $\psi  _{c}$,  $\rho _{cdm}$  and
$\rho_{\phi_c}$ are given by
\begin{eqnarray}
\ddot{\psi  _{c}}+3\frac{\dot{R}}{R}\dot{\psi _{c}}&=&-\frac{2}{\omega
} e^{-2\psi _{c}}\zeta \rho _{cdm} \label{psidotcluster1} \\ \dot\rho
_{cdm} &=& - 3 \frac{\dot R}{R} \rho_{cdm}
\label{rhocluster1}
\\  \dot\rho_{\phi_c}&=&-3  \frac{\dot  R}{R}  \left( 1  +  w_{\phi_c}
\right) \rho_{\phi_c}
\label{dark1}
\end{eqnarray}

\bigskip

Once again we will evolve  our overdensity from a redshift well inside
the radiation  era until its  virialisation occurs.  In order  to know
when  the  cluster becomes  virialised  we  will  use the  {\it{Virial
theorem}}, equation (\ref{virialtheorem}). Again  it will be useful to
write the  condition for virialisation  to occur, using  the potential
energies associated to the different components of the overdensity.

Since we are  now considering a dark energy fluid  with a general, but
constant, equation  of state $w_{\phi_c}$,  we need to  generalise the
potential   energy   associated    to   the   cosmological   constant,
$U_{\Lambda}$.  The  generalisation of equation  (\ref{potential}), in
order  to incorporate  a fluid  with a  general equation  of  state is
trivial.      Using     equations     (\ref{generalpotential})     and
(\ref{potential}), we  obtain the  potential energy associated  to the
dark energy fluid
\begin{eqnarray}
U_{\phi_c        }&=&-\frac{3}{5}         G        M_{\phi_c}        M
R^{2-3\left(1+w_{\phi_c}\right)}
\label{darkenergypotential}
\end{eqnarray}
where $M_{\phi_c}$ is the mass associated to that fluid, given by
\[
M_{\phi_c}= \frac{4\pi}{3}\rho_{\phi_c}R^{3\left(1+w_{\phi_c}\right)}.
\]

Using now  $U_{\phi_c}$ in equation  (\ref{virialtheorem}), instead of
$U_{\Lambda}$  we obtain  a  condition for  virialisation, similar  to
equation (\ref{virialcond}), in terms of the known variables.

\bigskip

The behaviour of the fine structure constant during the evolution of a
cluster  can  now  be  obtained numerically  evolving  the  background
equations (\ref{psidot}) and  (\ref{fried1}) and the cluster equations
(\ref {rcluster1})-(\ref{dark1}) until the virialisation holds.

Once again  the initial  conditions for  $\psi $ are  chosen so  as to
obtain our  measured laboratory value  of $ \alpha $  at virialisation
$\alpha _{c}(z_{v})\equiv\alpha_v=\alpha _{0}$ and to match the latest
observations     \cite{murphylast}:  $\Delta   \alpha   /\alpha  \approx
-5.4\times 10^{-6}$ for regions  at $3.5\geq |z-z_{v}|\geq 0.5$, where
$z_{v}$ will be the redshift at which our inhomogeneity virialised.

\section{Dependence on the Dark-Energy Equation of Sate}

Non-linear models of structure formation will present different features
depending on the equation of state of the universe \cite{lahav,wang}. The main
difference is the way the parameter $\Delta _{c}=\rho _{cdm}(z_{c})/\rho
_{b}(z_{v})$ evolves with the redshift. The evolution of $\Delta _{c}$
depends on the equation of state of the dark-energy component which
dominates the expansion dynamics. For instance, in a $\Lambda CDM$ model,
the density contrast, $\Delta _{c}$ increases as the redshift decreases. At
high redshifts, the density contrast at virialisation becomes asymptotically
constant in standard ($\Lambda =0$) $CDM$, with $\Delta _{c}\approx 178$ at
collapse or $\Delta _{c}\approx 148$ at virialisation. This behaviour is
common to other dark-energy models of structure formation ($wCDM$), where
the major difference is in the magnitude and the rate of change of $\Delta
_{c}$ at low redshifts.

The local value of the fine structure constant will be a function of the
redshift and will be dependent on the density of the region of the universe
we measure it, according to whether it is in the background or an
overdensity \cite{mota3}. The density contrast of the virialised clusters
depends on the dark-energy equation of state parameter, $w_{\phi }$ . Hence,
the evolution of $\alpha $ will be dependent on $w_{\phi }$ as well.

\bigskip

What difference we would expect to see in $\alpha $ if we compared two bound
systems, like two clusters of galaxies? And how does the difference depend
on the cosmological model of structure formation? These questions can be
answered by looking at the time and spatial variations of $\alpha $ at the
time of virialisation. The space variations will be tracked using a
``spatial'' density contrast, 
\begin{equation}
\frac{\delta \alpha }{\alpha }\equiv \frac{\alpha _{c}-\alpha _{b}}{\alpha
_{b}}  \label{deltaspatialalpha}
\end{equation}%
which is computed at virialisation (where $\alpha _{c}(z=z_{v})=\alpha _{v}$%
). 

\bigskip

Since the main dependence of $\alpha $ is on the density of the clusters and
the redshift of virialisation, we will only study dark-energy models where $%
w_{\phi }$ is a constant. Any effect contributed by a time-varying equation
of state should be negligible, since the important feature is the average
equation of state of the universe. This may not be the case for the models
where the scalar field responsible for the variations in $\alpha $ is
coupled to dark energy \cite{parkinson,copeland,anchordoqui}.

In order to have a qualitative behaviour of the evolution of the fine
structure constant, at the virialisation of an overdensity, we will then
compare the standard Cold Dark Matter model, $SCDM$, to the dark-energy Cold
Dark Matter, $wCDM$, models. In particular, we will examine the
representative cases of $w_{\phi }=-1,-0.8,-0.6$. All models will be
normalised to have $\alpha _{v}=\alpha _{0}$ at $z=0$ and to satisfy the
quasar observations \cite{murphylast}, as discussed above. This
normalisation, although unrealistic (Earth did not virialise today), give us
some indication of the dependence of the time and spatial evolution in $%
\alpha $ on the different models. In reality, this approximation will not
affect the order of magnitude of the spatial and time variations in $\alpha $
for the cases of virialisation at low redshift \cite{mota3}.

\subsection{Time-shifts and the evolution of $\protect\alpha $}

\begin{figure}[p]
\centering
\epsfig{file=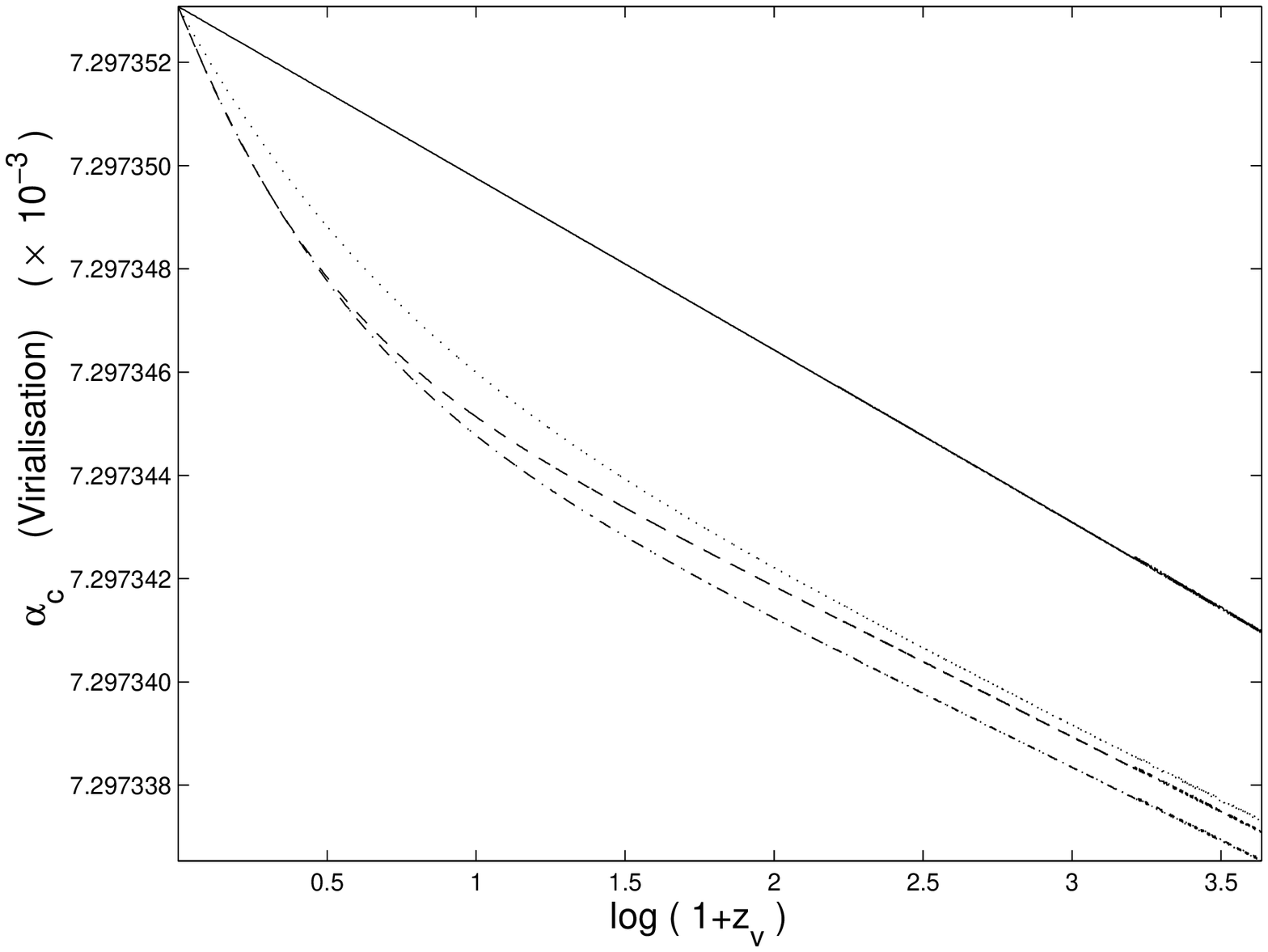,height=7cm,width=10cm}
\epsfig{file=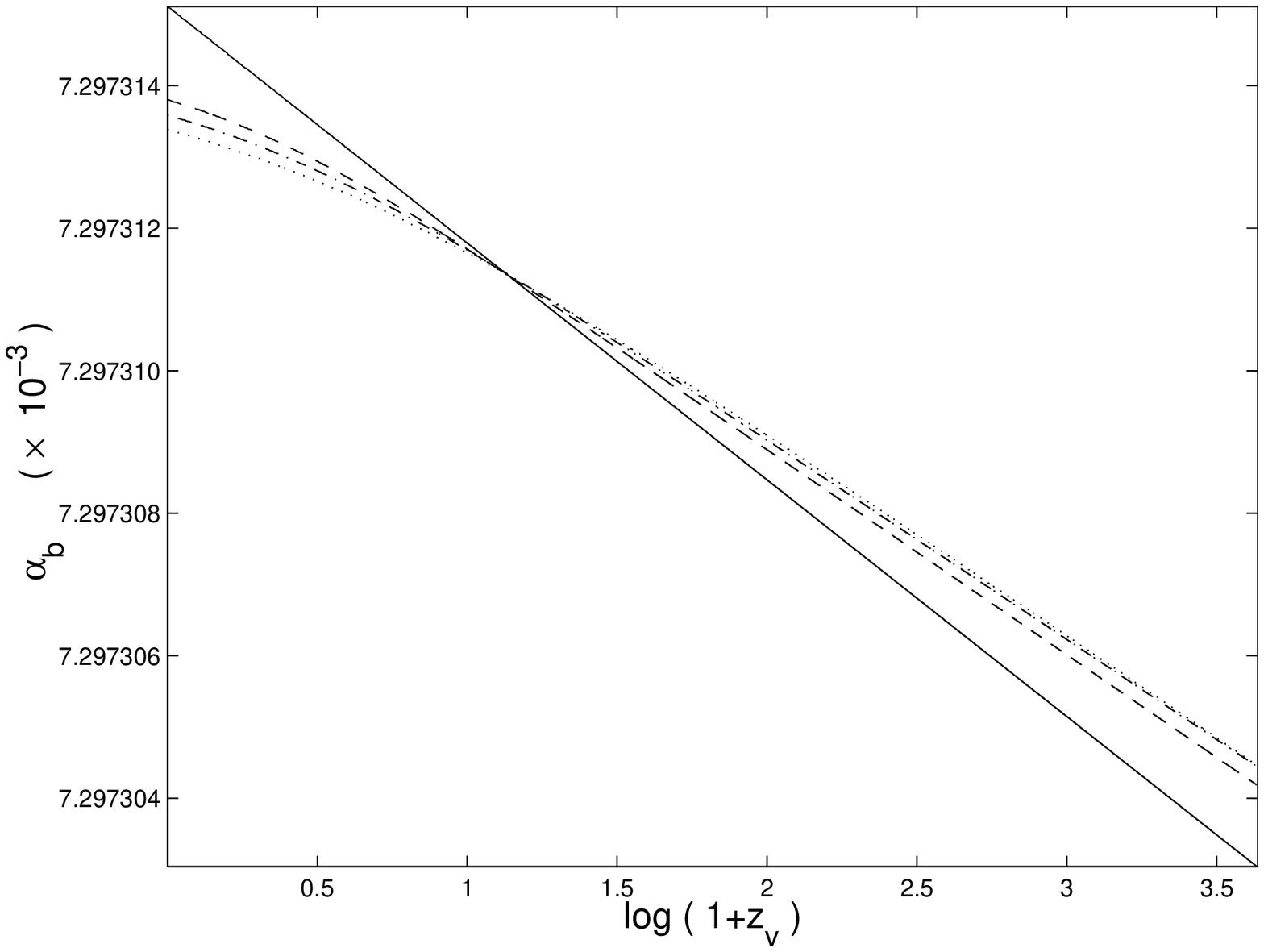,height=7cm,width=10cm}
\caption{\textit{Value of $\protect\alpha $ in the background and inside clusters as a function of 
$\log (1+z_{v})$, the epoch of virialisation. Solid line corresponds to the
standard ($\Lambda =0$) $CDM$, dashed line is the $\Lambda CDM$,
dashed-dotted corresponds to a $wCDM$ model with $w_{\protect\phi }=-0.8$,
dotted-line corresponds to a $wCDM$ model with $w_{\protect\phi }=-0.6$. In
all the models, the initial condition were set in order to have $\protect%
\alpha _{c}(z=0)=\protect\alpha _{0}$ and to satisfy $\Delta \protect\alpha /%
\protect\alpha \approx -5.4\times 10^{-6}$ at $3.5\geq |z-z_{v}|\geq 0.5$.}}
\label{modelsvirial}
\end{figure}

The final value of $\alpha $ inside virialised overdensities and its
evolution in the background is shown in figure \ref{modelsvirial}. 
From this plot, a feature common to all
the models stands out: the fine structure constant in the background
regions has a lower value than inside the virialised overdensities. Also,
its eventual local value depends on the redshift at which the overdensity
virialises. 

As expected, the equation of state of the dark energy affects the evolution
of $\alpha ,$ both in the overdensities and the background. A major
difference arises if we compare the $SCDM$ and $wCDM$ models. In a $SCDM$
model, the fine structure constant is always a growing function, both in
the background and inside the overdensities, and the growth rate is almost
constant. In a $wCDM$ model, the evolution of $\alpha $ will depend strongly
on whether one is inside a cluster or in the background. In the $wCDM$
background, $\alpha _{b}$ becomes constant (independent of the redshift) as
the universe enters the phase of accelerated expansion. Inside the clusters, 
$\alpha _{v}$ will always grow and its value now will depend on the redshift
at which virialisation occurred. The cumulative effect of this growth
increases as we consider overdensities which virialise at increasingly lower
redshift.

These differences arise due to the dependence of the fine structure
constant on the equation of state of the universe and the density of the
regions we are measuring it. In a $SCDM$ model, we will always live in a
dust-dominated era. The fine structure constant will then be an
ever-increasing logarithmic function of time, $\alpha \propto \ln (t)$ \cite%
{bsbm,mota1}. The growth rates of $\alpha _{b}$ and $\alpha _{c}$ will be
constant, since $\Delta _{c}$ is independent of the redshift in a $SCDM$
model. In a $wCDM$ model, dark energy plays an important role at low
redshifts. As we reach low redshifts, where dark energy dominates the
universal expansion, $\alpha _{b}$ becomes a constant \cite{bsbm,mota1}, but 
$\Delta _{c}$ continues to increase, as will $\alpha _{c}$ \cite{mota3}. The
growth of $\alpha _{v}$ becomes steeper as we go from a dark-energy fluid
with $w_{\phi }=-0.6$ to the $\Lambda $-like case of $w_{\phi }=-1$. The
intermediate situation is where $w_{\phi }=-0.8$, due to the dependence of $%
\Delta _{c}$ on $w_{\phi }$.

\begin{figure}[Hp]
\centering
\epsfig{file=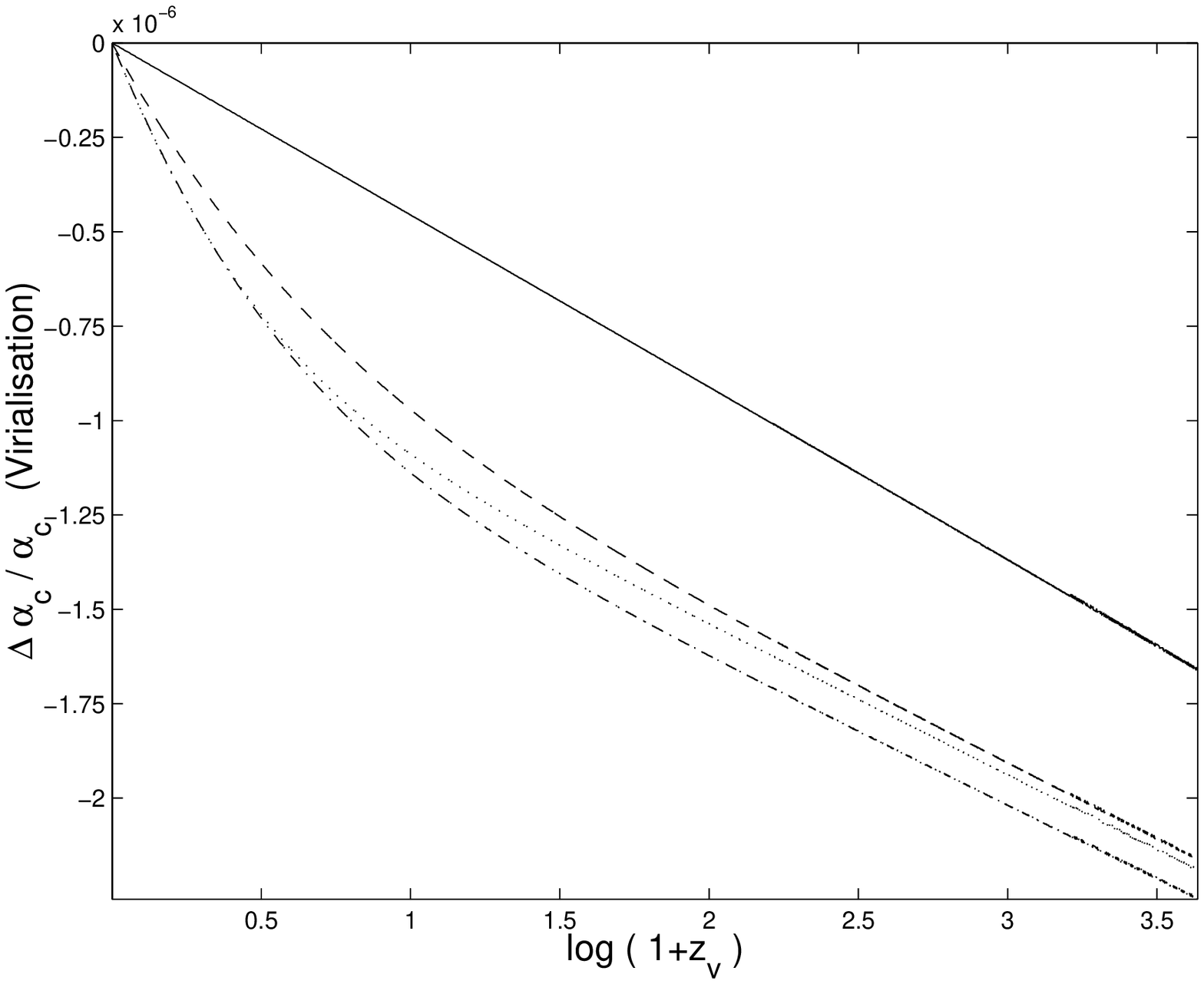,height=7cm,width=10cm}
\epsfig{file=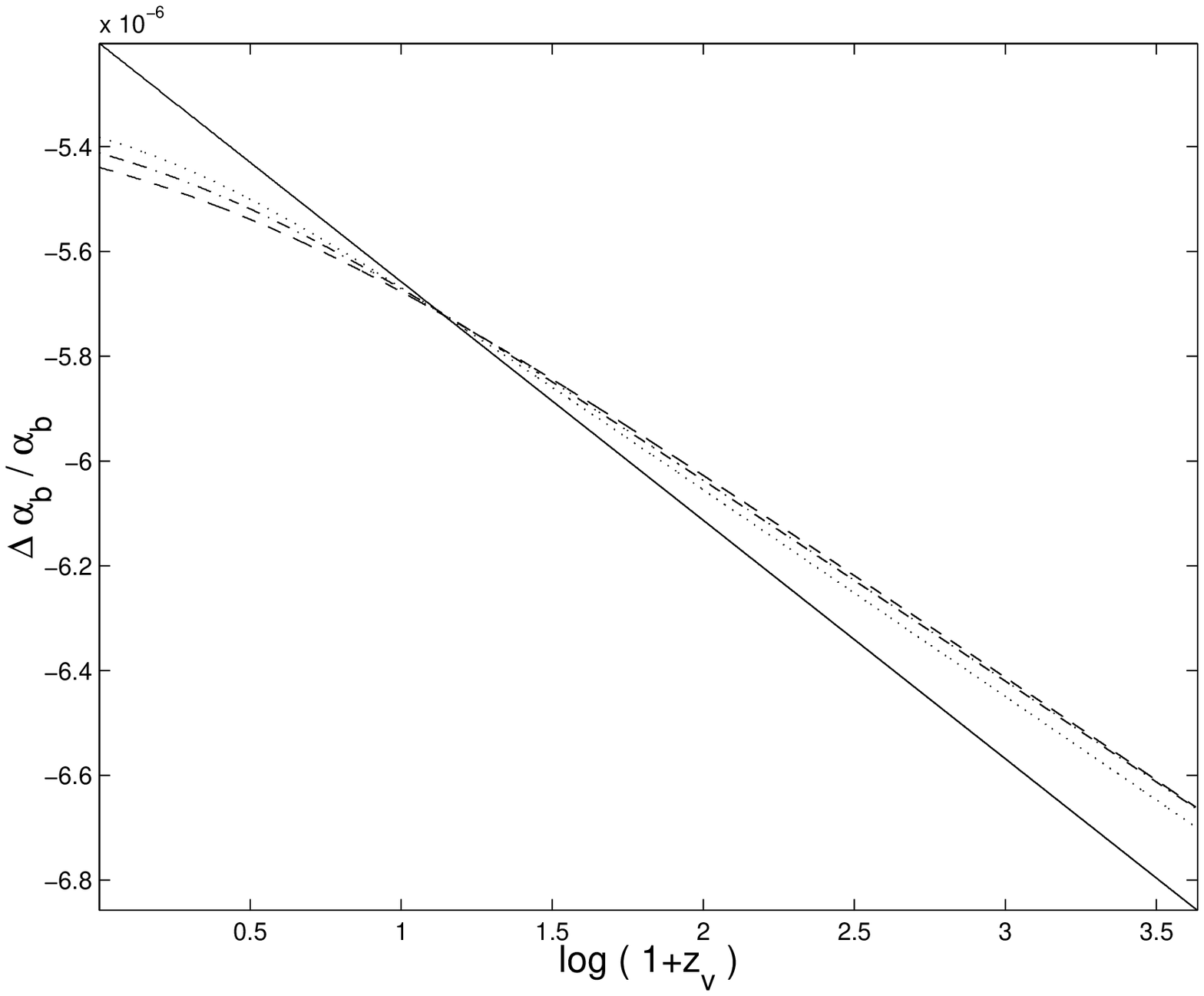,height=7cm,width=10cm}
\caption{\textit{Variation of $\Delta \protect\alpha /\protect\alpha $
    in the background and 
inside clusters as a function of the epoch of virialisation, $\log (1+z_{v})$%
. The solid line corresponds to the standard ($\Lambda =0$) $CDM$, the
dashed line is the $\Lambda CDM$, the dashed-dotted line corresponds to a $%
wCDM$ model with $w_{\protect\phi }=-0.8$, the dotted line corresponds to a $%
wCDM$ model with $w_{\protect\phi }=-0.6$. In all models, the initial
conditions were set in order to have $\protect\alpha _{c}(z=0)=\protect%
\alpha _{0}$ and to satisfy $\Delta \protect\alpha /\protect\alpha \approx
-5.4\times 10^{-6}$ at $3.5\geq |z-z_{v}|\geq 0.5$.}}
\label{modelsdeltatimevirial}
\end{figure}

Similar conclusions can be drawn with respect to the ``time shift'' of the
fine structure constant, ($\Delta \alpha /\alpha $), at virialisation, see
figure \ref{modelsdeltatimevirial} 
\clearpage

\subsection{Spatial variations in $\protect\alpha$}

\begin{figure}\centering
\epsfig{file=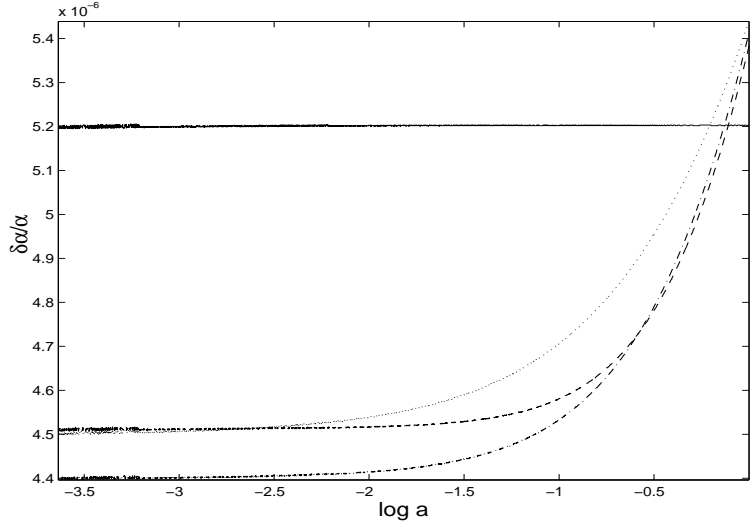,height=7cm,width=10cm}
\caption{\textit{Variation of $\protect\delta \protect\alpha /\protect\alpha 
$ as a function of $\log (1+z_{v})$. Solid line corresponds to the standard (%
$\Lambda =0$) $CDM$, dashed line is the $\Lambda CDM$, dashed-dotted line
corresponds to a $wCDM$ model with $w_{\protect\phi }=-0.8$, dotted line
corresponds to a $wCDM$ model with $w_{\protect\phi }=-0.6$. In all the
models, the initial conditions were set in order to have $\protect\alpha %
_{c}(z=0)=\protect\alpha _{0}$ and to satisfy $\Delta \protect\alpha /%
\protect\alpha \approx -5.4\times 10^{-6}$ at $3.5\geq |z-z_{v}|\geq 0.5$.}}
\label{modelsdeltaa}
\end{figure}

Spatial variations in $\alpha $ will be dramatically different when
comparing the standard $CDM$ model with the $wCDM$ models. In the $SCDM$
model, the difference between the fine structure constant in a virialised
cluster ($\alpha _{v}$) and in the background ($\alpha _{b}$) will always be
the same, $\delta \alpha /\alpha \approx 5.2\times 10^{-6}$, independently
of the redshift at which we measure it, figure \ref{modelsdeltaa}. 
Again, this is because, in a $SCDM$ model, $\Delta _{c}$ is always a
constant independent of the redshift at which virialisation occurs. The
constancy of $\delta \alpha /\alpha $ is a signature of the $SCDM$
structure-formation model, and it may even provide a means to rule out the $%
SCDM$ model completely if, when comparing the value of $\delta \alpha
/\alpha $ in two different clusters, we do not find the same value,
independently of $z_{v}$. 
A similar result is found for the case of a dark-energy structure formation
model at high redshifts where $\delta \alpha /\alpha $ will be constant.
This behaviour is expected since at high redshift any $wCDM$ model is
asymptotically equivalent to standard $CDM$.

\begin{figure}\centering
\epsfig{file=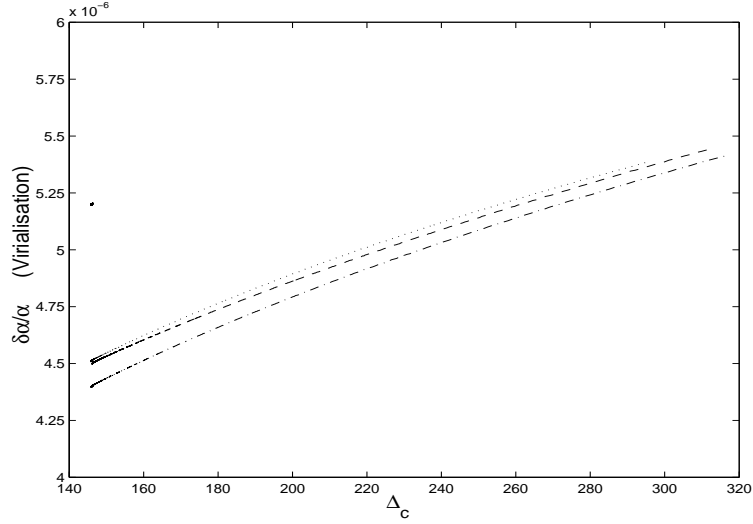,height=7cm,width=10cm}
\caption{\textit{Variation of $\protect\delta \protect\alpha /\protect\alpha 
$ as a function as a function of $\Delta _{c}$. The single-point corresponds
to the standard ($\Lambda =0$) $CDM$, the dashed line is the $\Lambda CDM$,
the dash-dot line corresponds to a $wCDM$ model with $w_{\protect\phi }=-0.8$%
, dotted-line corresponds to a $wCDM$ model with $w_{\protect\phi }=-0.6$.
In all the models, the initial conditions were set in order to have $\protect%
\alpha _{c}(z=0)=\protect\alpha _{0}$ and to satisfy $\Delta \protect\alpha /%
\protect\alpha \approx -5.4\times 10^{-6}$ at $3.5\geq |z-z_{v}|\geq 0.5$.}}
\label{modelsdeltarho}
\end{figure}

\begin{figure}\centering
\epsfig{file=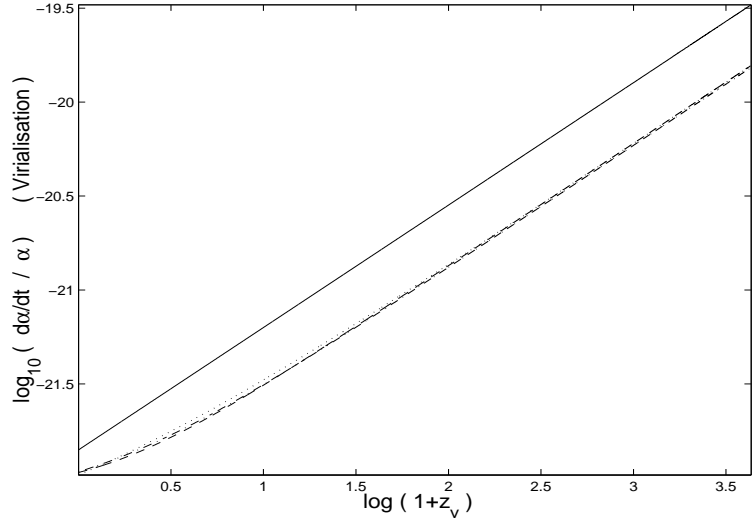,height=7cm,width=10cm}
\caption{\textit{Variation of $log_{10}(\dot{\protect\alpha}/\protect\alpha )
$ as a function of $log(1+z_{v})$. Solid line corresponds to the Standard ($%
\Lambda =0$) $CDM$, dashed-line is the $\Lambda CDM$, dashed-dotted
corresponds to a $wCDM$ model with $w_{\protect\phi }=-0.8$, dotted line
corresponds to a $wCDM$ model with $w_{\protect\phi }=-0.6$. In all the
models, the initial conditions were set in order to have $\protect\alpha %
_{c}(z=0)=\protect\alpha _{0}$ and to satisfy $\Delta \protect\alpha /%
\protect\alpha \approx -5.4\times 10^{-6}$ at $3.5\geq |z-z_{v}|\geq 0.5$.}}
\label{modelspsidot}
\end{figure}

As expected, it is at low redshifts that the difference between $wCDM$ and $%
SCDM$ emerges. When comparing virialised regions at low redshifts, $\delta
\alpha /\alpha $ will increase in a $wCDM$ model as we approach $z=0$. This
is due to an increase of the density contrast of the virialised regions, $%
\Delta _{c}$, and the approach to a constant value of $\alpha _{b}$, see
figure \ref{modelsdeltarho}. 
In general, the growth will be steeper for smaller values of $w_{\phi }$,
although there will be parameter degeneracies between the behaviour of
different models, which ensure that there is no simple relation between $%
\delta \alpha /\alpha $ and the dark-energy equation of state. Note that,
independently of the structure formation model we use, $\dot{\alpha}%
_{c}/\alpha _{c}$ at virialisation is always a decreasing function of time,
as shown in figure \ref{modelspsidot}. 
%

\section{Dependence on the Coupling of $\protect\alpha $ to
the Matter Fields}

The evolution of $\alpha $ in the background and inside clusters depends
mainly on the dominant equation of state of the universe and the \textit{sign%
} of the coupling constant $\zeta /\omega ,$ which is determined by the
theory and the dark matter's identity. %
%
As was shown in chapter \ref{qualitative}, $\alpha _{b}$ will be nearly constant for
an accelerated expansion and also during the radiation era far from the
initial singularity (where the kinetic term, $\rho _{\psi }$, can dominate). 
\emph{\ }Slow evolution of $\alpha $ will occur during the dust-dominated
epoch, where $\alpha $ increases logarithmically in time for $\zeta <0$.
When $\zeta $ is negative, $\alpha $ will be a slowly growing function of
time but $\alpha $ will fall rapidly (even during a curvature-dominated era)
for $\zeta $ positive \cite{bsbm}. A similar behaviour is found for the
evolution of the fine structure constant inside overdensities. Thus, we
see that a slow change in $\alpha $, cut off by the accelerated expansion at
low redshift, that may be required by the data, demands that $\zeta <0$ in
the cosmological background.

The sign of $\zeta $ is determined by the physical character of the matter
that carries electromagnetic coupling. If it is dominated by magnetic energy
then $\zeta <0$, if not then $\zeta >0.$ Baryons will usually have a
positive $\zeta $ (although Bekenstein has argued for negative baryonic $%
\zeta $ in references \cite{bek1,bek2}, but see \cite{damour1}), in particular $\zeta \approx
10^{-4}$ for neutrons and protons. Dark matter may have negative values of $%
\zeta $, for instance superconducting cosmic strings have$\ \zeta \approx -1$.

In the previous section, we have chosen the sign of $\zeta $ to be negative
so $\alpha $ is a slowly-growing function in time during the era of dust
domination. This was done in order to match the latest observations which
suggest that $\alpha $ had a smaller value in the past \cite{murphy,murphy1,webb,webb1}. This
is a good approximation, since we have been studying the cosmological
evolution of $\alpha $ during large-scale structure formation, when dark
matter dominates. However, we know that on sufficiently small scales the
dark matter will become dominated by a baryonic contribution for which $%
\zeta >0.$ The transition in the dominant form of total density, from
non-baryonic to baryonic as one goes from large to small scales requires a
significant evolution in the magnitude and sign of $\zeta /\omega $. This
inhomogeneity will create distinctive behaviours in the evolution of the
fine structure constant and will not be studied in this thesis. It is clear
that a change in the sign of $\zeta /\omega $ will lead to a completely
different type of evolution for $\alpha $, although the expected variations
in the sign of $\zeta /\omega $ will occur on scales much smaller than those
to which we are applying the spherical collapse model here. Hence, we will
only investigate the effects of changing the absolute value of the coupling, 
$|\zeta /\omega |$, for the evolution of the fine structure constant.


\bigskip

\begin{figure}\centering
\epsfig{file=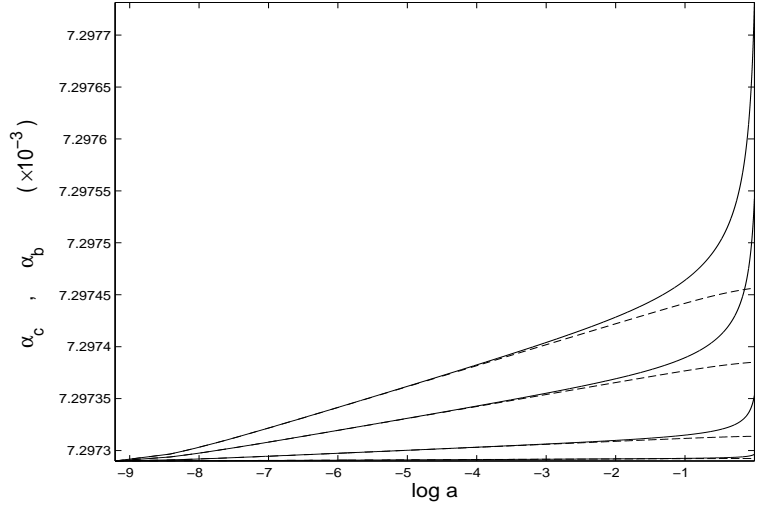,height=7cm,width=10cm}
\caption{\textit{Evolution of $\protect\alpha $ in the background
(dashed-line) and inside a cluster (Solid line) as a function of $\log (1+z)$.
For $\protect\zeta /\protect\omega =-10^{-4},-4\times 10^{-4},-7\times
10^{-4}$ and $-10^{-5}$. The lower curves correspond to a lower value of $%
\protect\zeta /\protect\omega $. The initial conditions were set in order to
have $\protect\alpha _{c}(z=0)=\protect\alpha _{0}$ and to satisfy $\Delta 
\protect\alpha /\protect\alpha =-5.4\times 10^{-6}$ at $3\geq |z-z_{v}|\geq 1
$, in the case where $\protect\zeta =-7\times 10^{-4}$. }}
\label{zetasalphacb}
\end{figure}

\begin{figure}\centering
\epsfig{file=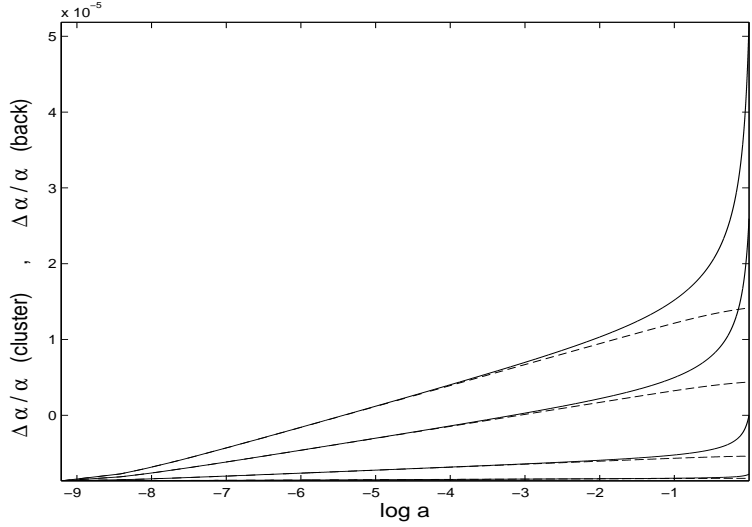,height=7cm,width=10cm}
\caption{\textit{Evolution of $\Delta \protect\alpha /\protect\alpha $ in
the background (dashed-line) and inside a cluster (Solid line) as a function
of $\log (1+z)$. For $\protect\zeta /\protect\omega =-10^{-4},-4\times
10^{-4},-7\times 10^{-4}$ and $-10^{-5}$. The lower curves correspond to a
lower value of $\protect\zeta /\protect\omega $. The initial conditions were
set in order to have $\protect\alpha _{c}(z=0)=\protect\alpha _{0}$ and to
satisfy $\Delta \protect\alpha /\protect\alpha =-5.4\times 10^{-6}$ at $%
3\geq |z-z_{v}|\geq 1$, in the case where $\protect\zeta =-7\times 10^{-4}$. 
}}
\label{zetasalphacbback}
\end{figure}

From figure \ref{zetasalphacb} it is clear that the rate of changes in $%
\alpha _{b}$ and $\alpha _{c}$ will be functions of the absolute value of $%
\zeta /\omega $. Smaller values of $|\zeta /\omega |$ lead to a slower
variation of $\alpha $. A similar behaviour is found for the time variations
in $\Delta \alpha /\alpha $, see figure \ref{zetasalphacbback}. 

\begin{figure}\centering
\epsfig{file=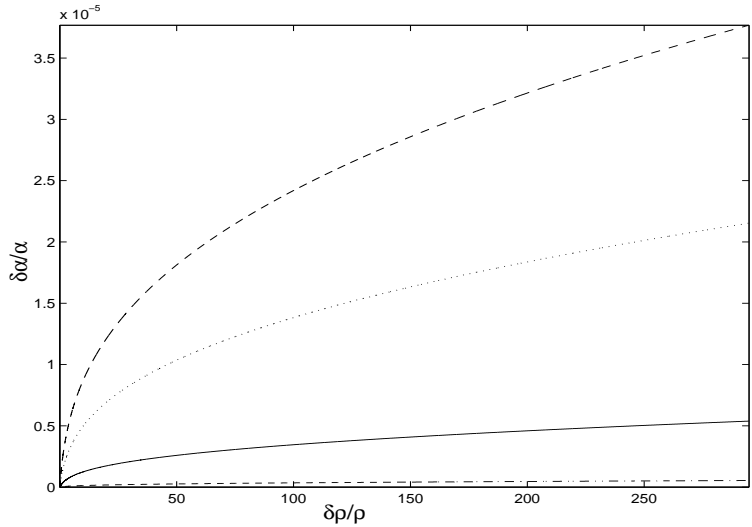,height=7cm,width=10cm}
\caption{\textit{Evolution of $\protect\delta \protect\alpha /\protect\alpha 
$ as a function $\protect\delta \protect\rho /\protect\rho $. For $\protect%
\zeta /\protect\omega =-10^{-4},-4\times 10^{-4},-7\times 10^{-4}$ and $%
-10^{-5}$. The lower curves correspond to a lower value of $\protect\zeta /%
\protect\omega $. The initial conditions were set in order to have $\protect%
\alpha _{c}(z=0)=\protect\alpha _{0}$ and to satisfy $\Delta \protect\alpha /%
\protect\alpha \approx -5.4\times 10^{-6}$ at $3\geq |z-z_{v}|\geq 1$, in
the case where $\protect\zeta =-7\times 10^{-4}$. }}
\label{zetasdeltaa}
\end{figure}

The faster variation in $\alpha $ and $\Delta \alpha /\alpha $ for higher
values of $|\zeta /\omega |$ is also a common feature for $\delta \alpha
/\alpha $, see figure \ref{zetasdeltaa}. This is expected. A stronger
coupling to the matter fields would naturally lead to a stronger dependence
on the matter inhomogeneities, and in particular on their density contrast, $%
\Delta _{c}$. 
%
%
%
%

\clearpage

In reality, the dependence of $\alpha $ on the coupling $\zeta /\omega $ has
a degeneracy with respect to the initial condition chosen for $\psi $. This
is clear from the scale invariance of equation (\ref{psidot}) under linear
shifts in the value of $\psi \rightarrow \psi +const$ and rescaling of $%
\zeta /\omega $ and $t$. It is always possible, to obtain the same evolution
(and rate of change) of $\alpha $ and $\Delta \alpha /\alpha $ for any other
value of $|\zeta /\omega |$; see for example Tables \ref{zeta1},\ref{zeta4},%
\ref{zeta6} and \ref{zeta10},  where we have
tabulated the shifts $\Delta \alpha /\alpha ,\delta \alpha /\alpha ,$ and
time change, $\dot{\alpha}/\alpha ,$ obtained in the clusters and in the
background for various numerical choices of $\zeta /\omega $ and
virialisation redshift $z_{v}$. The observed Oklo and $\beta $-decay
constraints on variations in $\alpha $ are highlighted in italic and
boldface, respectively.

\begingroup

\begin{table}[htb!] 
\begin{center}
\begin{tabular}{|c|c|c|c|c|c|c|}
\hline $z_v$ & $\frac{\Delta\alpha}{\alpha} \vert_{b}\times 10^{-6}$ &
 $\frac{\Delta\alpha}{\alpha}         \vert_{c}\times         10^{-7}$
 &$\frac{\dot\alpha}{\alpha}      \vert_{b}\times      10^{-24}$     &
 $\frac{\dot\alpha}{\alpha}          \vert_{c}\times         10^{-22}$
 &$\frac{\delta\alpha}{\alpha}\times      10^{-6}$      \\      \hline
 $0.00$&$-5.381$&$     0.000$&$      0.361$&$     1.068$&$     5.38$\\
 ${\it{0.09}}$&${\it{-5.397}}$&${\it{ -1.618}}$&${\it{ 0.427}}$&${\it{
 1.128}}$&${\it{   5.24}}$\\   $0.18$&$-5.412$&$  -2.889$&$   0.493$&$
 1.192$&$   5.12$\\  $0.26$&$-5.427$&$  -3.915$&$   0.558$&$  1.260$&$
 5.04$\\   $0.34$&$-5.442$&$  -4.769$&$   0.623$&$   1.329$&$  4.97$\\
 ${\bf{0.41}}$&${\bf{-5.456}}$&$ {\bf{-5.485}}$&${\bf{ 0.688}}$&${\bf{
 1.401}}$&${\bf{   4.91}}$\\   $0.49$&$-5.470$&$  -6.097$&$   0.753$&$
 1.475$&$   4.86$\\  $0.51$&$-5.474$&$  -6.271$&$   0.773$&$  1.498$&$
 4.85$\\   $0.61$&$-5.492$&$  -6.941$&$   0.860$&$   1.601$&$  4.80$\\
 $0.70$&$-5.509$&$     -7.510$&$     0.948$&$     1.708$&$     4.76$\\
 $0.80$&$-5.526$&$     -8.001$&$     1.036$&$     1.816$&$     4.73$\\
 $0.89$&$-5.542$&$     -8.427$&$     1.124$&$     1.928$&$     4.70$\\
 $0.95$&$-5.552$&$     -8.684$&$     1.184$&$     2.005$&$     4.68$\\
 $1.07$&$-5.572$&$     -9.139$&$     1.303$&$     2.161$&$     4.66$\\
 $1.19$&$-5.590$&$     -9.534$&$     1.424$&$     2.321$&$     4.64$\\
 $1.30$&$-5.608$&$     -9.882$&$     1.546$&$     2.484$&$     4.62$\\
 $1.36$&$-5.617$&$-10.048$&$       1.608$&$      2.569$&$      4.61$\\
 $1.51$&$-5.639$&$-10.428$&$       1.773$&$      2.793$&$      4.60$\\
 $1.66$&$-5.660$&$-10.767$&$       1.940$&$      3.024$&$      4.58$\\
 $1.76$&$-5.674$&$-10.973$&$       2.053$&$      3.181$&$      4.58$\\
 $1.95$&$-5.699$&$-11.333$&$       2.278$&$      3.496$&$      4.57$\\
 $2.14$&$-5.722$&$-11.658$&$       2.508$&$      3.821$&$      4.56$\\
 $2.15$&$-5.723$&$-11.670$&$       2.519$&$      3.838$&$      4.56$\\
 $2.40$&$-5.751$&$-12.035$&$       2.824$&$      4.272$&$      4.55$\\
 $2.54$&$-5.767$&$-12.237$&$       3.009$&$      4.535$&$      4.54$\\
 $2.85$&$-5.799$&$-12.624$&$       3.417$&$      5.121$&$      4.54$\\
 $2.92$&$-5.807$&$-12.711$&$       3.521$&$      5.272$&$      4.54$\\
 $3.31$&$-5.843$&$-13.129$&$       4.055$&$      6.043$&$      4.53$\\
 $3.69$&$-5.876$&$-13.495$&$       4.612$&$      6.852$&$      4.53$\\
 $4.07$&$-5.906$&$-13.835$&$       5.190$&$      7.690$&$      4.52$\\
 $4.45$&$-5.935$&$-14.137$&$       5.788$&$      8.562$&$      4.52$\\
 $4.83$&$-5.961$&$-14.415$&$ 6.407$&$ 9.466$&$ 4.52$\\
\hline
\end{tabular}
\caption{\label{zeta1}$\frac{\protect\zeta }{\protect\omega }=-10^{-1}$: Time and space
variations in $\protect\alpha $ obtained for the corresponding redshifts of
virialisation, $z_{v}$, for $\protect\zeta /\protect\omega =-10^{-1}$. We
have assumed a $\Lambda CDM$ model. The indexes $^{\prime }b^{\prime }$ and $%
^{\prime }c^{\prime }$, stand for background and cluster respectively. The
italic and bold entries correspond approximately to the level of the Oklo
and $\protect\beta $-decay rate constraints, respectively. The quasar
absorption spectra observations correspond to the values of $\Delta \protect%
\alpha /\protect\alpha |_{b}$. The initial conditions were set in order to
have $\protect\alpha _{c}(z=0)=\protect\alpha _{0}$.}
\end{center}
\end{table}
\endgroup
\begingroup

\begin{table}[htb!] 
\begin{center}
\begin{tabular}{|c|c|c|c|c|c|c|}
\hline $z_v$ & $\frac{\Delta\alpha}{\alpha} \vert_{b}\times 10^{-6}$ &
 $\frac{\Delta\alpha}{\alpha}         \vert_{c}\times         10^{-7}$
 &$\frac{\dot\alpha}{\alpha}      \vert_{b}\times      10^{-24}$     &
 $\frac{\dot\alpha}{\alpha}          \vert_{c}\times         10^{-22}$
 &$\frac{\delta\alpha}{\alpha}\times      10^{-6}$      \\      \hline
 $0.00$&$-5.382$&$     0.000$&$      0.361$&$     1.068$&$     5.38$\\
 ${\it{0.09}}$&${\it{-5.398}}$&${\it{ -1.618}}$&${\it{ 0.427}}$&${\it{
 1.128}}$&${\it{   5.24}}$\\   $0.18$&$-5.414$&$  -2.889$&$   0.493$&$
 1.193$&$   5.12$\\  $0.26$&$-5.429$&$  -3.916$&$   0.558$&$  1.260$&$
 5.04$\\   $0.34$&$-5.443$&$  -4.769$&$   0.623$&$   1.329$&$  4.97$\\
 ${\bf{0.41}}$&${\bf{-5.458}}$&${\bf{ -5.486}}$&${\bf{ 0.688}}$&${\bf{
 1.401}}$&${\bf{   4.91}}$\\   $0.49$&$-5.471$&$  -6.098$&$   0.753$&$
 1.475$&$   4.86$\\  $0.51$&$-5.476$&$  -6.272$&$   0.773$&$  1.498$&$
 4.85$\\   $0.61$&$-5.493$&$  -6.943$&$   0.860$&$   1.602$&$  4.80$\\
 $0.70$&$-5.511$&$     -7.511$&$     0.948$&$     1.708$&$     4.76$\\
 $0.80$&$-5.527$&$     -8.003$&$     1.036$&$     1.817$&$     4.73$\\
 $0.89$&$-5.543$&$     -8.428$&$     1.124$&$     1.929$&$     4.70$\\
 $0.95$&$-5.553$&$     -8.685$&$     1.184$&$     2.006$&$     4.68$\\
 $1.07$&$-5.573$&$     -9.141$&$     1.303$&$     2.161$&$     4.66$\\
 $1.19$&$-5.592$&$     -9.535$&$     1.424$&$     2.321$&$     4.64$\\
 $1.30$&$-5.610$&$     -9.884$&$     1.546$&$     2.485$&$     4.62$\\
 $1.36$&$-5.619$&$-10.050$&$       1.609$&$      2.569$&$      4.61$\\
 $1.51$&$-5.641$&$-10.430$&$       1.773$&$      2.794$&$      4.60$\\
 $1.66$&$-5.662$&$-10.769$&$       1.940$&$      3.025$&$      4.58$\\
 $1.76$&$-5.675$&$-10.976$&$       2.053$&$      3.182$&$      4.58$\\
 $1.95$&$-5.700$&$-11.336$&$       2.278$&$      3.497$&$      4.57$\\
 $2.14$&$-5.723$&$-11.660$&$       2.508$&$      3.822$&$      4.56$\\
 $2.15$&$-5.725$&$-11.672$&$       2.520$&$      3.839$&$      4.56$\\
 $2.40$&$-5.753$&$-12.038$&$       2.825$&$      4.273$&$      4.55$\\
 $2.54$&$-5.769$&$-12.239$&$       3.009$&$      4.536$&$      4.54$\\
 $2.85$&$-5.801$&$-12.626$&$       3.417$&$      5.122$&$      4.54$\\
 $2.92$&$-5.808$&$-12.713$&$       3.522$&$      5.273$&$      4.54$\\
 $3.31$&$-5.844$&$-13.132$&$       4.056$&$      6.045$&$      4.53$\\
 $3.69$&$-5.877$&$-13.498$&$       4.613$&$      6.853$&$      4.53$\\
 $4.07$&$-5.908$&$-13.837$&$       5.191$&$      7.692$&$      4.52$\\
 $4.45$&$-5.936$&$-14.140$&$       5.790$&$      8.565$&$      4.52$\\
 $4.83$&$-5.962$&$-14.418$&$ 6.408$&$ 9.469$&$ 4.52$\\
\hline
\end{tabular}
\caption{\label{zeta4}$\frac{\protect\zeta }{\protect\omega }=-10^{-4}$: Time and space
variations in $\protect\alpha $ obtained for the corresponding redshifts of
virialisation, $z_{v}$, for $\protect\zeta /\protect\omega =-10^{-4}$. We
have assumed a $\Lambda CDM$ model. The indexes $^{\prime }b^{\prime }$ and $%
^{\prime }c^{\prime }$, stand for background and cluster respectively. The
italic and bold entries correspond approximately to the level of the Oklo
and $\protect\beta $-decay rate constraints, respectively. The quasar
absorption spectra observations correspond to the values of $\Delta \protect%
\alpha /\protect\alpha |_{b}$. The initial conditions were set in order to
have $\protect\alpha _{c}(z=0)=\protect\alpha _{0}$.}
\end{center}
\end{table}
\endgroup
\begingroup

\begin{table}[htb!] 
\begin{center}
\begin{tabular}{|c|c|c|c|c|c|c|}
\hline $z_v$ & $\frac{\Delta\alpha}{\alpha} \vert_{b}\times 10^{-6}$ &
 $\frac{\Delta\alpha}{\alpha}         \vert_{c}\times         10^{-7}$
 &$\frac{\dot\alpha}{\alpha}      \vert_{b}\times      10^{-24}$     &
 $\frac{\dot\alpha}{\alpha}          \vert_{c}\times         10^{-22}$
 &$\frac{\delta\alpha}{\alpha}\times      10^{-6}$      \\      \hline
 $0.00$&$-5.381$&$     0.000$&$      0.361$&$     1.068$&$     5.38$\\
 ${\it{0.09}}$&${\it{-5.397}}$&${\it{ -1.617}}$&${\it{ 0.427}}$&${\it{
 1.128}}$&${\it{   5.24}}$\\   $0.18$&$-5.413$&$  -2.889$&$   0.493$&$
 1.192$&$   5.12$\\  $0.26$&$-5.428$&$  -3.915$&$   0.558$&$  1.260$&$
 5.04$\\   $0.34$&$-5.442$&$  -4.768$&$   0.623$&$   1.329$&$  4.97$\\
 ${\bf{0.41}}$&${\bf{-5.457}}$&${\bf{ -5.485}}$&${\bf{ 0.688}}$&${\bf{
 1.401}}$&${\bf{   4.91}}$\\   $0.49$&$-5.470$&$  -6.097$&$   0.753$&$
 1.475$&$   4.86$\\  $0.51$&$-5.475$&$  -6.271$&$   0.773$&$  1.498$&$
 4.85$\\   $0.61$&$-5.492$&$  -6.941$&$   0.860$&$   1.601$&$  4.80$\\
 $0.70$&$-5.510$&$     -7.510$&$     0.948$&$     1.708$&$     4.76$\\
 $0.80$&$-5.526$&$     -8.001$&$     1.036$&$     1.817$&$     4.73$\\
 $0.89$&$-5.542$&$     -8.427$&$     1.124$&$     1.929$&$     4.70$\\
 $0.95$&$-5.552$&$     -8.684$&$     1.184$&$     2.006$&$     4.68$\\
 $1.07$&$-5.572$&$     -9.139$&$     1.303$&$     2.161$&$     4.66$\\
 $1.19$&$-5.591$&$     -9.534$&$     1.424$&$     2.321$&$     4.64$\\
 $1.30$&$-5.609$&$     -9.882$&$     1.546$&$     2.485$&$     4.62$\\
 $1.36$&$-5.618$&$-10.048$&$       1.608$&$      2.569$&$      4.61$\\
 $1.51$&$-5.640$&$-10.428$&$       1.773$&$      2.794$&$      4.60$\\
 $1.66$&$-5.661$&$-10.767$&$       1.940$&$      3.024$&$      4.58$\\
 $1.76$&$-5.674$&$-10.973$&$       2.053$&$      3.181$&$      4.58$\\
 $1.95$&$-5.699$&$-11.334$&$       2.278$&$      3.497$&$      4.57$\\
 $2.14$&$-5.722$&$-11.658$&$       2.508$&$      3.822$&$      4.56$\\
 $2.15$&$-5.724$&$-11.670$&$       2.519$&$      3.838$&$      4.56$\\
 $2.40$&$-5.752$&$-12.036$&$       2.824$&$      4.272$&$      4.55$\\
 $2.54$&$-5.768$&$-12.237$&$       3.009$&$      4.535$&$      4.54$\\
 $2.85$&$-5.800$&$-12.624$&$       3.417$&$      5.121$&$      4.54$\\
 $2.92$&$-5.807$&$-12.711$&$       3.521$&$      5.272$&$      4.54$\\
 $3.31$&$-5.843$&$-13.129$&$       4.056$&$      6.044$&$      4.53$\\
 $3.69$&$-5.876$&$-13.495$&$       4.612$&$      6.852$&$      4.53$\\
 $4.07$&$-5.907$&$-13.835$&$       5.190$&$      7.690$&$      4.52$\\
 $4.45$&$-5.935$&$-14.137$&$       5.789$&$      8.563$&$      4.52$\\
 $4.83$&$-5.961$&$-14.415$&$ 6.407$&$ 9.467$&$ 4.52$\\
\hline
\end{tabular}
\caption{\label{zeta6}$\frac{\protect\zeta }{\protect\omega }=-10^{-6}$: Time and space
variations in $\protect\alpha $ obtained for the corresponding redshifts of
virialisation, $z_{v}$, for $\protect\zeta /\protect\omega =-10^{-6}$. We
have assumed a $\Lambda CDM$ model. The indexes $^{\prime }b^{\prime }$ and $%
^{\prime }c^{\prime }$, stand for background and cluster respectively. The
italic and bold entries correspond approximately to the level of the Oklo
and $\protect\beta $-decay rate constraints, respectively. The quasar
absorption spectra observations correspond to the values of $\Delta \protect%
\alpha /\protect\alpha |_{b}$. The initial conditions were set in order to
have $\protect\alpha _{c}(z=0)=\protect\alpha _{0}$.}
\end{center}
\end{table}
\endgroup
\begingroup

\begin{table}[htb!] 
\begin{center}
\begin{tabular}{|c|c|c|c|c|c|c|}
\hline $z_v$ & $\frac{\Delta\alpha}{\alpha} \vert_{b}\times 10^{-6}$ &
 $\frac{\Delta\alpha}{\alpha}         \vert_{c}\times         10^{-7}$
 &$\frac{\dot\alpha}{\alpha}      \vert_{b}\times      10^{-24}$     &
 $\frac{\dot\alpha}{\alpha}          \vert_{c}\times         10^{-22}$
 &$\frac{\delta\alpha}{\alpha}\times      10^{-6}$      \\      \hline
 $0.00$&$-5.385$&$     0.000$&$      0.361$&$     1.068$&$     5.38$\\
 ${\it{0.09}}$&${\it{-5.401}}$&${\it{ -1.619}}$&${\it{ 0.428}}$&${\it{
 1.129}}$&${\it{   5.24}}$\\   $0.18$&$-5.416$&$  -2.891$&$   0.493$&$
 1.193$&$   5.13$\\  $0.26$&$-5.431$&$  -3.918$&$   0.559$&$  1.261$&$
 5.04$\\   $0.34$&$-5.446$&$  -4.772$&$   0.624$&$   1.330$&$  4.97$\\
 ${\bf{0.41}}$&${\bf{-5.460}}$&${\bf{ -5.489}}$&${\bf{ 0.689}}$&${\bf{
 1.402}}$&${\bf{   4.91}}$\\   $0.49$&$-5.474$&$  -6.101$&$   0.753$&$
 1.476$&$   4.86$\\  $0.51$&$-5.478$&$  -6.275$&$   0.773$&$  1.499$&$
 4.85$\\   $0.61$&$-5.496$&$  -6.946$&$   0.861$&$   1.602$&$  4.80$\\
 $0.70$&$-5.513$&$     -7.515$&$     0.948$&$     1.709$&$     4.76$\\
 $0.80$&$-5.530$&$     -8.007$&$     1.036$&$     1.818$&$     4.73$\\
 $0.89$&$-5.546$&$     -8.432$&$     1.125$&$     1.930$&$     4.70$\\
 $0.95$&$-5.556$&$     -8.690$&$     1.185$&$     2.007$&$     4.69$\\
 $1.07$&$-5.576$&$     -9.145$&$     1.304$&$     2.162$&$     4.66$\\
 $1.19$&$-5.595$&$     -9.540$&$     1.425$&$     2.322$&$     4.64$\\
 $1.30$&$-5.613$&$     -9.889$&$     1.547$&$     2.486$&$     4.62$\\
 $1.36$&$-5.621$&$-10.055$&$       1.610$&$      2.571$&$      4.62$\\
 $1.51$&$-5.644$&$-10.435$&$       1.774$&$      2.795$&$      4.60$\\
 $1.66$&$-5.664$&$-10.774$&$       1.941$&$      3.026$&$      4.59$\\
 $1.76$&$-5.678$&$-10.981$&$       2.054$&$      3.183$&$      4.58$\\
 $1.95$&$-5.703$&$-11.341$&$       2.279$&$      3.499$&$      4.57$\\
 $2.14$&$-5.726$&$-11.666$&$       2.510$&$      3.824$&$      4.56$\\
 $2.15$&$-5.727$&$-11.678$&$       2.521$&$      3.841$&$      4.56$\\
 $2.40$&$-5.756$&$-12.044$&$       2.826$&$      4.275$&$      4.55$\\
 $2.54$&$-5.771$&$-12.245$&$       3.011$&$      4.538$&$      4.55$\\
 $2.85$&$-5.803$&$-12.633$&$       3.419$&$      5.125$&$      4.54$\\
 $2.92$&$-5.811$&$-12.719$&$       3.523$&$      5.276$&$      4.54$\\
 $3.31$&$-5.847$&$-13.138$&$       4.058$&$      6.048$&$      4.53$\\
 $3.69$&$-5.880$&$-13.505$&$       4.615$&$      6.857$&$      4.53$\\
 $4.07$&$-5.911$&$-13.844$&$       5.193$&$      7.695$&$      4.53$\\
 $4.45$&$-5.939$&$-14.147$&$       5.792$&$      8.569$&$      4.52$\\
 $4.83$&$-5.965$&$-14.425$&$ 6.412$&$ 9.473$&$ 4.52$\\
\hline
\end{tabular}
\caption{\label{zeta10}$\frac{\protect\zeta }{\protect\omega }=-10^{-10}$: Time and space
variations in $\protect\alpha $ obtained for the corresponding redshifts of
virialisation, $z_{v}$, for $\protect\zeta /\protect\omega =-10^{-10}$. We
have assumed a $\Lambda CDM$ model. The indexes $^{\prime }b^{\prime }$ and $%
^{\prime }c^{\prime }$, stand for background and cluster respectively. The
italic and bold entries correspond approximately to the level of the Oklo
and $\protect\beta $-decay rate constraints, respectively. The quasar
absorption spectra observations correspond to the values of $\Delta \protect%
\alpha /\protect\alpha |_{b}$. The initial conditions were set in order to
have $\protect\alpha _{c}(z=0)=\protect\alpha _{0}$.}
\end{center}
\end{table}
\endgroup

In the plots of this section, we have chosen the value $\zeta /\omega
=-7\times 10^{-4}$ to be the one which satisfies the current observations,
but we could have used any value of $|\zeta /\omega |$ because of the
invariance under rescalings. However, once we set the initial condition for
a given value of $\zeta /\omega $, any deviation from that value leads to
quite different future variations in $\alpha $. This can occur if there are
regions of the universe where the dominant matter has a different nature,
and do possesses a different value (and even sign) of $\zeta /\omega $
to that in our solar system. The evolution of $\alpha $ in those regions may
then be different from the one that led to the value of $\alpha
_{c}(z=z_{v})=\alpha _{0}$ on Earth. 

It is important to note that, due to the degeneracy between the initial
condition and the coupling to the matter fields, there is no way to avoid
evolving differences in $\alpha $ variations between the overdensities and
the background. The difference will be of the same order of magnitude as the
effects indicated by the recent quasar absorption-line data. This is not a
coincidence and it is related to the fact that we have normalised $\alpha
(z=z_{v})=\alpha _{0}$ on Earth and $\Delta \alpha /\alpha =5.4\times
10^{-6} $ at $3.5\geq |z-z_{v}|\geq 0.5$. So, any varying-$\alpha $ model
that uses these normalisations will create a similar difference in $\alpha $
evolution between the overdensities and the background, independently of the
coupling $\zeta /\omega $.

\bigskip

We might ask: how model-independent are these results? In this connection it
is interesting to note that even with a zero coupling to the matter fields
(which is unrealistic), $\zeta =0$, there is no way to avoid difference
arising between the evolution of $\alpha $ in the background and in the
cluster overdensities. While the background\ expansion has a
monotonically-increasing scale factor, $a(t)$, the overdensities will have a
scale radius, $R(t)$, which will eventually collapse at a finite time. For
instance, in the case where $\zeta =0$, equations (\ref{psidot}) and (\ref%
{psidotcluster1}) can be automatically integrated to give: 
\begin{eqnarray}
\dot{\psi} &\propto &a^{-3} \\
\dot{\psi}_{c} &\propto &R^{-3}
\end{eqnarray}%
The difference between those two solutions clearly increases, especially
after the turnaround of the overdensity, when that region starts to
collapse. As the collapse proceeds, the bigger will be the difference
between the background and the overdensity. Variations in $\alpha $ between
the background and the overdensities are therefore\ quite natural although
they have always been ignored in studies of varying constants in cosmology. 

\section{Discussion of the Results and Observational Constraints}

The development of matter inhomogeneities in our universe affects the
cosmological evolution of the fine structure constant \cite{mota2,mota3}.
Therefore, variations in $\alpha $ depend on nature of its coupling to the
matter fields and the detailed large-scale structure formation model.
Large-scale structure formation models depend in turn on the dark-energy
equation of state. This dependence is particularly strong at low redshifts,
when dark energy dominates the density of the universe \cite{lahav,wang}. Using
the BSBM varying-$\alpha $ theory, and the simplest spherical collapse
model, we have studied the effects of the dark-energy equation of state and
the coupling to the matter fields on the evolution of the fine structure
constant. We have compared the evolution of $\alpha $ inside virialised
overdensities, using the standard ($\Lambda =0$) $CDM$ model of structure
formation and dark-energy modification ($wCDM$). It was shown that,
independently of the model of structure formation one considers, there is
always a spatial contrast, $\delta \alpha /\alpha $, between $\alpha $ in an
overdensity and in the background. In a $SCDM$ model, $\delta \alpha /\alpha 
$ is always a constant, independent of the virialisation redshift, see Table %
\ref{darkscdm}. In the case of a $wCDM$ model, especially at low redshifts,
the spatial contrast depends on the time when virialisation occurs and the
equation of state of the dark energy. At high redshifts, when the $wCDM$
model becomes asymptotically equivalent to the $SCDM$ one, $\delta \alpha
/\alpha $ is a constant. At low redshifts, when dark energy starts to
dominate, the difference between $\alpha $ in a cluster and in the
background grows. The growth rate is proportional to $|w_{\phi }|$, see
Tables \ref{darkw6cdm}, \ref{darkw8cdm} and \ref{darklcdm}. These
differences in the behaviour of the fine structure constant, its ``time
shift density contrast'' ($\Delta \alpha /\alpha $) and its ``spatial density
contrast'' ($\delta \alpha /\alpha $) could help us to distinguish among
different dark-energy models of structure formation at low redshifts. %

\begingroup

\begin{table}[htb!] 
\begin{center}
\begin{tabular}{|c|c|c|c|c|c|c|}
\hline $z_v$ & $\frac{\Delta\alpha}{\alpha} \vert_{b}\times 10^{-6}$ &
 $\frac{\Delta\alpha}{\alpha}         \vert_{c}\times         10^{-7}$
 &$\frac{\dot\alpha}{\alpha}      \vert_{b}\times      10^{-24}$     &
 $\frac{\dot\alpha}{\alpha}          \vert_{c}\times         10^{-22}$
 &$\frac{\delta\alpha}{\alpha}\times      10^{-6}$      \\      \hline
 $0.00$&$-5.203$&$     0.000$&$      0.959$&$     1.409$&$     5.20$\\
 $0.06$&$-5.229$&$-0.266$&$       1.047$&$       1.537$&$      5.20$\\
 ${\it{0.12}}$&${\it{-5.254}}$&${\it{-0.517}}$&${\it{   1.137}}$&${\it{
 1.670}}$&${\it{    5.20}}$\\    $0.18$&$-5.278$&$-0.754$&$   1.230$&$
 1.806$&$ 5.20$\\ $0.24$&$-5.301$&$-0.981$&$ 1.325$&$ 1.945$&$ 5.20$\\
 $0.30$&$-5.322$&$-1.196$&$       1.422$&$       2.088$&$      5.20$\\
 $0.36$&$-5.343$&$-1.401$&$       1.522$&$       2.235$&$      5.20$\\
 $0.38$&$-5.349$&$-1.463$&$       1.553$&$       2.281$&$      5.20$\\
 ${\bf{0.46}}$&${\bf{-5.374}}$&${\bf{-1.717}}$&${\bf{   1.690}}$&${\bf{
 2.481}}$&${\bf{    5.20}}$\\    $0.54$&$-5.399$&$-1.962$&$   1.830$&$
 2.687$&$ 5.20$\\ $0.62$&$-5.422$&$-2.192$&$ 1.974$&$ 2.898$&$ 5.20$\\
 $0.70$&$-5.443$&$-2.408$&$       2.121$&$       3.116$&$      5.20$\\
 $0.76$&$-5.459$&$-2.568$&$       2.235$&$       3.283$&$      5.20$\\
 $0.86$&$-5.486$&$-2.830$&$       2.438$&$       3.580$&$      5.20$\\
 $0.97$&$-5.510$&$-3.079$&$       2.645$&$       3.885$&$      5.20$\\
 $1.07$&$-5.534$&$-3.313$&$       2.859$&$       4.198$&$      5.20$\\
 $1.14$&$-5.548$&$-3.457$&$       2.996$&$       4.399$&$      5.20$\\
 $1.27$&$-5.576$&$-3.740$&$       3.285$&$       4.824$&$      5.20$\\
 $1.41$&$-5.603$&$-4.000$&$       3.583$&$       5.262$&$      5.20$\\
 $1.52$&$-5.623$&$-4.200$&$       3.827$&$       5.620$&$      5.20$\\
 $1.69$&$-5.653$&$-4.506$&$       4.230$&$       6.211$&$      5.20$\\
 $1.86$&$-5.681$&$-4.792$&$       4.646$&$       6.822$&$      5.20$\\
 $1.89$&$-5.686$&$-4.840$&$       4.723$&$       6.936$&$      5.20$\\
 $2.12$&$-5.720$&$-5.175$&$       5.273$&$       7.744$&$      5.20$\\
 $2.27$&$-5.742$&$-5.402$&$       5.679$&$       8.340$&$      5.20$\\
 $2.55$&$-5.779$&$-5.768$&$       6.417$&$       9.426$&$      5.20$\\
 $2.65$&$-5.792$&$-5.899$&$       6.693$&$       9.830$&$      5.20$\\
 $3.00$&$-5.834$&$-6.315$&$          7.668$&$11.261$&$         5.20$\\
 $3.03$&$-5.837$&$-6.345$&$          7.760$&$11.399$&$         5.20$\\
 $3.41$&$-5.878$&$-6.760$&$ 8.879$&$13.039$&$ 5.20$\\
$3.79$&$-5.916$&$-7.136$&$10.046$&$14.754$&$ 5.20$\\
$4.17$&$-5.950$&$-7.476$&$11.261$&$16.542$&$                    5.20$\\
$4.54$&$-5.982$&$-7.803$&$12.521$&$18.389$&$                    5.20$\\
$4.92$&$-6.012$&$-8.102$&$13.824$&$20.305$&$ 5.20$\\
\hline
\end{tabular}
\caption{\label{darkscdm}$(\Lambda =0)$ CDM: Time and space variations in $\protect\alpha $
for the corresponding redshifts of virialisation, $z_{v}$, in the $(\Lambda
=0)$ standard Cold Dark Matter model. The indexes $^{\prime }b^{\prime }$
and $^{\prime }c^{\prime }$, stand for background and cluster respectively.
The italic and bold entries correspond approximately to the Oklo and to the $%
\protect\beta $-decay rate constraint, respectively. The quasar absorption
spectra observations correspond to the values of $\Delta \protect\alpha /%
\protect\alpha |_{b}$. The initial conditions were set for $\protect\zeta /%
\protect\omega =-2\times 10^{-4}$, in order to have $\protect\alpha %
_{c}(z=0)=\protect\alpha _{0}$.}
\end{center}
\end{table}
\endgroup
\begingroup

\begin{table}[htb!] 
\begin{center}
\begin{tabular}{|c|c|c|c|c|c|c|}
\hline $z_v$ & $\frac{\Delta\alpha}{\alpha} \vert_{b}\times 10^{-6}$ &
 $\frac{\Delta\alpha}{\alpha}         \vert_{c}\times         10^{-7}$
 &$\frac{\dot\alpha}{\alpha}      \vert_{b}\times      10^{-24}$     &
 $\frac{\dot\alpha}{\alpha}          \vert_{c}\times         10^{-22}$
 &$\frac{\delta\alpha}{\alpha}\times      10^{-6}$      \\      \hline
 $0.00$&$-5.440$&$     0.000$&$      0.332$&$     1.036$&$     5.44$\\
 $0.08$&$-5.453$&$     -1.130$&$     0.385$&$     1.121$&$     5.34$\\
 ${\it{0.17}}$&${\it{-5.466}}$&${\it{ -2.077}}$&${\it{ 0.440}}$&${\it{
 1.207}}$&${\it{   5.26}}$\\   $0.25$&$-5.478$&$  -2.886$&$   0.497$&$
 1.293$&$   5.19$\\  $0.32$&$-5.490$&$  -3.591$&$   0.554$&$  1.381$&$
 5.13$\\     ${\bf{0.40}}$&${\bf{-5.502}}$&${\bf{     -4.208}}$&${\bf{
 0.613}}$&${\bf{ 1.469}}$&${\bf{ 5.08}}$\\ $0.48$&$-5.513$&$ -4.759$&$
 0.673$&$   1.558$&$  5.04$\\  $0.50$&$-5.517$&$   -4.914$&$  0.691$&$
 1.586$&$   5.03$\\  $0.60$&$-5.532$&$  -5.564$&$   0.776$&$  1.710$&$
 4.98$\\   $0.70$&$-5.546$&$  -6.128$&$   0.862$&$   1.837$&$  4.93$\\
 $0.80$&$-5.560$&$     -6.631$&$     0.950$&$     1.966$&$     4.90$\\
 $0.90$&$-5.574$&$     -7.078$&$     1.040$&$     2.097$&$     4.87$\\
 $0.95$&$-5.582$&$     -7.317$&$     1.091$&$     2.172$&$     4.85$\\
 $1.08$&$-5.599$&$     -7.828$&$     1.217$&$     2.354$&$     4.82$\\
 $1.21$&$-5.616$&$     -8.283$&$     1.345$&$     2.540$&$     4.79$\\
 $1.34$&$-5.632$&$     -8.691$&$     1.476$&$     2.729$&$     4.76$\\
 $1.39$&$-5.638$&$     -8.828$&$     1.524$&$     2.798$&$     4.75$\\
 $1.55$&$-5.658$&$     -9.297$&$     1.704$&$     3.059$&$     4.73$\\
 $1.72$&$-5.677$&$     -9.715$&$     1.889$&$     3.326$&$     4.71$\\
 $1.80$&$-5.687$&$     -9.910$&$     1.984$&$     3.464$&$     4.70$\\
 $2.02$&$-5.710$&$-10.372$&$       2.238$&$      3.831$&$      4.67$\\
 $2.22$&$-5.731$&$-10.741$&$       2.471$&$      4.168$&$      4.66$\\
 $2.49$&$-5.758$&$-11.218$&$       2.821$&$      4.675$&$      4.64$\\
 $2.62$&$-5.770$&$-11.415$&$       2.983$&$      4.909$&$      4.63$\\
 $2.98$&$-5.802$&$-11.922$&$       3.458$&$      5.597$&$      4.61$\\
 $3.02$&$-5.806$&$-11.980$&$       3.518$&$      5.684$&$      4.61$\\
 $3.42$&$-5.839$&$-12.467$&$ 4.076$&$ 6.493$&$ 4.59$\\
$3.81$&$-5.869$&$-12.899$&$ 4.655$&$ 7.333$&$ 4.58$\\
$4.20$&$-5.898$&$-13.275$&$       5.255$&$       8.207$&$      4.57$\\
$4.59$&$-5.924$&$-13.626$&$ 5.876$&$ 9.107$&$ 4.56$\\
\hline
\end{tabular}
\caption{\label{darkw6cdm}$w=-0.6$ CDM: Time and space variations in $\protect\alpha $ for
the corresponding redshifts of virialisation, $z_{v}$, in the $w=-0.6$ CDM
model . The indices $^{\prime }b^{\prime }$ and $^{\prime }c^{\prime }$,
stand for background and cluster respectively. The italic and bold entries
correspond approximately to the Oklo and $\protect\beta $-decay rate
constraints, respectively. The quasar absorption spectra observations
correspond to the values of $\Delta \protect\alpha /\protect\alpha |_{b}$.
The initial conditions were set for $\protect\zeta /\protect\omega =-2\times
10^{-4}$, in order to have $\protect\alpha _{c}(z=0)=\protect\alpha _{0}$.}
\end{center}
\end{table}
\endgroup
\begingroup

\begin{table}[htb!] 
\begin{center}
\begin{tabular}{|c|c|c|c|c|c|c|}
\hline $z_v$ & $\frac{\Delta\alpha}{\alpha} \vert_{b}\times 10^{-6}$ &
 $\frac{\Delta\alpha}{\alpha}         \vert_{c}\times         10^{-7}$
 &$\frac{\dot\alpha}{\alpha}      \vert_{b}\times      10^{-24}$     &
 $\frac{\dot\alpha}{\alpha}          \vert_{c}\times         10^{-22}$
 &$\frac{\delta\alpha}{\alpha}\times      10^{-6}$      \\      \hline
 $0.00$&$-5.412$&$     0.000$&$      0.340$&$     1.078$&$     5.41$\\
 ${\it{0.09}}$&${\it{-5.427}}$&${\it{ -1.587}}$&${\it{ 0.401}}$&${\it{
 1.148}}$&${\it{   5.27}}$\\   $0.18$&$-5.441$&$  -2.857$&$   0.462$&$
 1.220$&$   5.16$\\  $0.26$&$-5.455$&$  -3.899$&$   0.524$&$  1.294$&$
 5.06$\\   $0.34$&$-5.468$&$  -4.777$&$   0.586$&$   1.370$&$  4.99$\\
 ${\bf{0.42}}$&${\bf{-5.481}}$&${\bf{ -5.527}}$&${\bf{ 0.649}}$&${\bf{
 1.447}}$&${\bf{   4.93}}$\\   $0.49$&$-5.494$&$  -6.175$&$   0.711$&$
 1.525$&$   4.88$\\  $0.52$&$-5.498$&$  -6.362$&$   0.731$&$  1.549$&$
 4.86$\\   $0.62$&$-5.514$&$  -7.088$&$   0.817$&$   1.659$&$  4.81$\\
 $0.72$&$-5.530$&$     -7.714$&$     0.905$&$     1.771$&$     4.76$\\
 $0.81$&$-5.545$&$     -8.252$&$     0.993$&$     1.885$&$     4.72$\\
 $0.91$&$-5.560$&$     -8.728$&$     1.082$&$     2.002$&$     4.69$\\
 $0.97$&$-5.569$&$     -9.000$&$     1.139$&$     2.077$&$     4.67$\\
 $1.09$&$-5.587$&$     -9.516$&$     1.260$&$     2.239$&$     4.64$\\
 $1.22$&$-5.605$&$     -9.967$&$     1.384$&$     2.404$&$     4.61$\\
 $1.34$&$-5.622$&$-10.365$&$       1.509$&$      2.574$&$      4.59$\\
 $1.39$&$-5.629$&$-10.534$&$       1.566$&$      2.652$&$      4.58$\\
 $1.55$&$-5.650$&$-10.975$&$       1.735$&$      2.884$&$      4.55$\\
 $1.71$&$-5.670$&$-11.364$&$       1.908$&$      3.122$&$      4.53$\\
 $1.80$&$-5.682$&$-11.577$&$       2.014$&$      3.270$&$      4.52$\\
 $2.01$&$-5.706$&$-11.997$&$       2.247$&$      3.595$&$      4.51$\\
 $2.20$&$-5.728$&$-12.363$&$       2.484$&$      3.928$&$      4.49$\\
 $2.46$&$-5.755$&$-12.779$&$       2.801$&$      4.376$&$      4.48$\\
 $2.60$&$-5.769$&$-12.986$&$       2.976$&$      4.624$&$      4.47$\\
 $2.92$&$-5.800$&$-13.421$&$       3.401$&$      5.232$&$      4.46$\\
 $2.99$&$-5.806$&$-13.503$&$       3.489$&$      5.358$&$      4.46$\\
 $3.38$&$-5.841$&$-13.948$&$ 4.023$&$ 6.125$&$ 4.45$\\
$3.77$&$-5.872$&$-14.342$&$ 4.578$&$ 6.925$&$ 4.44$\\
$4.15$&$-5.901$&$-14.689$&$       5.153$&$       7.757$&$      4.43$\\
$4.54$&$-5.928$&$-15.012$&$       5.748$&$       8.617$&$      4.43$\\
$4.92$&$-5.953$&$-15.297$&$ 6.362$&$ 9.510$&$ 4.42$\\
\hline
\end{tabular}
\caption{\label{darkw8cdm}$w=-0.8$ CDM: Time and space variations in $\protect\alpha $ for
the corresponding redshifts of virialisation, $z_{v},$ for the $w=-0.8$ CDM
model. The indices $^{\prime }b^{\prime }$ and $^{\prime }c^{\prime }$,
stand for background and cluster respectively. The italic and bold entries
correspond approximately to the Oklo and $\protect\beta $-decay rate
constraints, respectively. The quasar absorption spectra observations
correspond to the values of $\Delta \protect\alpha /\protect\alpha |_{b}$.
The initial conditions were set for $\protect\zeta /\protect\omega =-2\times
10^{-4}$, in order to have $\protect\alpha _{c}(z=0)=\protect\alpha _{0}$.}
\end{center}
\end{table}
\endgroup
\begingroup

\begin{table}[htb!] 
\begin{center}
\begin{tabular}{|c|c|c|c|c|c|c|}
\hline $z_v$ & $\frac{\Delta\alpha}{\alpha} \vert_{b}\times 10^{-6}$ &
 $\frac{\Delta\alpha}{\alpha}         \vert_{c}\times         10^{-7}$
 &$\frac{\dot\alpha}{\alpha}      \vert_{b}\times      10^{-24}$     &
 $\frac{\dot\alpha}{\alpha}          \vert_{c}\times         10^{-22}$
 &$\frac{\delta\alpha}{\alpha}\times      10^{-6}$      \\      \hline
 $0.00$&$-5.382$&$     0.000$&$      0.361$&$     1.068$&$     5.38$\\
 ${\it{0.09}}$&${\it{-5.398}}$&${\it{ -1.618}}$&${\it{ 0.427}}$&${\it{
 1.128}}$&${\it{   5.24}}$\\   $0.18$&$-5.414$&$  -2.889$&$   0.493$&$
 1.193$&$   5.12$\\  $0.26$&$-5.429$&$  -3.916$&$   0.558$&$  1.260$&$
 5.04$\\   $0.34$&$-5.443$&$  -4.769$&$   0.623$&$   1.329$&$  4.97$\\
 ${\bf{0.41}}$&${\bf{-5.458}}$&$    {\bf{-5.486}}$&$   {\bf{0.688}}$&$
 {\bf{1.401}}$&$  {\bf{4.91}}$\\ $0.49$&$-5.471$&$  -6.098$&$ 0.753$&$
 1.475$&$   4.86$\\  $0.51$&$-5.476$&$  -6.272$&$   0.773$&$  1.498$&$
 4.85$\\   $0.61$&$-5.493$&$  -6.943$&$   0.860$&$   1.602$&$  4.80$\\
 $0.70$&$-5.511$&$     -7.511$&$     0.948$&$     1.708$&$     4.76$\\
 $0.80$&$-5.527$&$     -8.003$&$     1.036$&$     1.817$&$     4.73$\\
 $0.89$&$-5.543$&$     -8.428$&$     1.124$&$     1.929$&$     4.70$\\
 $0.95$&$-5.553$&$     -8.685$&$     1.184$&$     2.006$&$     4.68$\\
 $1.07$&$-5.573$&$     -9.141$&$     1.303$&$     2.161$&$     4.66$\\
 $1.19$&$-5.592$&$     -9.535$&$     1.424$&$     2.321$&$     4.64$\\
 $1.30$&$-5.610$&$     -9.884$&$     1.546$&$     2.485$&$     4.62$\\
 $1.36$&$-5.619$&$-10.050$&$       1.609$&$      2.569$&$      4.61$\\
 $1.51$&$-5.641$&$-10.430$&$       1.773$&$      2.794$&$      4.60$\\
 $1.66$&$-5.662$&$-10.769$&$       1.940$&$      3.025$&$      4.58$\\
 $1.76$&$-5.675$&$-10.976$&$       2.053$&$      3.182$&$      4.58$\\
 $1.95$&$-5.700$&$-11.336$&$       2.278$&$      3.497$&$      4.57$\\
 $2.14$&$-5.723$&$-11.660$&$       2.508$&$      3.822$&$      4.56$\\
 $2.15$&$-5.725$&$-11.672$&$       2.520$&$      3.839$&$      4.56$\\
 $2.40$&$-5.753$&$-12.038$&$       2.825$&$      4.273$&$      4.55$\\
 $2.54$&$-5.769$&$-12.239$&$       3.009$&$      4.536$&$      4.54$\\
 $2.85$&$-5.801$&$-12.626$&$       3.417$&$      5.122$&$      4.54$\\
 $2.92$&$-5.808$&$-12.713$&$       3.522$&$      5.273$&$      4.54$\\
 $3.31$&$-5.844$&$-13.132$&$       4.056$&$      6.045$&$      4.53$\\
 $3.69$&$-5.877$&$-13.498$&$       4.613$&$      6.853$&$      4.53$\\
 $4.07$&$-5.908$&$-13.837$&$       5.191$&$      7.692$&$      4.52$\\
 $4.45$&$-5.936$&$-14.140$&$       5.790$&$      8.565$&$      4.52$\\
 $4.83$&$-5.962$&$-14.418$&$ 6.408$&$ 9.469$&$ 4.52$\\
\hline
\end{tabular}
\caption{\label{darklcdm}$\Lambda $ CDM: Time and space variations in $\protect\alpha $ for
the corresponding redshifts of virialisation, $z_{v}$, in the $\Lambda CDM$
model. The indices $^{\prime }b^{\prime }$ and $^{\prime }c^{\prime }$,
stand for background and cluster respectively. The italic and bold entries
correspond approximately to the Oklo and $\protect\beta $-decay rate
constraints, respectively. The quasar absorption spectra observations
correspond to the values of $\Delta \protect\alpha /\protect\alpha |_{b}$.
The initial conditions were set for $\protect\zeta /\protect\omega =-2\times
10^{-4}$, in order to have $\protect\alpha _{c}(z=0)=\protect\alpha _{0}$.}
\end{center}
\end{table}
\endgroup

\bigskip

Variations in $\alpha $ also depend on the value and sign of the coupling, $%
\zeta /\omega $, of the scalar field responsible for variations in $\alpha $%
, to the matter fields. A higher value of $|\zeta /\omega |$, leads to a
stronger dependence on the density contrast of the matter inhomogeneities.
If the value or sign of $\zeta /\omega $ changes in space, then spatial
inhomogeneities in $\alpha $ occur. This could happen if we take into
account that on small enough scales, baryons will dominate the dark matter
density. The sign and value of $\zeta /\omega $ will change, and variations
in $\alpha $ will evolve differently on different scales. If there are no
variations in the sign and value of $\zeta /\omega $, then the only spatial
variations in $\alpha $ are the ones resulting from the dependence of the
fine structure constant on the density contrast of the region in which one
is measuring $\alpha $. At first sight, one might conclude that the
difference between $\alpha _{c}$ and $\alpha _{b}$, is only a consequence of
the coupling of $\alpha $ to matter. In reality this is not so. It is always
possible to obtain the same results for any value $|\zeta /\omega |$ with a
suitable choice of initial conditions, as can be seen from the results in
Tables \ref{zeta1}, \ref{zeta4}, \ref{zeta6} and \ref{zeta10}.

\bigskip

The results of this chapter confirm the natural explanation, given in
chapter \ref{realistic}, for why any experiment
carried out on Earth \cite{fuj, prestage,sortais}, or in our local solar system \cite%
{olive}, gives constraints on possible time-variation in $\alpha $ that are
much stronger than the magnitude of variation that is consistent with the
quasar observations \cite{murphylast} on extragalactic scales. The value of
the fine structure constant on Earth, and most probably in our local
cluster, differs from that in the background universe because of the
different histories of these regions. It can be seen from the Tables \ref%
{darkscdm}, \ref{darkw6cdm}, \ref{darkw8cdm} and \ref{darklcdm} that inside
a virialised overdensity we expect $\Delta \alpha /\alpha \approx -10^{-7}$
for $z_{v}\leq 1,$ while in the background we have $\Delta \alpha /\alpha
\approx -10^{-6}$, independently of the structure formation model used. The
same conclusions arise independently of the absolute value of the coupling $%
\zeta /\omega $, see Tables \ref{zeta1}, \ref{zeta4}, \ref{zeta6} and \ref%
{zeta10}.

The dependence of $\alpha $ on the matter-field perturbations is much less
important when one is studying effects of varying $\alpha $ on the early
universe, for example on the last scattering of the CMB or the course of
primordial nucleosynthesis \cite{martins,martins1,martins2}. In the linear regime of the
cosmological perturbations, small perturbations in $\alpha $ will decay or
become constant in the radiation era \cite{mota2}. At the redshift of last
scattering, $z=1100$, it was found in \cite{mota3} that $\Delta \alpha
/\alpha \leq 10^{-5}$. In the background, the fine structure constant,
will be a constant during the radiation era \cite{bsbm} so long as the
kinetic term is negligible. A growth in value of $\alpha $ will occur only
in the matter-dominated era \cite{mota1}. Hence, the early-universe
constraints, coming from the CMB and primordial nucleosynthesis, are
comparatively weak, $\Delta \alpha /\alpha \leq 10^{-2}$, and are easily
satisfied. 

%% file: conclusion.tex
\chapter{Conclusions and Future Directions}

\begin{flushright}
{\it {\small Enquanto n\~ao desvendarmos este paradoxo do fim,}}\\
{\it {\small  por um lado o desejo de  ter um t\'elos,}}\\  
{\it {\small e  por  outro  o  desejo  de  n\~ao  ter  um  fim, de  ser  eterno,}}\\
{\it {\small continuaremos agrilhoados sobre a verdade...}}\\
{\it {\small -- Ernesto Mota --}} 
\end{flushright}

\bigskip

The aim of this thesis was to study the influence of  matter field
inhomogeneities on the evolution of the fine structure constant.

In chapter \ref{qualitative}, using a phase plane analysis, we
studied the cosmological evolution of a
time-varying fine-structure constant. We considered the case of a
power-law $a(t)=t^{n}$ for the  scale factor of the background universe. We have shown that in general $%
\alpha $ increases with time or asymptotes to a constant value at late
times. We have  found
general asymptotic solutions for all the different and possible behaviours
via the analysis of the critical points of the system that determines the
evolution of $\alpha $. In particular, we have also found asymptotic solutions
for the dust, radiation and curvature-dominated FRW universe which also
generalise the asymptotes found in \cite{bsm1}. These solutions correspond to
late-time attractors that describes the $\psi $ and $\alpha $ evolution in
time and will be useful to study the linear regime of the cosmological perturbations.

In chapter \ref{gaugeinvariant}, by applying the gauge-invariant formalism of ref. \cite{mukhanov},
we have determined the evolution of
small inhomogeneities in $\delta \alpha /\alpha $ in the presence of small
adiabatic density inhomogeneities. The evolution of the
perturbations of the metric and of the stress-energy filling the
universe  behave as in cosmological models with constant $\alpha $ and
we can to determine the behaviour of small inhomogeneities in $\delta \alpha
/\alpha $ in the gravitational fields created by the density and metric
perturbations.
In a flat radiation-dominated universe we find that inhomogeneous
perturbations in $\alpha $ decay on large scales while on scales
smaller than the Hubble radius they  undergo bounded oscillations. 
In a flat dust-dominated universe, linear perturbation theory, indicate
that small inhomogeneities in $\alpha $ will
become constant on large scales at late times while on small scales they
will increase as $t^{2/3}$. In reality, the small-scale evolution will be
made more complicated by the breakdown of the assumptions underlying the
perturbation analysis and the development of local deviations from the FRW
behaviour.
In an accelerated phase of our universe, as is the case for an early
inflationary epoch, or during a $\Lambda $- or quintessence-dominated late
phase of evolution, we show that inhomogeneous perturbations in $\alpha $
will decrease on all scales. This result complements the earlier discovery 
\cite{bsbm}, \cite{bsm1}, that $\alpha $ tends to a constant with
exponential rapidity in Friedmann universes dominated by a
cosmological constant. Any pre-existing inhomogeneities will be frozen
in but their scale will be exponentially increased by the de Sitter
expansion.

In chapter \ref{realistic}, we have shown that spatial variations in $%
\alpha $ inevitably occur because of the development of non-linear
inhomogeneities in the universe.
We used
the spherical collapse model to study the space-time variations in $\alpha $
in the non-linear regime. Strong differences arise between the value of $%
\alpha $ inside a cluster and its background value and also between
clusters. Variations in $\alpha $ depend on the matter density contrast of
the cluster region and the redshift at which it virialised. If the overdense
regions are still contracting and have not yet virialised, the value of 
$\alpha $ within them will continue to change. Variations in $\alpha $ will
cease when the cluster virialises so long as it does so at moderate
redshift. This leads to larger values of $\alpha $ in the overdense regions
than in the cosmological background and means that time variations in $
\alpha $ will turn off in virialised overdensities even though they continue
in the background universe. 
The fact that local $\alpha $ values 'freeze in'
at virialisation, means we would observe no time or spatial variations in $%
\alpha $ on Earth, or elsewhere in our Galaxy, even though time-variations
in $\alpha $ might still be occurring on extragalactic scales. For a
cluster, the value of $\alpha $ today will be the value of $\alpha $ at the
virialisation time of the cluster. We should observe significant differences
in $\alpha $ only when comparing clusters which virialised at quite
different redshifts. Differences will arise within the same bound system
only if it has not reached viral equilibrium. Hence, variations in $\alpha $
using geochemical methods could easily give a value that is $100$ times
smaller than is inferred from quasar spectra.

In chapter \ref{darkenergy}, we study the  dependence of the fine structure constant
on the equation of state of dark energy and on the coupling to matter,$\zeta/\omega$
.  It was shown that independently of the model of
structure formation we consider or the coupling to matter, there will
be always a spatial contrast, $\delta\alpha/\alpha$, between $\alpha$ in 
an overdensity and in the background. 
When comparing (already) virialised overdensities, in the Standard Cold Dark
Matter model ($SCDM$),
spatial variations of the fine structure constant are independent of
the redshift we measure them, and are constant. However,
that is not the case with time-variations which grow in time.
A similar behaviour is found, at high redshifts, in a $\Lambda CDM$ or
other dark-energy $CDM$ models. At low redshifts, spatial
variations of the fine
structure constant become a function of the virialisation time,  the density
of the clusters and the equation of state of the dark energy
component.
We conclude the chapter, showing that  the natural possibility of 
occurring,  time or spatial inhomogeneities in $\zeta/\omega$, would
lead to spatial and time variations in the fine structure constant. 

\bigskip

The independence of the results above, in particular the effect of the
matter inhomogeneities on the cosmological evolution of the fine
structure constant, with respect to the model used to
drive variations in $\alpha$, is an important issue. All the results
we have obtained were only performed in
the case of the BSBM theory.  In spite of being a quite simple and elegant
theory, the BSBM model but does not include the effects coming from grand unification
theories. These may be important, since in many situations the field
responsible for the variations of $\alpha$ is coupled to all the other
matter fields. The coupling to the other matter fields might result on
effects which are more important than the matter inhomogeneities, and
could avoid the spatial inhomogeneity $\delta\alpha/\alpha$.
Until similar studies are performed to
other models, which are more complex and ``realistic'', for instance
low energy string-theory models,  there will be always doubts about
the universality of the results.

Another important result, which requires a more detailed
investigation, is our claim that
$\alpha$ ``freeze in'' after virialisation occurs. That conclusion was
taken only based on qualitative arguments.  Hence, a quantitative analysis of
the process of virialisation of the overdensities, and the behaviour
of the fine structure constant inside the overdensity, is of major
importance. These can be obtained 
performing a more complete numerical simulation where many more
variables and features of the structure formation models are
involved. 

If the results above are indeed confirmed in the future, and for other
theoretical models,  the
consideration of the evolution of inhomogeneities, notably the one inside
which we live, is essential if we are to make meaningful comparisons of
different pieces of astronomical and terrestrial evidence for the constancy
of $\alpha$. 